\PassOptionsToPackage{unicode}{hyperref}
\PassOptionsToPackage{hyphens}{url}
\PassOptionsToPackage{dvipsnames,svgnames,x11names}{xcolor}
\documentclass[
  letterpaper,
  DIV=11,
  numbers=noendperiod]{scrartcl}

\usepackage{amsmath,amssymb}
\usepackage{lmodern}
\usepackage[noend]{algpseudocode} 
\usepackage[linesnumbered,ruled]{algorithm2e}
\usepackage{caption}
\usepackage{iftex}
\ifPDFTeX
  \usepackage[T1]{fontenc}
  \usepackage[utf8]{inputenc}
  \usepackage{textcomp} 
\else 
  \usepackage{unicode-math}
  \defaultfontfeatures{Scale=MatchLowercase}
  \defaultfontfeatures[\rmfamily]{Ligatures=TeX,Scale=1}
\fi
\IfFileExists{upquote.sty}{\usepackage{upquote}}{}
\IfFileExists{microtype.sty}{
  \usepackage[]{microtype}
  \UseMicrotypeSet[protrusion]{basicmath} 
}{}
\makeatletter
\@ifundefined{KOMAClassName}{
  \IfFileExists{parskip.sty}{%
    \usepackage{parskip}
  }{
    \setlength{\parindent}{0pt}
    \setlength{\parskip}{6pt plus 2pt minus 1pt}}
}{
  \KOMAoptions{parskip=half}}
\makeatother
\usepackage{xcolor}
\ifx\paragraph\undefined\else
  \let\oldparagraph\paragraph
  \renewcommand{\paragraph}[1]{\oldparagraph{#1}\mbox{}}
\fi
\ifx\subparagraph\undefined\else
  \let\oldsubparagraph\subparagraph
  \renewcommand{\subparagraph}[1]{\oldsubparagraph{#1}\mbox{}}
\fi

\usepackage{longtable,booktabs,array}
\usepackage{calc} 
\usepackage{etoolbox}
\makeatletter
\patchcmd\longtable{\par}{\if@noskipsec\mbox{}\fi\par}{}{}
\makeatother
\IfFileExists{footnotehyper.sty}{\usepackage{footnotehyper}}{\usepackage{footnote}}
\makesavenoteenv{longtable}
\usepackage{graphicx}
\makeatletter
\def\maxwidth{\ifdim\Gin@nat@width>\linewidth\linewidth\else\Gin@nat@width\fi}
\def\maxheight{\ifdim\Gin@nat@height>\textheight\textheight\else\Gin@nat@height\fi}
\makeatother
\setkeys{Gin}{width=\maxwidth,height=\maxheight,keepaspectratio}
\makeatletter
\def\fps@figure{htbp}
\makeatother
\newlength{\cslhangindent}
\newlength{\csllabelwidth}
\newlength{\cslentryspacingunit} 
\newenvironment{CSLReferences}[2] 
 {
  \setlength{\parindent}{0pt}
  \ifodd #1
  \let\oldpar\par
  \def\par{\hangindent=\cslhangindent\oldpar}
  \fi
  \setlength{\parskip}{#2\cslentryspacingunit}
 }%
 {}
\usepackage{calc}

\KOMAoption{captions}{tableheading}
\makeatletter
\@ifpackageloaded{caption}{}{\usepackage{caption}}
\AtBeginDocument{%
\ifdefined\contentsname
  \renewcommand*\contentsname{Table of contents}
\else
  \newcommand\contentsname{Table of contents}
\fi
\ifdefined\listfigurename
  \renewcommand*\listfigurename{List of Figures}
\else
  \newcommand\listfigurename{List of Figures}
\fi
\ifdefined\listtablename
  \renewcommand*\listtablename{List of Tables}
\else
  \newcommand\listtablename{List of Tables}
\fi
\ifdefined\figurename
  \renewcommand*\figurename{Figure}
\else
  \newcommand\figurename{Figure}
\fi
\ifdefined\tablename
  \renewcommand*\tablename{Table}
\else
  \newcommand\tablename{Table}
\fi
}
\@ifpackageloaded{float}{}{\usepackage{float}}
\floatstyle{ruled}
\@ifundefined{c@chapter}{\newfloat{codelisting}{h}{lop}}{\newfloat{codelisting}{h}{lop}[chapter]}
\floatname{codelisting}{Listing}

\makeatother
\makeatletter
\@ifpackageloaded{caption}{}{\usepackage{caption}}
\@ifpackageloaded{subcaption}{}{\usepackage{subcaption}}
\makeatother
\makeatletter
\@ifpackageloaded{tcolorbox}{}{\usepackage[many]{tcolorbox}}
\makeatother
\makeatletter
\@ifundefined{shadecolor}{\definecolor{shadecolor}{rgb}{.97, .97, .97}}
\makeatother
\ifLuaTeX
  \usepackage{selnolig}  
\fi
\IfFileExists{bookmark.sty}{\usepackage{bookmark}}{\usepackage{hyperref}}
\IfFileExists{xurl.sty}{\usepackage{xurl}}{} 
\hypersetup{
  pdftitle={Solving multiphysics-based inverse problems with learned surrogates and constraints},
  pdfauthor={Ziyi Yin*; Rafael Orozco; Mathias Louboutin; Felix J. Herrmann},
  colorlinks=true,
  linkcolor={blue},
  filecolor={Maroon},
  citecolor={Blue},
  urlcolor={Blue},
  pdfcreator={LaTeX via pandoc}}

\title{Solving multiphysics-based inverse problems with learned
surrogates and constraints}
\author{Ziyi Yin* \and Rafael Orozco \and Mathias Louboutin \and Felix
J. Herrmann}
\date{}

\begin{document}
\maketitle
\ifdefined\Shaded\renewenvironment{Shaded}{\begin{tcolorbox}[enhanced, sharp corners, borderline west={3pt}{0pt}{shadecolor}, boxrule=0pt, breakable, interior hidden, frame hidden]}{\end{tcolorbox}}\fi

Affiliation: Georgia Institute of Technology

Address: 756 West Peachtree Street NW, Atlanta, Georgia, USA 30308

Emails:

\begin{itemize}
\item
  Ziyi Yin (corresponding author):
  \href{mailto:ziyi.yin@gatech.edu}{\nolinkurl{ziyi.yin@gatech.edu}}
\item
  Rafael Orozco:
  \href{mailto:rorozco@gatech.edu}{\nolinkurl{rorozco@gatech.edu}}
\item
  Mathias Louboutin:
  \href{mailto:mlouboutin3@gatech.edu}{\nolinkurl{mlouboutin3@gatech.edu}}
\item
  Felix J. Herrmann:
  \href{mailto:felix.herrmann@gatech.edu}{\nolinkurl{felix.herrmann@gatech.edu}}
\end{itemize}

\hypertarget{abstract}{%
\subsection{Abstract}\label{abstract}}

Solving multiphysics-based inverse problems for geological carbon
storage monitoring can be challenging when multimodal time-lapse data
are expensive to collect and costly to simulate numerically. We overcome
these challenges by combining computationally cheap learned surrogates
with learned constraints. Not only does this combination lead to vastly
improved inversions for the important fluid-flow property, permeability,
it also provides a natural platform for inverting multimodal data
including well measurements and active-source time-lapse seismic data.
By adding a learned constraint, we arrive at a computationally feasible
inversion approach that remains accurate. This is accomplished by
including a trained deep neural network, known as a normalizing flow,
which forces the model iterates to remain in-distribution, thereby
safeguarding the accuracy of trained Fourier neural operators that act
as surrogates for the computationally expensive multiphase flow
simulations involving partial differential equation solves. By means of
carefully selected experiments, centered around the problem of
geological carbon storage, we demonstrate the efficacy of the proposed
constrained optimization method on two different data modalities, namely
time-lapse well and time-lapse seismic data. While permeability
inversions from both these two modalities have their pluses and minuses,
their joint inversion benefits from either, yielding valuable superior
permeability inversions and CO\textsubscript{2} plume predictions near,
and far away, from the monitoring wells.

\textbf{Keywords}: Fourier neural operators, normalizing flows,
multiphysics, deep learning, learned surrogates, learned constraints,
inverse problems

\hypertarget{introduction}{%
\subsection{Introduction}\label{introduction}}

In this paper, we introduce a novel learned inversion algorithm designed
to address inverse problems based on partial differential equations
(PDEs). These problems can be represented using the following general
form:

\begin{equation}\protect\hypertarget{eq-inv}{}{
\mathbf{d} = \mathcal{H}\circ\mathcal{S}(\mathbf{K}) + \boldsymbol \epsilon.
}\label{eq-inv}\end{equation}

In this expression, the nonlinear operator \(\mathcal{S}\) represents
the solution operator of a nonlinear parametric PDE mapping the
coefficients \(\mathbf{K}\) to the solution. Given numerical solutions
of the PDE, partially observed data, collected in the vector
\(\mathbf{d}\), are modeled by compounding the solution operator with
the measurement operator, \(\mathcal{H}\), followed by adding the noise
term \(\boldsymbol{\epsilon}\) with noise level of \(\sigma\)---i.e.,
\(\boldsymbol{\epsilon}\sim \mathcal{N}(\mathbf{0}, \sigma^2 \mathbf{I})\).
This problem is quite general and pertinent to various physical
applications, including geophysical exploration (Tarantola 1984, 2005),
medical imaging (Arridge 1999), and experimental design (Alexanderian
2021).

Without loss of generality, we focus on time-lapse seismic monitoring of
geological carbon storage (GCS), which involves underground storage of
supercritical CO\textsubscript{2} captured from the atmosphere or from
industrial smoke stacks (Furre et al. 2017). We consider GCS in saline
aquifers, which involves multiphase flow physics where
CO\textsubscript{2} replaces brine in the porous rocks (Nordbotten and
Celia 2011). In this context, the PDE solution operator,
\(\mathcal{S}\), serves as the multiphase flow simulator, which takes
the gridded spatially varying permeability in the reservoir,
\(\mathbf{K}\), as input and produces \(n_t\) time-varying
CO\textsubscript{2} saturation snapshots,
\(\mathbf{c}=[\mathbf{c}_1, \mathbf{c}_2, \cdots,\mathbf{c}_{n_t}]\), as
output. The governing equations for the multiphase flow involve Darcy's
and the mass conservation law. Detailed information on the governing
equations, initial and boundary conditions, and numerical solution
schemes can be found in (Rasmussen et al. 2021) and the references
therein. To ensure safety, conformance, and containment of GCS projects,
various kinds of time-lapse data are collected to monitor
CO\textsubscript{2} plumes. These different data modalities include
measurements in wells (Freifeld et al. 2009; Nogues, Nordbotten, and
Celia 2011), and the collection of gravity (Nooner et al. 2007; Alnes et
al. 2011), electromagnetic (Carcione et al. 2012; Zhdanov et al. 2013),
and seismic time-lapse data (Arts et al. 2004; Lumley 2010; Yin et al.
2023) that can be used to follow the plume and invert for reservoir
properties such as the permeability, \(\mathbf{K}\). The latter is the
property of interest in this exposition.

Overall, solving for the reservoir model parameter, \(\mathbf{K}\),
poses significant challenges for two primary reasons:

\begin{itemize}
\item
  the forward modeling operator, \(\mathcal{H}\circ\mathcal{S}\), can be
  ill-posed, resulting in multiple model parameters that fit the
  observed data equally well. This necessitates the use of regularizers
  (Golub, Hansen, and O'Leary 1999; Tarantola 2005) in the form of
  penalties or constraints (Peters, Smithyman, and Herrmann 2019).
\item
  The PDE modeling operator \(\mathcal{S}\), and the sensitivity
  calculations with respect to the model parameters can be
  computationally expensive for large-scale problems, limiting the
  efficacy of iterative methods such as gradient-based (D. C. Liu and
  Nocedal 1989) or Markov chain Monte Carlo (Cowles and Carlin 1996)
  methods.
\end{itemize}

To overcome the second challenge, numerous attempts have been made to
replace computationally expensive PDE solves with more affordable
approximate alternatives (Razavi, Tolson, and Burn 2012; Asher et al.
2015), which include the use of radial basis functions to learn the
complex models from few sample points (Powell 1985) or reduced-order
modeling where the dimension of the model space is reduced (Schilders,
Van der Vorst, and Rommes 2008; K. Lu et al. 2019). More recently,
various deep learning techniques have emerged as cheap alternatives to
numerical PDE solves (L. Lu, Jin, and Karniadakis 2019; Pestourie et al.
2020; Qian et al. 2020; Karniadakis et al. 2021; Kovachki et al. 2021;
Rahman, Ross, and Azizzadenesheli 2022; Hijazi, Freitag, and Landwehr
2023). After incurring initial training costs, these neural operators
lead to vastly improved computation of PDE solves. Data-driven methods
have also been used successfully to learn coarse-to-fine grid mappings
of PDEs solves. Because of their advertised performance on approximating
solution operators of the multiphase flow in porous media (G. Wen et al.
2022, 2023; Grady et al. 2023; Philipp A. Witte, Redmond, et al. 2022;
Philipp A. Witte, Hewett, et al. 2022; Philipp A. Witte et al. 2023), we
will consider Fourier neural operators (FNOs, Z. Li et al. 2020;
Kovachki, Lanthaler, and Mishra 2021) in this work even though
alternative choices can be made. Once trained, FNOs produce approximate
PDE solutions orders of magnitude faster than traditional solvers (Z. Li
et al. 2020, 2021; Grady et al. 2023; De Hoop et al. 2022). In addition,
Yin et al. (2022), Louboutin et al. (2022) and Louboutin, Yin, et al.
(2023) demonstrated that trained FNOs can replace PDE solution operators
during inversion. This latest development is especially beneficial to
applications such as GCS where trained FNOs can be used in lieu of
numerically costly flow simulators (Lie 2019; Rasmussen et al. 2021;
Gross and Mazuyer 2021). However, despite their promising results,
unconstrained inversion formulations offer little to no guarantees that
the model iterates remain within the statistical distribution on which
the FNO was trained initially during inversion. As a consequence, FNOs
may no longer produce accurate fluid-flow simulations throughout the
iterations, which can lead to erroneous inversion results when the
errors become too large, possibly rendering surrogate modeling by FNOs
ineffective. To avoid this situation, we propose a constrained
formulation where a trained normalizing flow (NF, Rezende and Mohamed
(2015)) is included as a learned constraint. This added constraint
guarantees that the model iterates remain within the desired statistical
distribution. Because our approach safeguards the FNO's accuracy, it
allows FNOs to act as reliable low-cost neural surrogates replacing
costly fluid-flow simulations and gradient calculations that rely on
numerically expensive PDE solves during inversion.

The organization of this paper is as follows: First, we introduce FNOs
and explore the possibility of replacing the forward modeling operator
with a trained FNO surrogate. Next, NFs are introduced. By means of a
motivational example, we demonstrate how these learned generative
networks control the prediction error of FNOs by ensuring that the model
iterates remain in distribution. Based on this motivational example, we
propose our novel method for using trained NFs as a learned constraint
to guarantee performance of FNO surrogates during inversion. Through
four synthetic experiments related to GCS monitoring, the efficacy of
our method will be demonstrated.

\hypertarget{fourier-neural-operators}{%
\subsection{Fourier neural operators}\label{fourier-neural-operators}}

There is an extensive literature on training deep neural networks to
serve as affordable alternatives to computationally expensive numerical
simulators (L. Lu, Jin, and Karniadakis 2019; Karniadakis et al. 2021;
Kovachki et al. 2021; Kontolati et al. 2023; Benitez et al. 2023).
Without loss of generality, we limit ourselves in this exposition to the
training of a special class of neural operators known as Fourier neural
operators (FNOs). These FNOs are designed to approximate numerical
solution operators of the PDE solution operator, \(\mathcal{S}\), by
minimizing the following objective:
\begin{equation}\protect\hypertarget{eq-fno-train}{}{
\underset{\boldsymbol{\theta}}{\operatorname{minimize}} \quad \, \frac{1}{N}\sum_{j=1}^{N}\|\mathcal{S}_{\boldsymbol{\theta}}(\mathbf{K}^{(j)})-\mathbf{c}^{(j)}\|_2^2\quad\text{where}\quad\mathbf{c}^{(j)}=\mathcal{S}(\mathbf{K}^{(j)}).
}\label{eq-fno-train}\end{equation}

Here, \(\mathcal{S}_{\boldsymbol{\theta}}\) denotes the FNO with network
weights \({\boldsymbol{\theta}}\). The optimization aims to minimize the
\(\ell_2\) misfit between numerically simulated PDE solutions,
\(\mathbf{c}^{(j)}\), and solutions approximated by the FNO, across
\(N\) training samples (permeability models),
\(\{\mathbf{K}^{(j)}\}_{j=1}^N\) compiled by domain experts. Once
trained, FNOs can generate approximate PDE solutions for unseen model
parameters orders of magnitude faster than numerical simulations (Grady
et al. 2023; De Hoop et al. 2022). For model parameters that fall within
the distribution used to train, approximation by FNOs are reasonably
accurate---i.e.,
\(\mathcal{S}_{\boldsymbol\theta^\ast}(\mathbf{K})\approx\mathcal{S}(\mathbf{K})\),
with \(\boldsymbol\theta^\ast\) being the minimizer of
Equation~\ref{eq-fno-train}. We refer to the numerical examples section
for details calculating these weights. Before studying the impact of
applying these surrogates on samples for the permeability that are out
of distribution, let us first consider an example where data is inverted
using surrogate modeling.

\hypertarget{inversion-with-learned-surrogates}{%
\subsection{Inversion with learned
surrogates}\label{inversion-with-learned-surrogates}}

Replacing PDE solutions by approximate solutions yielded by trained FNO
surrogates has two main advantages when solving inverse problems. First,
as mentioned earlier, FNOs are orders of magnitude faster than numerical
PDE solves, which allows for many simulations at negligible costs
(Chandra et al. 2020; Lan, Li, and Shahbaba 2022). Second, existing
softwate for multiphase flow simulations may not always support
computationally efficient calculations of sensitivity, e.g.~via
adjoint-state calculations (Cao et al. 2003; Plessix 2006; Jansen 2011)
of the simulations with respect to their input. In such cases, FNO
surrogates are favorable because automatic differentiation on the
trained network (Griewank et al. 1989; Louboutin et al. 2022; Yin et al.
2022; Yang et al. 2023; Louboutin, Yin, et al. 2023) readily provides
access to gradients with respect to model parameters. As a result, the
PDE solver, \(\mathcal{S}\), in Equation~\ref{eq-inv} can be replaced by
trained surrogate, \(\mathcal{S}_{\boldsymbol\theta^\ast}\)---i.e., we
have

\begin{equation}\protect\hypertarget{eq-inv-fno}{}{
\underset{\mathbf{K}}{\operatorname{minimize}} \quad \|\mathbf{d} - \mathcal{H}\circ\mathcal{S}_{\boldsymbol\theta^\ast}(\mathbf{K})\|_2^2
}\label{eq-inv-fno}\end{equation}

where \(\boldsymbol\theta^\ast\) represent the optimized weights
minimizing Equation~\ref{eq-fno-train}. While the above formulation in
terms of trained surrogates has been applied successfully during
permeability inversion from time-lapse seismic data (D. Li et al. 2020;
Yin et al. 2022; Louboutin, Yin, et al. 2023), this type of inversion is
only valid as long as the (intermediate) permeabilities remain within
distribution during the inversion. Practically, this means two things.
First, the data need to be in the range of permeability models that are
in distribution. This means that there can not be too much noise neither
can the observed data be the result of an out-of-distribution
permeability. Second, there are no guarantees that the permeability
model iterates remain in distribution during inversion even though some
bias of the gradients of the surrogate towards in-distribution
permeabilities may be expected. To overcome this challenge, we propose
to add a learned constraint to Equation~\ref{eq-inv-fno} that offers
guarantees that the model iterates remain in distribution.

\hypertarget{learned-constraints-with-normalizing-flows}{%
\subsection{Learned constraints with normalizing
flows}\label{learned-constraints-with-normalizing-flows}}

As demonstrated by Peters and Herrmann (2017), Esser et al. (2018),
Peters, Smithyman, and Herrmann (2019) regularization of non-linear
inverse problems, such as full-waveform inversion, with constraints,
e.g., total-variation (Esser et al. 2018) or transform-domain sparsity
with \(\ell_1\)-norms (X. Li et al. 2012), offers distinct advantages
over regularizations based adding these norms as penalties. Even though
constraint and penalty formulations are equivalent for linear inverse
problems for the appropriate choice of the Lagrange multiplier,
minimizing the constraint formulation leads to completely different
solution paths compared to adding a penalty term to the data misfit
objective (Hennenfent et al. 2008). In the constrained formulation, the
model iterates remain at all times within the constraint set while model
iterates yielded by the penalty formulation does not offer these
guarantees. Peters and Herrmann (2017) demonstrated this importance
difference for the non-convex problem of full-waveform inversion. For
this problem, it proved essential to work with a homotopy where the
intersection of multiple handcrafted constraints (intersection of box
and size of total-variation-norm ball constraints) are relaxed slowly
during the inversion, so the model iterates remain physically feasible
and local minima are avoided.

Motivated by these results, we propose a similar approach but now for
``data-driven'' learned constraints based on normalizing flows (NFs,
Rezende and Mohamed (2015)). NFs are powerful deep generative neural
networks capable of learning to generate samples from complex
distributions (Dinh, Sohl-Dickstein, and Bengio 2016; Siahkoohi et al.
2023b; Orozco, Louboutin, et al. 2023; Louboutin, Yin, et al. 2023).
Designed to be invertible, these NFs require the latent and model spaces
to share identical dimensions, which confers several advantages:

\begin{itemize}
\item
  unlike variational autoencoders (Kingma and Welling 2013) or
  generative adversarial networks (GANs, Goodfellow et al. 2014), which
  both have a lower-dimensional latent space, NFs do not impose any
  intrinsic dimensionality constraints. This flexibility lets NFs
  capture model space characteristics across high dimensions (Kobyzev,
  Prince, and Brubaker 2020). Relevantly, concurrent literature has
  delved into the intrinsic dimensionality of NFs, indicating the
  potential to using NFs to generate models with inherently lower
  dimensions (Horvat and Pfister 2022).
\item
  NFs' inherent invertibility negates the need to store state variables
  during gradient calculations, enabling memory-efficient training and
  inversion in large-scale 3D applications, such as in geophysics (Zhang
  and Curtis 2020, 2021; Siahkoohi et al. 2021, 2023b; Siahkoohi and
  Herrmann 2021; Zhao, Curtis, and Zhang 2021; Lensink, Peters, and
  Haber 2022) and ultrasound imaging (Orozco et al. 2021, 2023; Orozco,
  Louboutin, and Herrmann 2022; Orozco, Louboutin, et al. 2023; Orozco,
  Siahkoohi, Louboutin, et al. 2023; J. Wen, Ahmad, and Schniter 2023).
\item
  because of their invertibility NFs guarantee unique latent codes for
  all model space samples, including out-of-distribution ones.
  Therefore, they can still be used to invert for out-of-distribution
  model parameters, while other methods like GANs may introduce bias
  (Asim et al. 2020).
\end{itemize}

Aside from being invertible, NFs are trained to map samples from a
target distribution in the physical space to samples from the standard
zero-mean Gaussian distribution noise in the latent space. After
training is completed, samples from the target distribution are
generated by running the NF in reverse on samples in the latent space
from the standard Gaussian distribution. Below, we will demonstrate how
NFs can be used to guarantee that the permeability remains in
distribution during the inversion.

\hypertarget{training-normalizing-flows}{%
\subsubsection{Training normalizing
flows}\label{training-normalizing-flows}}

Given samples from the permeability distribution,
\(\{\mathbf{K}^{(j)}\}_{j=1}^N\), training NFs entails minimizing the
Kullback-Leibler divergence between the base and target distributions
(Ardizzone et al. 2018). This involves solving the following variational
problem: \begin{equation}\protect\hypertarget{eq-train-nf}{}{
\underset{\mathbf{w}}{\operatorname{minimize}} \quad \frac{1}{N}\sum_{i=1}^{N} \left(\frac{1}{2}\|\mathcal{G}^{-1}_{\mathbf{w}}(\mathbf{K}_i)\|_2^2-\log\left|\det\mathrm{J}_{\mathcal{G}^{-1}_{\mathbf{w}}}(\mathbf{K}_i)\right|\right).
}\label{eq-train-nf}\end{equation}

\begin{figure}

\begin{minipage}[t]{0.33\linewidth}

{\centering 

\raisebox{-\height}{

\includegraphics{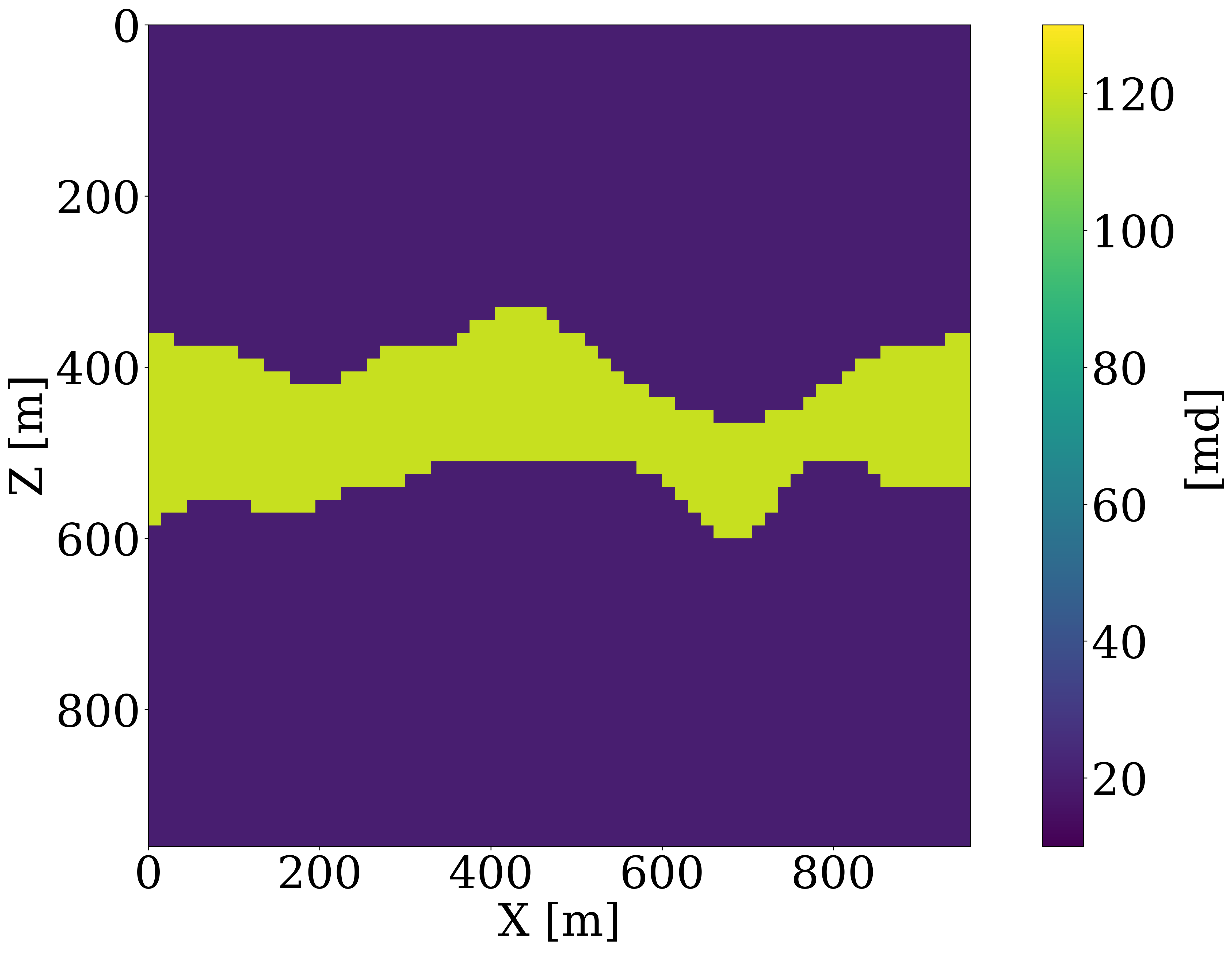}

}

}

\end{minipage}%
\begin{minipage}[t]{0.33\linewidth}

{\centering 

\raisebox{-\height}{

\includegraphics{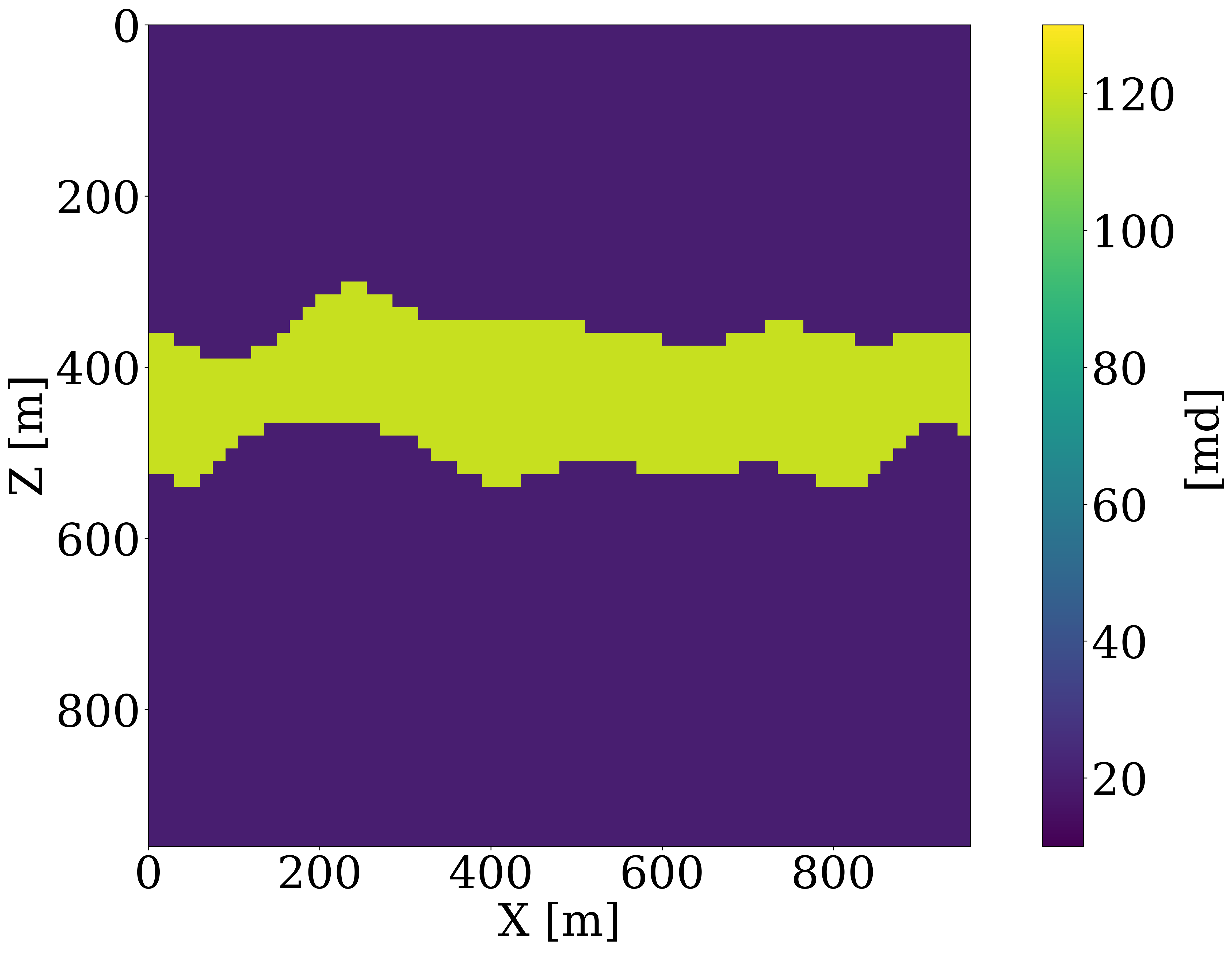}

}

}

\end{minipage}%
\begin{minipage}[t]{0.33\linewidth}

{\centering 

\raisebox{-\height}{

\includegraphics{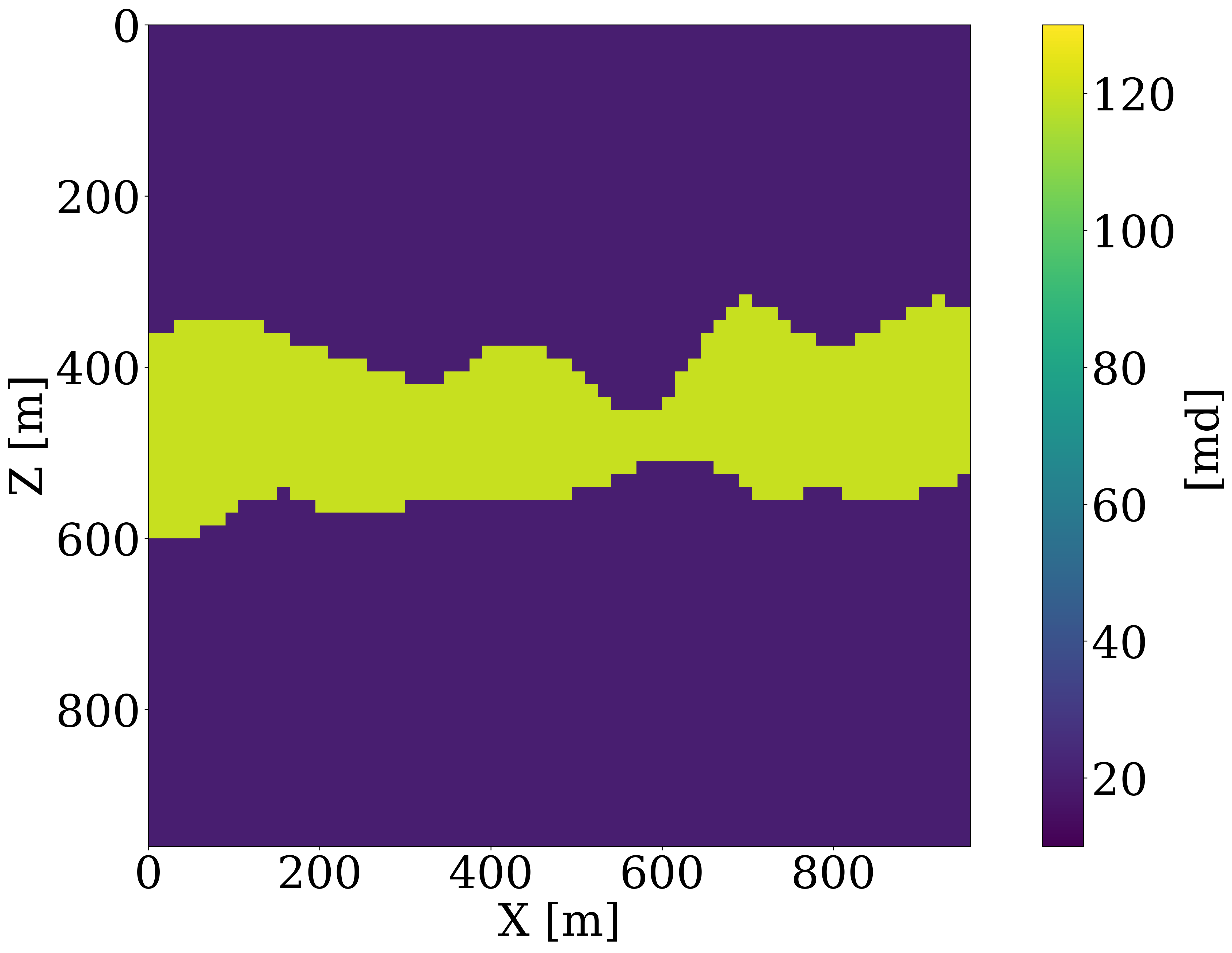}

}

}

\end{minipage}%
\newline
\begin{minipage}[t]{0.33\linewidth}

{\centering 

\raisebox{-\height}{

\includegraphics{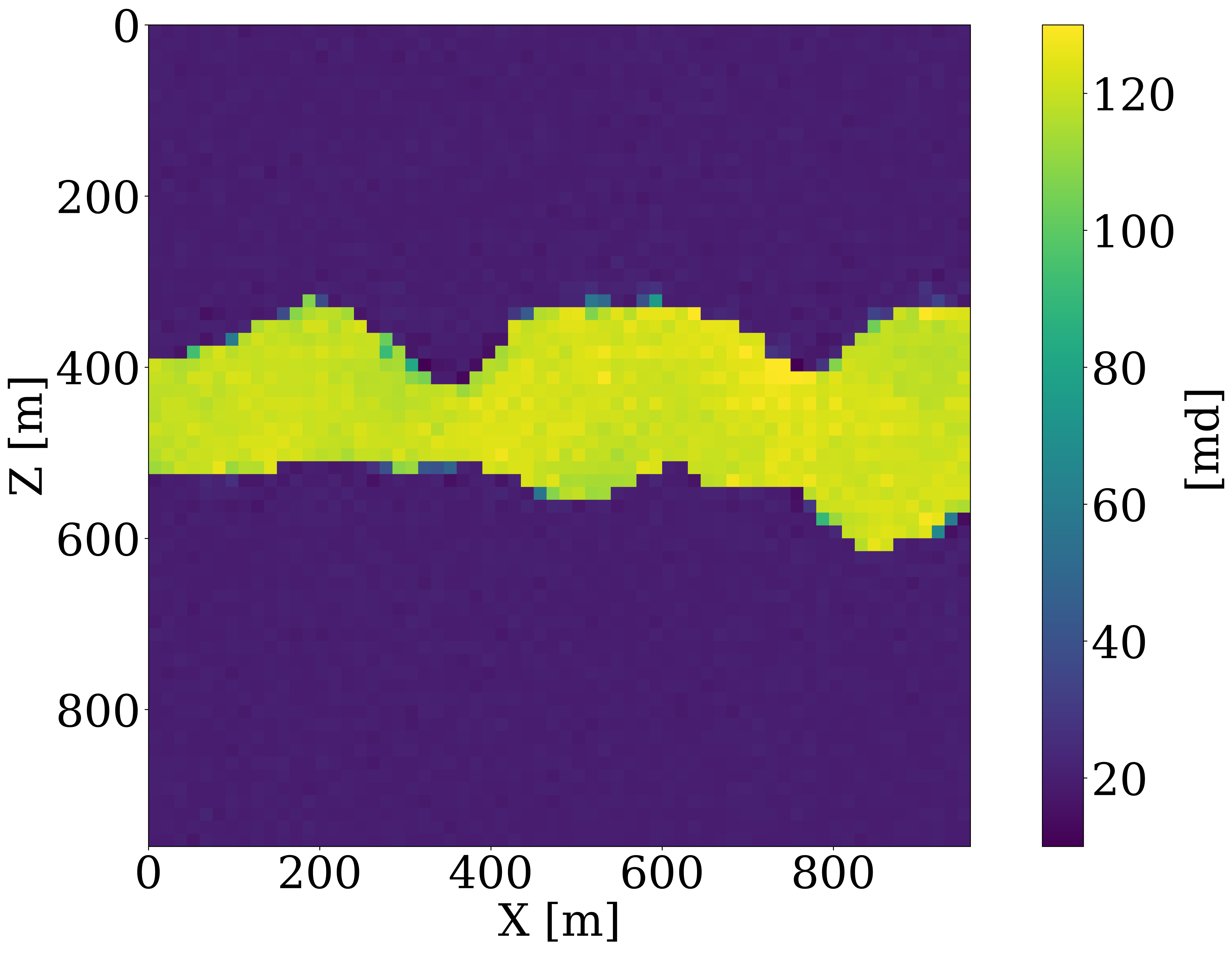}

}

}

\end{minipage}%
\begin{minipage}[t]{0.33\linewidth}

{\centering 

\raisebox{-\height}{

\includegraphics{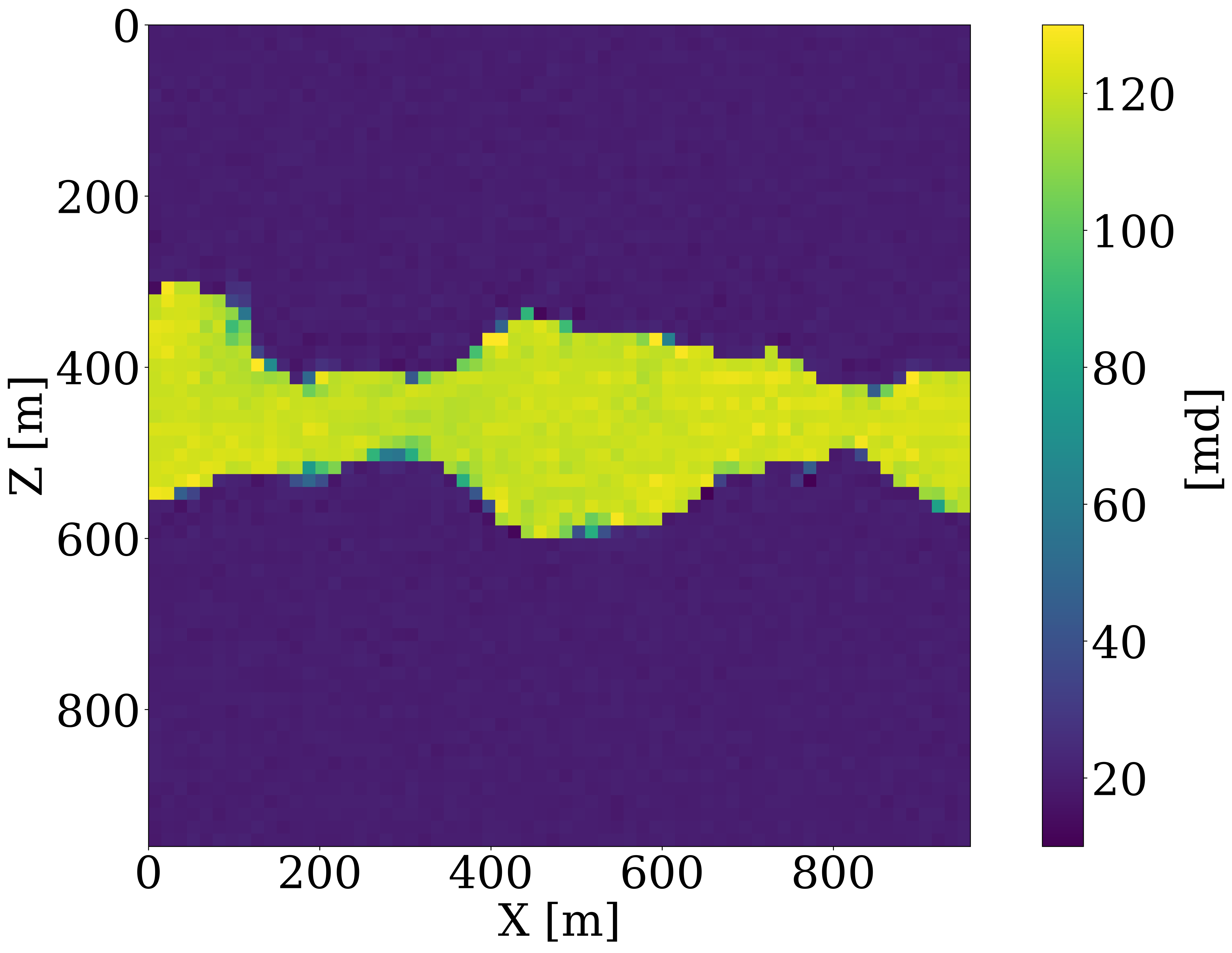}

}

}

\end{minipage}%
\begin{minipage}[t]{0.33\linewidth}

{\centering 

\raisebox{-\height}{

\includegraphics{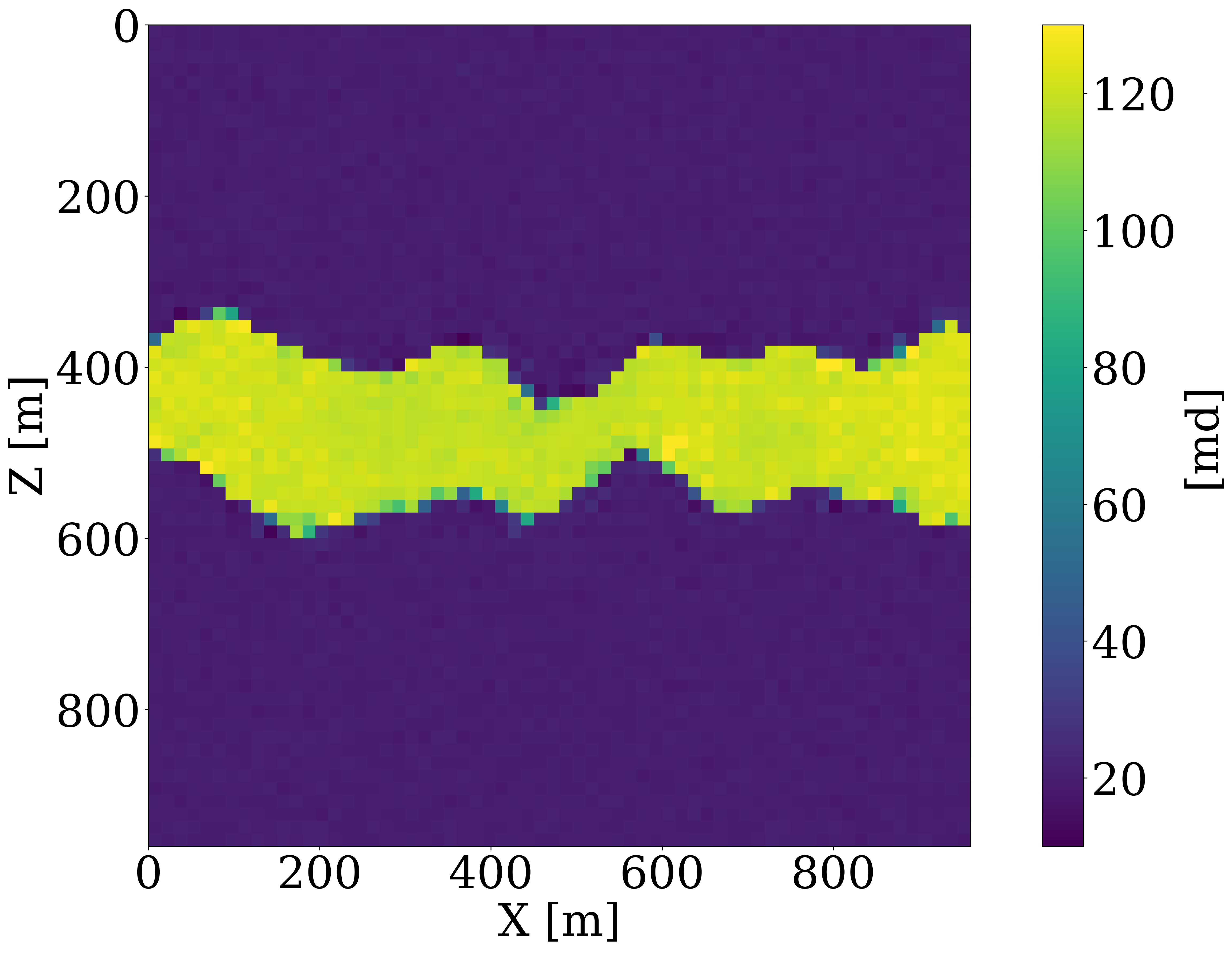}

}

}

\end{minipage}%

\caption{\label{fig-trainingsample}Permeability models. First row shows
the realistic permeability samples for FNO and NF training. Second row
shows the generative samples from the trained NF.}

\end{figure}

In this optimization problem, \(\mathcal{G}^{-1}_{\mathbf{w}}\)
represents the NF, which is parameterized by its network weights
\(\mathbf{w}\), while \(\mathrm{J}_{\mathcal{G}^{-1}_{\mathbf{w}}}\)
denotes its Jacobian. By minimizing the \(\ell_2\)-norm, the objective
imposes a Gaussian distribution on the network's output and the second
\(\log\det\) term prevents trivial solutions, i.e., cases where
\(\mathcal{G}^{-1}_{\mathbf{w}}\) produces zeros. To ensure alignment
between the permeability distributions, Equation~\ref{eq-fno-train} and
Equation~\ref{eq-train-nf} are trained on the same dataset consisting of
2000 permeability models examples of which are included in
Figure~\ref{fig-trainingsample}. Each \(64\times64\) permeability model
of consists of a randomly generated highly permeable channels
(\(120\,\mathrm{mD}\)) in a low-permeable background of
\(20\,\mathrm{mD}\), where \(\,\mathrm{mD}\) denotes millidarcy.
Generative examples produced by the trained NF are included in the
second row of Figure~\ref{fig-trainingsample}, which confirm the NF's
ability to learn distributions from examples. Aside from generating
samples from the learned distribution, trained NFs are also capable of
carrying out density calculations, an ability we will exploit below.

\hypertarget{trained-normalizing-flows-as-constraints}{%
\subsubsection{Trained normalizing flows as
constraints}\label{trained-normalizing-flows-as-constraints}}

As we mentioned before, adding constraints to the solution of non-convex
optimization problems offers guarantees that model iterates remain
within constrained sets. When solving inverse problems with learned
surrogates, it is important that model iterates remain ``in
distribution'', which can be achieved by recasting the optimization
problem in Equation Equation~\ref{eq-inv-fno} into the following
constrained form:

\begin{equation}\protect\hypertarget{eq-inv-fno-nf}{}{
\underset{\mathbf{z}}{\operatorname{minimize}} \quad \|\mathbf{d} - \mathcal{H}\circ\mathcal{S}_{\boldsymbol\theta^\ast}\circ\mathcal{G}_{\mathbf{w}^\ast}(\mathbf{z})\|_2^2\quad\text{subject to}\quad\|\mathbf{z}\|_2\leq \tau.
}\label{eq-inv-fno-nf}\end{equation}

To arrive at this constrained optimization problem, two important
changes were made. First, the permeability \(\mathbf{K}\) is replaced by
the output of a trained NF with trained weights \(\mathbf{w}^\ast\)
obtained by minimizing Equation~\ref{eq-train-nf}. This
reparameterization in terms of the latent variable, \(\mathbf{z}\),
produces permeabilities that are in distribution as long as
\(\mathbf{z}\) remains distributed according to the standard normal
distribution. Second, we added a constraint on this latent space
variable in Equation~\ref{eq-inv-fno-nf}, which ensures that the latent
variable \(\mathbf{z}\) remains within an \(\ell_2\)-norm ball of size
\(\tau\).

\begin{figure}

\begin{minipage}[t]{0.50\linewidth}

{\centering 

\raisebox{-\height}{

\includegraphics{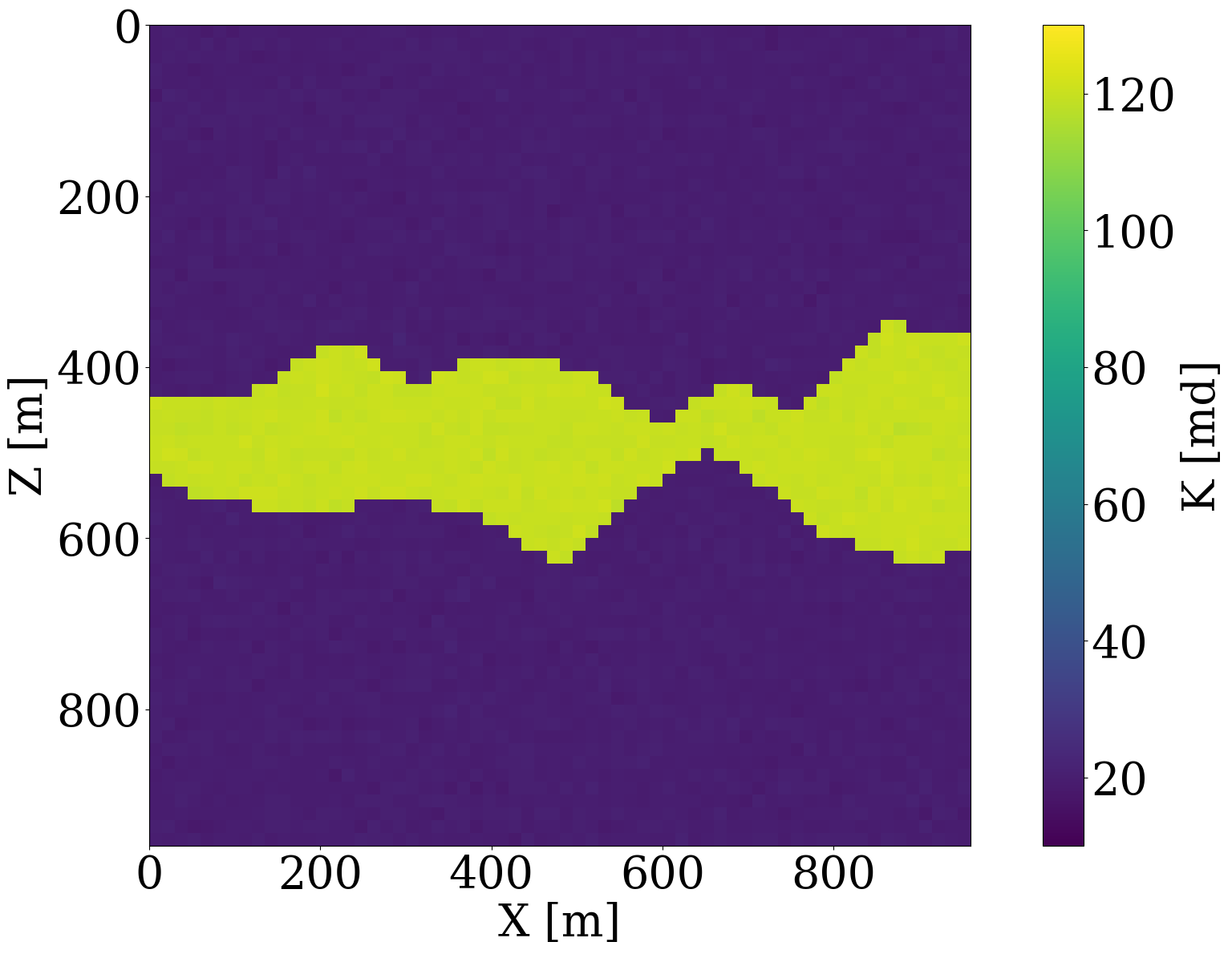}

}

}

\subcaption{\label{fig-ind-sample}}
\end{minipage}%
\begin{minipage}[t]{0.50\linewidth}

{\centering 

\raisebox{-\height}{

\includegraphics{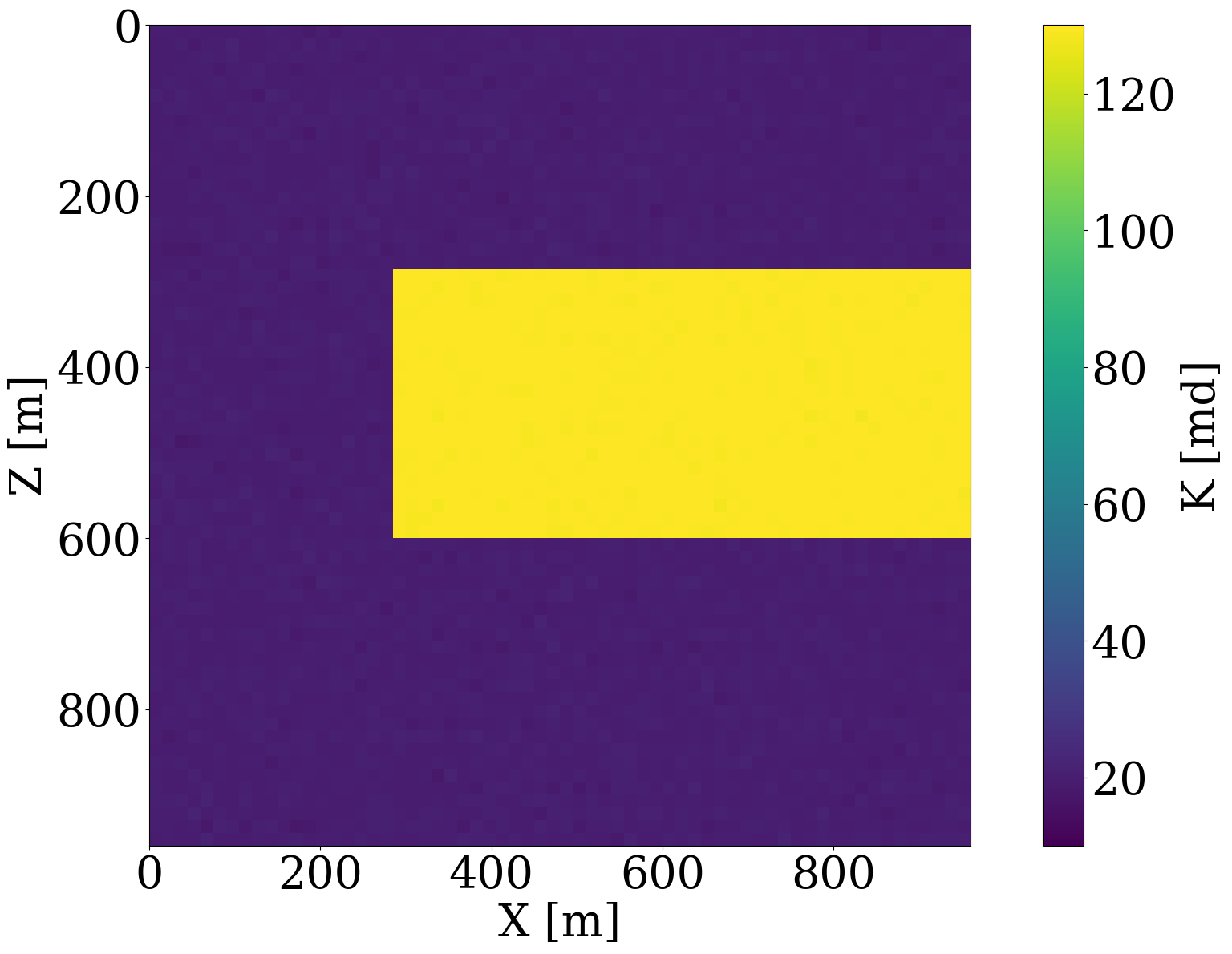}

}

}

\subcaption{\label{fig-ood-sample}}
\end{minipage}%
\newline
\begin{minipage}[t]{0.50\linewidth}

{\centering 

\raisebox{-\height}{

\includegraphics{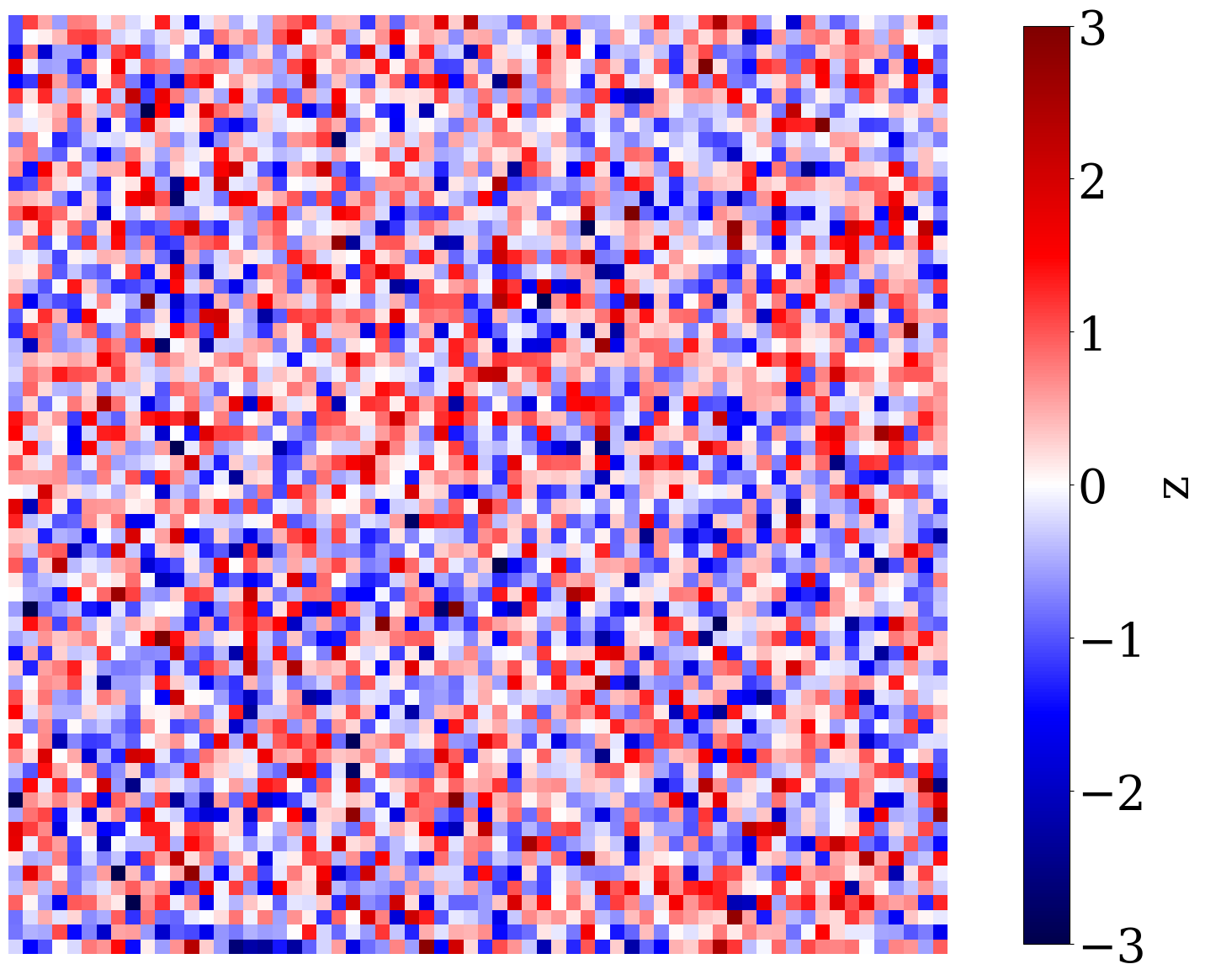}

}

}

\subcaption{\label{fig-ind-latent}}
\end{minipage}%
\begin{minipage}[t]{0.50\linewidth}

{\centering 

\raisebox{-\height}{

\includegraphics{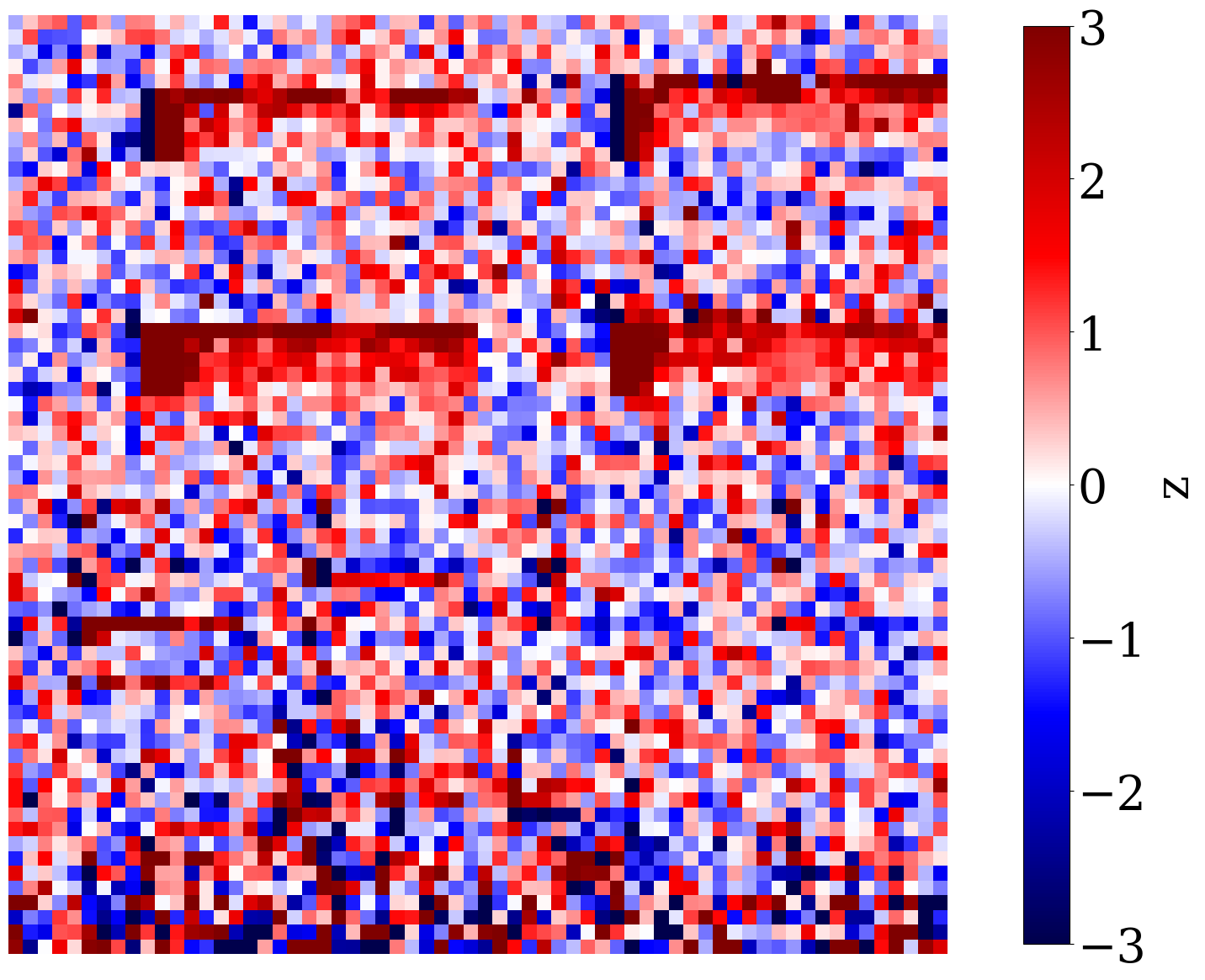}

}

}

\subcaption{\label{fig-ood-latent}}
\end{minipage}%

\caption{\label{fig-ood-samples}Sample permeability models in the
physical and latent space. \emph{(a)} An in-distribution permeability
model. \emph{(b)} An out-of-distribution permeability model. \emph{(c)}
An in-distribution permeability model in the latent space. \emph{(d)} An
out-of-distribution permeability model in the latent space.}

\end{figure}

\begin{figure}

\begin{minipage}[t]{0.20\linewidth}

{\centering 

\raisebox{-\height}{

\includegraphics{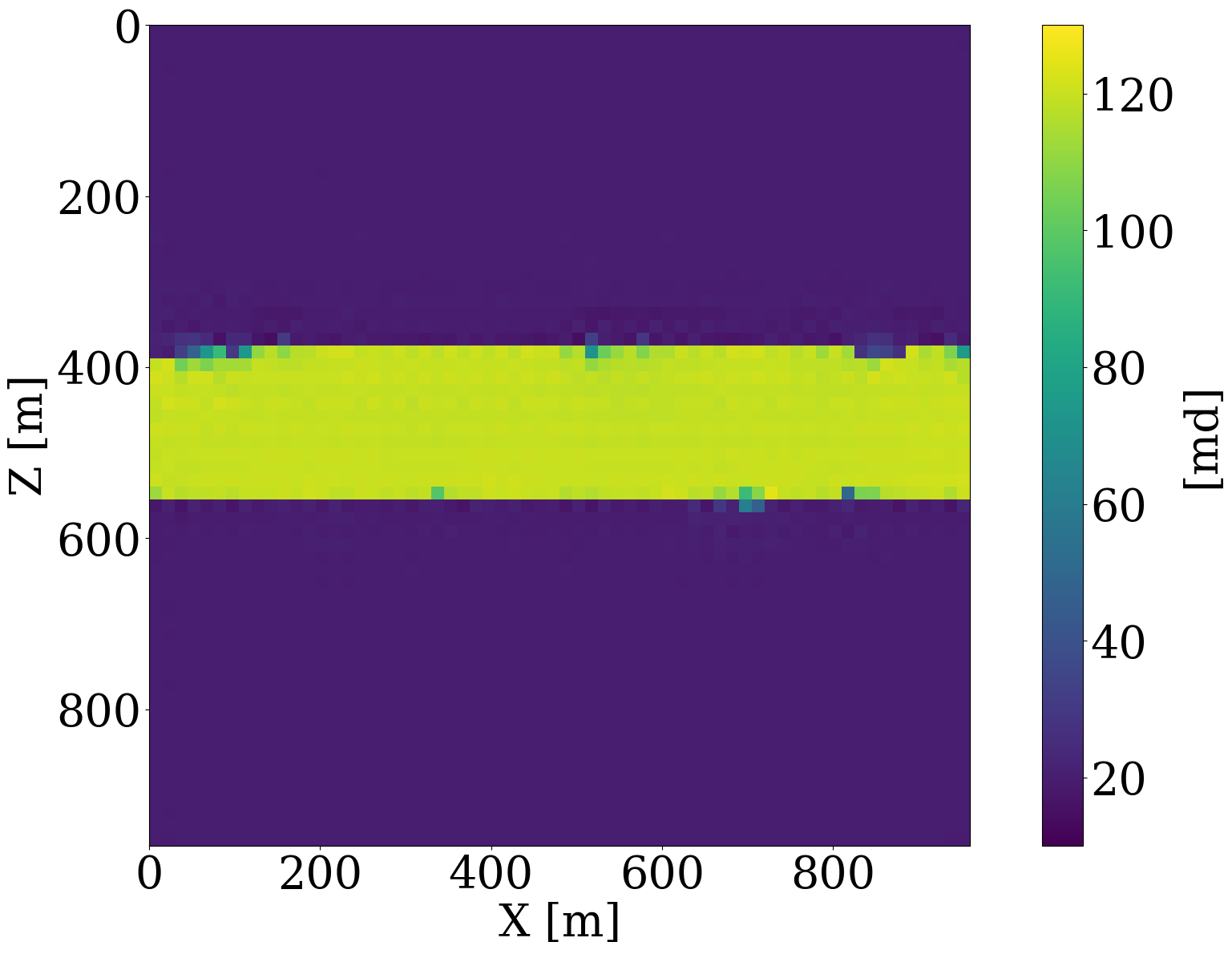}

}

}

\subcaption{\label{fig-ind-sample-ho1}}
\end{minipage}%
\begin{minipage}[t]{0.20\linewidth}

{\centering 

\raisebox{-\height}{

\includegraphics{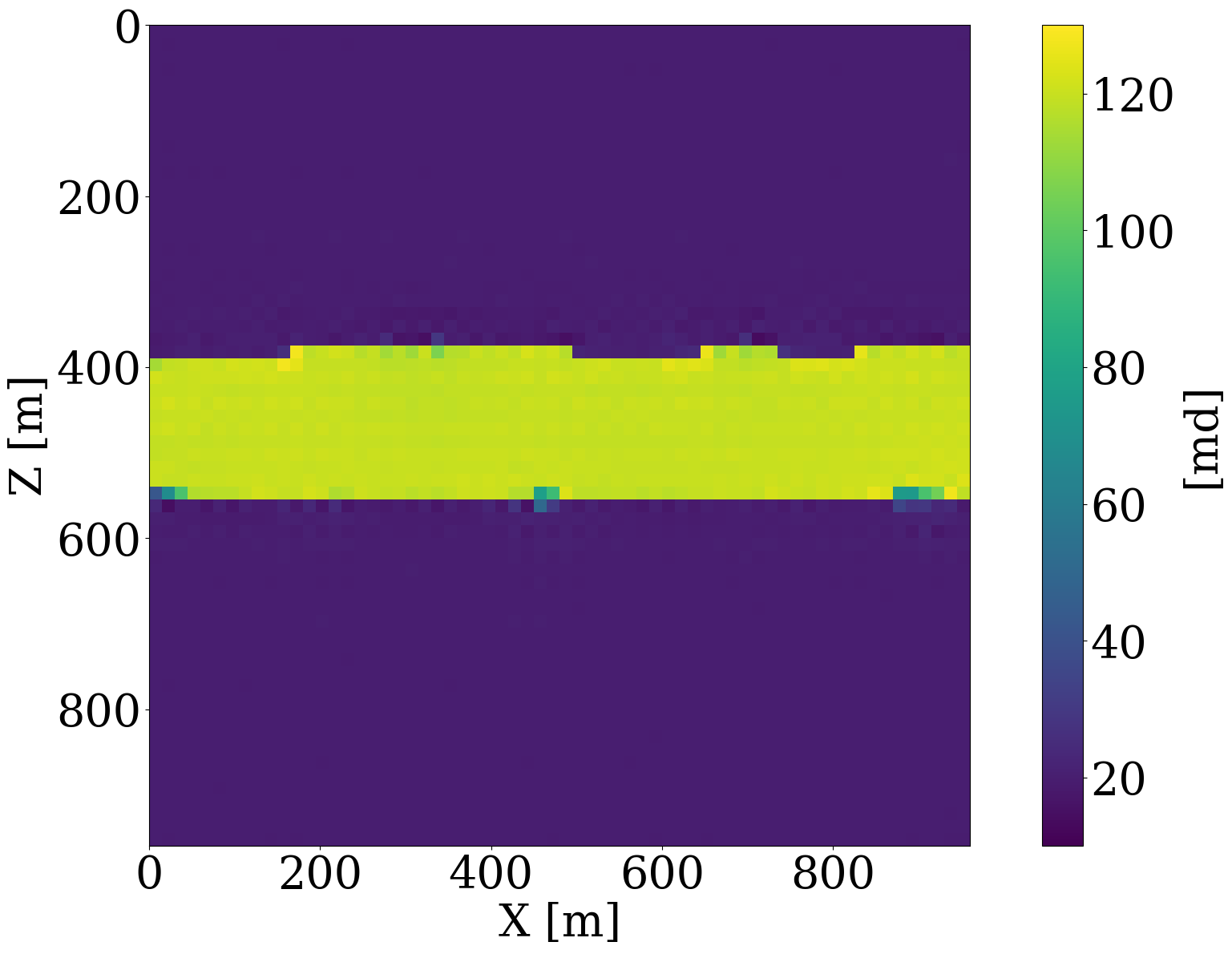}

}

}

\subcaption{\label{fig-ind-sample-ho2}}
\end{minipage}%
\begin{minipage}[t]{0.20\linewidth}

{\centering 

\raisebox{-\height}{

\includegraphics{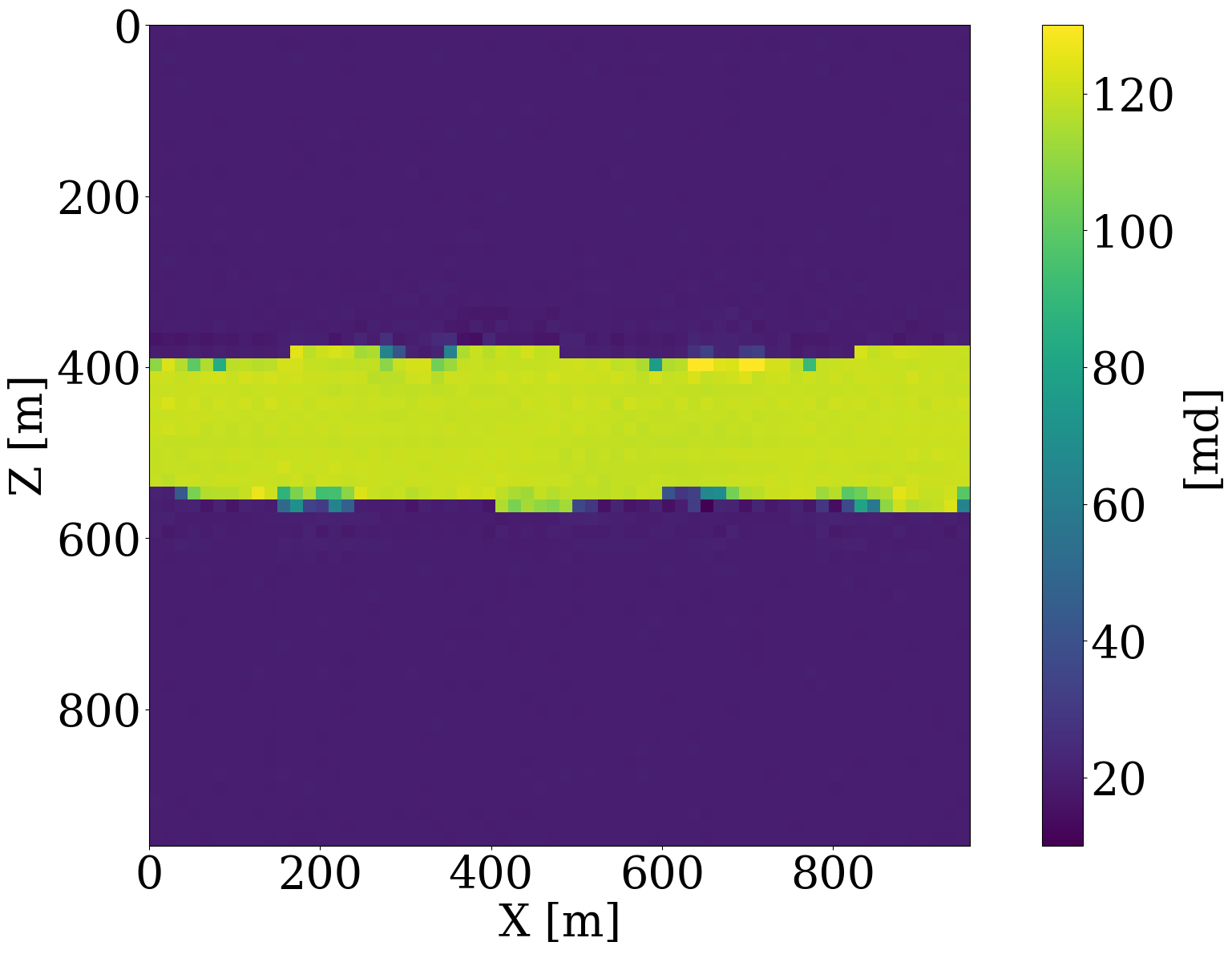}

}

}

\subcaption{\label{fig-ind-sample-ho3}}
\end{minipage}%
\begin{minipage}[t]{0.20\linewidth}

{\centering 

\raisebox{-\height}{

\includegraphics{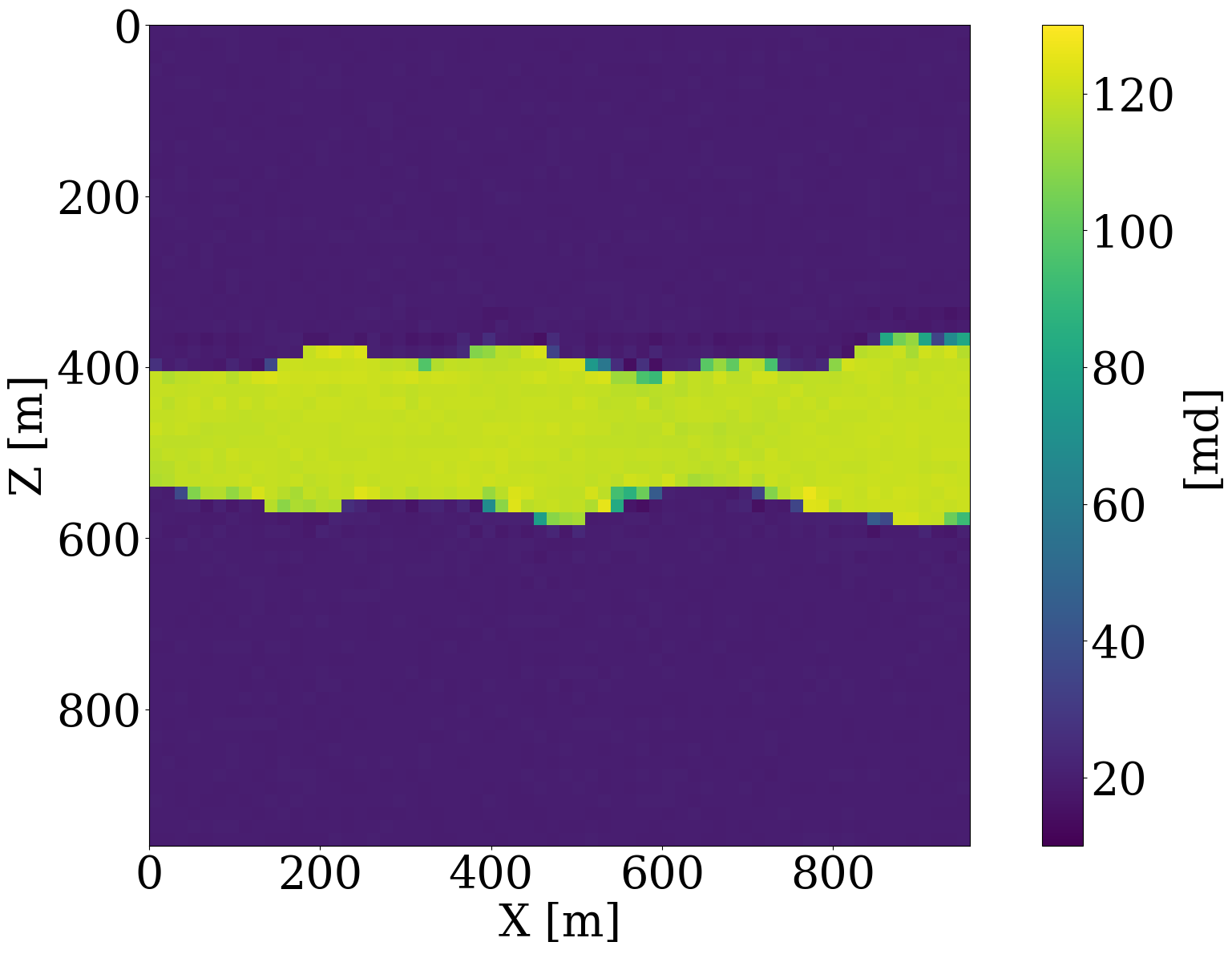}

}

}

\subcaption{\label{fig-ind-sample-ho4}}
\end{minipage}%
\begin{minipage}[t]{0.20\linewidth}

{\centering 

\raisebox{-\height}{

\includegraphics{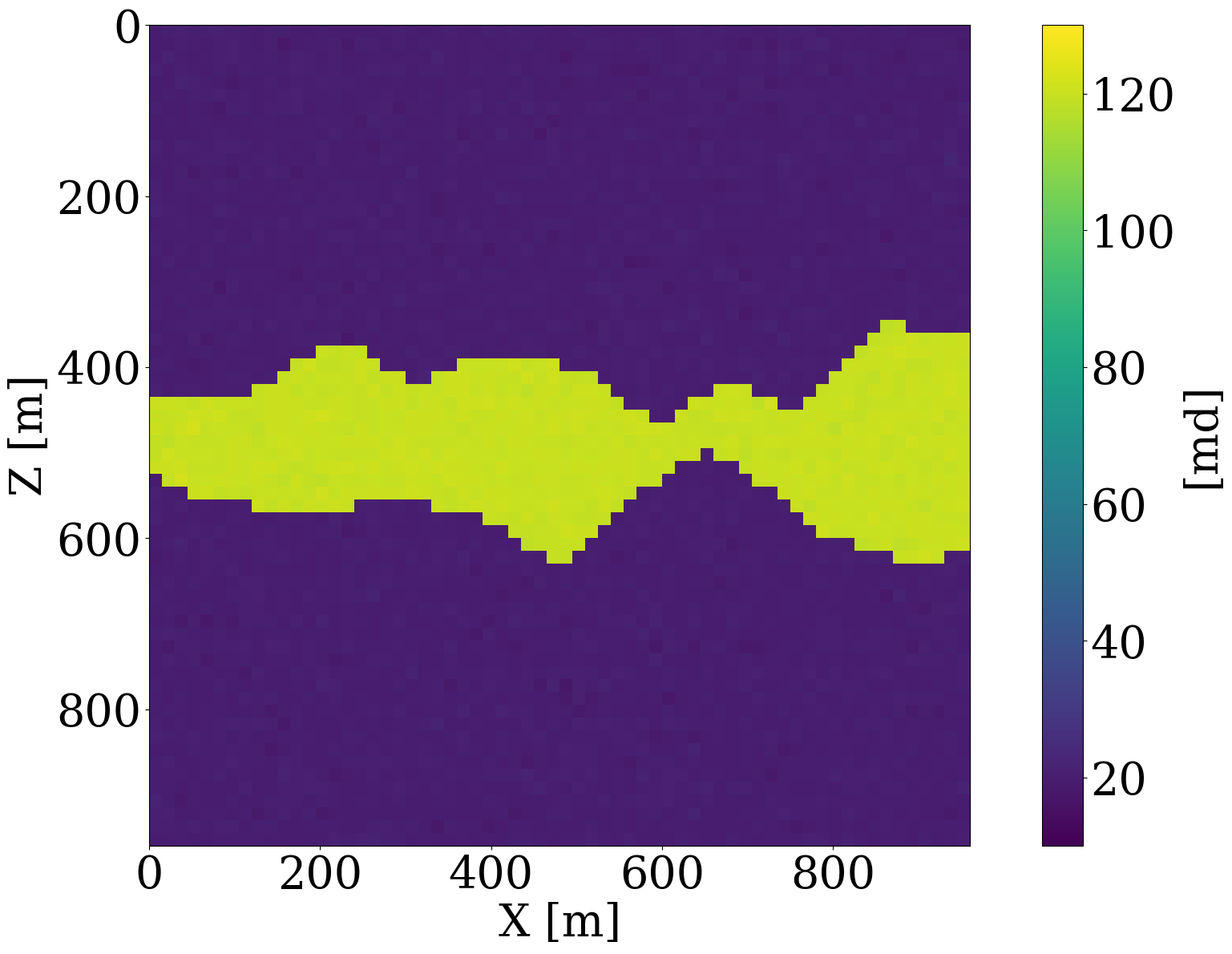}

}

}

\subcaption{\label{fig-ind-sample-ho5}}
\end{minipage}%
\newline
\begin{minipage}[t]{0.20\linewidth}

{\centering 

\raisebox{-\height}{

\includegraphics{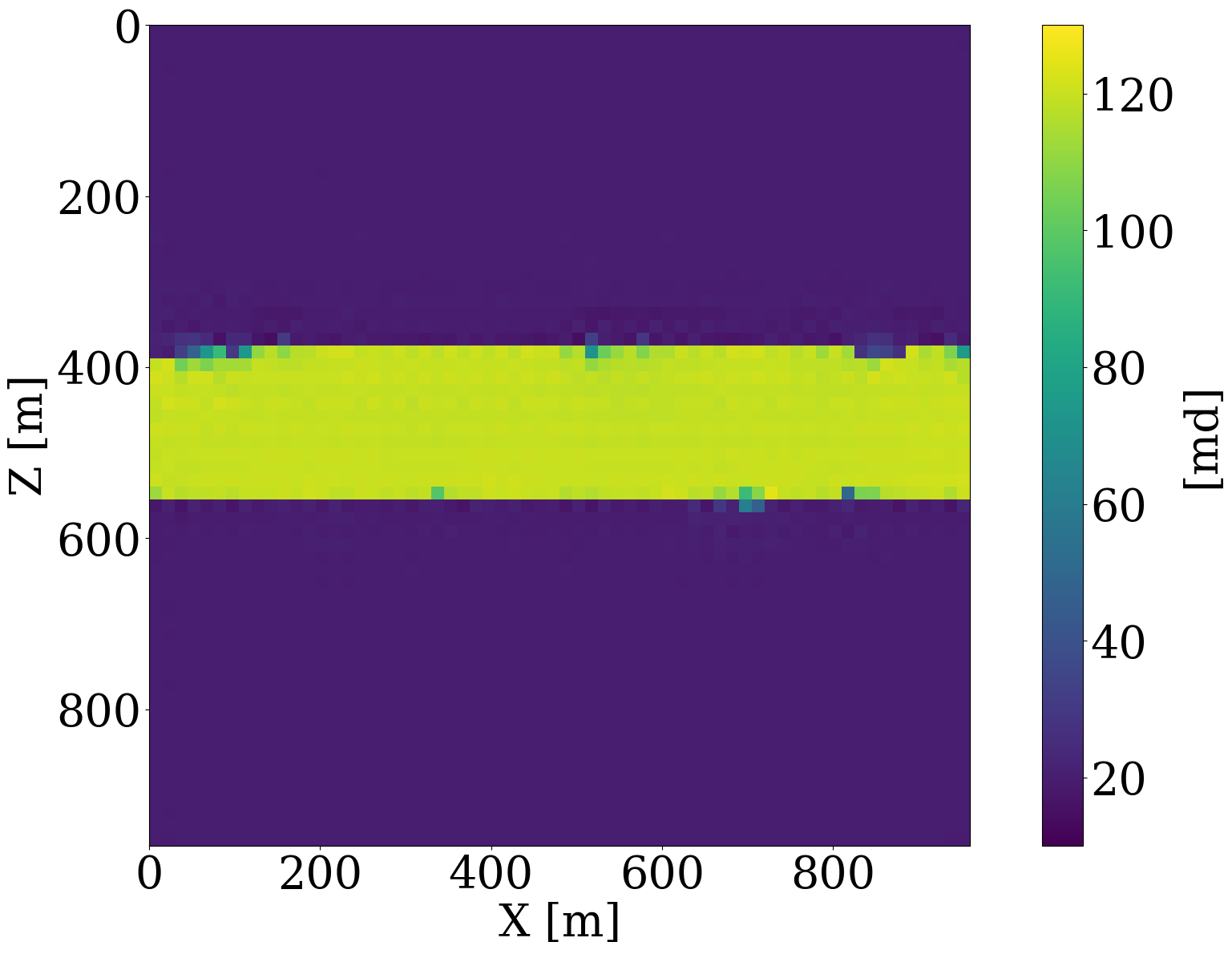}

}

}

\subcaption{\label{fig-ood-sample-ho1}}
\end{minipage}%
\begin{minipage}[t]{0.20\linewidth}

{\centering 

\raisebox{-\height}{

\includegraphics{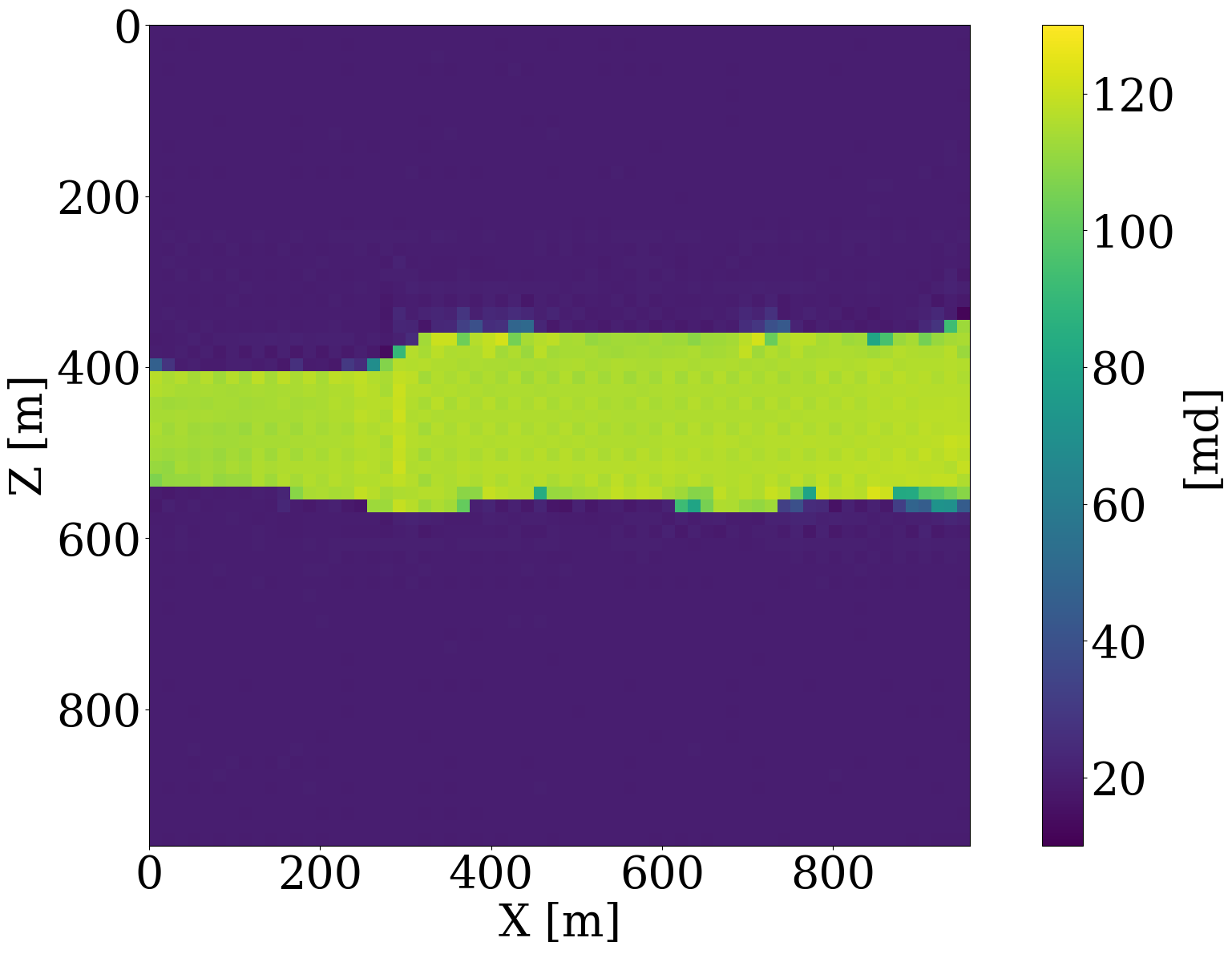}

}

}

\subcaption{\label{fig-ood-sample-ho2}}
\end{minipage}%
\begin{minipage}[t]{0.20\linewidth}

{\centering 

\raisebox{-\height}{

\includegraphics{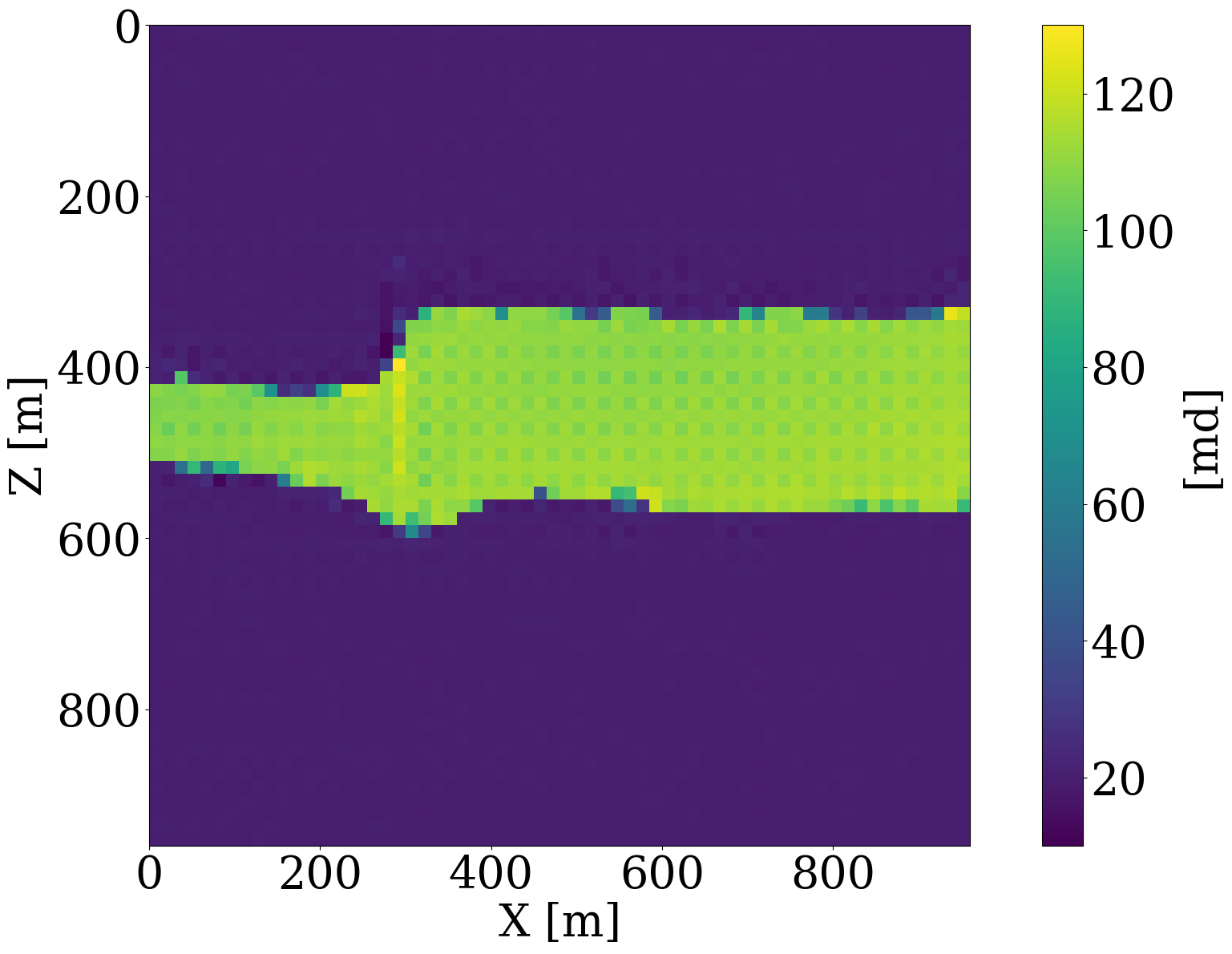}

}

}

\subcaption{\label{fig-ood-sample-ho3}}
\end{minipage}%
\begin{minipage}[t]{0.20\linewidth}

{\centering 

\raisebox{-\height}{

\includegraphics{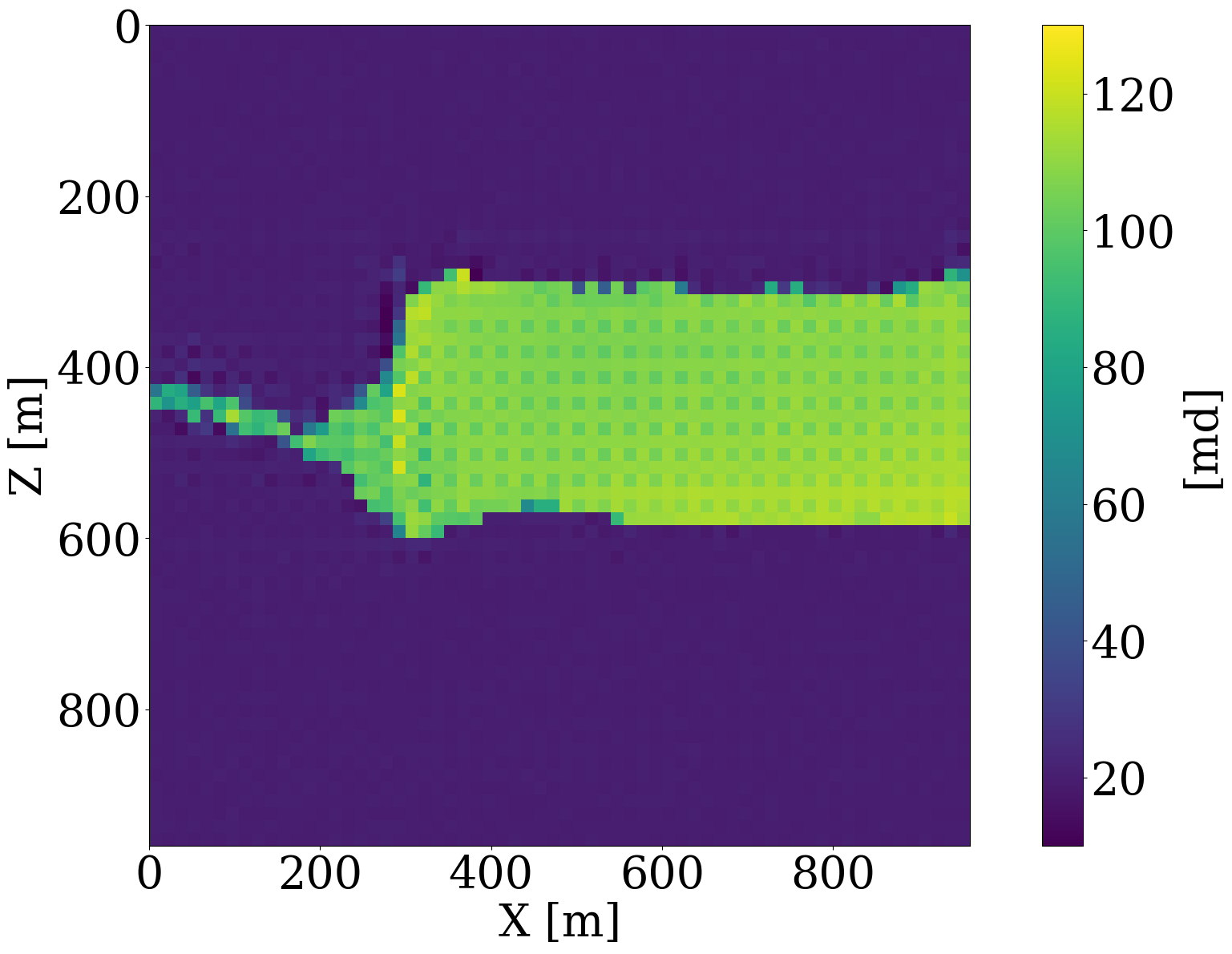}

}

}

\subcaption{\label{fig-ood-sample-ho4}}
\end{minipage}%
\begin{minipage}[t]{0.20\linewidth}

{\centering 

\raisebox{-\height}{

\includegraphics{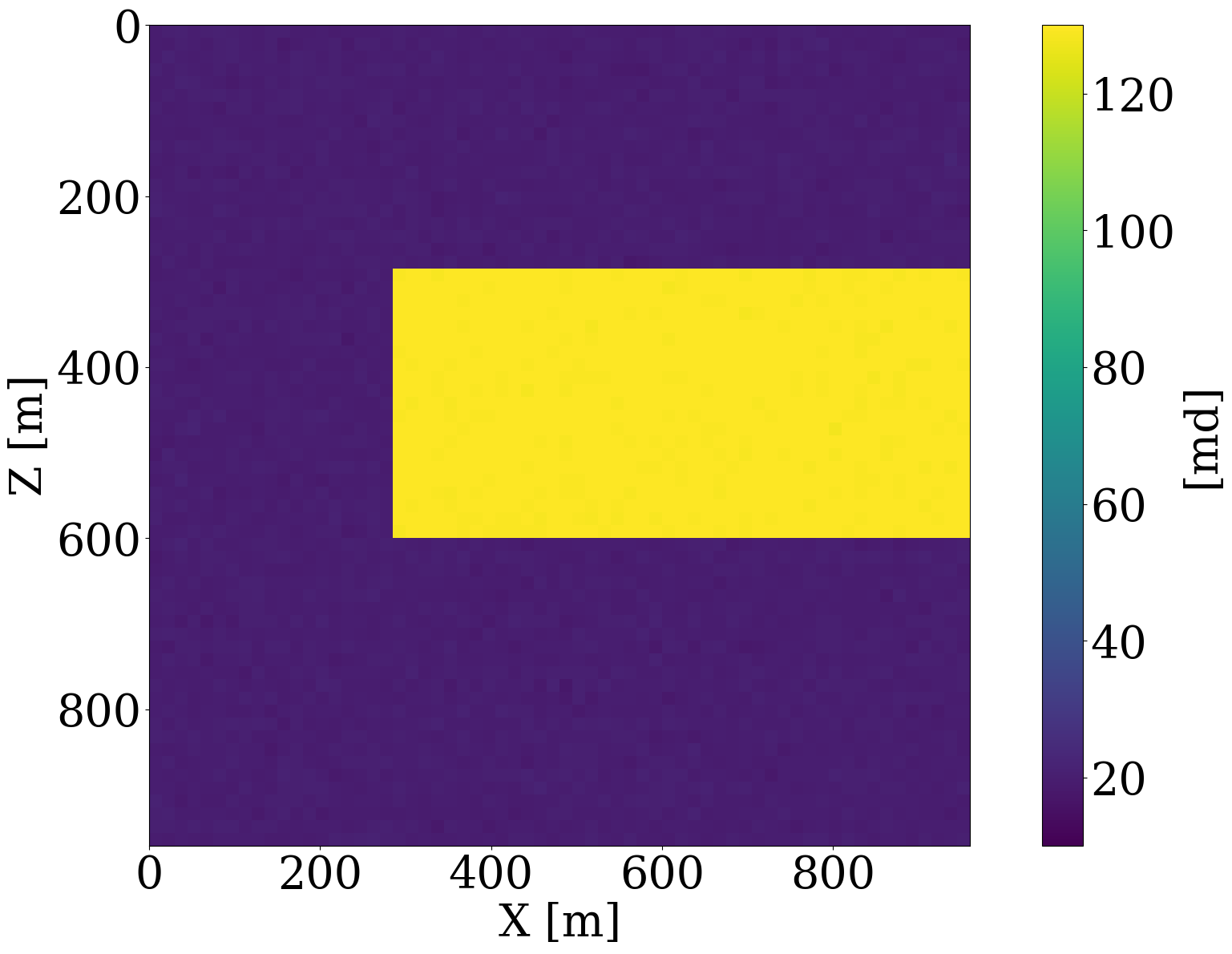}

}

}

\subcaption{\label{fig-ood-sample-ho5}}
\end{minipage}%

\caption{\label{fig-homotopy}Projections onto increasing \(\ell_2\)-norm
balls for the in- and out-of-distribution examples of
Figure~\ref{fig-ood-samples}. Top row: projections of in-distribution
sample. Bottom row: projections of out-of-distribution sample. Each
column corresponds to setting \(\alpha=0,0.1,0.2,0.4,1\) in
Equation~\ref{eq-shrinkage}.}

\end{figure}

\begin{figure}

\begin{minipage}[t]{0.50\linewidth}

{\centering 

\raisebox{-\height}{

\includegraphics{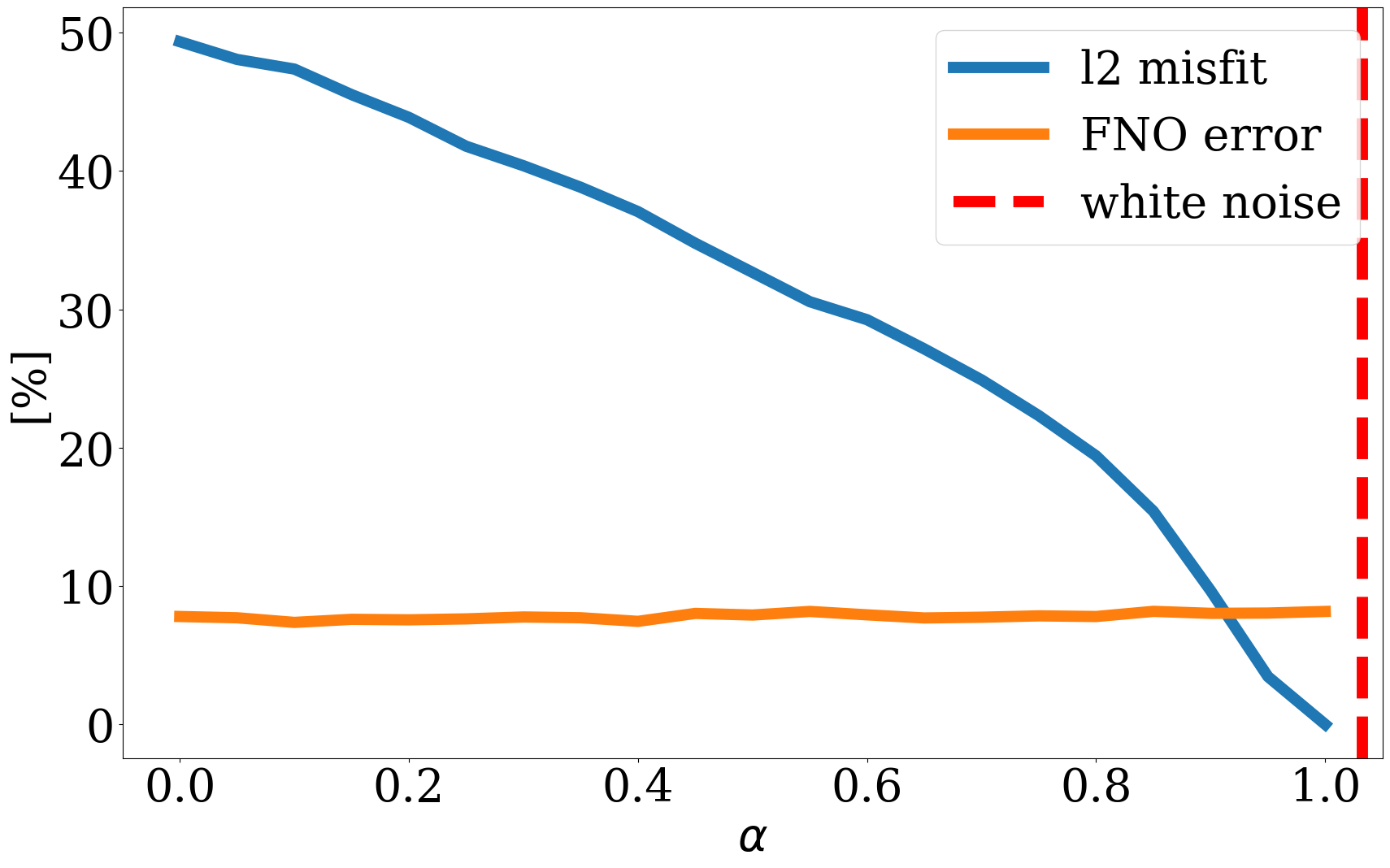}

}

}

\subcaption{\label{fig-ind-alpha}}
\end{minipage}%
\begin{minipage}[t]{0.50\linewidth}

{\centering 

\raisebox{-\height}{

\includegraphics{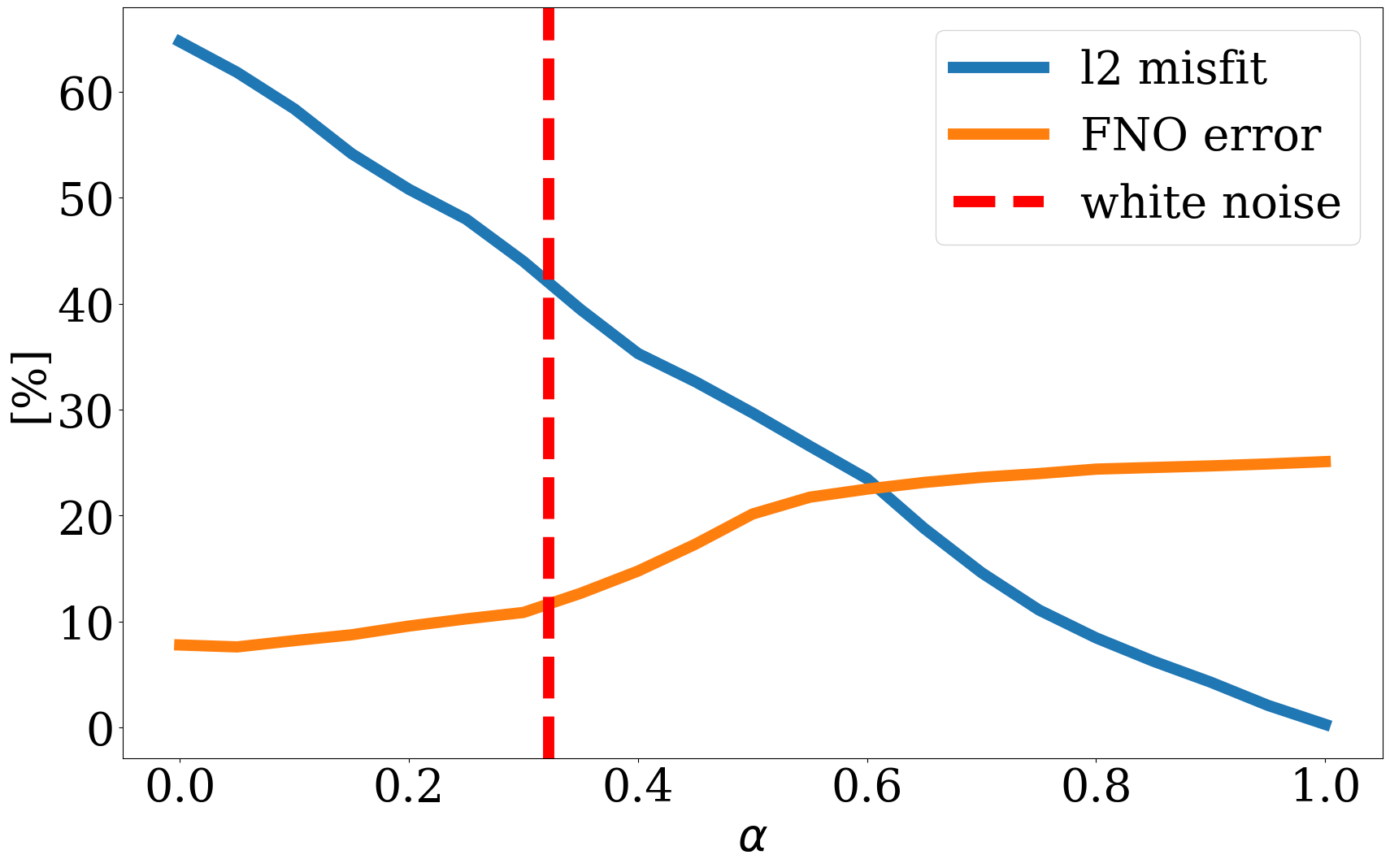}

}

}

\subcaption{\label{fig-ood-alpha}}
\end{minipage}%

\caption{\label{fig-ood}Latent space projection experiments.\emph{(a)}
Relative \(\ell_2\) reconstruction error and FNO prediction error for
in-distribution sample. \emph{(b)} The same but for out-of-distribution
sample. The blue curve shows the relative \(\ell_2\) misfit between the
permeability models before and after latent space shrinkage. The orange
curve shows the FNO prediction error on the permeability model after
shrinking the \(\ell_2\)-norm ball. The red dashed line denotes the
amplitude of standard Gaussian noise.}

\end{figure}

\begin{figure}

{\centering 

\includegraphics[width=0.5\textwidth,height=\textheight]{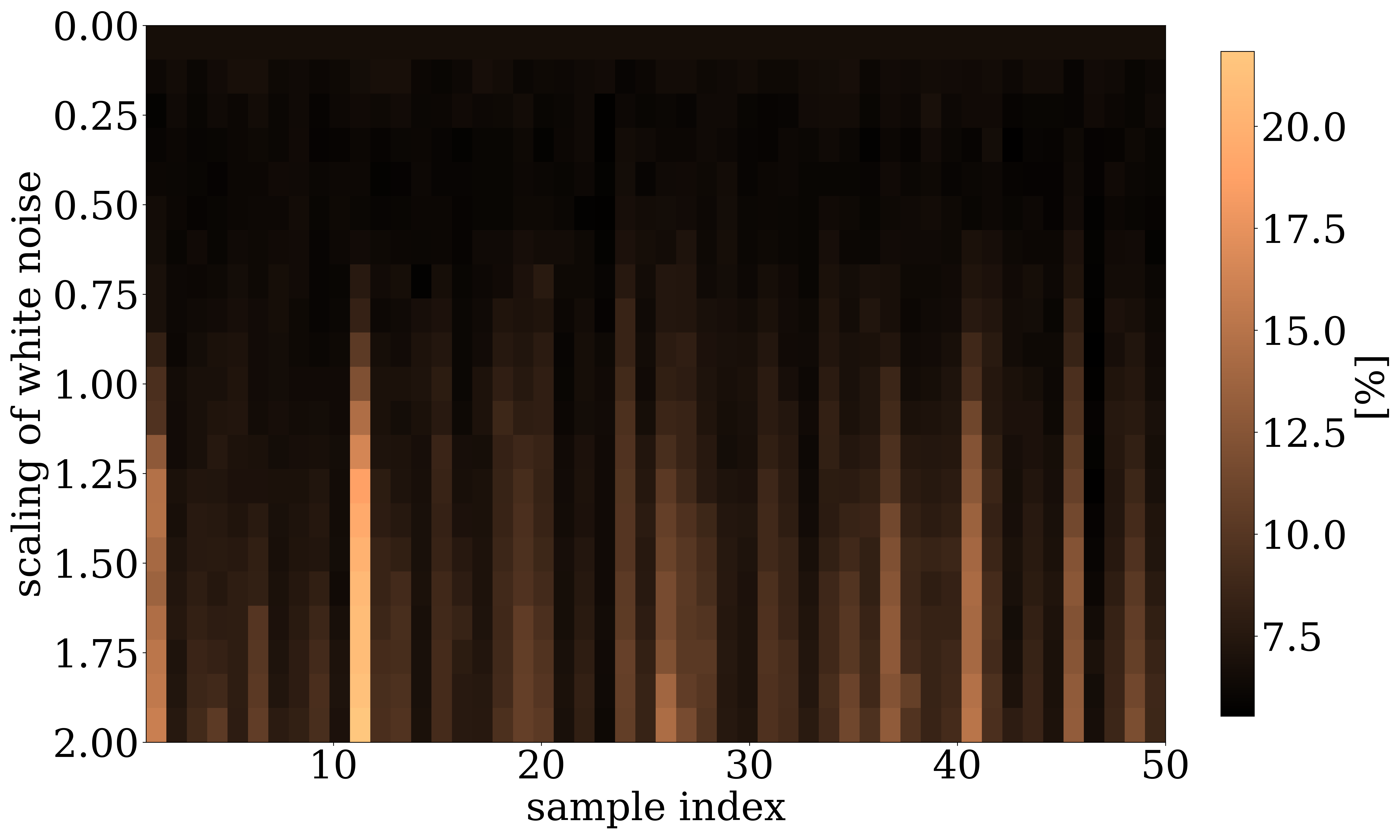}

}

\caption{\label{fig-init-tau}FNO prediction errors for the latent space
shrinkage experiment in Equation~\ref{eq-shrinkage} for 50 random
realizations of standard Gaussian noise.}

\end{figure}

To better understand the behavior of a trained normalizing flow in
conjunction with the \(\ell_2\)-norm constraint for in- and
out-of-distribution examples, we include Figure~\ref{fig-ood-samples}
and Figure~\ref{fig-homotopy}. In the latter Figure, nonlinear
projections (via latent space shrinkage),
\begin{equation}\protect\hypertarget{eq-shrinkage}{}{
\widetilde{\mathbf{K}} = \mathcal{G}_{\mathbf{w}^\ast}\left(\alpha\mathbf{z}\right)\quad\text{where}\quad \mathbf{z}=\mathcal{G}_{\mathbf{w}^\ast}^{-1}(\mathbf{K})
}\label{eq-shrinkage}\end{equation}

are plotted as a function of increasing \(\alpha\). We also plot in
Figure~\ref{fig-ood} the NF's relative nonlinear approximation error,
\(\|\mathbf{\widetilde{K}}-\mathbf{K}\|_2/\|\mathbf{K}\|_2\), and the
corresponding relative FNO prediction
error,\(\|\mathcal{S}_{\boldsymbol\theta^\ast}(\widetilde{\mathbf{K}})-\mathcal{S}(\widetilde{\mathbf{K}})\|_2/\|\mathcal{S}(\widetilde{\mathbf{K}})\|_2\)
as a function of increasing \(0\leq\alpha\leq 1\). From these plots, we
can make the following observations. First, the latent representations
(Figure~\ref{fig-ind-latent} and Figure~\ref{fig-ood-latent}) of the in-
and out-of-distribution samples (Figure~\ref{fig-ind-sample} and
Figure~\ref{fig-ood-sample} ) clearly show that NF applied to
out-of-distribution samples produces a latent variable far from the
standard normal distribution, while the latent variable corresponding to
the in-distribution example is close to being white Gaussian noise.
Quantitatively, the \(\ell_2\) norm of the latent variables are
\(0.99\|\mathcal{N}(0,\mathbf{I})\|_2\) and
\(3.11\|\mathcal{N}(0,\mathbf{I})\|_2\), respectively, where
\(\|\mathcal{N}(0,\mathbf{I})\|_2\) corresponds to the \(\ell_2\)-norm
of the standard normal distribution. Second, we observe from
Figure~\ref{fig-homotopy} that for small \(\ell_2\)-norm balls
(\(\alpha\ll 1\)) the projected solutions tend to be close to the most
probable sample, which is a flat permeability channel in the middle.
This is true for both the in- and out-of-distribution example. As
\(\alpha\) increases, the in-distribution example is reconstructed
accurately when the \(\ell_2\) norm of the scaled latent variable,
\(\|\alpha\mathbf{z}\|_2\), is close to the
\(\|\mathcal{N}(0,\mathbf{I})\|_2\). Clearly, this is not the case for
the out-of-distribution example. When
\(\|\alpha\mathbf{z}\|_2\approx\|\mathcal{N}(0,\mathbf{I})\|_2\), the
reconstruction still looks like an in-distribution permeability sample
and is not close to the out-of-distribution sample. However, if
\(\alpha=1\), which makes \(\|\alpha\mathbf{z}\|_2\) well beyond the
norm of the standard normal distribution, the out-of-distribution
example is recovered accurately by virtue of the invertibility of NFs,
irrespective on their input and what they have been trained on. Third,
the relative FNO prediction error for the in-distribution example
(Figure~\ref{fig-ind-alpha}) remains flat while the error of the FNO
surrogate increases as soon as \(\alpha\approx 0.25\). At that value for
\(\alpha\), the projection, \(\widetilde{\mathbf{K}}\), is gradually
transitioning from being in-distribution to out-of-distribution, which
occurs at a non-linear approximation error of about 45\%. As expected
the plots in Figure~\ref{fig-ood} also show a monotonous decay of the
nonlinear approximation error as a function of increasing \(\alpha\). To
further analyze the effects of the nonlinear projections in
Equation~\ref{eq-shrinkage}, we draw 50 random realizations from the
standard normal distribution, scale each of them by
\(0\leq\alpha\leq 2\), and calculate the FNO prediction errors on these
samples. Figure~\ref{fig-init-tau} includes the results of this
excercise where each column represents the FNO prediction error
calculated for \(0\leq \alpha \leq 2\). From these experiments, we make
the following two observations. First, when \(\alpha<0.8\), the FNO
consistently makes accurate predictions for all projected samples.
Second, as expected, the FNO starts to make less accurate predictions
for \(\alpha>1\) with errors that increase as the size of the
\(\ell_2\)-norm ball of the latent space expands, demarcating the
transition from being in distribution to being out-of-distribution.

In summary, the experiments of Figure~\ref{fig-ood-samples} to
Figure~\ref{fig-ood} indicate that FNO errors remain small and
relatively constant for the in-distribution example. Irrespective of the
value of \(\alpha\), the generated samples remain in distribution while
moving from the most likely---i.e., a flat high-permeability channel in
the middle, to the in-distribution sample as \(\alpha\) increases.
Conversely, the projection of the out-of-distribution example morphs
from being in distribution to being out-of-distribution for
\(\alpha \geq 0.25\). The FNO prediction errors also increase during
this transition from an in-distribution sample to an out-of-distribution
sample. Therefore, shrinkage in the latent space by multiplying with a
small \(\alpha\) can serve as an effective projection that ensures
relatively low FNO prediction errors. We will use this unique ability to
control the distribution during inversion.

\hypertarget{inversion-with-progressively-relaxed-learned-constraints}{%
\subsubsection{Inversion with progressively relaxed learned
constraints}\label{inversion-with-progressively-relaxed-learned-constraints}}

Our main objective is to perform inversions where the multiphase flow
equations are replaced with pretrained FNO surrogates. To make sure the
learned surrogates remain accurate, we propose working with a
continuation scheme where the learned constraint in
Equation~\ref{eq-inv-fno-nf} is steadily relaxed by increasing the size
of the \(\ell_2\)-norm ball constraint. Compared to the more common
penalty formulation, where regularization entails adding a
Lagrange-multiplier weighted \(\ell_2\)-norm squared, constrained
formulations offer guarantees that the model iterates for the latent
variable, \(\mathbf{z}\), remain within the constraint set---i.e.,
within the \(\ell_2\)-norm ball of size \(\tau\). Using the argument of
the previous section, this implies that permeability distributions
generated by the trained NF remain in distribution as long as the size
of the initial \(\ell_2\)-norm ball, \(\tau_{\mathrm{init}}\), is small
enough (e.g., smaller than \(0.6\|\mathcal{N}(0,\mathbf{I})\|_2\),
following the observations from Figure~\ref{fig-init-tau}). Taking
advantage of this trained NF in a homotopy, we propose the following
algorithm:

\hypertarget{algorithm-1-inversion-with-relaxed-learned-constraints}{%
\paragraph{Algorithm 1: Inversion with relaxed learned
constraints}\label{algorithm-1-inversion-with-relaxed-learned-constraints}}

\begin{algorithm}
\textbf{Input:} initial model parameter $\mathbf{K}_0\in \mathbb{R}^N$, observed data $\mathbf{d}$, noise level $\sigma$

\textbf{Input:} trained FNO $\mathcal{S}_{\boldsymbol{\theta}^\ast}$, trained NF $\mathcal{G}_{\mathbf{w}^\ast}$

\textbf{Input:} number of inner-loop iterations $maxiter$

\textbf{Input:} initial $\ell_2$ ball size $\tau_{\mathrm{init}}$, multiplier $\beta>1$, final $\ell_2$ ball size $\tau_{\mathrm{final}}$

$\mathbf{z} = \mathcal{G}_{\mathbf{w}^\ast}^{-1}(\mathbf{K}_0)$

$\tau = \tau_{\mathrm{init}}$

\While {$\|\mathbf{d} - \mathcal{H}\circ\mathcal{S}_{\boldsymbol\theta^\ast}\circ\mathcal{G}_{\mathbf{w}^\ast}(\mathbf{z})\|_2 > \sigma\|\mathcal{N}(0,\mathbf{I})\|_2$ $\mathrm{and}$ $\tau\leq\tau_{\mathrm{final}}$}{
\For{$iter = 1:maxiter$}{ $\displaystyle\mathbf{g}=\nabla_{\mathbf{z}}\|\mathbf{d} - \mathcal{H}\circ\mathcal{S}_{\boldsymbol\theta^\ast}\circ\mathcal{G}_{\mathbf{w}^\ast}(\mathbf{z})\|_2^2$

$\mathbf{z} = \mathcal{P}_{\tau}(\mathbf{z} - \gamma\mathbf{g})$
}
$\tau = \beta\tau$
}
\textbf{Output:} inverted model parameter $\mathbf{K}=\mathcal{G}_{\mathbf{w}^\ast}(\mathbf{z})$
\caption{Inversion with relaxed learned constraints}
\end{algorithm}

Given observed data, \(\mathbf{d}\), trained networks,
\(\mathcal{S}_{\boldsymbol{\theta}^\ast}\) and
\(\mathcal{G}_{\mathbf{w}^\ast}\), the initial guess for the
permeability distribution, \(\mathbf{K} _0\), the initial size of the
\(\ell_2\)-norm ball, \(\tau_{\mathrm{init}}\), and the final size of
the \(\ell_2\)-norm ball, \(\tau_{\mathrm{final}}\), Algorithm 1
proceeds by solving a series of constrained optimization problems where
the size of the constraint set is increased by a factor of \(\beta\)
after each iteration (cf.~line 12 in Algorithm 1). The constrained
optimization subproblems themselves (cf.~line 8 to 11 of Algorithm 1)
are solved with projected gradient descent (Beck 2014). Each iteration
of the projected gradient descent method first calculates the gradient
(cf.~line 9 of Algorithm 1), followed by the much cheaper projection of
the updated latent variable back onto the \(\ell_2\)-norm ball of size
\(\tau\) via the projection operator \(\mathcal{P}_{\tau}\) (cf.~line 10
in Algorithm 1). This projection is a trivial scaling operation if the
updated latent variable \(\ell_2\)-norm exceeds the constraint --- i.e.,
\begin{equation}\protect\hypertarget{eq-proj}{}{
\mathcal{P}_{\tau}(\mathbf{z}) = 
\begin{cases} 
\mathbf{z} & \text{if } \|\mathbf{z}\|_2 \leq \tau \\
\tau\mathbf{z}/\|\mathbf{z}\|_2 & \text{if } \|\mathbf{z}\|_2 > \tau 
\end{cases}
}\label{eq-proj}\end{equation}

A line search determines the steplength \(\gamma\) (Stanimirović and
Miladinović 2010) for each iteration shown in line 8 to 11. As is common
in continuation methods, the relaxed gradient-descent iterations are
warm-started with the optimization result from the previous iteration,
which at the first iteration is initialized by the latent representation
of the initial permeability model, \(\mathbf{K}_0\) (cf.~line 5 in
Algorithm 1). Practically, each subproblem does not need to be fully
solved, but only need a few iterations instead. The number of iterations
to solve each subproblem is denoted by \(maxiter\) in line 8 of
Algorithm 1. This continuation strategy serves two purposes. First, for
small \(\tau\)'s it makes sure the model iterates remain in
distribution, so accuracy of the learned surrogate is preserved. Second,
by relaxing the constraint slowly, the data residual is gradually
allowed to decrease, bringing in more and more features derived from the
data. By slowly relaxing the constraint, we find a careful balance
between these two purposes as long as progress is made towards the
solution when solving the subproblem (cf.~line 8 to 11 in Algorithm 1).
One notable distinction of the surrogate-assisted inversion, compared to
the conventional inversion with relaxed constraints (Esser et al. 2018),
is that the size of the \(\ell_2\)-norm projection ball cannot increase
far beyond the \(\ell_2\)-norm of the standard Gaussian white noise on
which the NFs are trained. Otherwise, there is no guarantee the learned
surrogate is accurate because the NF may generate samples that are
out-of-distribution (cf. Figure~\ref{fig-init-tau}). This is explicitly
incorporated into the stopping criteria,
\(\tau\leq\tau_{\mathrm{final}}\), in line 7 of Algorithm 1.

\hypertarget{numerical-experiments}{%
\subsection{Numerical Experiments}\label{numerical-experiments}}

To showcase the advocasy of the proposed optimization method with
relaxed learned constraints, a series of carefully chosen experiments of
increasing complexity are conducted. These experiments are designed to
be relevant to GCS, which in its ultimate form involves coupling of
multiphase flow with the wave equation to perform end-to-end inversion
for the permeability given multimodal data. To convince ourselves of the
validity of our approach, at all times comparisons will be made between
inversion results involving numerical solves of the multiphase equations
and inversions yielded by approximations with our learned surrogate.

\begin{figure}

\begin{minipage}[t]{0.50\linewidth}

{\centering 

\raisebox{-\height}{

\includegraphics{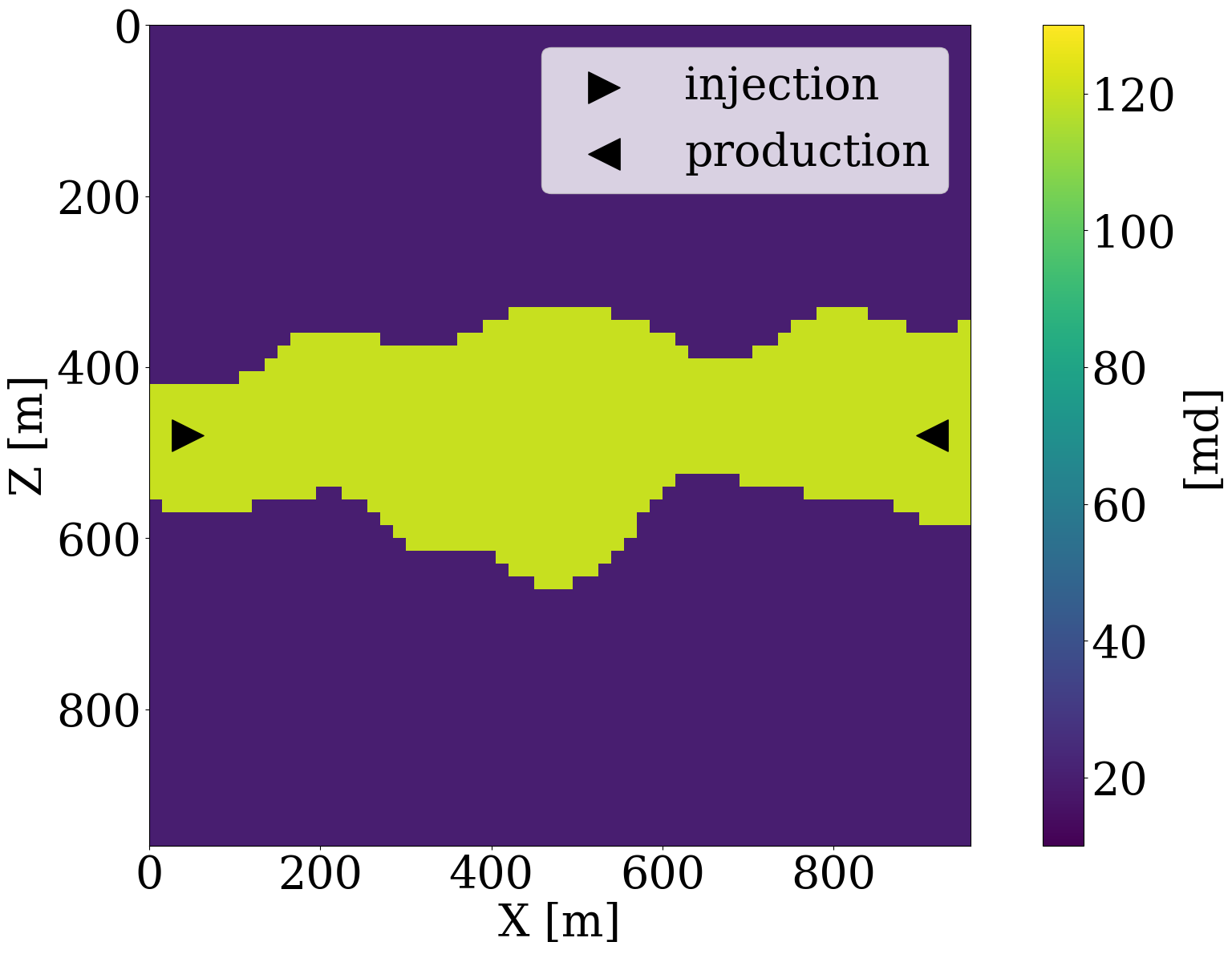}

}

}

\subcaption{\label{fig-true-perm}}
\end{minipage}%
\begin{minipage}[t]{0.50\linewidth}

{\centering 

\raisebox{-\height}{

\includegraphics{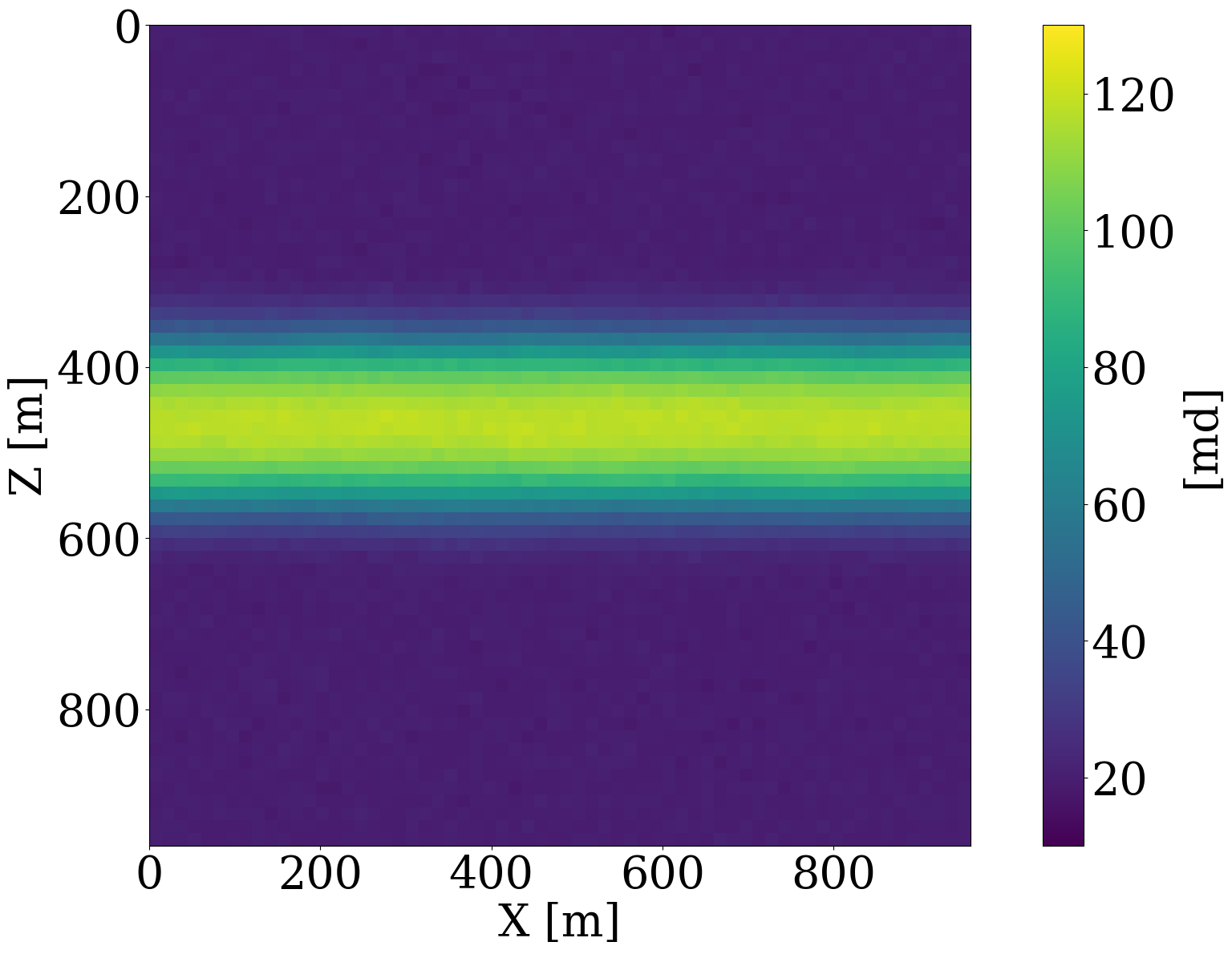}

}

}

\subcaption{\label{fig-init-perm}}
\end{minipage}%

\caption{\label{fig-perm-init-true}Permeability models. \emph{(a)}
unknown ``ground-truth'' permeability model from unseen test set, where
the symbols \(\blacktriangleright\) and \(\blacktriangleleft\) denote
the CO\textsubscript{2} injection and brine production location,
respectively; \emph{(b)} initial permeability model, \(\mathbf{K}_0\).}

\end{figure}

For all numerical experiments, the ``ground-truth'' permeability model
will be selected from the unseen test set and is shown in
Figure~\ref{fig-true-perm}. The inversions will be initiated with the
smooth permeability model depicted in Figure~\ref{fig-init-perm}. This
initial model, \(\mathbf{K}_0\), represents the arithmetic mean of all
permeability samples in the training dataset. To ensure that the model
iterates remain in distribution, we set the starting \(\ell_2\)-norm
ball size to
\(\tau_{\mathrm{init}}=0.6 \|\mathcal{N}(0,\mathbf{I})\|_2\)---i.e.,
\(0.6\times\) the \(\ell_2\)-norm of standard white Gauss noise
realizations for the discrete permeability model of 64 by 64 gridpoints.
To gradually relax the learned constraint, the multiplier of the
projection ball size is taken to be \(\beta=1.2\), and we set the
ultimate projection ball size \(\tau_{\mathrm{final}}\) in Algorithm 1
to be \(1.2\) times the norm of standard white noise. To limit
computational costs of solving the subproblems, we allow each
constrained subproblem (cf.~line 8 to 11 in Algorithm 1) to perform 8
iterations of projected gradient descent to solve for the latent
variable. From practical experience, we found that the proposed
inversions are not very sensitive to the choice of these
hyperparameters.

To simulate the evolution of injected CO\textsubscript{2} plumes, we
make use of the open-source software package
\href{https://github.com/sintefmath/Jutul.jl}{Jutul.jl} (Møyner et al.
2023; Møyner, Bruer, and Yin 2023; Yin, Bruer, and Louboutin 2023),
which for each permeability model, \(\mathbf{K}^{(j)}\), solves the
immiscible and compressible two-phase flow equations for the
CO\textsubscript{2} and brine saturation. As shown in
Figure~\ref{fig-true-perm}, an injection well is set up on the left-hand
side of the model, which injects supercritical CO\textsubscript{2} with
density 700 \(\mathrm{kg}/\mathrm{m}^3\) at a constant rate of 0.005
\(\mathrm{m}^3/\mathrm{s}\). To relieve pressure, a production well is
included on the right-hand side of the model, which produces brine with
density 1000 \(\mathrm{kg}/\mathrm{m}^3\) with a constant rate of also
0.005 \(\mathrm{m}^3/\mathrm{s}\). This finally results in approximately
a \(6\%\) storage capacity after 800 days of CO\textsubscript{2}
injection. From these simulations, we collect eight snapshots for the
CO\textsubscript{2} concentration,
\(\mathbf{c}=[\mathbf{c}_1, \mathbf{c}_2, \cdots,\mathbf{c}_{n_t}]\)
with \(n_t=8\) the number of snapshots that cover a total time period of
800 days. The last five snapshots of these simulations are included in
the top row of Figure~\ref{fig-true-perm}. Due to buoyancy effects and
well control, the CO\textsubscript{2} plume gradually moves from the
left to the right and upwards.

\begin{figure}

\begin{minipage}[t]{0.20\linewidth}

{\centering 

\raisebox{-\height}{

\includegraphics{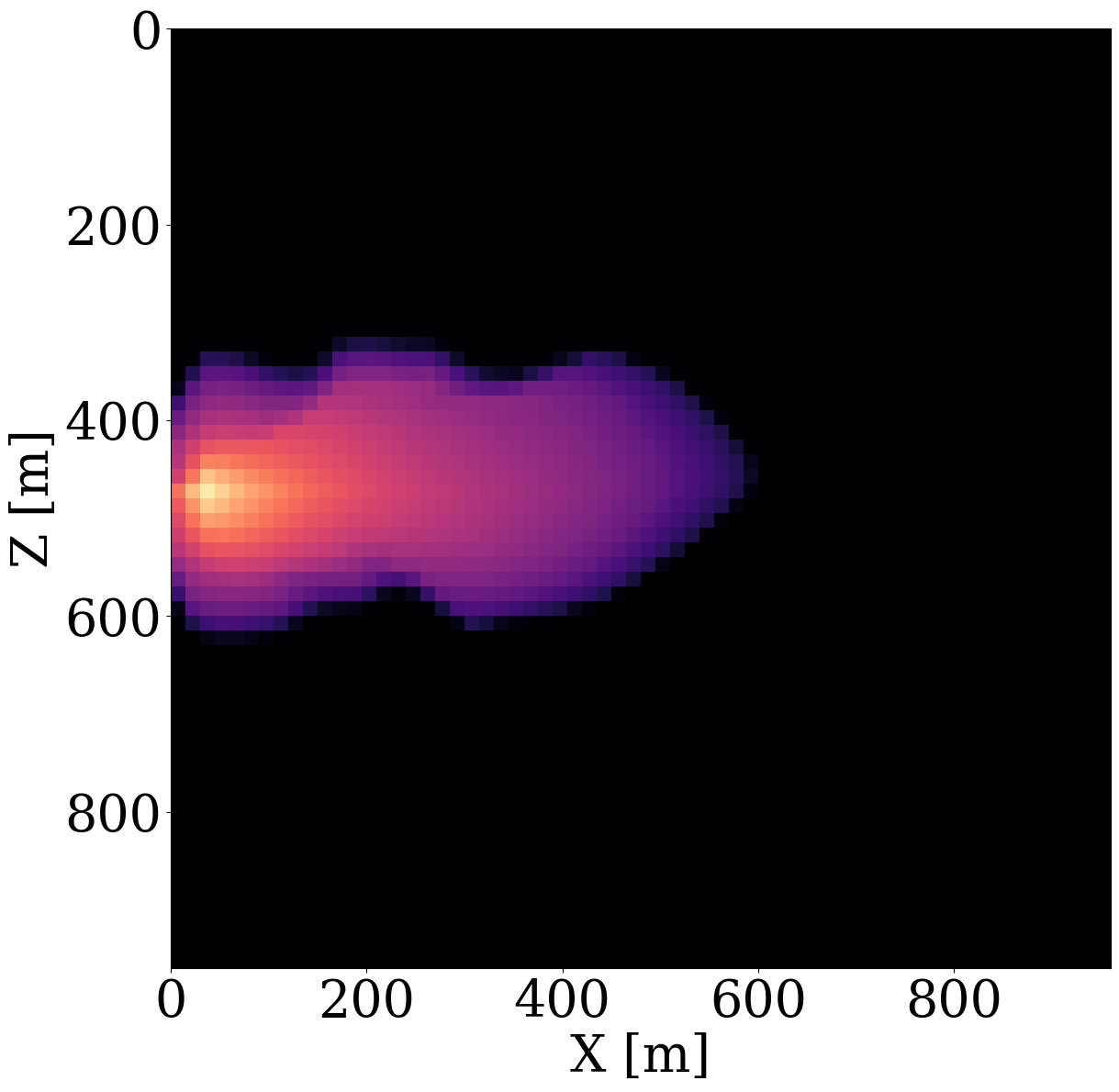}

}

}

\subcaption{\label{fig-co2-true-1}}
\end{minipage}%
\begin{minipage}[t]{0.20\linewidth}

{\centering 

\raisebox{-\height}{

\includegraphics{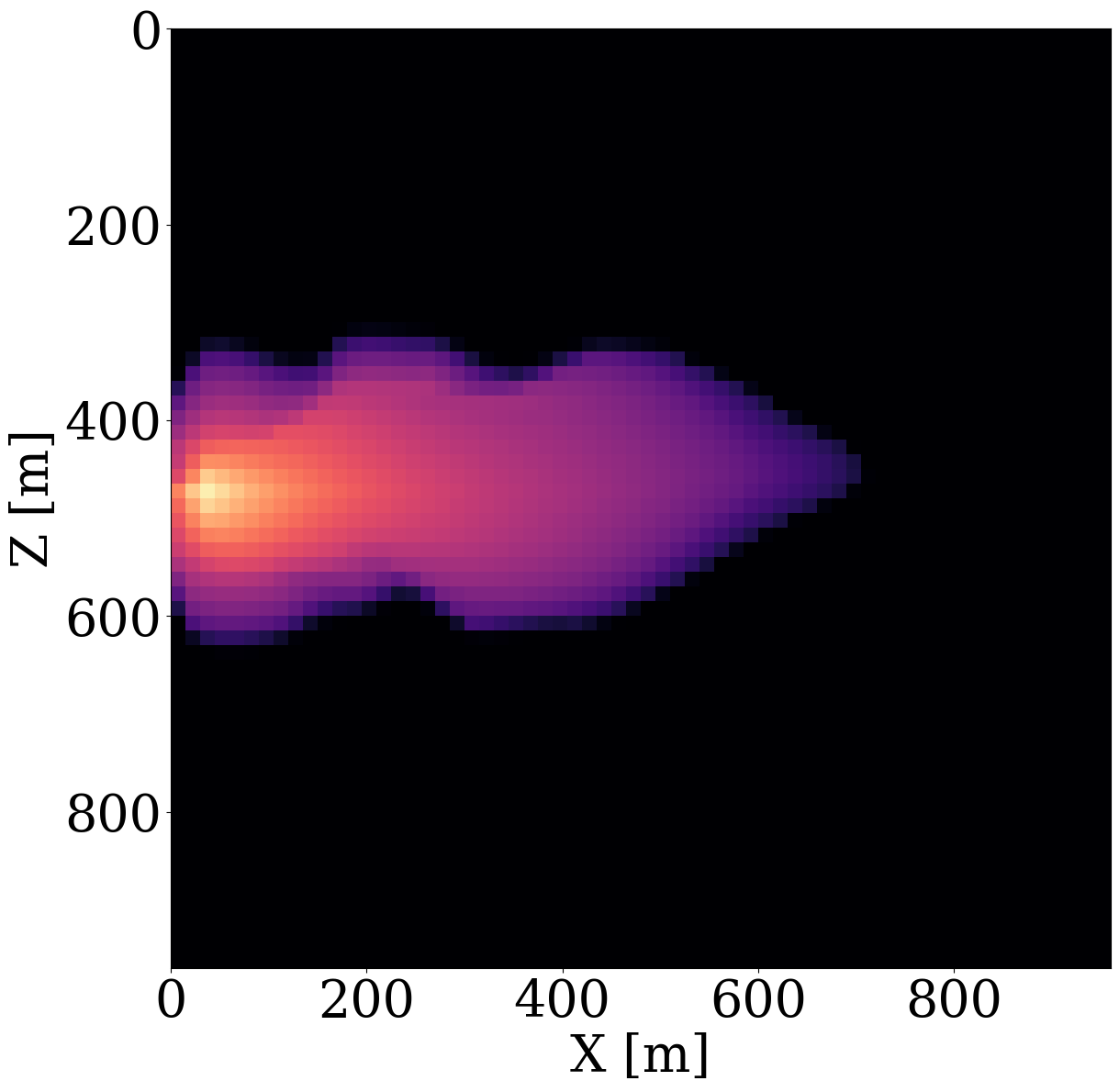}

}

}

\subcaption{\label{fig-co2-true-2}}
\end{minipage}%
\begin{minipage}[t]{0.20\linewidth}

{\centering 

\raisebox{-\height}{

\includegraphics{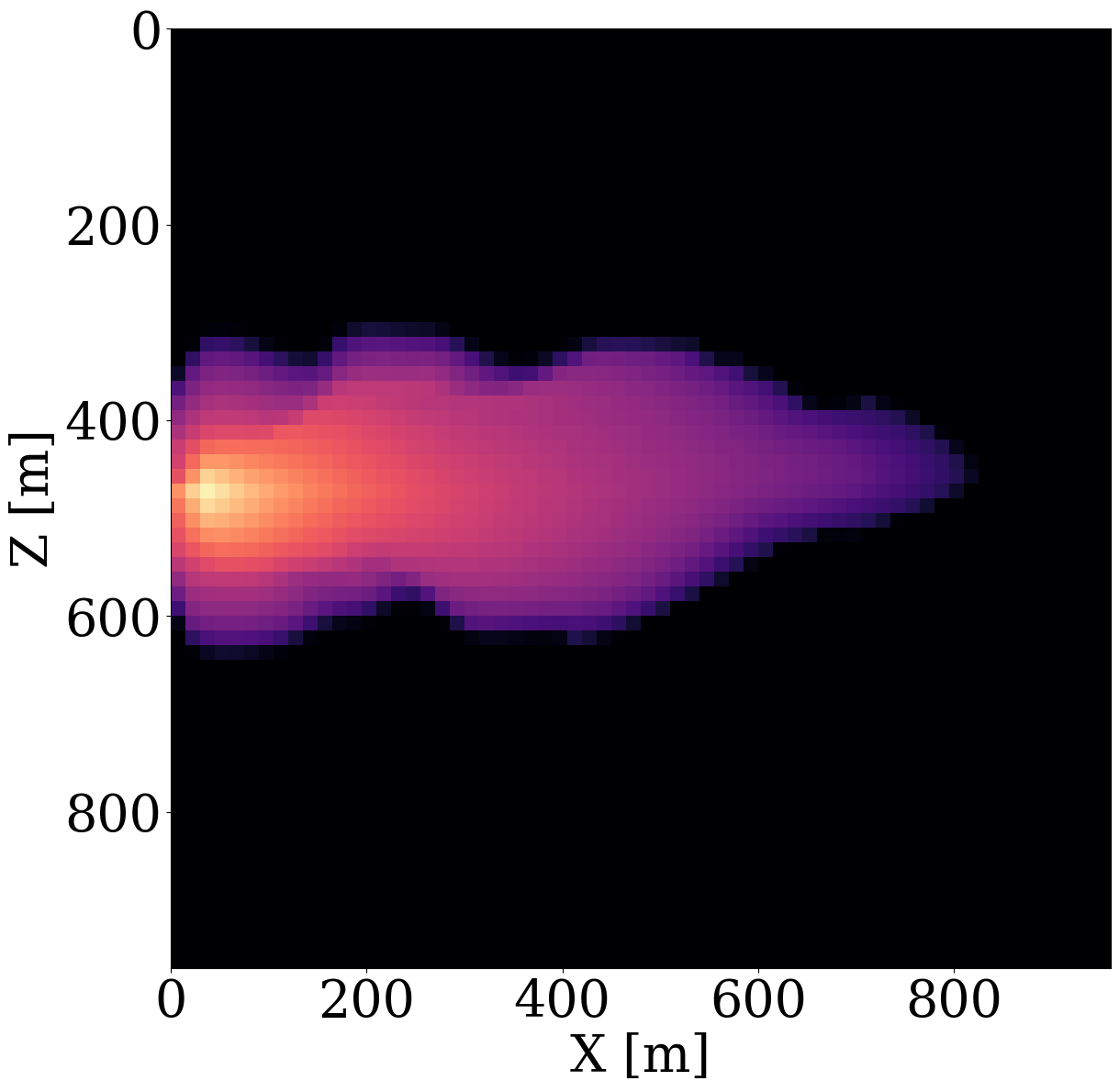}

}

}

\subcaption{\label{fig-co2-true-3}}
\end{minipage}%
\begin{minipage}[t]{0.20\linewidth}

{\centering 

\raisebox{-\height}{

\includegraphics{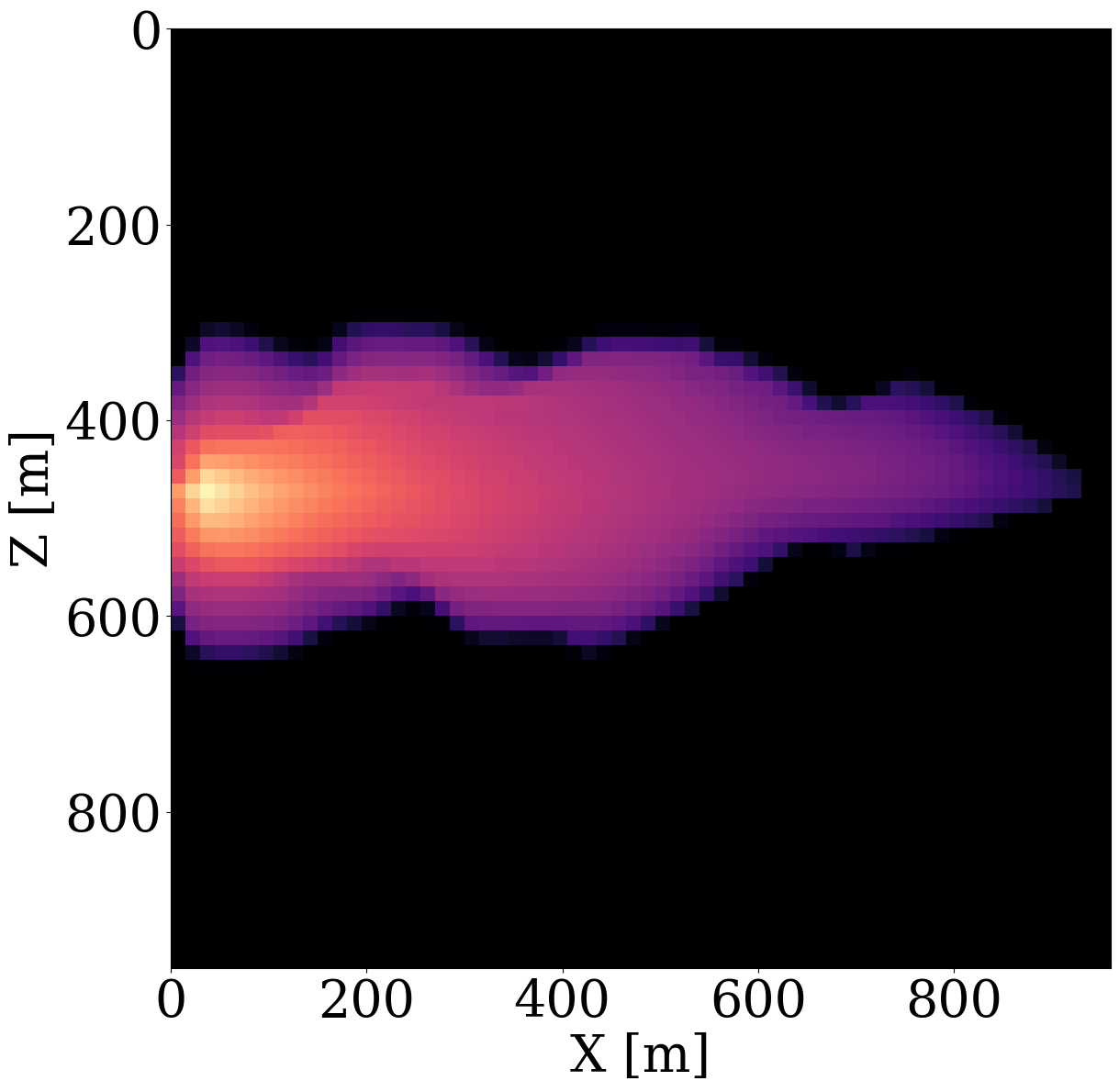}

}

}

\subcaption{\label{fig-co2-true-4}}
\end{minipage}%
\begin{minipage}[t]{0.20\linewidth}

{\centering 

\raisebox{-\height}{

\includegraphics{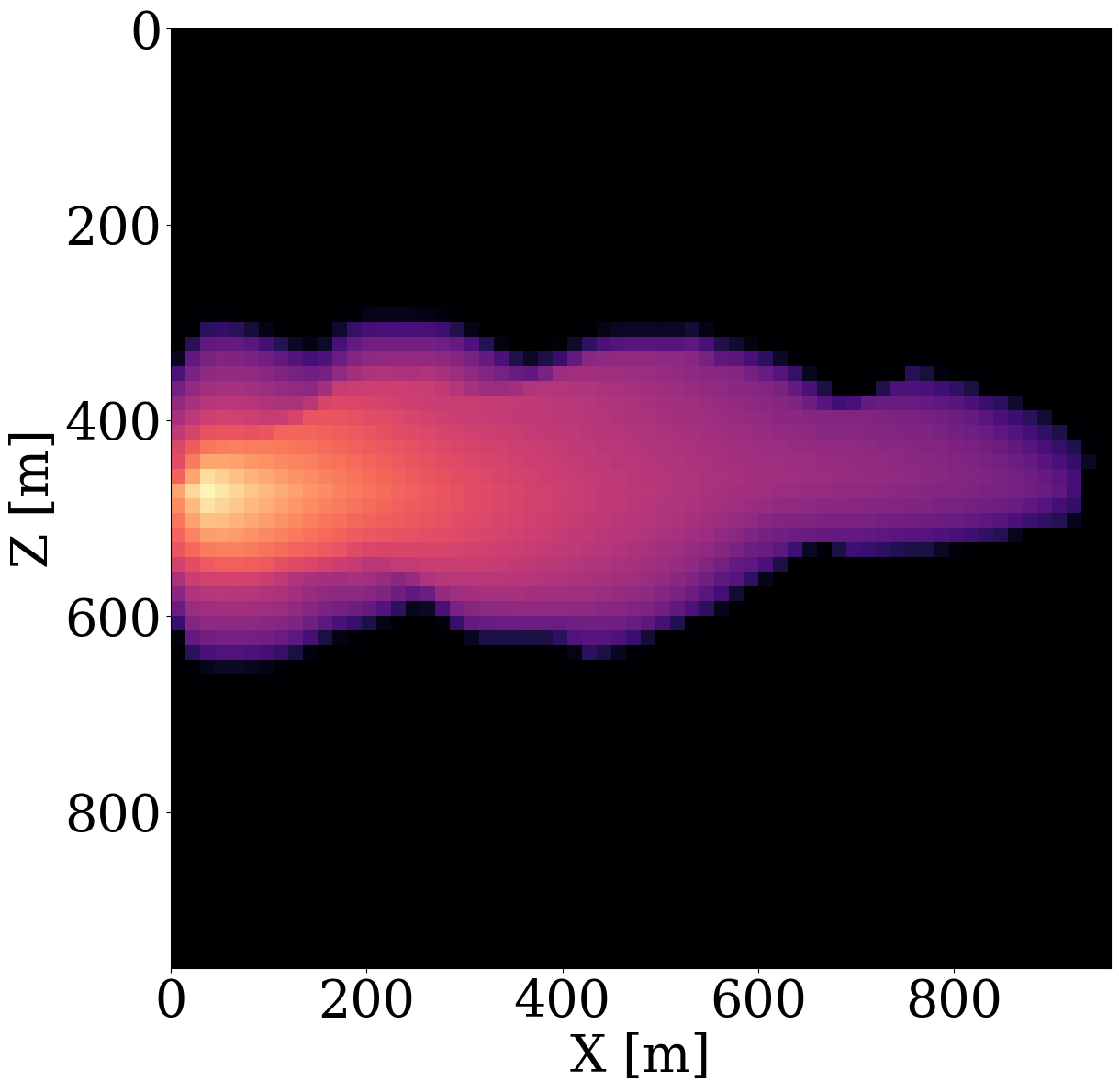}

}

}

\subcaption{\label{fig-co2-true-5}}
\end{minipage}%
\newline
\begin{minipage}[t]{0.20\linewidth}

{\centering 

\raisebox{-\height}{

\includegraphics{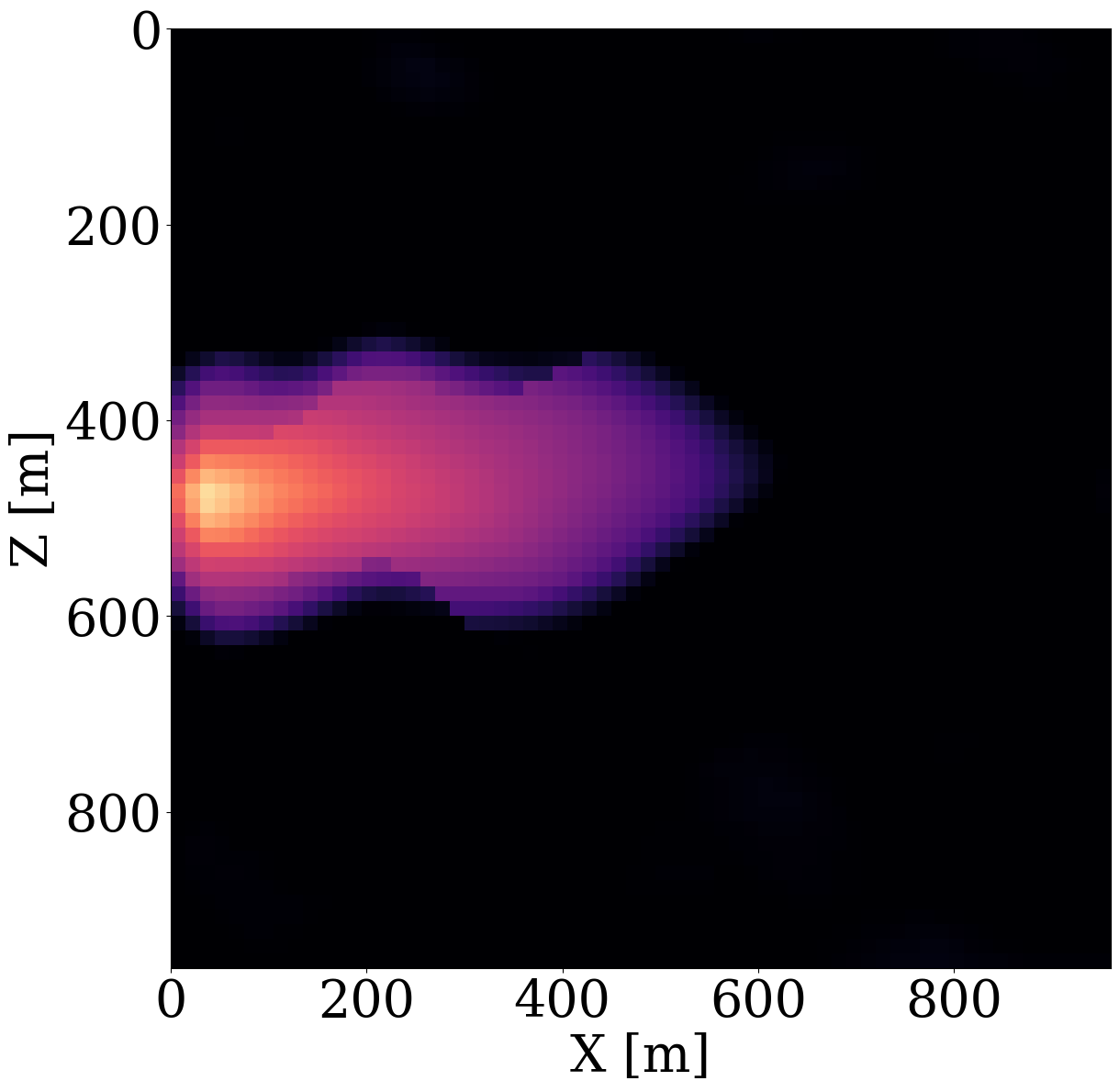}

}

}

\subcaption{\label{fig-co2-fno-1}}
\end{minipage}%
\begin{minipage}[t]{0.20\linewidth}

{\centering 

\raisebox{-\height}{

\includegraphics{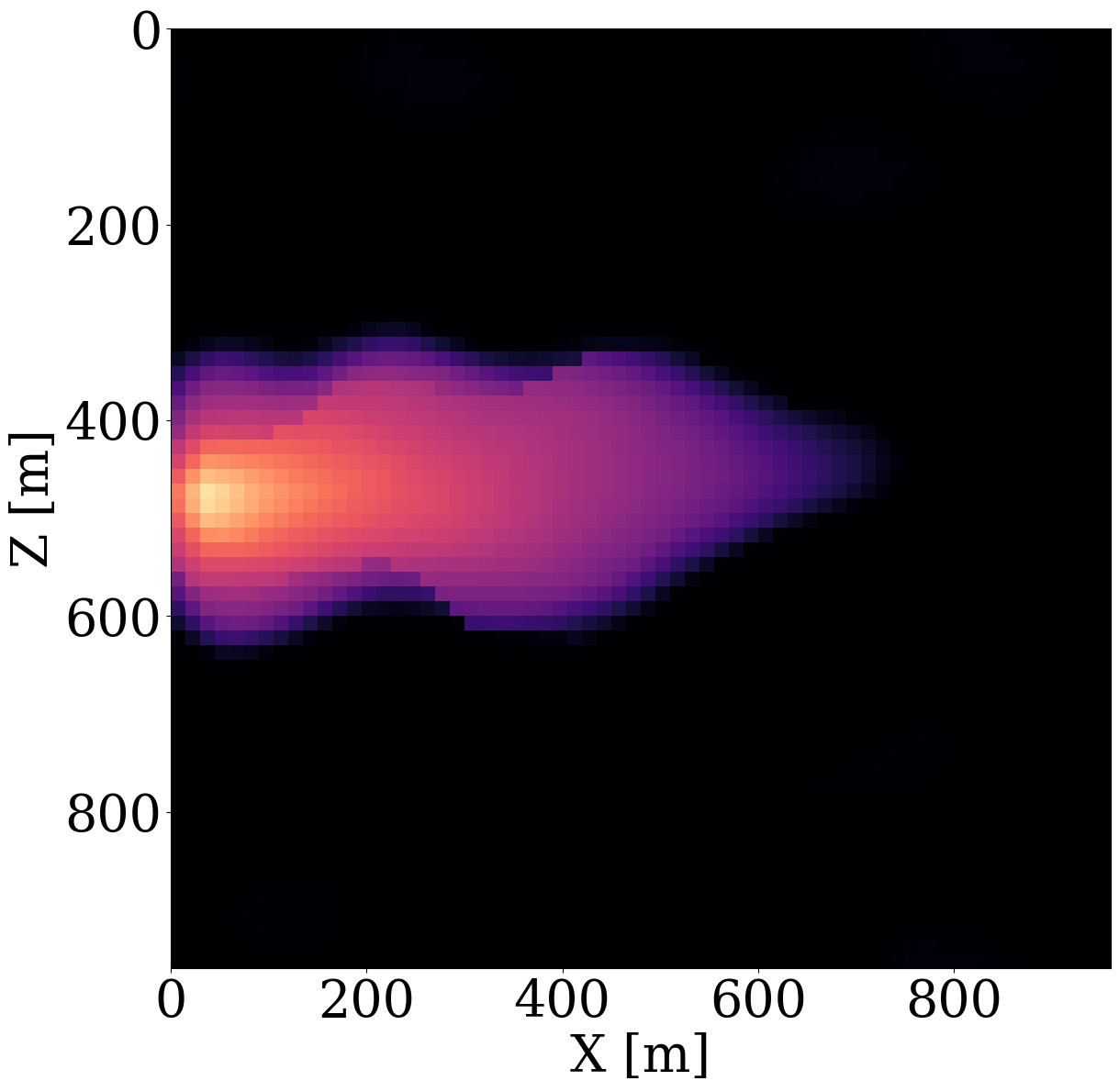}

}

}

\subcaption{\label{fig-co2-fno-2}}
\end{minipage}%
\begin{minipage}[t]{0.20\linewidth}

{\centering 

\raisebox{-\height}{

\includegraphics{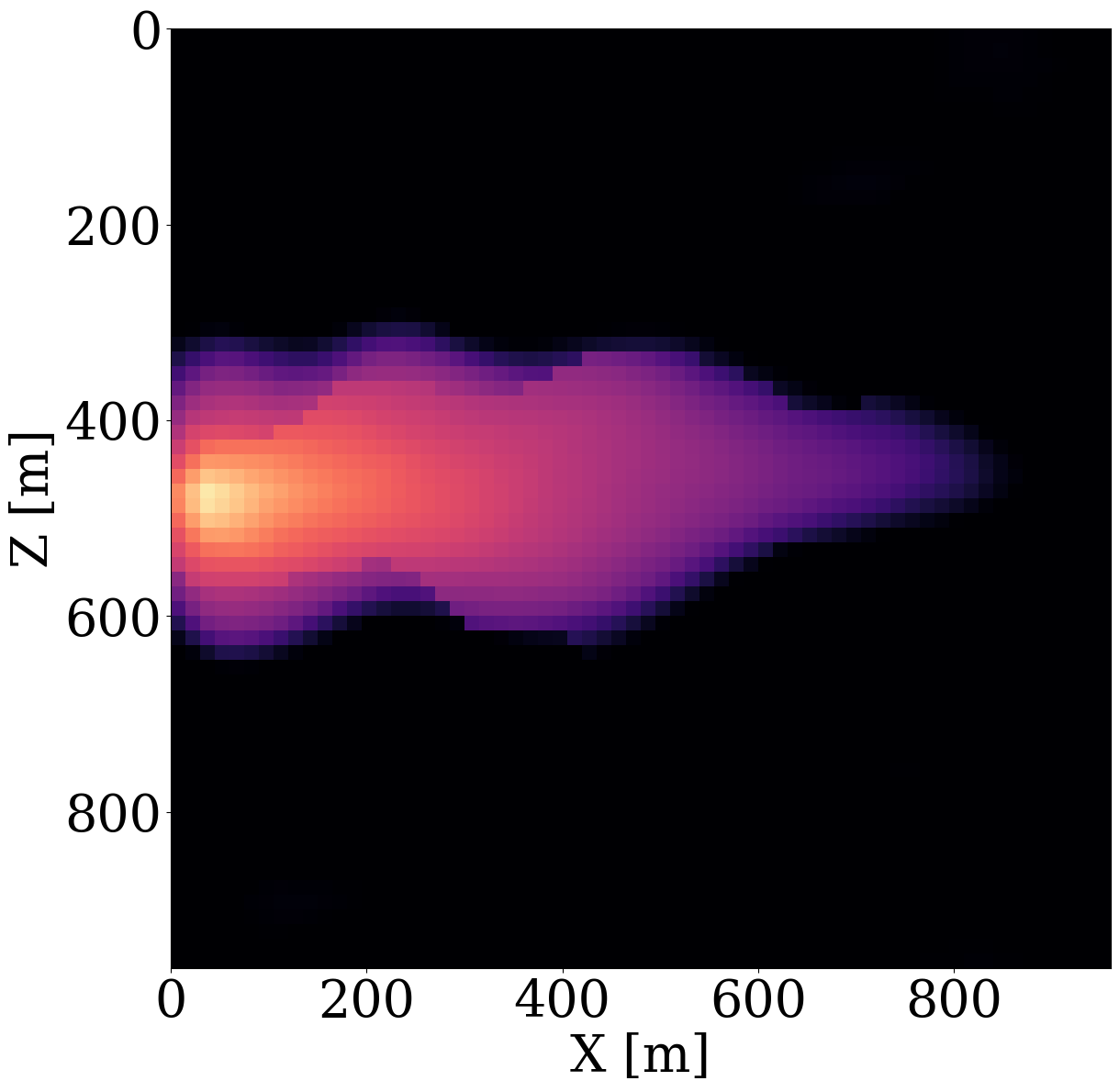}

}

}

\subcaption{\label{fig-co2-fno-3}}
\end{minipage}%
\begin{minipage}[t]{0.20\linewidth}

{\centering 

\raisebox{-\height}{

\includegraphics{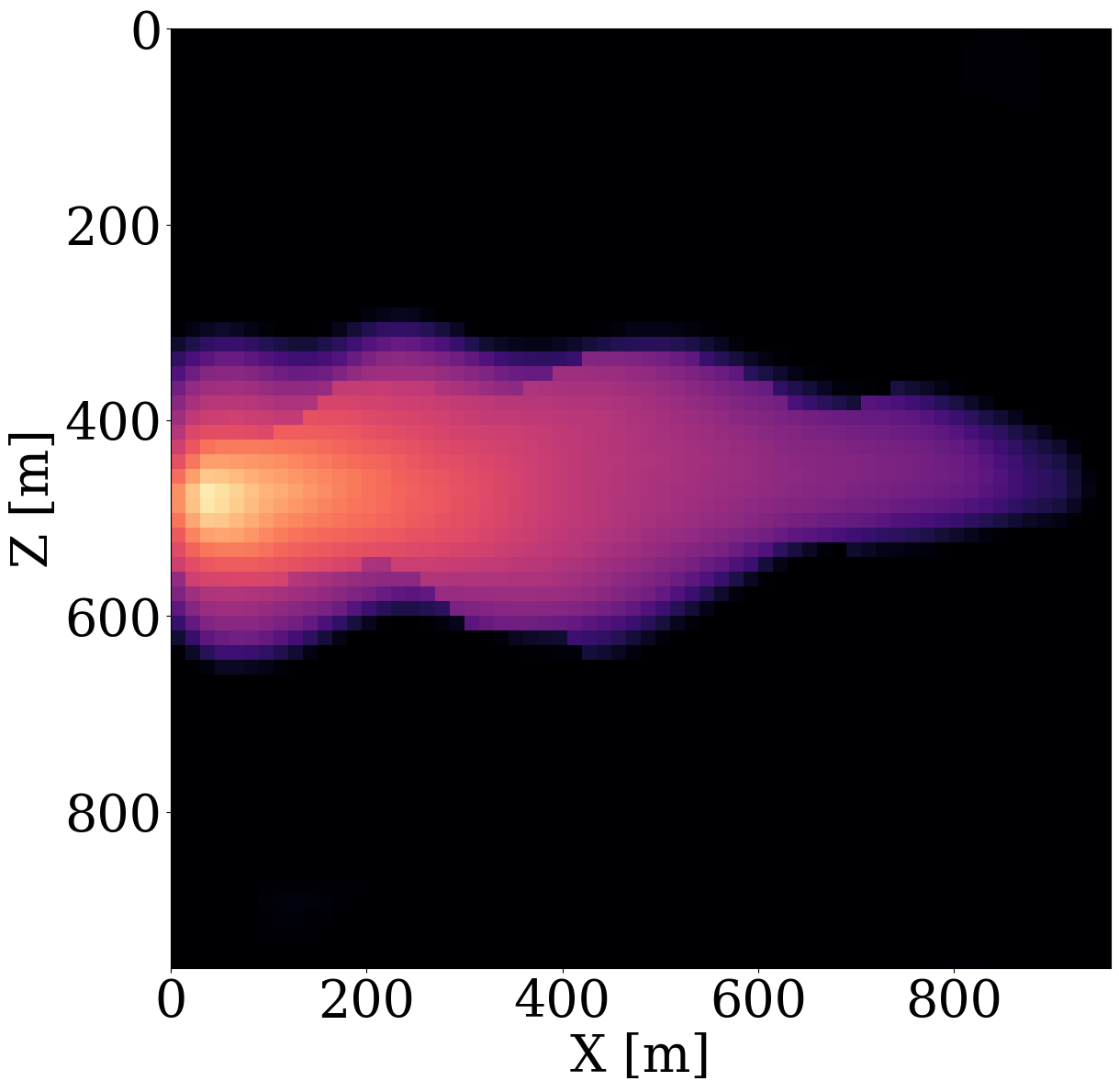}

}

}

\subcaption{\label{fig-co2-fno-4}}
\end{minipage}%
\begin{minipage}[t]{0.20\linewidth}

{\centering 

\raisebox{-\height}{

\includegraphics{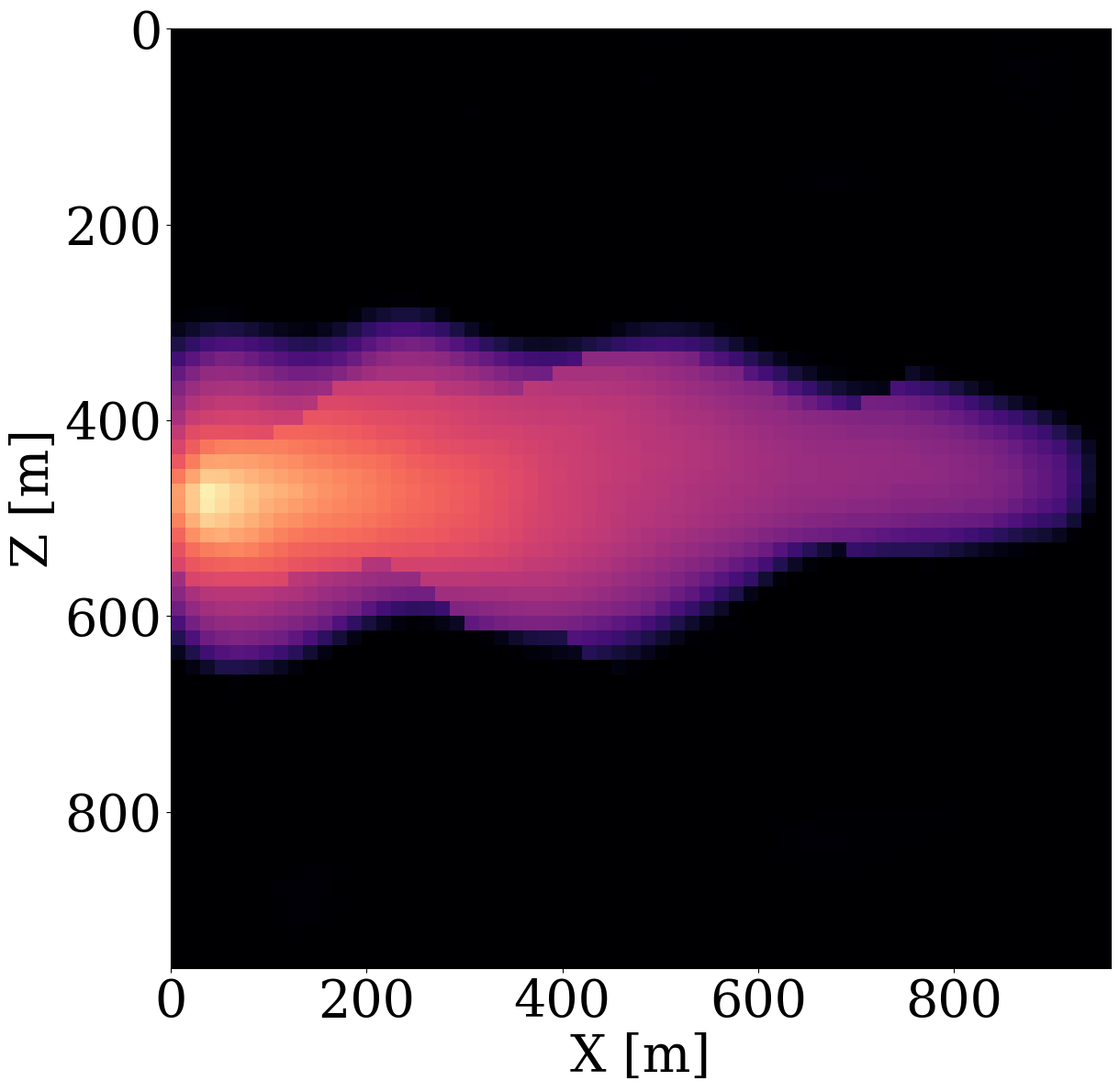}

}

}

\subcaption{\label{fig-co2-fno-5}}
\end{minipage}%
\newline
\begin{minipage}[t]{0.20\linewidth}

{\centering 

\raisebox{-\height}{

\includegraphics{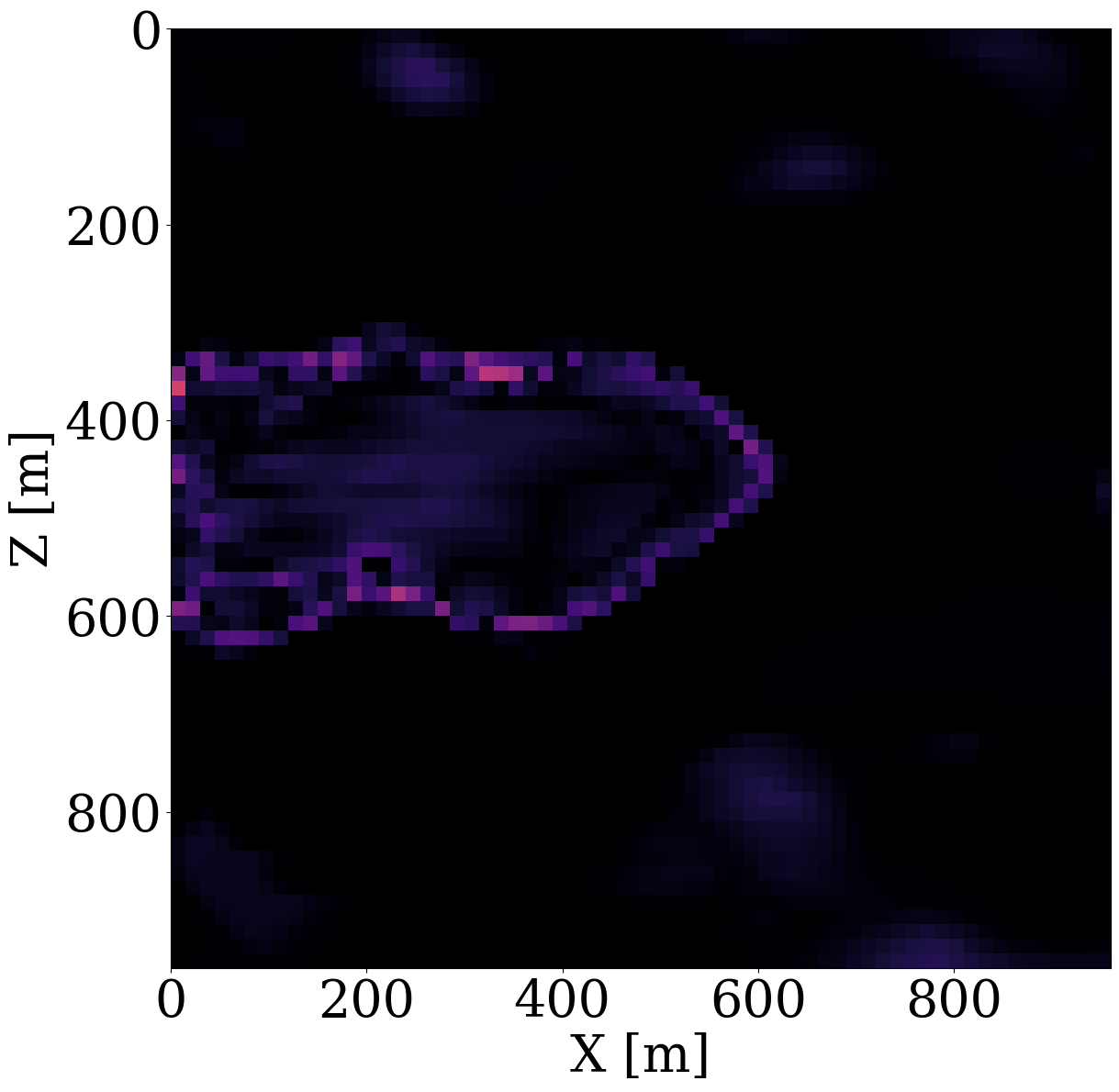}

}

}

\subcaption{\label{fig-co2-diff-1}}
\end{minipage}%
\begin{minipage}[t]{0.20\linewidth}

{\centering 

\raisebox{-\height}{

\includegraphics{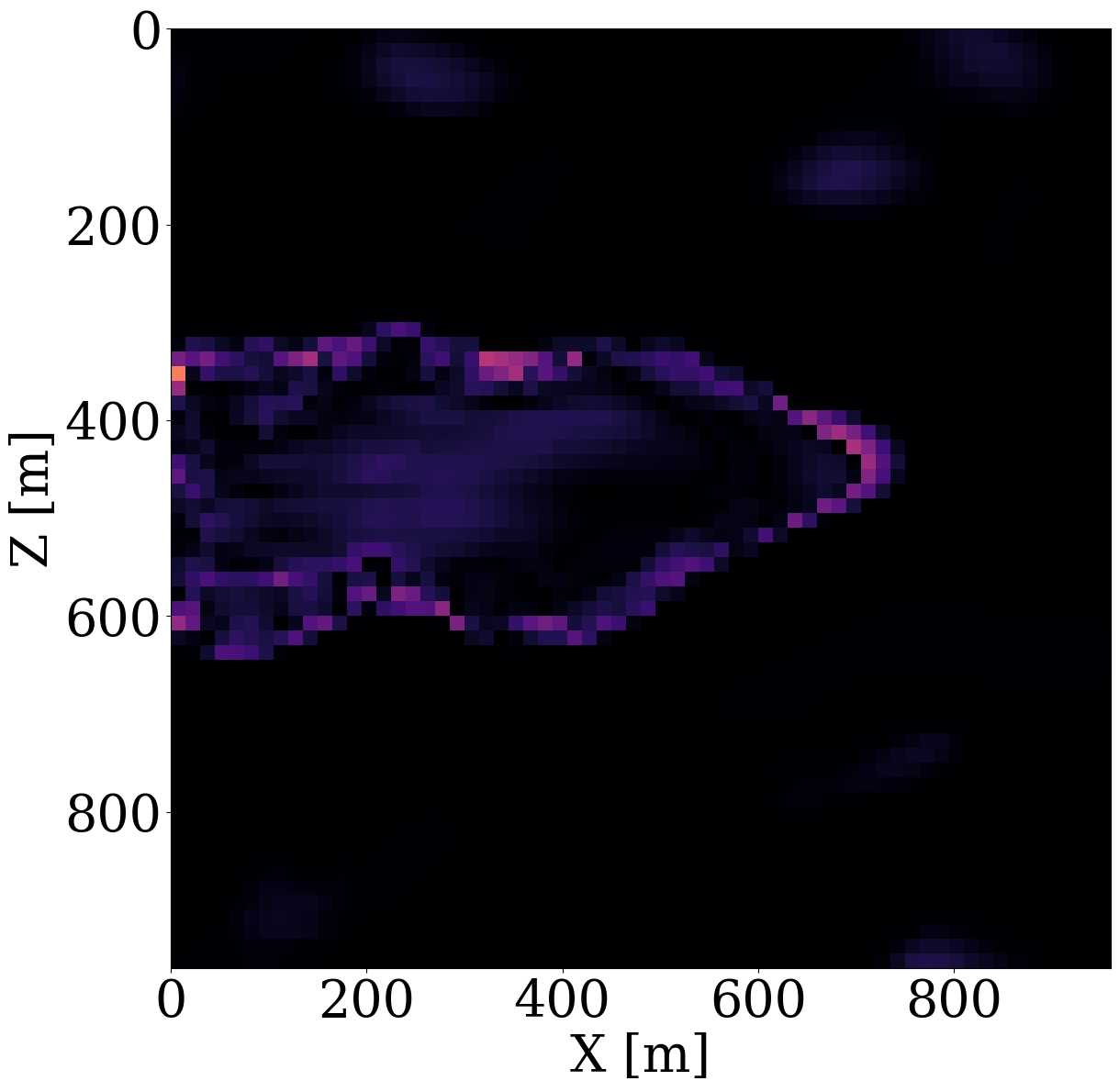}

}

}

\subcaption{\label{fig-co2-diff-2}}
\end{minipage}%
\begin{minipage}[t]{0.20\linewidth}

{\centering 

\raisebox{-\height}{

\includegraphics{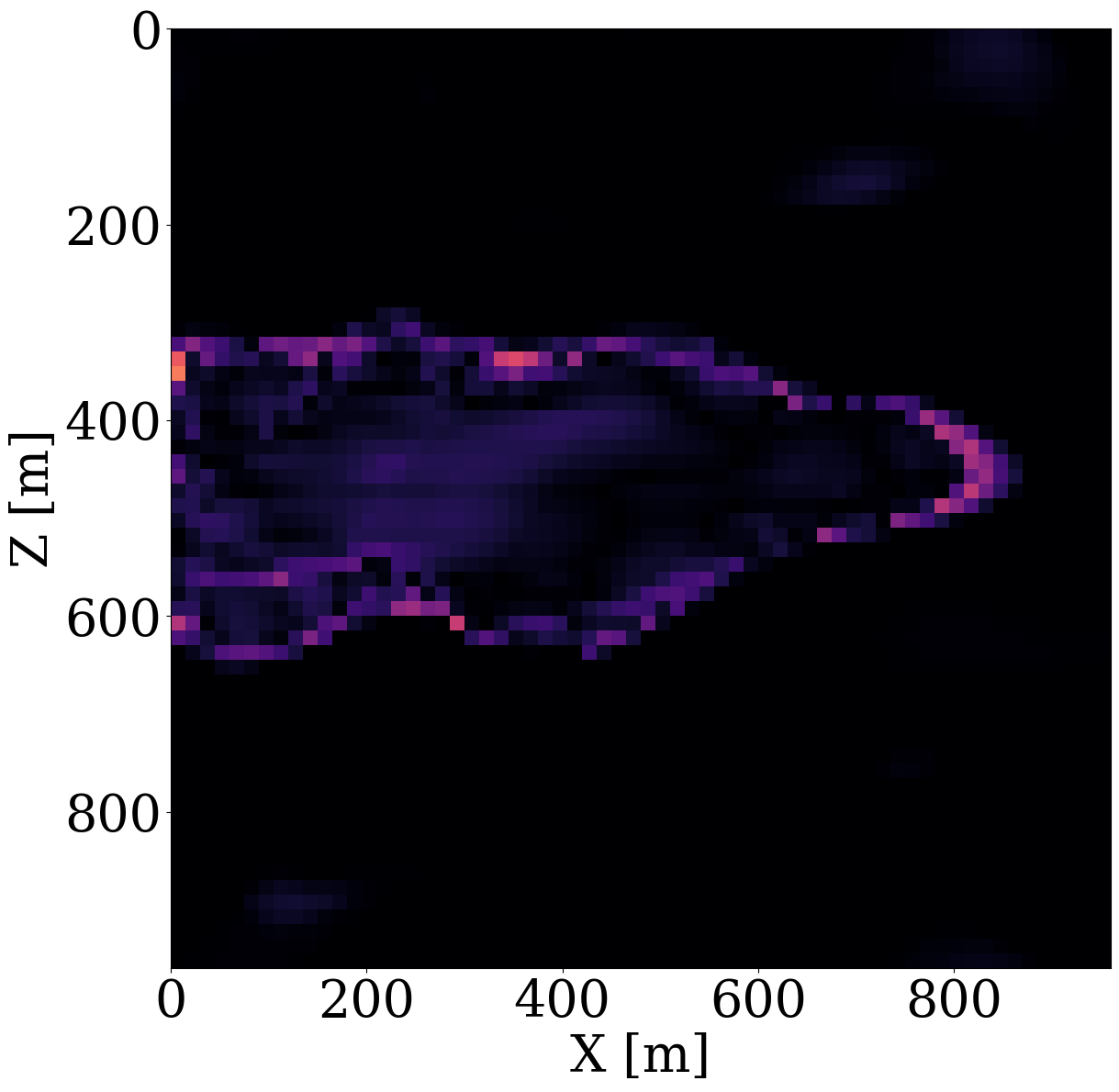}

}

}

\subcaption{\label{fig-co2-diff-3}}
\end{minipage}%
\begin{minipage}[t]{0.20\linewidth}

{\centering 

\raisebox{-\height}{

\includegraphics{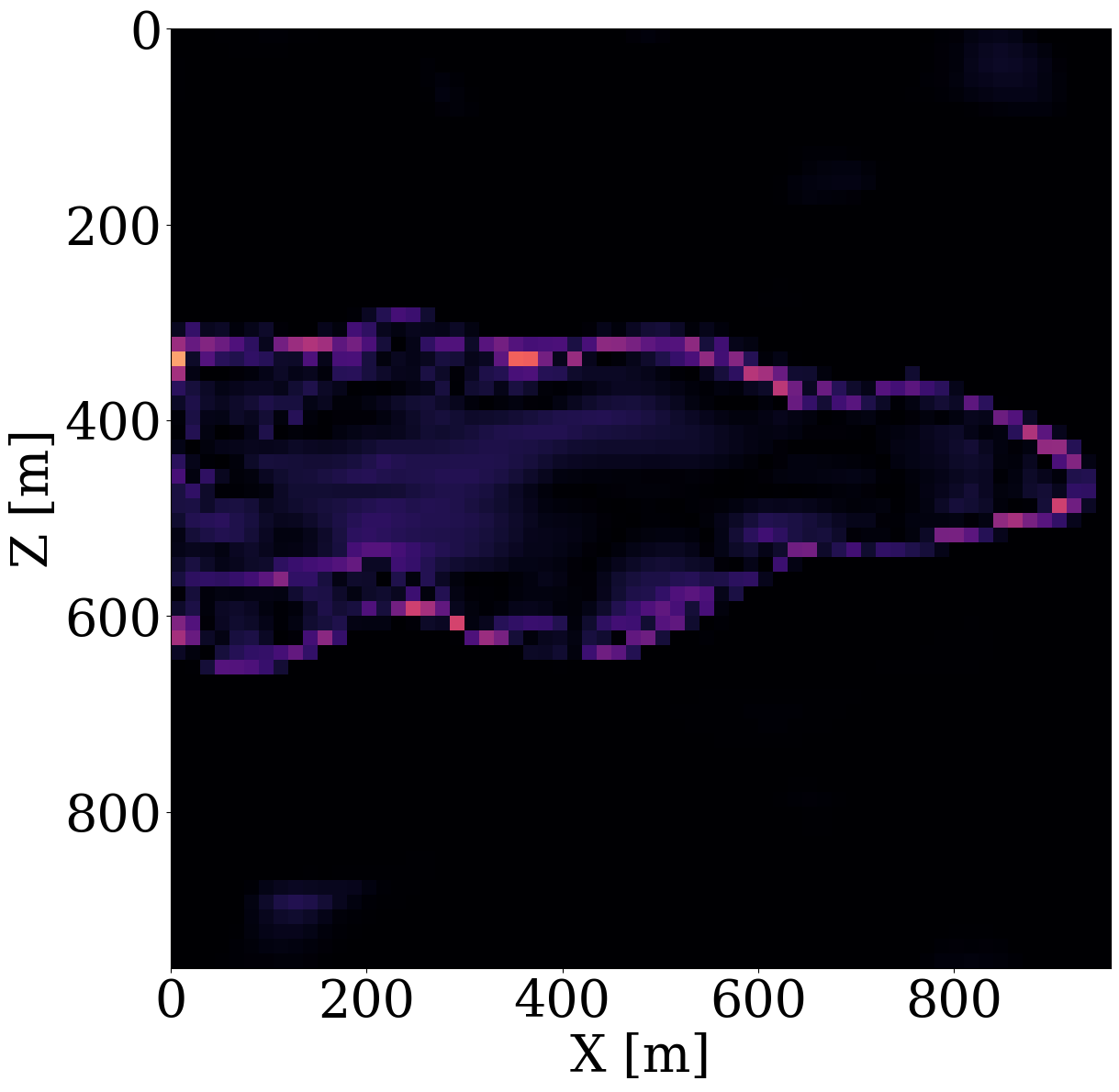}

}

}

\subcaption{\label{fig-co2-diff-4}}
\end{minipage}%
\begin{minipage}[t]{0.20\linewidth}

{\centering 

\raisebox{-\height}{

\includegraphics{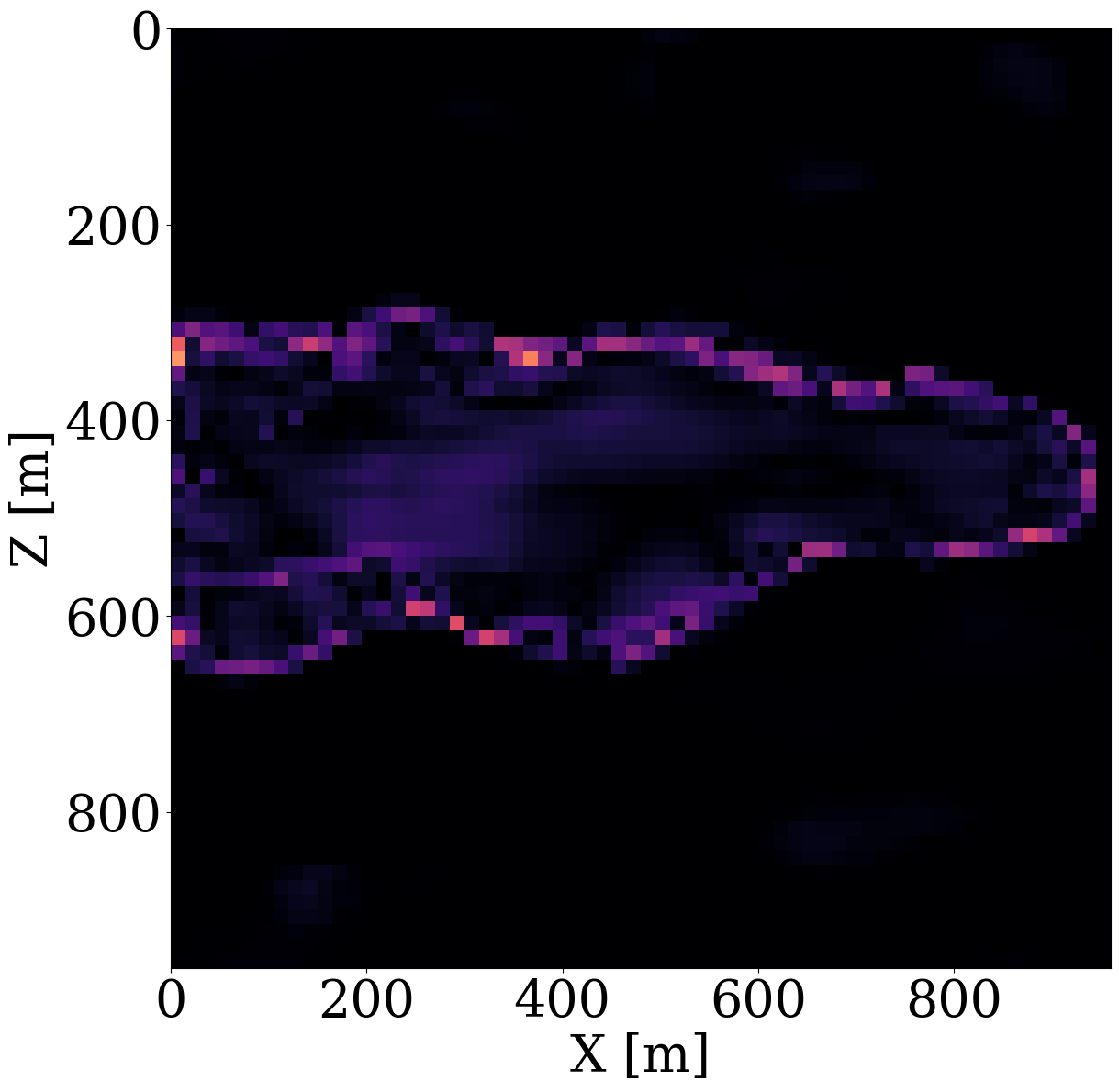}

}

}

\subcaption{\label{fig-co2-diff-5}}
\end{minipage}%

\caption{\label{fig-fno-train-plume}Five CO\textsubscript{2} saturation
snapshots after 400, 500, 600, 700, and 800 days. First row shows the
CO\textsubscript{2} saturation simulated by the PDE. Second row shows
the CO\textsubscript{2} saturation predicted by the trained FNO. Third
row shows the \(5\times\) difference between the first row and the
second row.}

\end{figure}

Given these simulated CO\textsubscript{2} concentrations, the optimized
weights, \(\mathbf{w}^\ast\), for the FNO surrogate are calculated by
minimizing Equation~\ref{eq-fno-train} for \(N=1900\) training pairs,
\(\{\mathbf{K}^{(j)},\, \mathbf{c}^{(j)}\}_{j=1}^N\). Another 100
training pairs are used for validation. After training with 350 epochs,
an average of 7\% prediction error is achieved for permeability samples
from the unseen test set. As observed from
Figure~\ref{fig-fno-train-plume}, the approximation errors of the FNO
are mostly concentrated at the leading edge of the CO\textsubscript{2}
plumes. The same permeability models are used to train the NF by
minimizing Equation~\ref{eq-train-nf} for 245 epochs using the
open-source software package
\href{https://github.com/slimgroup/InvertibleNetworks.jl}{InvertibleNetworks.jl}
(P. Witte et al. 2023). We use the HINT network structure (Kruse et al.
2021) for the NF. Three generative samples are shown in the second row
of Figure~\ref{fig-trainingsample}. From these examples, we can see that
the trained NF is capable of generating random permeability models that
resemble the ones in the training samples closely, despite minor noisy
artifacts.

\hypertarget{unconstrainedconstrained-permeability-inversion-from-co2-saturation-data}{%
\subsubsection{\texorpdfstring{Unconstrained/constrained permeability
inversion from CO\textsubscript{2} saturation
data}{Unconstrained/constrained permeability inversion from CO2 saturation data}}\label{unconstrainedconstrained-permeability-inversion-from-co2-saturation-data}}

\begin{figure}

\begin{minipage}[t]{0.50\linewidth}

{\centering 

\raisebox{-\height}{

\includegraphics{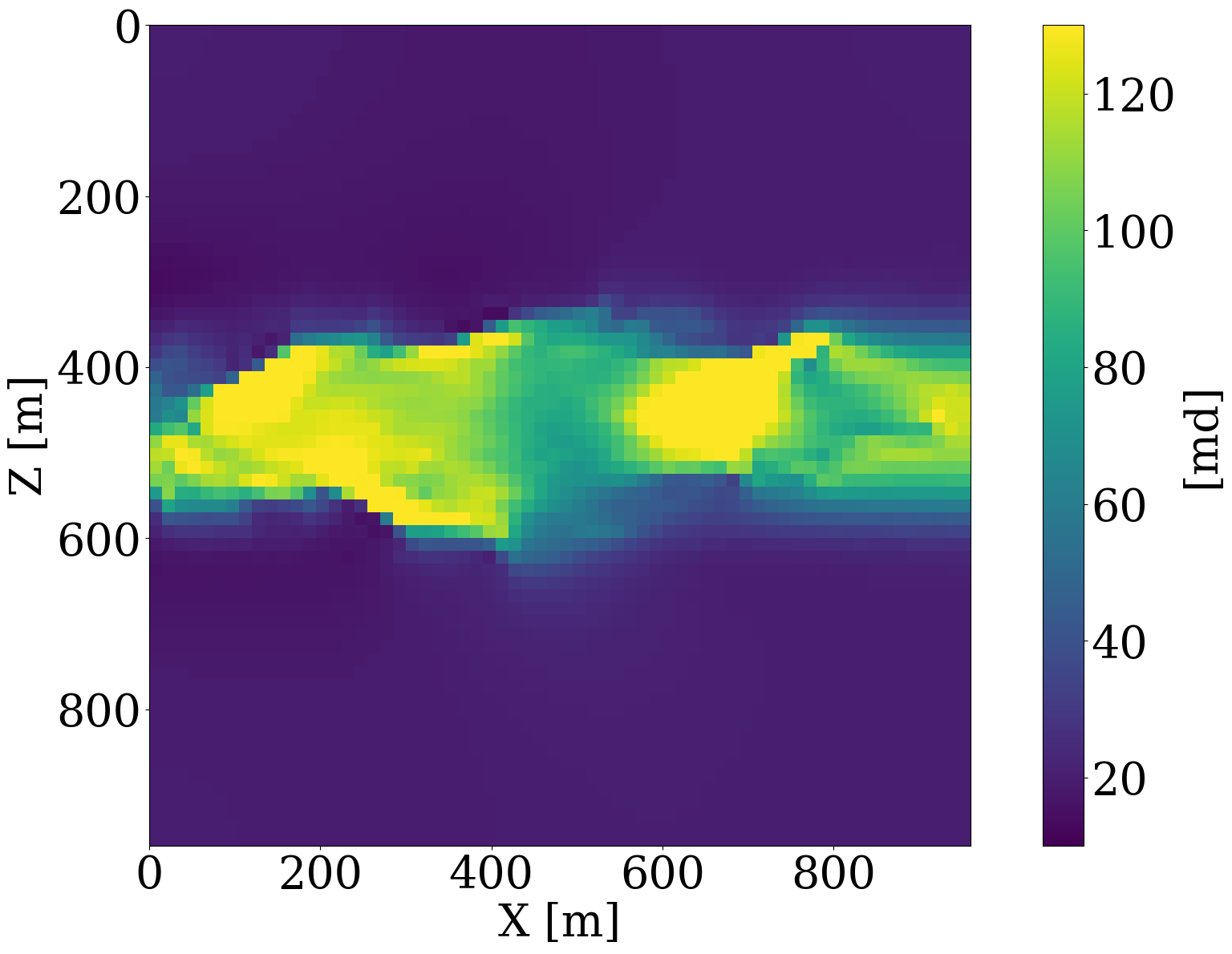}

}

}

\subcaption{\label{fig-flow-inv-jutul-all}}
\end{minipage}%
\begin{minipage}[t]{0.50\linewidth}

{\centering 

\raisebox{-\height}{

\includegraphics{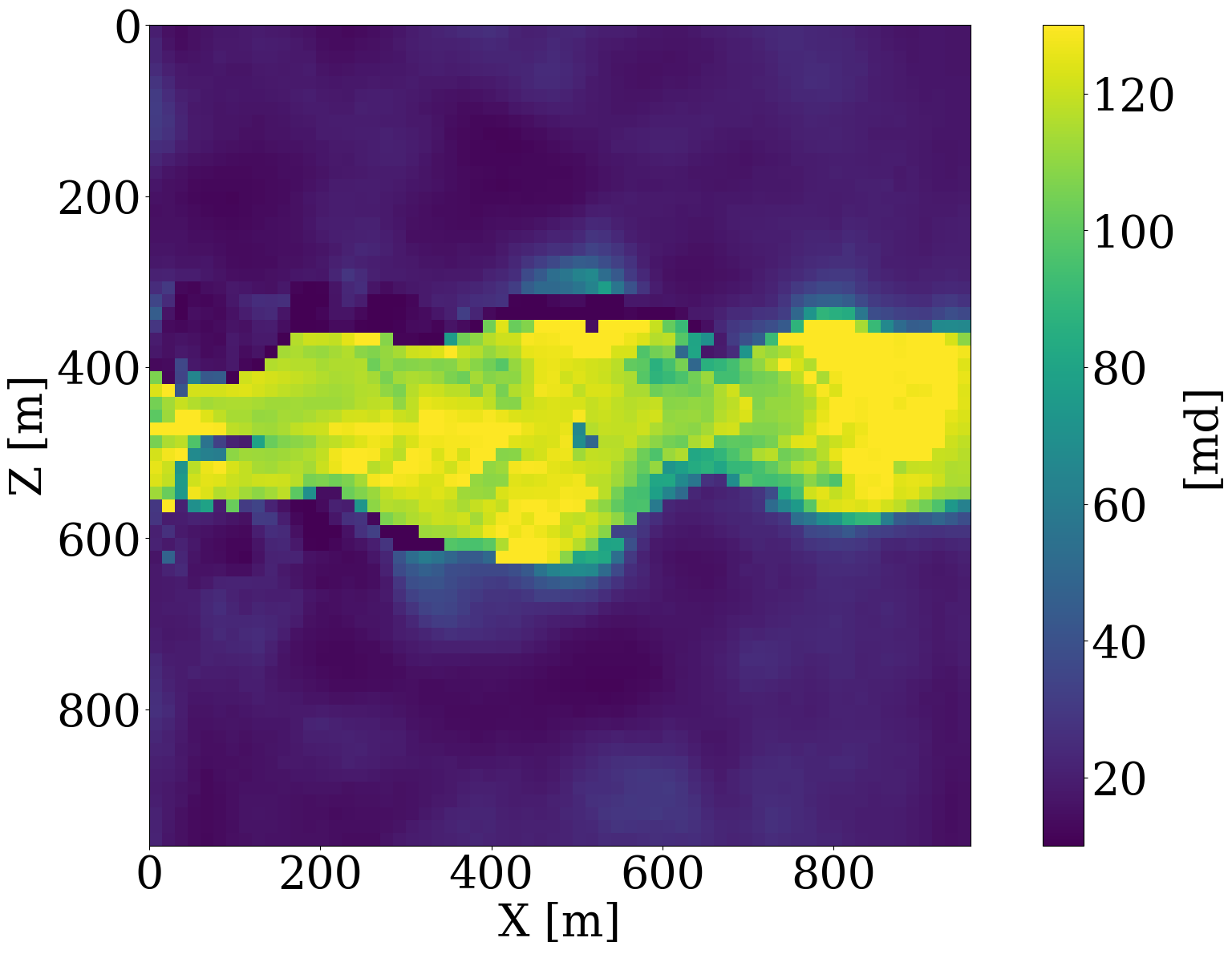}

}

}

\subcaption{\label{fig-flow-inv-fno-all}}
\end{minipage}%
\newline
\begin{minipage}[t]{0.50\linewidth}

{\centering 

\raisebox{-\height}{

\includegraphics{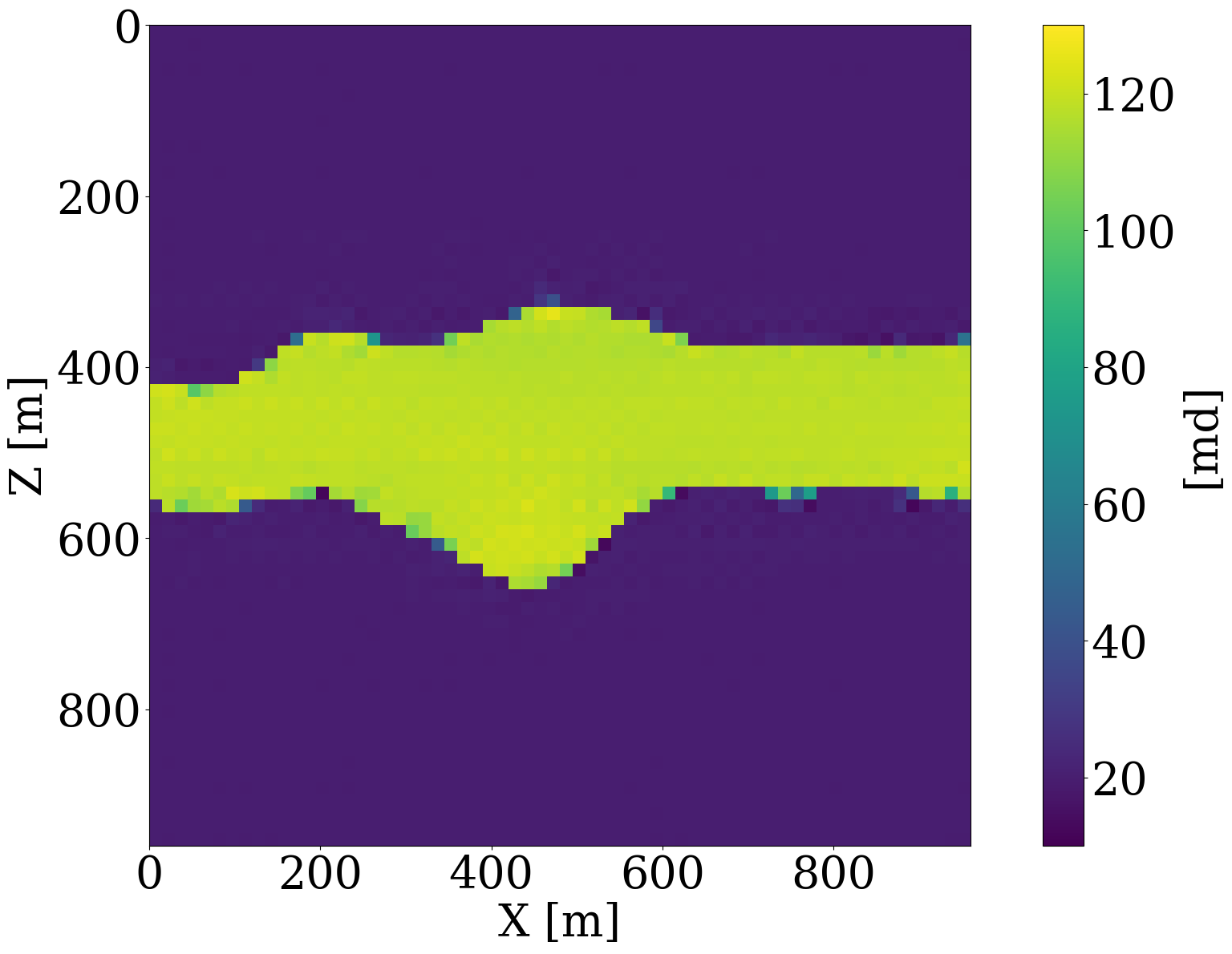}

}

}

\subcaption{\label{fig-flow-inv-jutul-nf-all}}
\end{minipage}%
\begin{minipage}[t]{0.50\linewidth}

{\centering 

\raisebox{-\height}{

\includegraphics{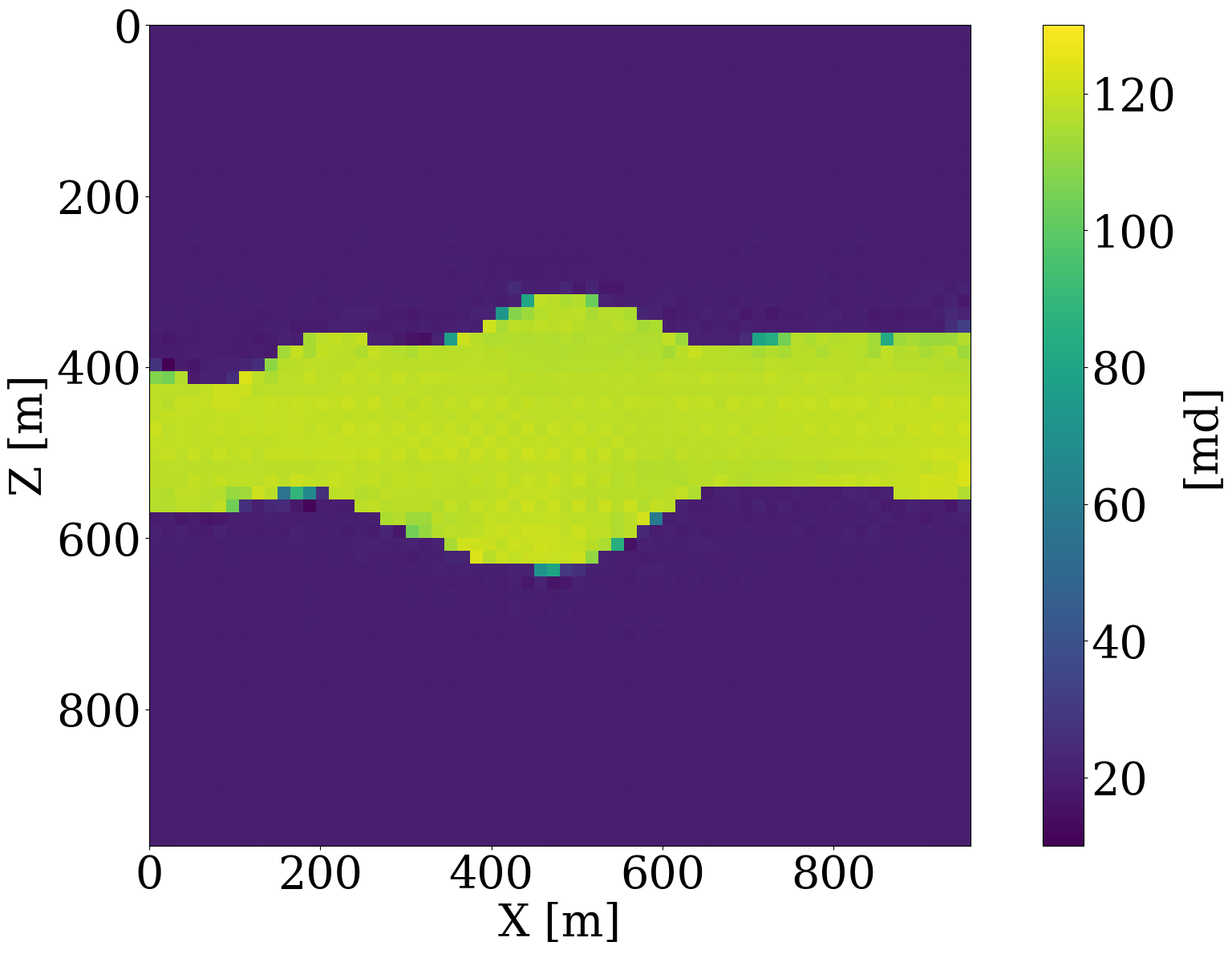}

}

}

\subcaption{\label{fig-flow-inv-fno-nf-all}}
\end{minipage}%
\newline
\begin{minipage}[t]{\linewidth}

{\centering 

\raisebox{-\height}{

\includegraphics{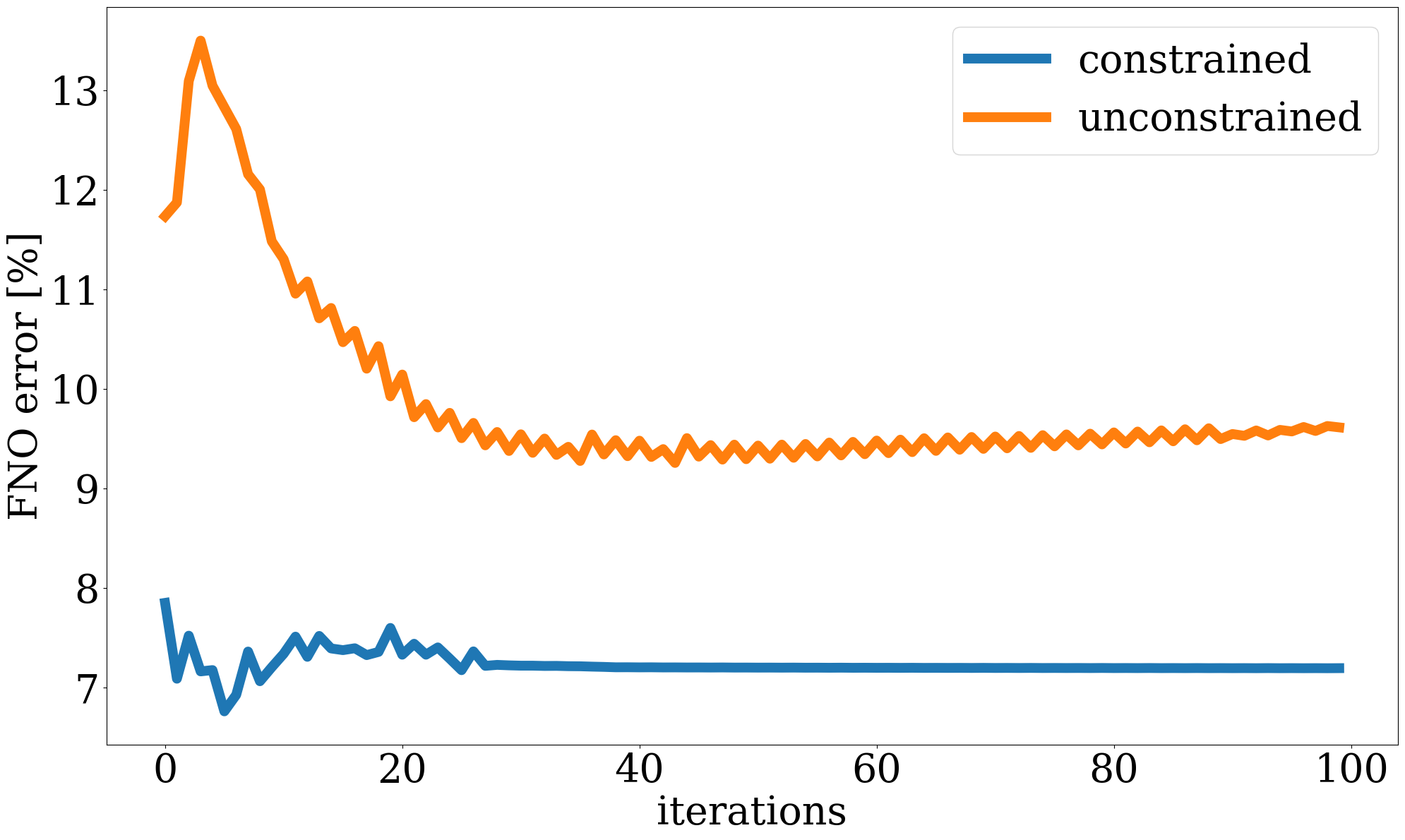}

}

}

\subcaption{\label{fig-flow-inv-fno-loss-all}}
\end{minipage}%

\caption{\label{fig-flow-inv-all}Permeability inversion from fully
observed time-lapse CO\textsubscript{2} saturations. \emph{(a)}
Inversion result with PDE solvers. \emph{(b)} The same but via the
approximate FNO surrogate. \emph{(c)} Same as \emph{(a)} but with NF
constraint. \emph{(d)} Same as \emph{(b)} but with NF constraint.
\emph{(e)} The FNO approximation errors as a function of the number of
iterations for the result plotted in \emph{(b)} and \emph{(d)}.}

\end{figure}

To demonstrate that permeability inversion with surrogates is indeed
feasible, we first consider the idealized, impossible in practice,
situation where we assume to have access to the time-lapse
CO\textsubscript{2} concentration,
\(\mathbf{c}=[\mathbf{c}_1, \mathbf{c}_2, \cdots,\mathbf{c}_{n_t}]\),
everywhere, and for all \(n_t=8\) timesteps. In that case, the
measurement operator, \(\mathcal{H}\) in Equation~\ref{eq-inv},
corresponds to the identity matrix. Given CO\textsubscript{2}
concentrations simulated from the ``ground-truth'' permeability
distribution plotted in Figure~\ref{fig-true-perm}, we invert for the
permeability by minimizing the unconstrained formulation (cf.
Equation~\ref{eq-inv-fno}) for the correct, yielded by the PDE, and
approximate fluid-flow physics, yielded by the trained FNO. The results
of these inversions after 100 iterations of gradient descent with
back-tracking linesearch (Stanimirović and Miladinović 2010) are plotted
in Figure~\ref{fig-flow-inv-jutul-all} and
Figure~\ref{fig-flow-inv-fno-all}. From these plots, we observe that the
inversion results using PDE solvers delineates most of the upper
boundary of the channel accurately. Because there is a null space in the
fluid-flow modeling---i.e., this null space mostly corresponds to
regions of the permeability model that are barely touched by the
CO\textsubscript{2} plume (e.g.~bottom and right-hand side of the
channel) --- artifacts are present in the high-permeability channel
itself. As expected, the reconstruction of the permeability is also not
perfect at the bottom and at the far right of the model. The inversion
result with the FNO surrogate is similar but introduces unrealistic
artifacts in the high-permeability channel and also outside the channel.
These more severe artifacts can be explained by the behavior of the FNO
approximation error plotted as the orange curve in
Figure~\ref{fig-flow-inv-fno-loss-all}. The error value increases
rapidly to 13\%, and finally saturates at 10\%. This behavior of the
error is a manifestation of out-of-distribution model iterates that
explain the erroneous behavior of the surrogate and its gradient with
respect to the permeability.

Inversions yielded by the relaxed constrained formulation with the
trained NF (see Algorithm 1), on the other hand, show virtually artifact
free inversion results (see Figure~\ref{fig-flow-inv-jutul-nf-all} and
Figure~\ref{fig-flow-inv-fno-nf-all}) that compare favorably with the
``ground-truth'' permeability plotted in
Figure~\ref{fig-perm-init-true}. While adding the NF as a constraint
obviously adds information, explaining the improved inversion for the
accurate physics (Figure~\ref{fig-flow-inv-jutul-nf-all}), it also
renders the approximate surrogates more accurate, as can be observed
from the blue curve in Figure~\ref{fig-flow-inv-fno-loss-all}, where the
FNO approximation error is controlled thanks to adding the constraint to
the inversion. This behavior underlines the importance of ensuring model
iterates to remain within distribution. It also demonstrates the
benefits of a solution strategy where we start with a small \(\tau\),
followed by relaxing the constraint slowly by increasing the size of the
constraint set gradually.

\hypertarget{unconstrainedconstrained-permeability-inversion-from-well-observations}{%
\subsubsection{Unconstrained/constrained permeability inversion from
well
observations}\label{unconstrainedconstrained-permeability-inversion-from-well-observations}}

\begin{figure}

\begin{minipage}[t]{0.25\linewidth}

{\centering 

\raisebox{-\height}{

\includegraphics{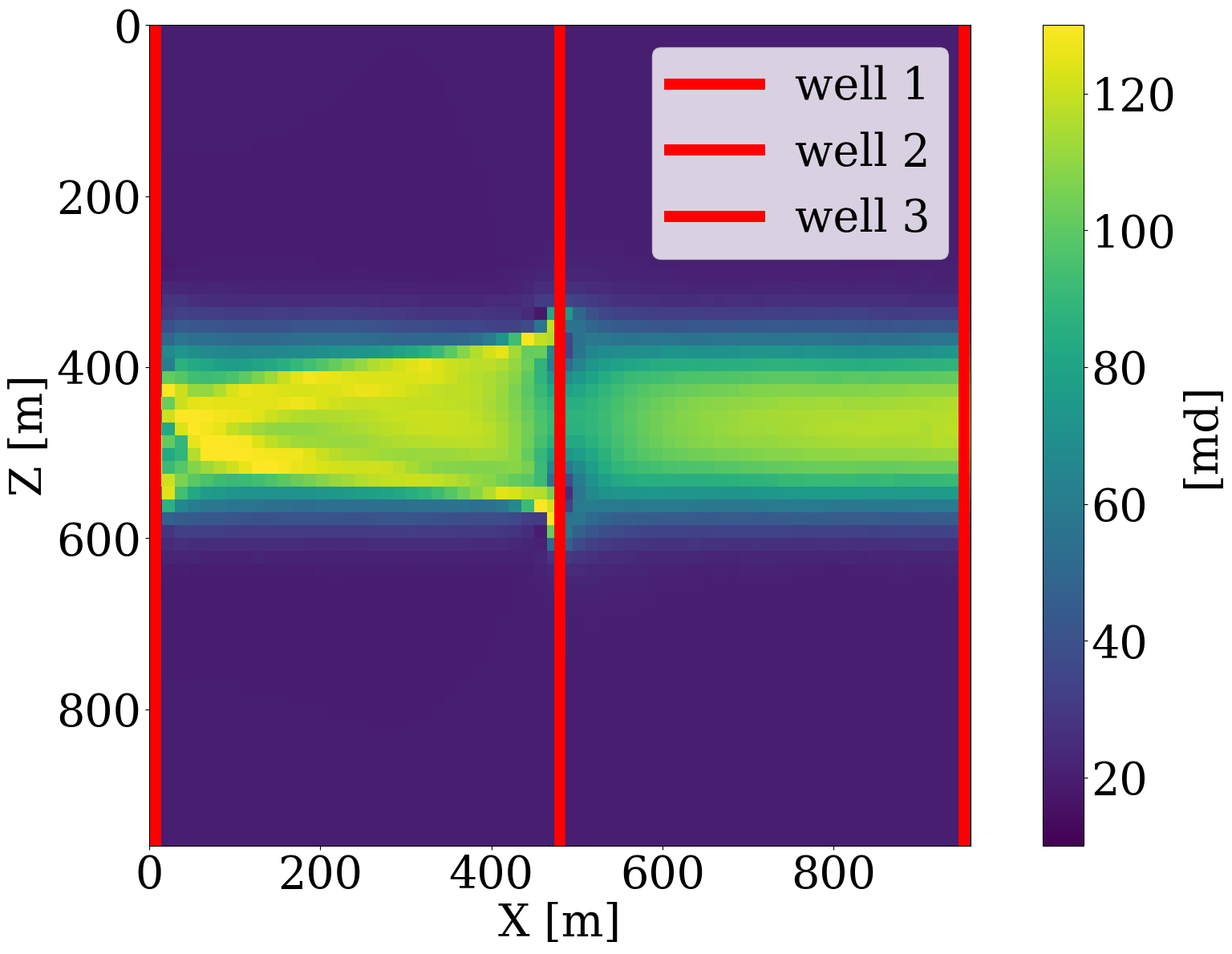}

}

}

\subcaption{\label{fig-flow-inv-jutul}}
\end{minipage}%
\begin{minipage}[t]{0.25\linewidth}

{\centering 

\raisebox{-\height}{

\includegraphics{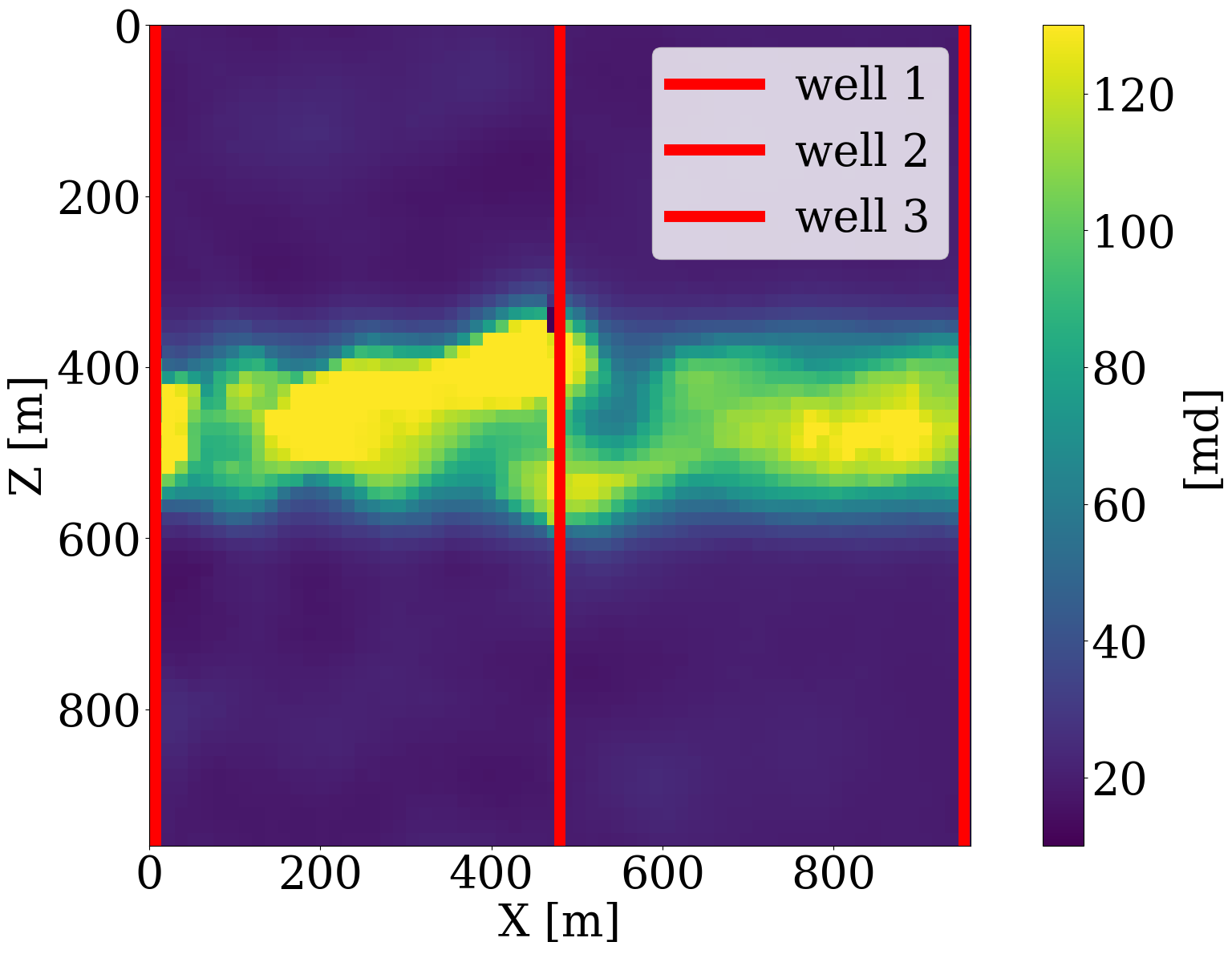}

}

}

\subcaption{\label{fig-flow-inv-fno}}
\end{minipage}%
\begin{minipage}[t]{0.25\linewidth}

{\centering 

\raisebox{-\height}{

\includegraphics{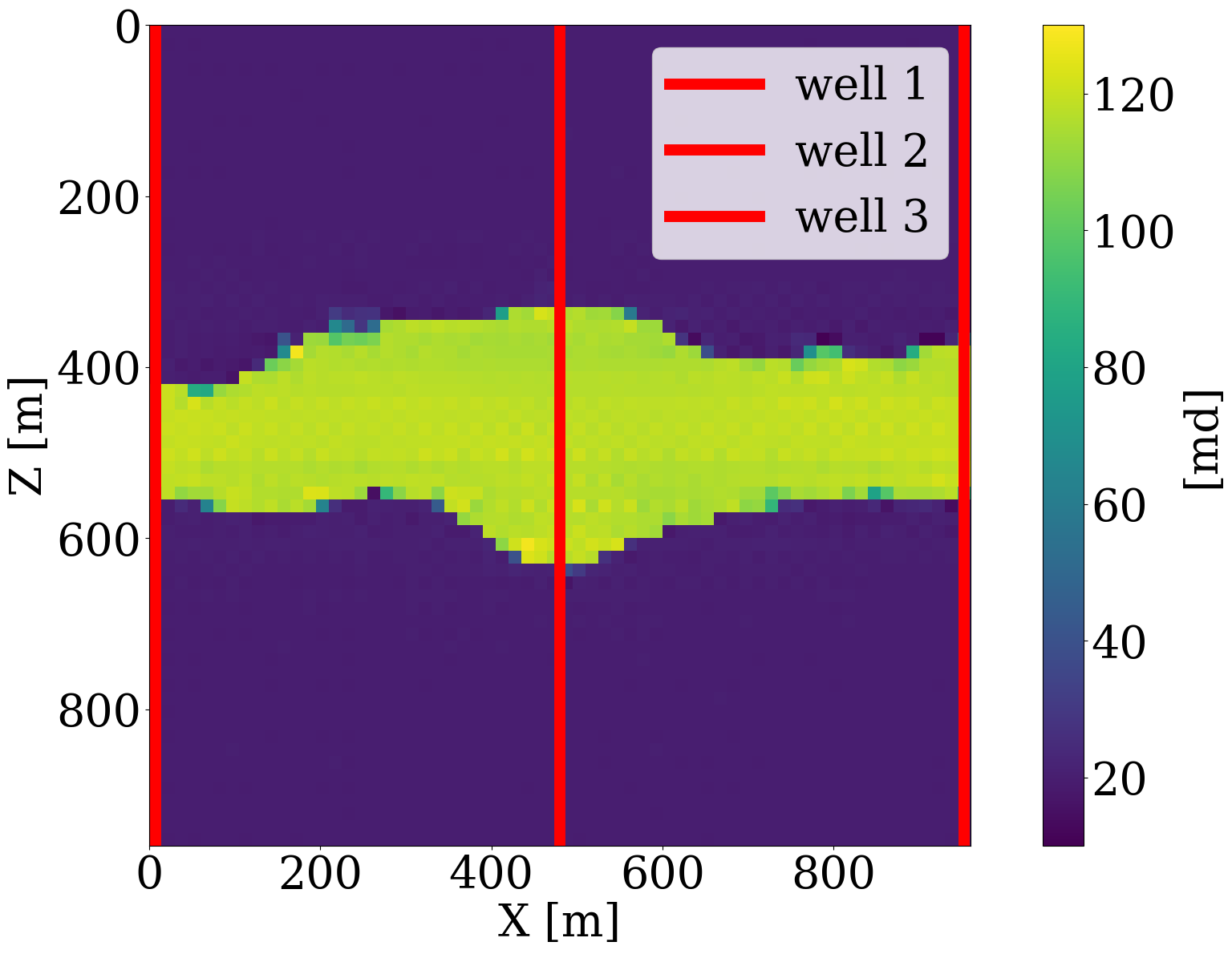}

}

}

\subcaption{\label{fig-flow-inv-jutul-nf}}
\end{minipage}%
\begin{minipage}[t]{0.25\linewidth}

{\centering 

\raisebox{-\height}{

\includegraphics{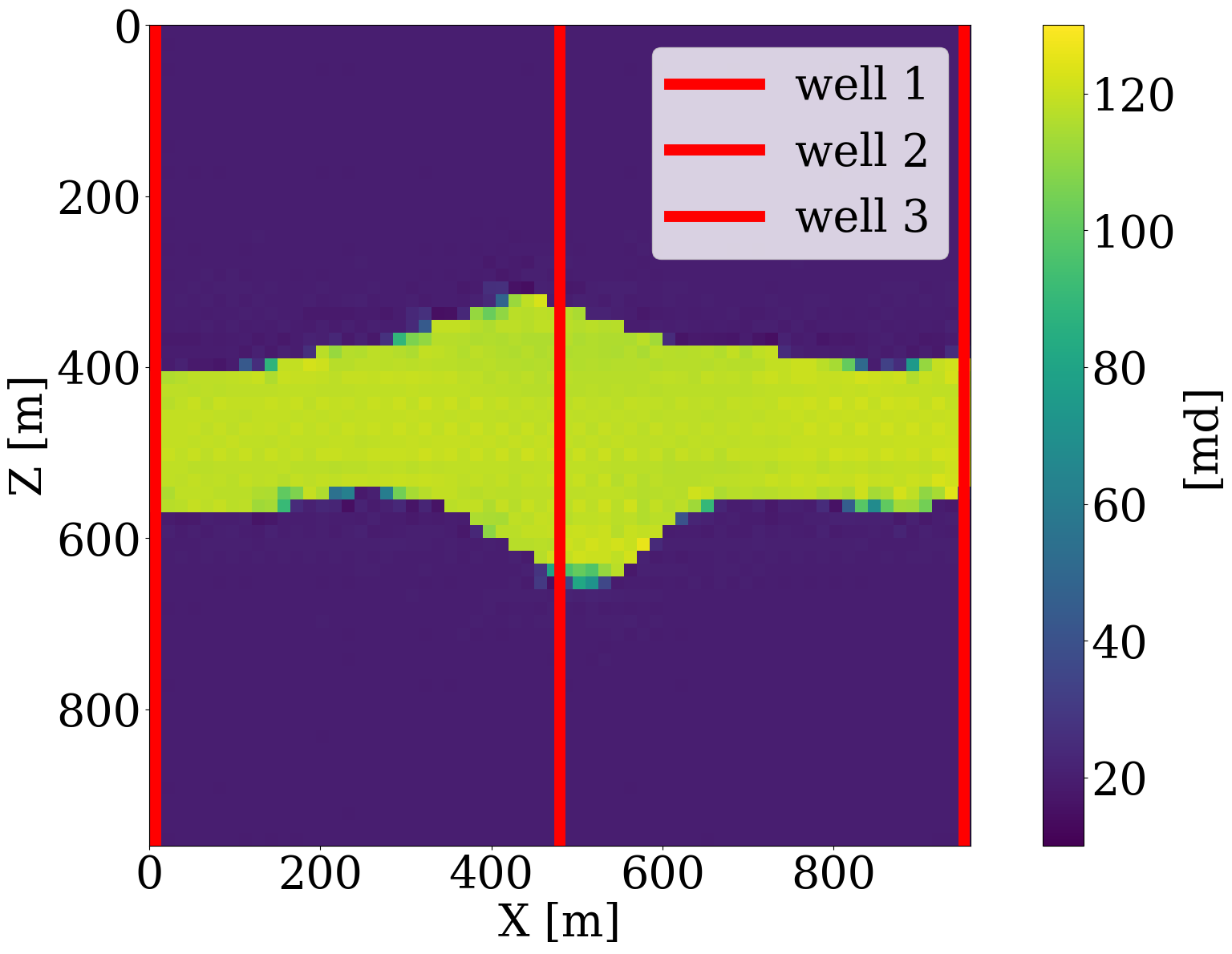}

}

}

\subcaption{\label{fig-flow-inv-fno-nf}}
\end{minipage}%
\newline
\begin{minipage}[t]{0.25\linewidth}

{\centering 

\raisebox{-\height}{

\includegraphics{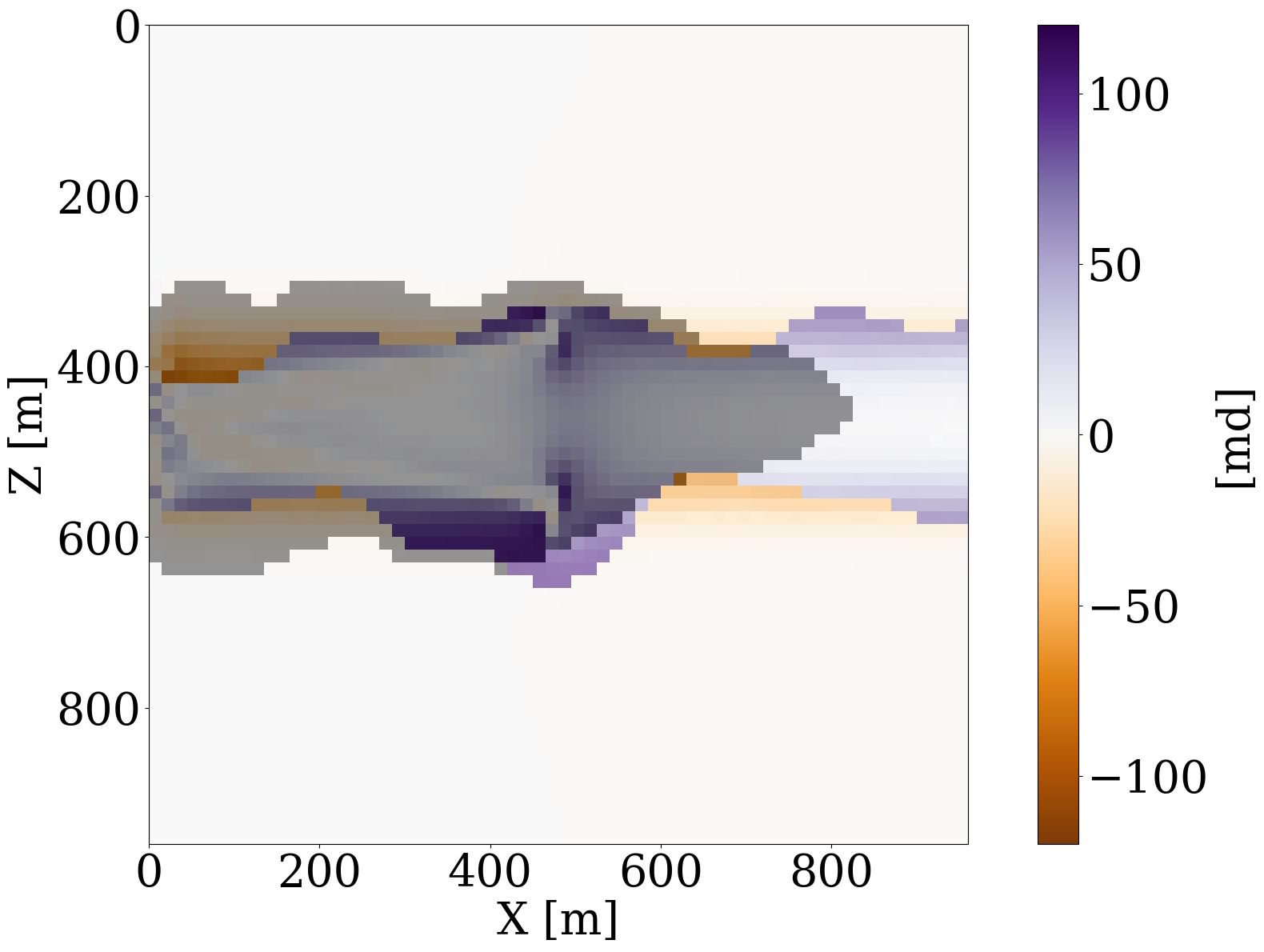}

}

}

\subcaption{\label{fig-diff-flow-inv-jutul}}
\end{minipage}%
\begin{minipage}[t]{0.25\linewidth}

{\centering 

\raisebox{-\height}{

\includegraphics{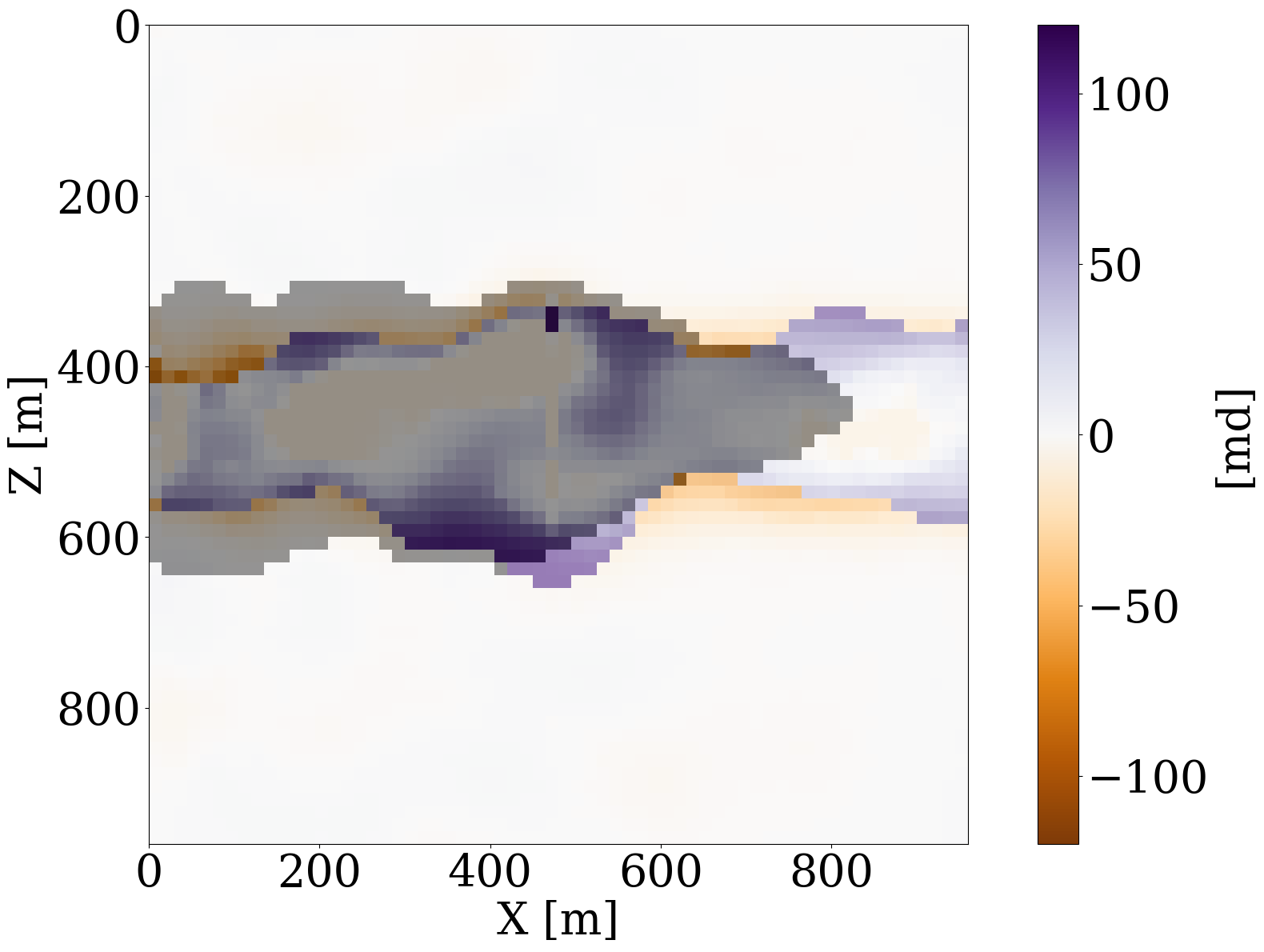}

}

}

\subcaption{\label{fig-diff-flow-inv-fno}}
\end{minipage}%
\begin{minipage}[t]{0.25\linewidth}

{\centering 

\raisebox{-\height}{

\includegraphics{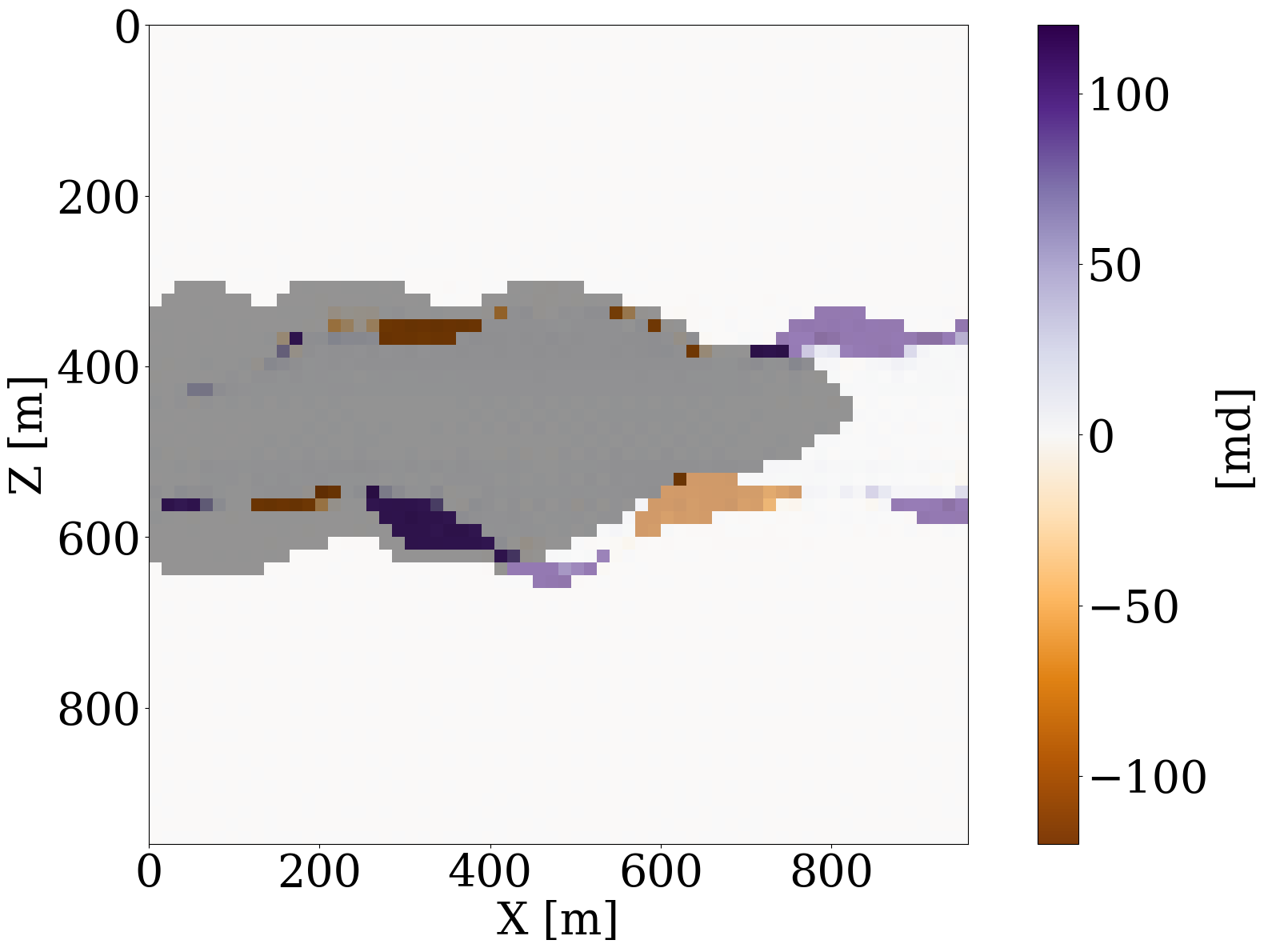}

}

}

\subcaption{\label{fig-diff-flow-inv-jutul-nf}}
\end{minipage}%
\begin{minipage}[t]{0.25\linewidth}

{\centering 

\raisebox{-\height}{

\includegraphics{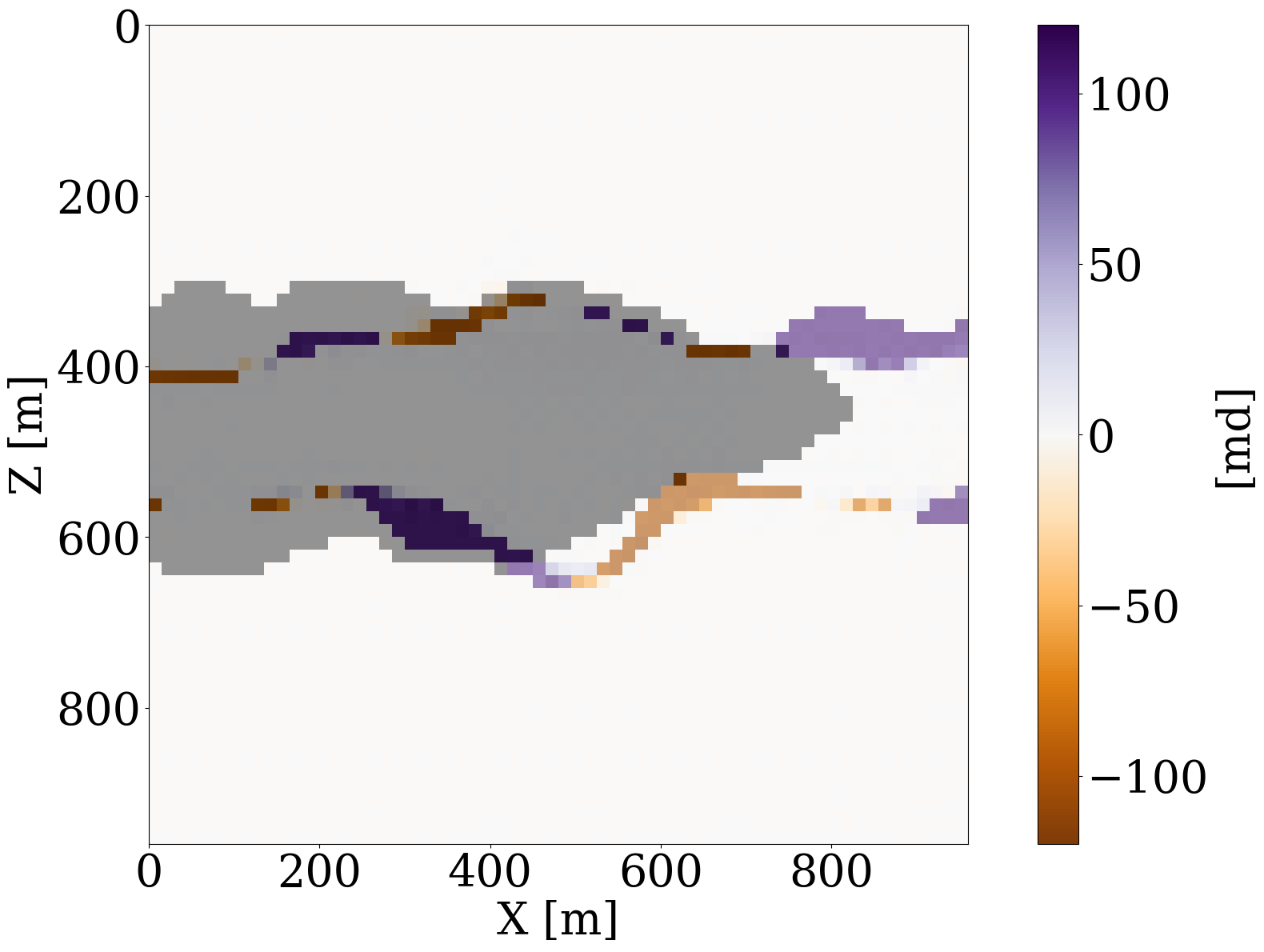}

}

}

\subcaption{\label{fig-diff-flow-inv-fno-nf}}
\end{minipage}%
\newline
\begin{minipage}[t]{\linewidth}

{\centering 

\raisebox{-\height}{

\includegraphics{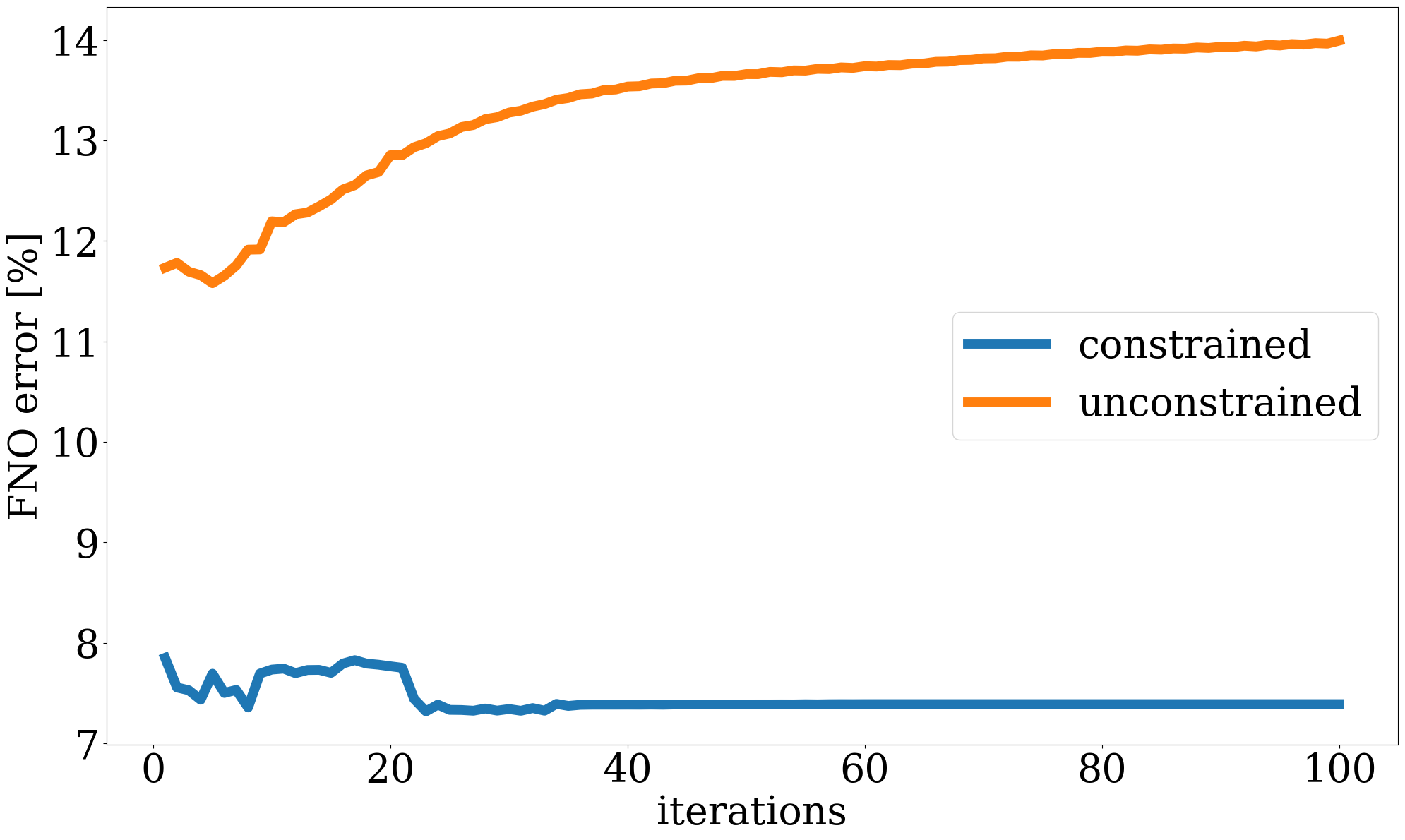}

}

}

\subcaption{\label{fig-flow-inv-fno-loss-well}}
\end{minipage}%

\caption{\label{fig-flow-inv}Permeability inversions from
CO\textsubscript{2} saturations sampled at three well locations at 6
early snapshots. The well locations are denoted by the red vertical
lines. \emph{(a)} Unconstrained inversion result based on PDE solves.
\emph{(b)} Same as \emph{(a)} but now with FNO surrogate approximation.
\emph{(c)} Constrained inversion result based on PDE solves. \emph{(d)}
Same as \emph{(c)} but now with FNO surrogate approximation.
\emph{(e)}-\emph{(h)} The error of the permeability inversion results in
\emph{(a)}-\emph{(d)} compared to the unseen ground truth shown in
Figure~\ref{fig-true-perm}. \emph{(i)} The FNO prediction errors as a
function of the number of iterations for \emph{(b)} and \emph{(d)}.}

\end{figure}

While the example of the previous section established feasibility of
constrained permeability inversion, it relied on having access to the
CO\textsubscript{2} saturation everywhere, which is unrealistic in
practice. To address this issue, we first consider permeability
inversion from CO\textsubscript{2} saturations, collected at three
equally spaced monitoring well locations, for only the first 6 timesteps
over the period of 600 days (Mosser, Dubrule, and Blunt 2019). In this
more realistic setting, the measurement operator, \(\mathcal{H}\) in
Equation~\ref{eq-inv}, corresponds to a restriction operator that
extracts simulated CO\textsubscript{2} saturations at each well location
in first six snapshots. The objective function reads

\begin{equation}\protect\hypertarget{eq-inv-well}{}{
\underset{\mathbf{z}}{\operatorname{minimize}} \quad \|\mathbf{d}_{\mathrm{w}}-\mathbf{M}\circ\mathcal{S}_{\boldsymbol\theta^\ast}\circ\mathcal{G}_{\mathbf{w}^\ast}(\mathbf{z})\|_2^2\quad\text{subject to}\quad\|\mathbf{z}\|_2\leq \tau,
}\label{eq-inv-well}\end{equation}

where \(\mathbf{d}_{\mathrm{w}}\) represents the well measurements
collected at three well locations through the linear restriction
operator \(\mathbf{M}\). The goal is to invert for the permeability by
minimizing the misfit of the well measurements of the
CO\textsubscript{2} saturation without and with constraints on the
\(\ell_2\)-norm ball in the latent space. The results of these numerical
experiments are included in the first row of Figure~\ref{fig-flow-inv},
where the differences with respect to the ground truth permeability
shown in Figure~\ref{fig-true-perm} are plotted in the second row.
Because the part of the permeability that is not touched by the
CO\textsubscript{2} plume lives in the null space, we highlight the
CO\textsubscript{2} plume in the difference plots by dark color and
focus on analyzing errors within the plume region. As expected, the
unconstrained inversions based on PDE solves
(Figure~\ref{fig-flow-inv-jutul}) and surrogate approximations
(Figure~\ref{fig-flow-inv-fno}) are both poorly resolved because of the
limited spatial information on the saturation. Contrasting these
unconstrained inversions with results for the constrained inversions for
the PDE (Figure~\ref{fig-flow-inv-jutul-nf}) and surrogate
(Figure~\ref{fig-flow-inv-fno-nf}) again shows the importance of adding
constraints to the inversion. Figure~\ref{fig-flow-inv-fno-loss-well}
clearly demonstrates that the FNO prediction errors remain relatively
constant during constrained inversion while the error continues to grow
during the unconstrained iterations eventually exceeding 14\%. Both
constrained results improve significantly, even though they converge to
different solutions in the end. This is because history matching is
typically an ill-posed problem with many distinctive solutions
(Canchumuni, Emerick, and Pacheco 2019). This observation further
motivates us to consider the experiment below, where time-lapse seismic
data are jointly inverted for the subsurface permeability.

\hypertarget{multiphysics-end-to-end-inversion}{%
\subsubsection{Multiphysics end-to-end
inversion}\label{multiphysics-end-to-end-inversion}}

Next, we consider the alternative setting for seismic monitoring of
geological carbon storage, where the dynamics of the CO\textsubscript{2}
plumes are indirectly observed from time-lapse seismic data. In this
case, the measurement operator, \(\mathcal{H}\), involves the
composition of the rock physics modeling operator, \(\mathcal{R}\),
which converts CO\textsubscript{2} saturations to decreases in the
compressional wavespeeds for rocks within the reservoir (Avseth,
Mukerji, and Mavko 2010), and the seismic modeling operator,
\(\mathcal{F}\), which generates time-lapse seismic data recorded at the
receiver locations and based on acoustic wave equation modeling (Sheriff
and Geldart 1995). The multiphysics end-to-end inversion process
estimates permeability from time-lapse seismic data via inversion of
these nested physics operators for the flow, rock physics, and waves (D.
Li et al. 2020). Following earlier work by Yin et al. (2022) and
Louboutin, Yin, et al. (2023), the fluid-flow PDE modeling is replaced
by the trained FNO (cf. Equation~\ref{eq-inv-fno-nf}), resulting in the
following optimization problem:

\begin{equation}\protect\hypertarget{eq-inv-end2end2end}{}{
\underset{\mathbf{z}}{\operatorname{minimize}} \quad \|\mathbf{d}_{\mathrm{s}} - \mathcal{F}\circ\mathcal{R}\circ\mathcal{S}_{\boldsymbol\theta^\ast}\circ\mathcal{G}_{\mathbf{w}^\ast}(\mathbf{z})\|_2^2 \quad\text{subject to}\quad\|\mathbf{z}\|_2\leq \tau,
}\label{eq-inv-end2end2end}\end{equation}

where \(\mathbf{d}_{\mathrm{s}}\) represents the observed time-lapse
seismic data. While this end-to-end inversion problem benefits from
having remote access to changes in the compressional wavespeed, it may
now suffer from null spaces associated with the flow,
\(\mathcal{S}_{\boldsymbol\theta^\ast}\), and the wave/rock physics,
\(\mathcal{F}\circ\mathcal{R}\). For instance, the latter suffers from
bandwidth limitation of the source function and from limited aperture.
Because important components are missing in the observed data, inversion
based on the data objective alone in Equation~\ref{eq-inv-end2end2end}
are likely to suffer from artifacts that can easily drive the
intermediate permeability model iterates out-of-distribution.

\begin{figure}

{\centering 

\includegraphics[width=0.5\textwidth,height=\textheight]{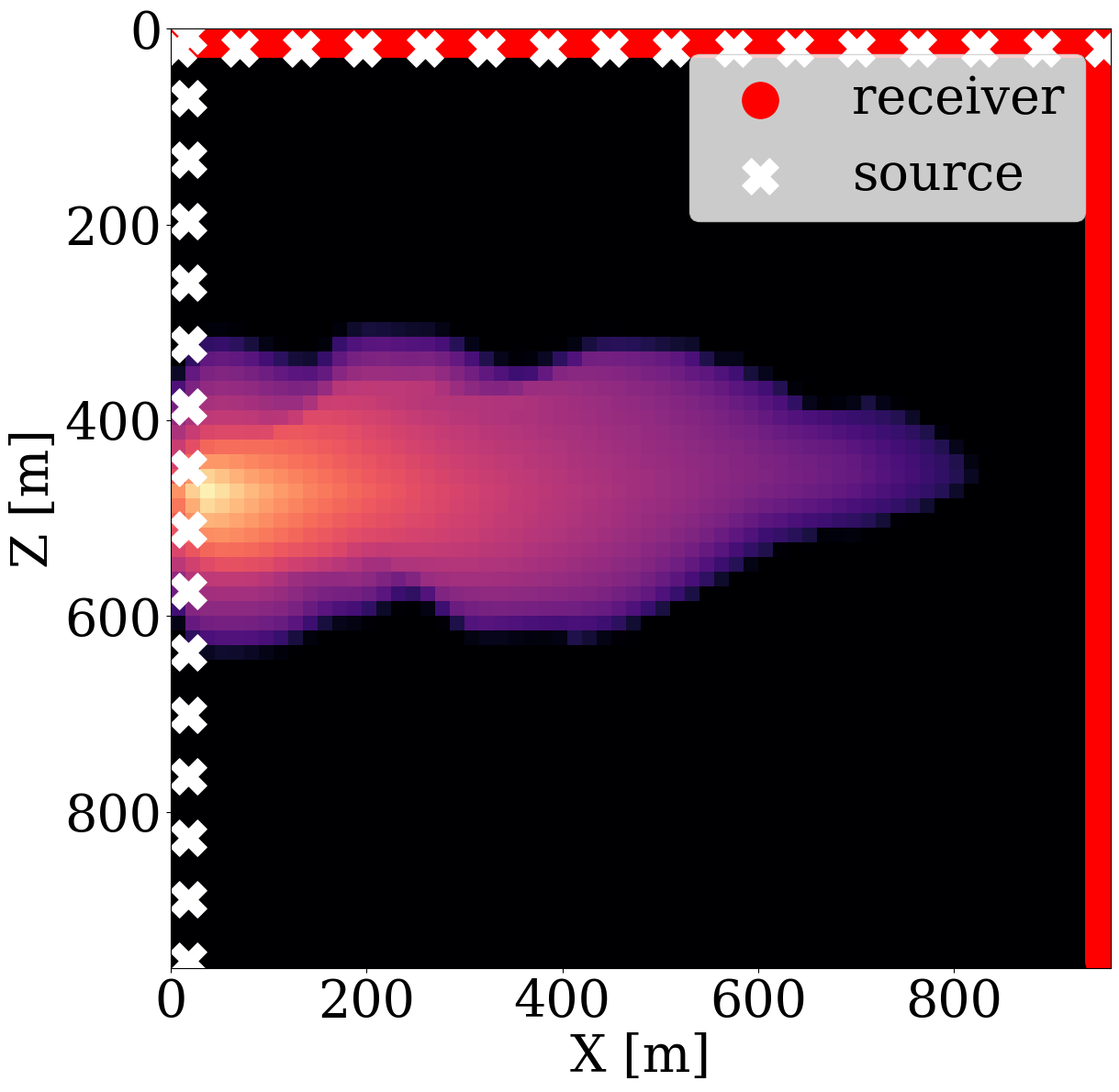}

}

\caption{\label{fig-seismic-acquisition}Seismic acquisition. The white
\(\boldsymbol{\times}\) represents the acoustic sources, and the red
lines represent the dense receivers. The CO\textsubscript{2} saturation
snapshot at day 600 is plotted in the middle, which is the last snapshot
that is monitored seismically.}

\end{figure}

To demonstrate capabilities of the proposed relaxed inversion procedure
with surrogates for the fluid flow, we assume the baseline to be
known---i.e, we assume the brine-filled reservoir with \(25\%\) porosity
to be acoustically homogeneous prior to CO\textsubscript{2} injection
with a compressional wavespeed of \(3500 \mathrm{m/s}\). We use the
patchy saturation model (Avseth, Mukerji, and Mavko 2010) to convert the
time-dependent CO\textsubscript{2} saturation resulting in
\(<300\mathrm{m/s}\) decreases in the wavespeed within the
CO\textsubscript{2} plumes. We collect six seismic surveys at the first
six snapshots for the CO\textsubscript{2} saturation from day 100 to day
600, which are the same snapshots as the ones used in the previous
experiment. For each time-lapse seismic survey, 16 active-seismic
sources are located within a well on the left-hand side of the model. We
also position 16 sources on the top of the model. Each active source
uses a Ricker wavelet with a central frequency of \(50\mathrm{Hz}\). The
transmitted and reflected wavefields are collected by 480 receivers on
the top and 480 receivers on the right-hand side of the model. The
seismic acquisition is shown in Figure~\ref{fig-seismic-acquisition},
where the plume at the last seismic vintage (at day 600) is plotted in
the middle.

To avoid numerical dispersion, the velocity model is upsampled by a
factor of two in both the horizontal and vertical directions, which
results in a \(7.5\mathrm{m}\) grid spacing. For the simulations, use is
made of the open-source software package
\href{https://github.com/slimgroup/JUDI.jl}{JUDI.jl} (Philipp A. Witte
et al. 2019; Louboutin, Witte, et al. 2023) to generate the time-lapse
seismic data at the first six snapshots. The fact that this software is
based on \href{https://www.devitoproject.org/}{Devito}'s wave
propagators (Louboutin et al. 2019; Luporini et al. 2020) allows us to
do this quickly. For realism, we add 10 dB Gaussian noise to the
time-lapse seismic data. Given these six time-lapse vintages, our goal
is to invert for the permeability in the reservoir by minimizing the
time-lapse seismic data misfit through the nested physics operators
shown in Equation~\ref{eq-inv-end2end2end}.

\begin{figure}

\begin{minipage}[t]{0.25\linewidth}

{\centering 

\raisebox{-\height}{

\includegraphics{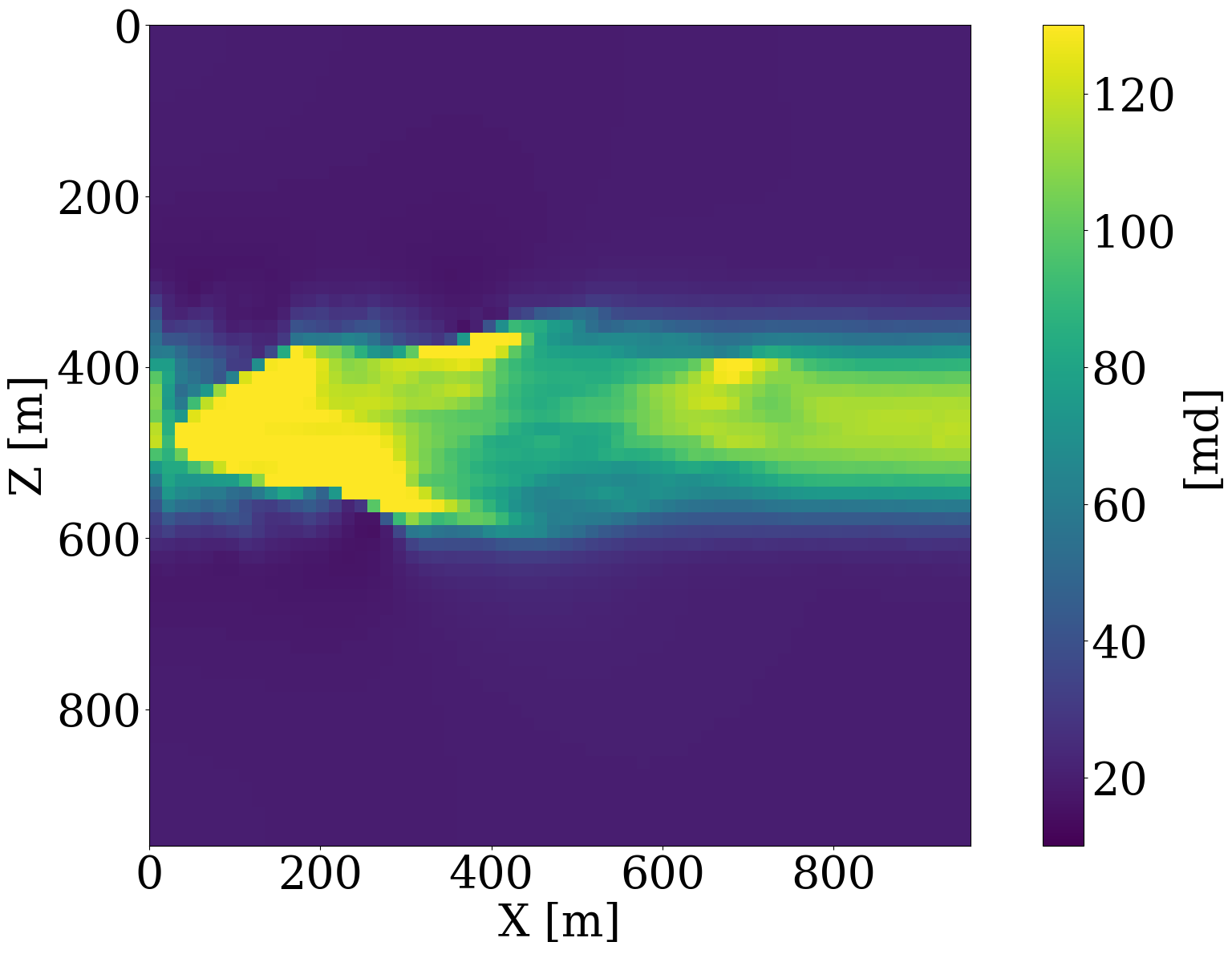}

}

}

\subcaption{\label{fig-end2end-inv-jutul}}
\end{minipage}%
\begin{minipage}[t]{0.25\linewidth}

{\centering 

\raisebox{-\height}{

\includegraphics{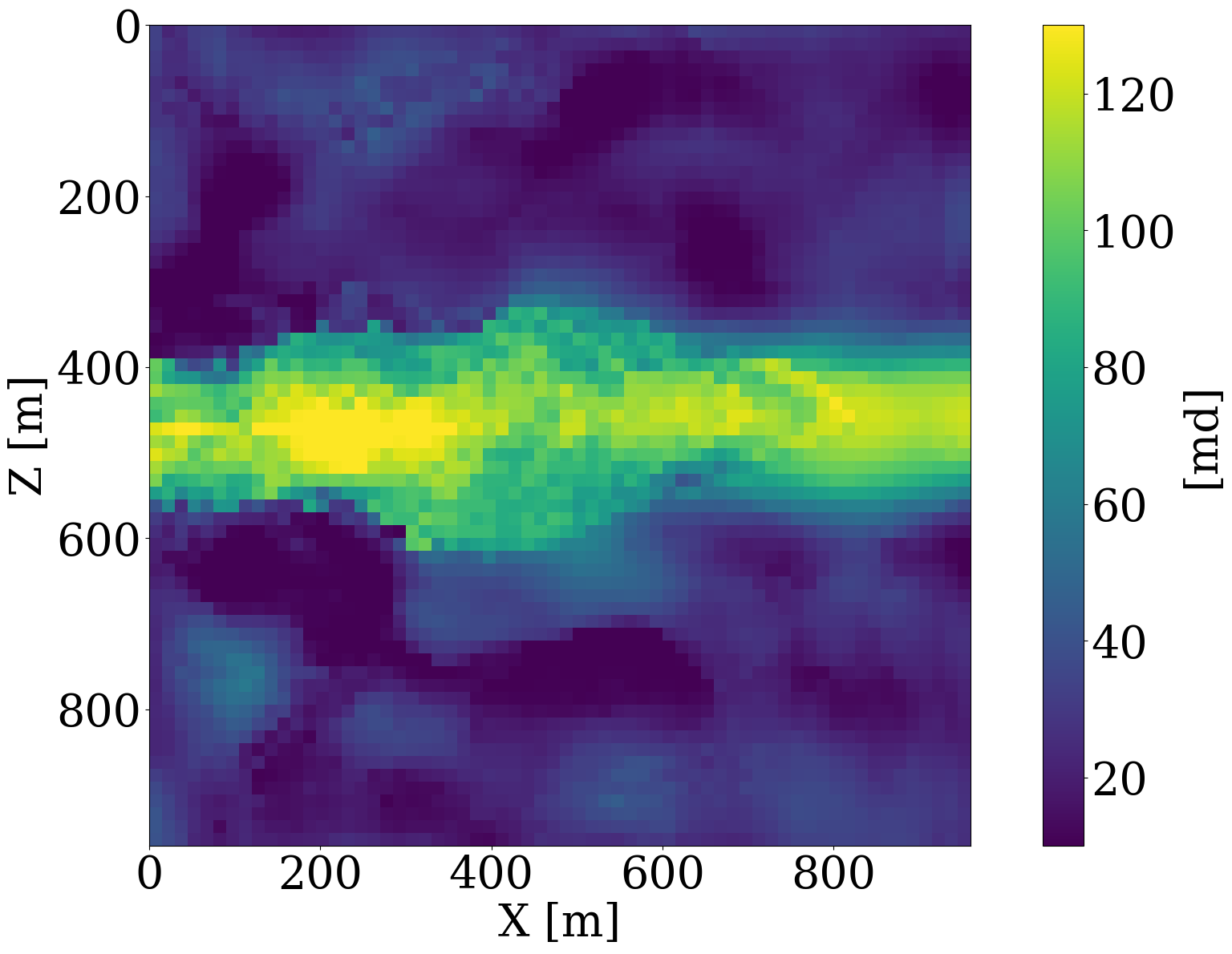}

}

}

\subcaption{\label{fig-end2end-inv-fno}}
\end{minipage}%
\begin{minipage}[t]{0.25\linewidth}

{\centering 

\raisebox{-\height}{

\includegraphics{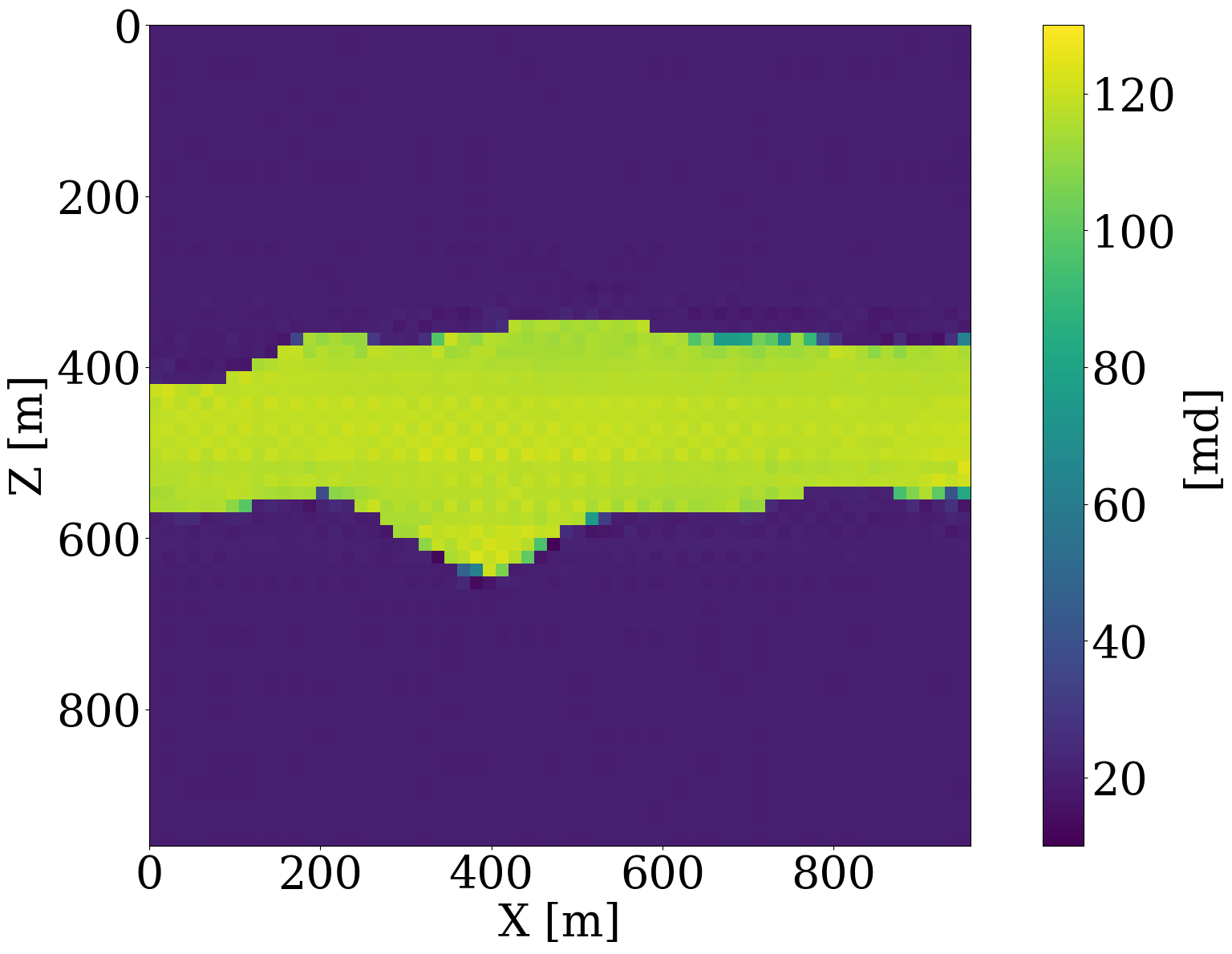}

}

}

\subcaption{\label{fig-end2end-inv-jutul-nf}}
\end{minipage}%
\begin{minipage}[t]{0.25\linewidth}

{\centering 

\raisebox{-\height}{

\includegraphics{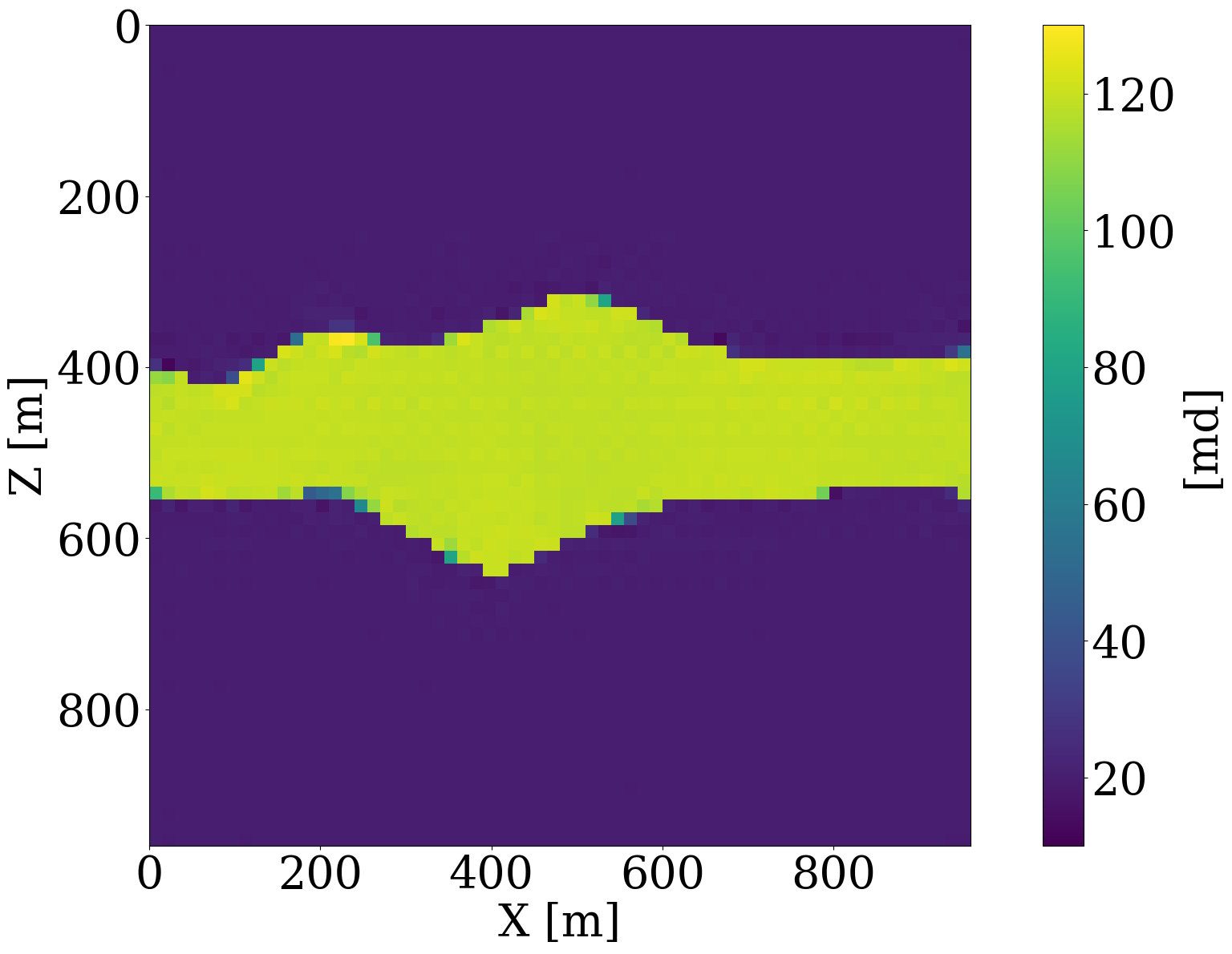}

}

}

\subcaption{\label{fig-end2end-inv-fno-nf}}
\end{minipage}%
\newline
\begin{minipage}[t]{0.25\linewidth}

{\centering 

\raisebox{-\height}{

\includegraphics{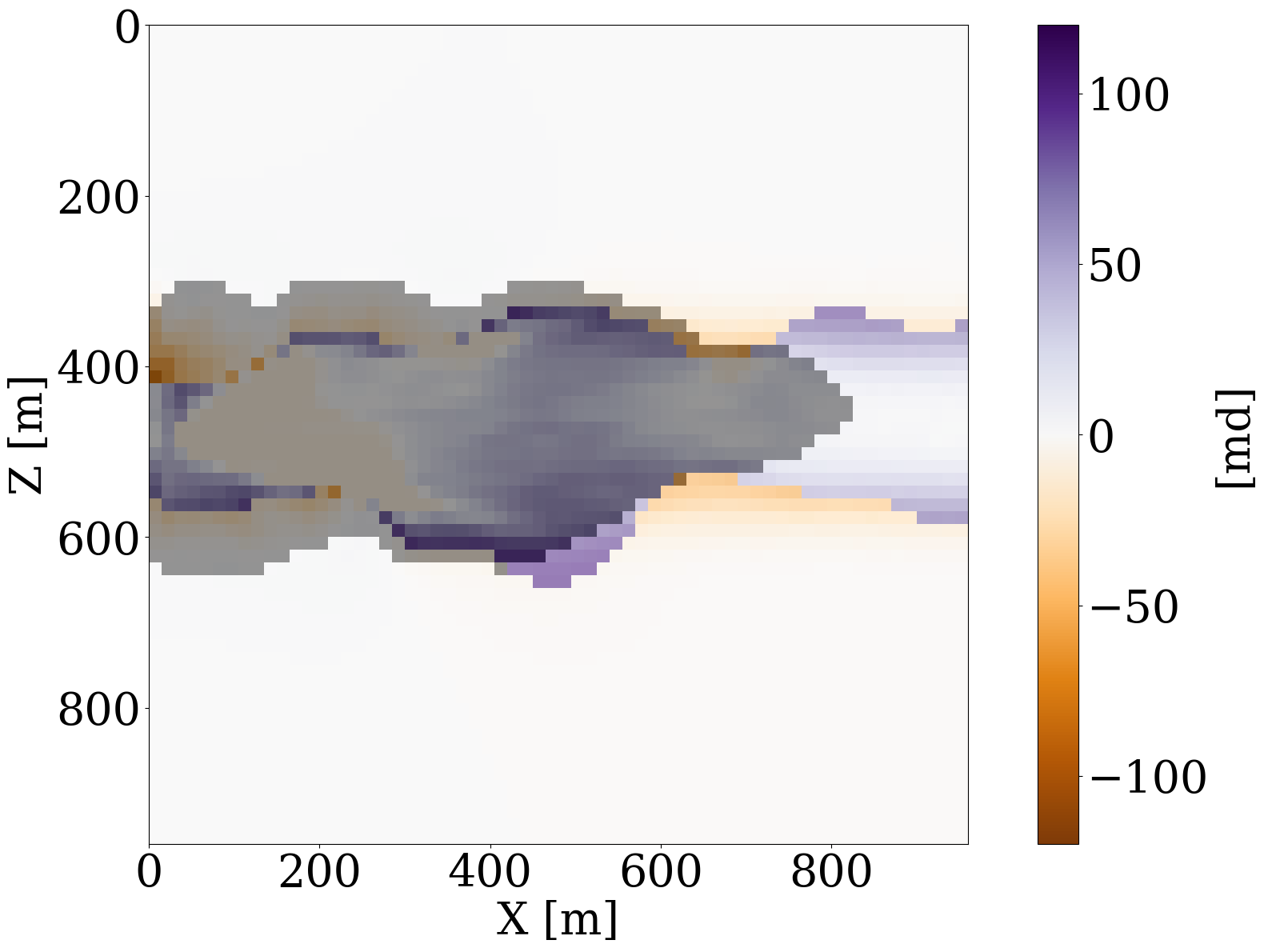}

}

}

\subcaption{\label{fig-diff-end2end-flow-inv-jutul}}
\end{minipage}%
\begin{minipage}[t]{0.25\linewidth}

{\centering 

\raisebox{-\height}{

\includegraphics{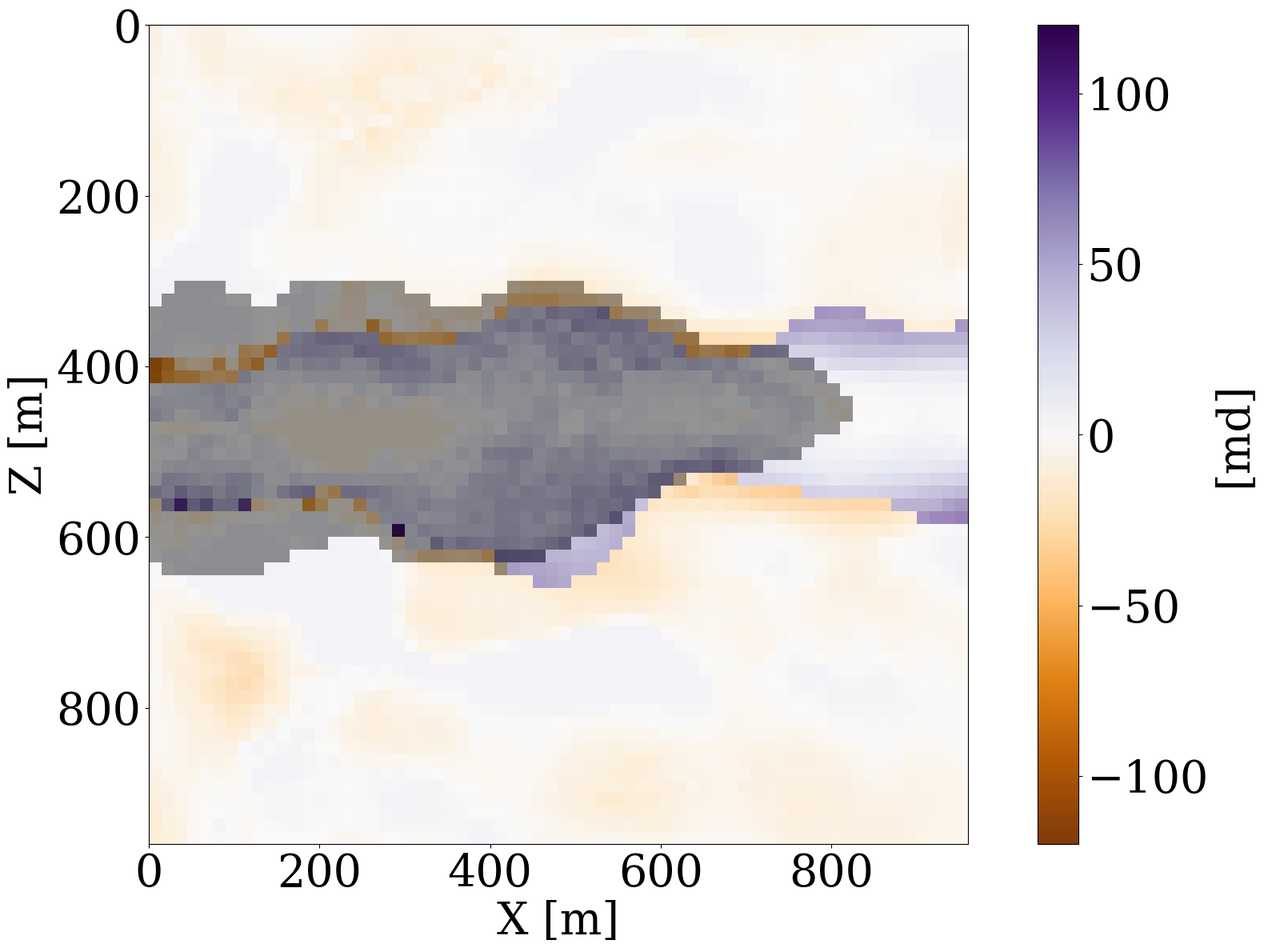}

}

}

\subcaption{\label{fig-diff-end2end-flow-inv-fno}}
\end{minipage}%
\begin{minipage}[t]{0.25\linewidth}

{\centering 

\raisebox{-\height}{

\includegraphics{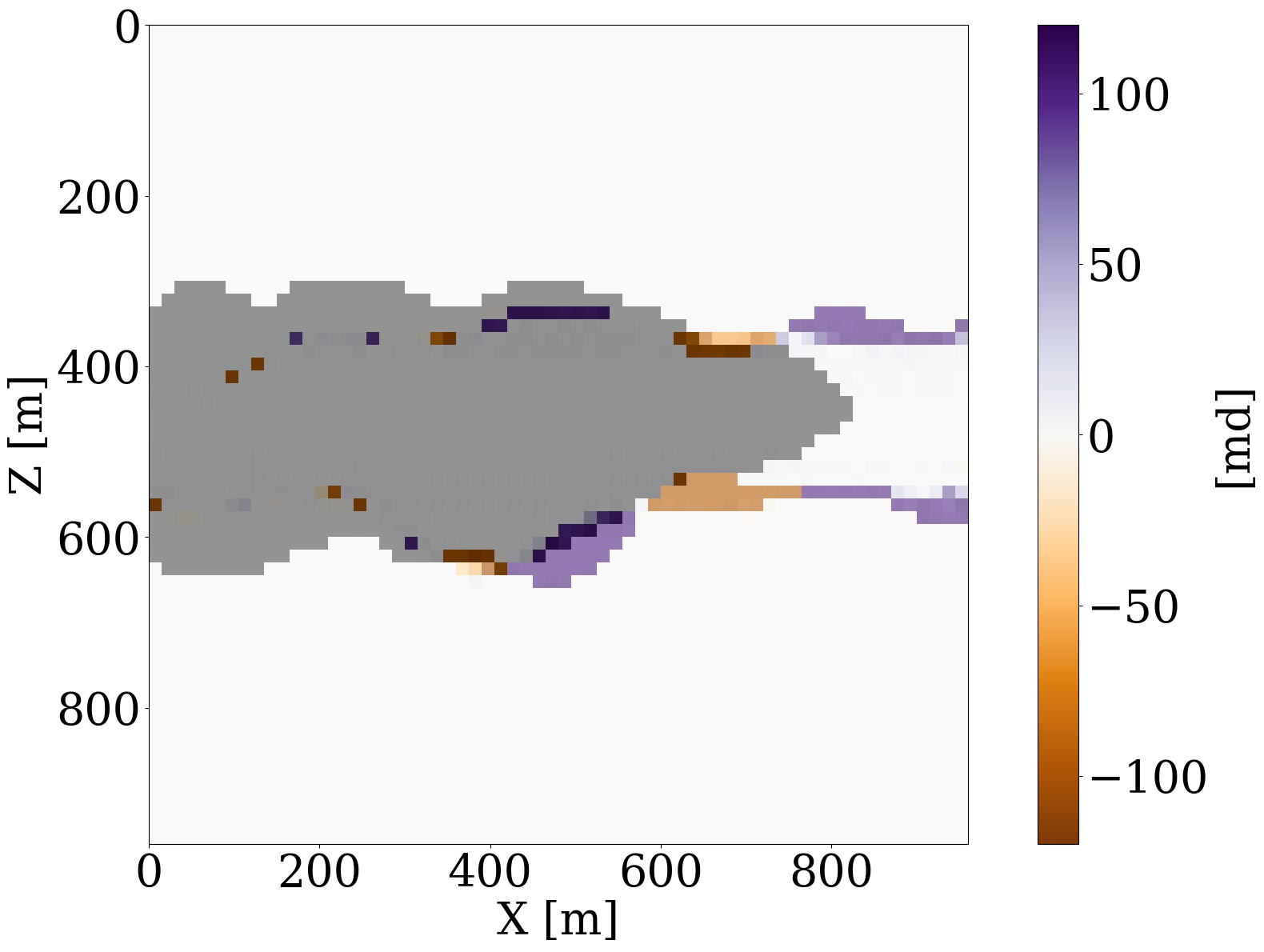}

}

}

\subcaption{\label{fig-diff-end2end-flow-inv-jutul-nf}}
\end{minipage}%
\begin{minipage}[t]{0.25\linewidth}

{\centering 

\raisebox{-\height}{

\includegraphics{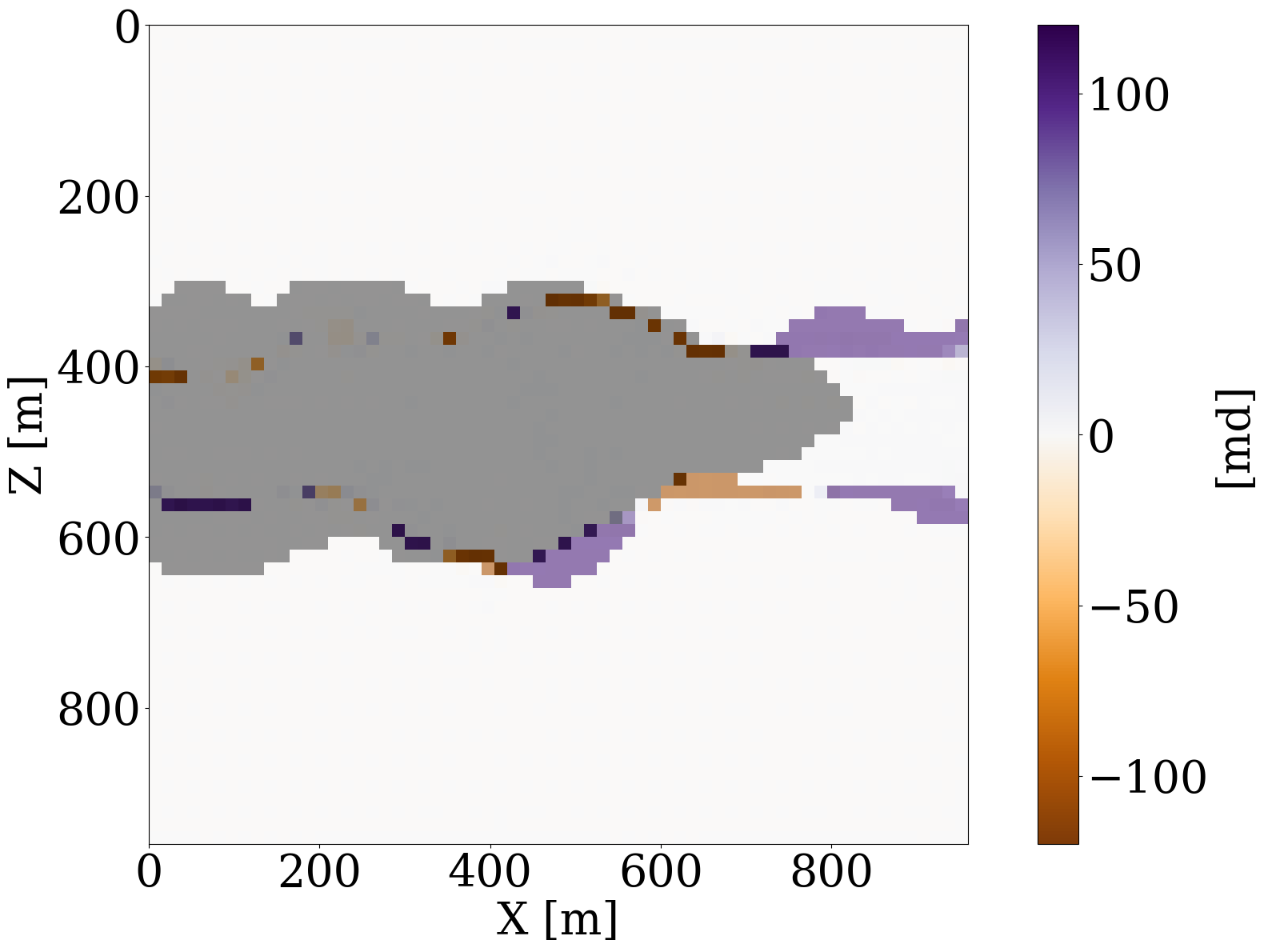}

}

}

\subcaption{\label{fig-diff-end2end-flow-inv-fno-nf}}
\end{minipage}%
\newline
\begin{minipage}[t]{\linewidth}

{\centering 

\raisebox{-\height}{

\includegraphics{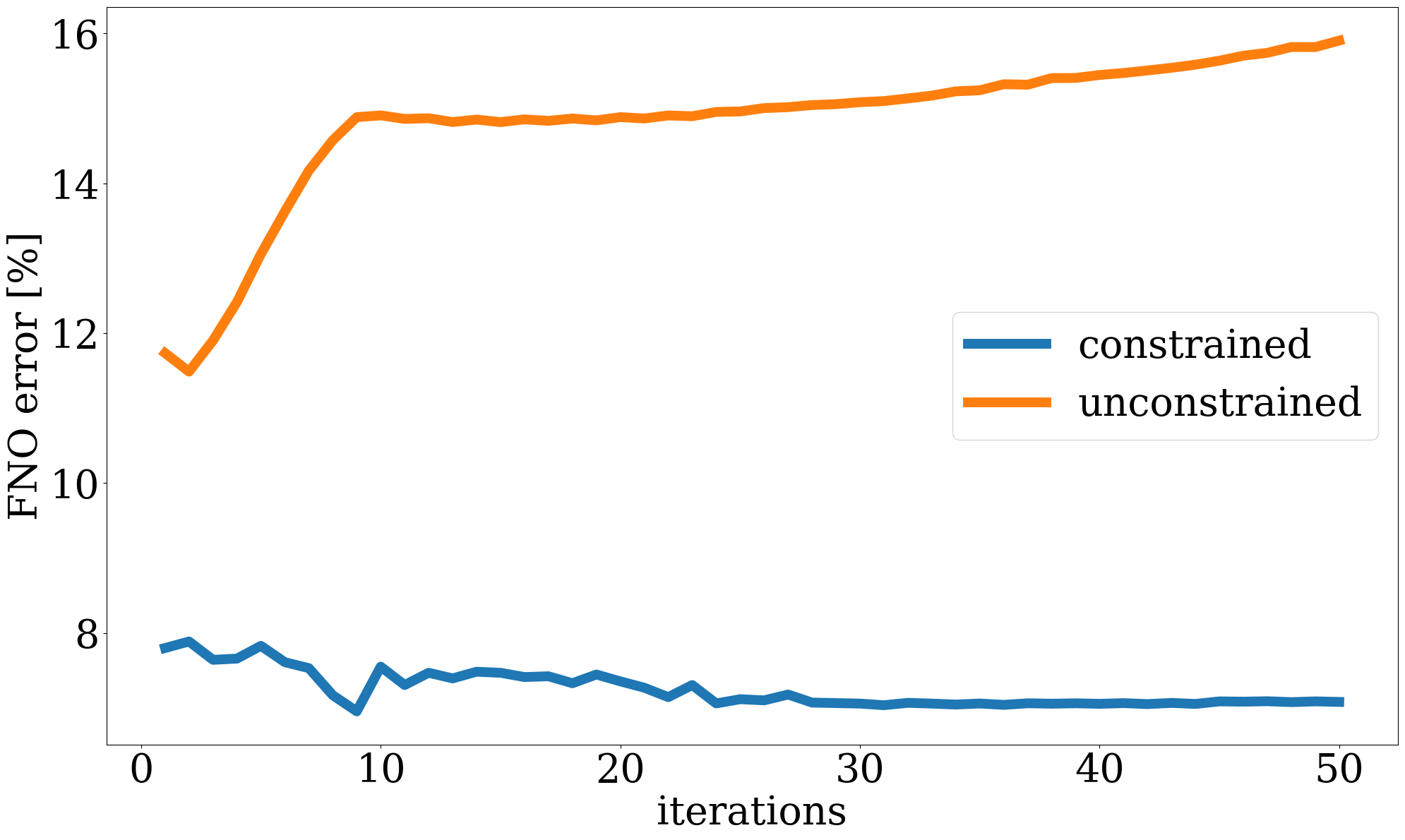}

}

}

\subcaption{\label{fig-end2end-inv-fno-loss}}
\end{minipage}%

\caption{\label{fig-end2end-inv}Permeability inversions from time-lapse
seismic data. \emph{(a)} Inversion result using PDE solvers. \emph{(b)}
The same as \emph{(a)} but for the FNO surrogate. \emph{(c)} The same as
\emph{(a)} but with the NF-based constraint. \emph{(d)} The same as
\emph{(a)} but now for the FNO surrogate with the NF-based constraint.
\emph{(e)}-\emph{(h)} The error of the permeability inversion results in
\emph{(a)}-\emph{(d)} compared to the unseen ground truth shown in
Figure~\ref{fig-true-perm}. \emph{(i)} The FNO prediction errors as a
function of the number of iterations for \emph{(b)} and \emph{(d)}.}

\end{figure}

Inversion results obtained by solving the PDEs for the fluid flow during
the inversion are shown in Figure~\ref{fig-end2end-inv-jutul} and
Figure~\ref{fig-end2end-inv-jutul-nf}. As before, the inversions benefit
majorly from adding the trained NF as a constraint. Remarkably, the
end-to-end inversion results shown in
Figure~\ref{fig-end2end-inv-jutul},
Figure~\ref{fig-end2end-inv-jutul-nf}, and
Figure~\ref{fig-end2end-inv-fno-nf} are close to the results plotted in
Figure~\ref{fig-flow-inv-jutul-all},
Figure~\ref{fig-flow-inv-jutul-nf-all}, and
Figure~\ref{fig-flow-inv-fno-nf-all}, which was obtained with access to
the CO\textsubscript{2} saturation everywhere. This reaffirms the notion
that time-lapse seismic can indeed provide useful spatial information
away from the monitoring wells to estimate the reservoir permeability,
which aligns with earlier observations by D. Li et al. (2020), Yin et
al. (2022), and Louboutin, Yin, et al. (2023). Juxtaposing the results
for the FNO surrogate without (Figure~\ref{fig-end2end-inv-fno}) and
with the constraint (Figure~\ref{fig-end2end-inv-fno-nf}) again
underlines the importance of adding constraints especially in situations
where the forward (wave) operator has a non-trivial nullspace. The
presence of this nullspace has a detrimental affect on the unconstrained
result obtained by the FNO. Contrary to solutions yielded by the PDE,
trained FNOs offer little to no control on the feasibility of the
solution, which explains the strong artifacts in
Figure~\ref{fig-end2end-inv-fno}. As we can see from
Figure~\ref{fig-end2end-inv-fno-loss}, these artifacts are mainly due to
the FNO-approximation errors that dominate and grow after a few
iterations. Conversely, the errors for the constrained case remain more
or less flatlined between \(7\%\) and \(8\%\). In contrast, using the
trained NF as a learned constraint yields better recovery where the
errors are minor within the plume region and mostly live on the edges,
shown in the second row of Figure~\ref{fig-end2end-inv}.

\hypertarget{jointly-inverting-time-lapse-seismic-data-and-well-measurements}{%
\subsubsection{Jointly inverting time-lapse seismic data and well
measurements}\label{jointly-inverting-time-lapse-seismic-data-and-well-measurements}}

Finally, we consider the most preferred scenario for GCS monitoring,
where multiple modalities of data are jointly inverted for the reservoir
permeability (Huang 2022; M. Liu et al. 2023). In our experiment, we
consider to jointly invert time-lapse seismic data and well measurements
by minimizing the following objective function:
\begin{equation}\protect\hypertarget{eq-inv-combined}{}{
\underset{\mathbf{z}}{\operatorname{minimize}} \quad \|\mathbf{d}_{\mathrm{s}} - \mathcal{F}\circ\mathcal{R}\circ\mathcal{S}_{\boldsymbol\theta^\ast}\circ\mathcal{G}_{\mathbf{w}^\ast}(\mathbf{z})\|_2^2 + \lambda \|\mathbf{d}_{\mathrm{w}}-\mathbf{M}\circ\mathcal{S}_{\boldsymbol\theta^\ast}\circ\mathcal{G}_{\mathbf{w}^\ast}(\mathbf{z})\|_2^2 \quad\text{subject to}\quad\|\mathbf{z}\|_2\leq \tau.
}\label{eq-inv-combined}\end{equation}

\begin{figure}

\begin{minipage}[t]{0.25\linewidth}

{\centering 

\raisebox{-\height}{

\includegraphics{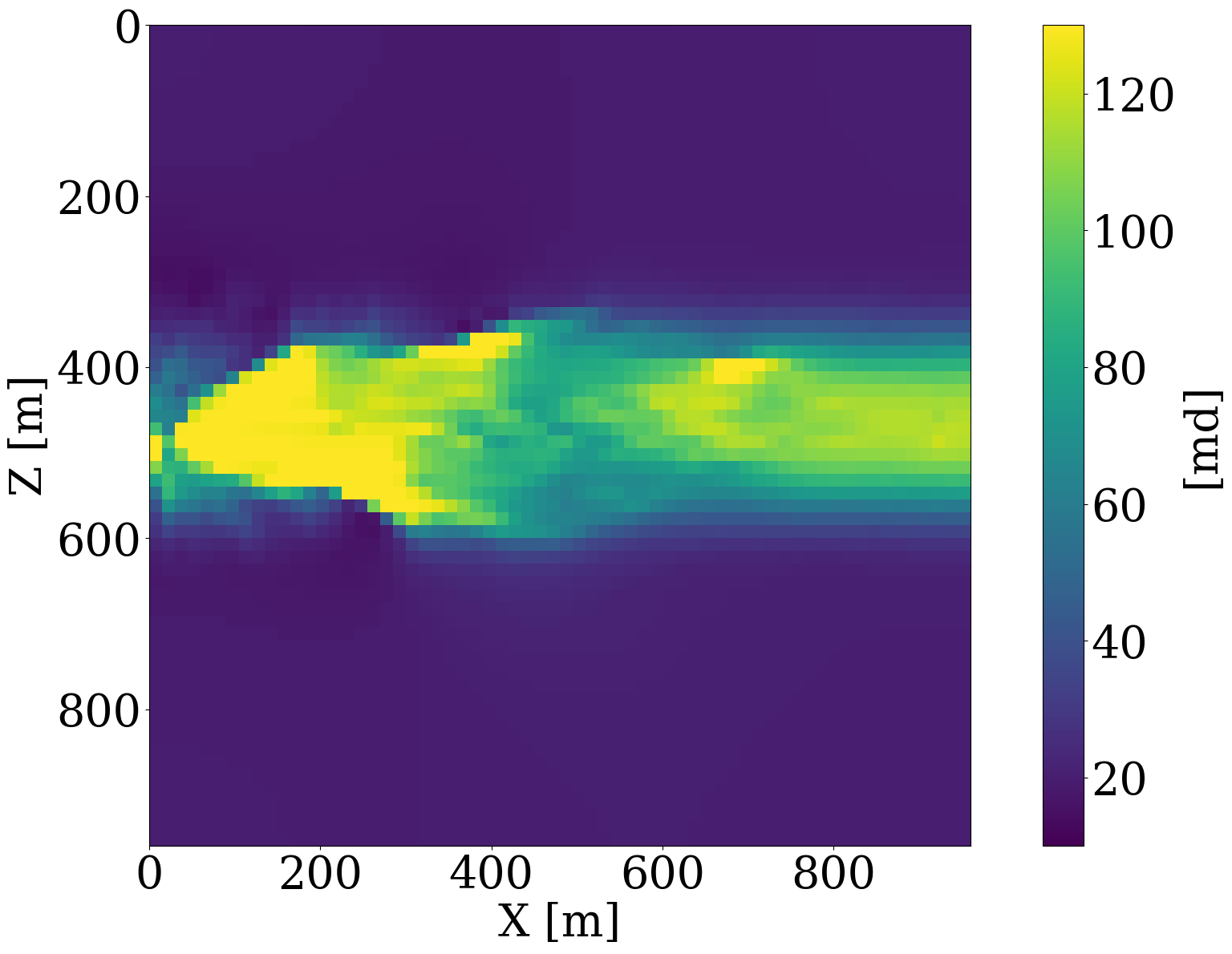}

}

}

\subcaption{\label{fig-combine-inv-jutul}}
\end{minipage}%
\begin{minipage}[t]{0.25\linewidth}

{\centering 

\raisebox{-\height}{

\includegraphics{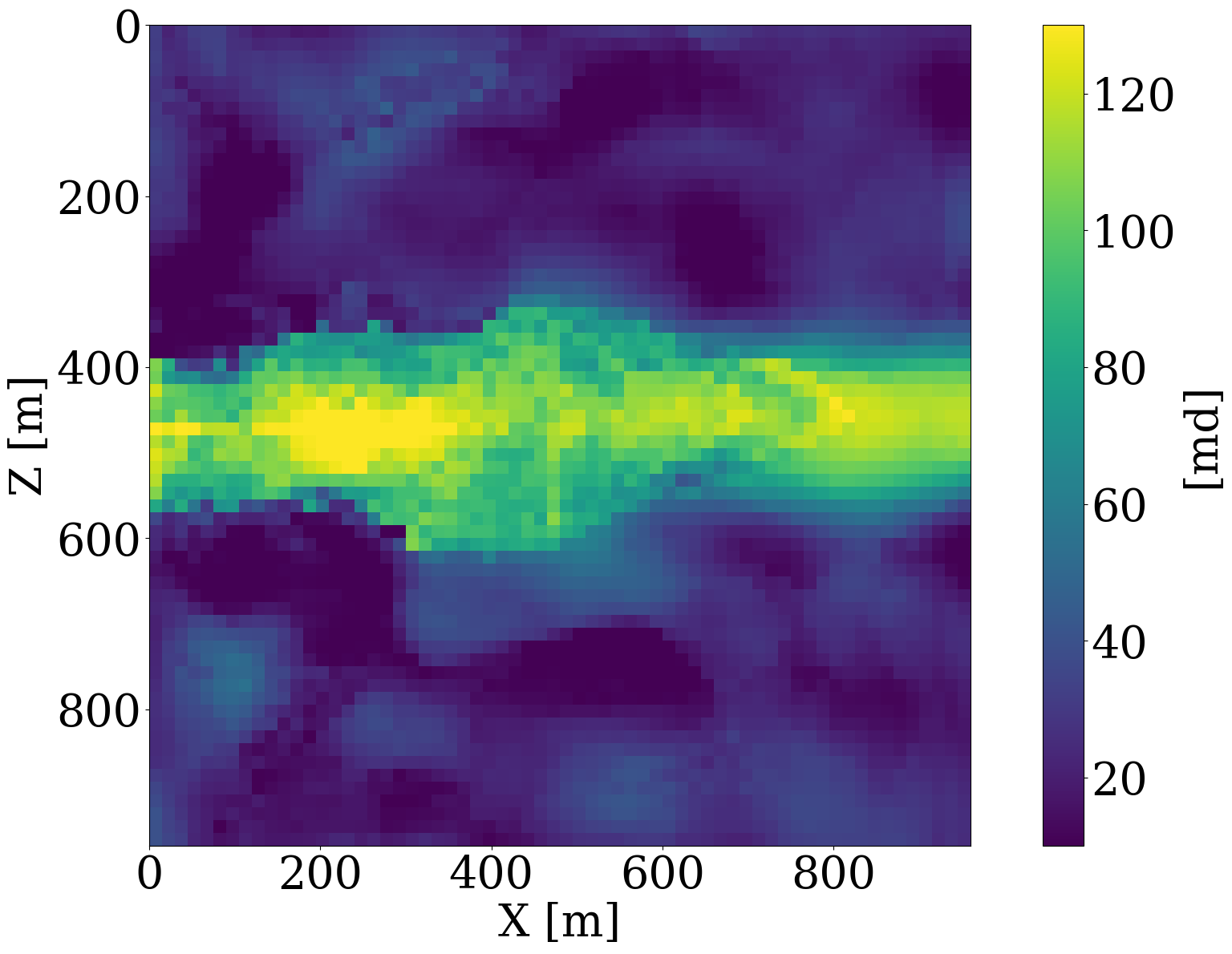}

}

}

\subcaption{\label{fig-combine-inv-fno}}
\end{minipage}%
\begin{minipage}[t]{0.25\linewidth}

{\centering 

\raisebox{-\height}{

\includegraphics{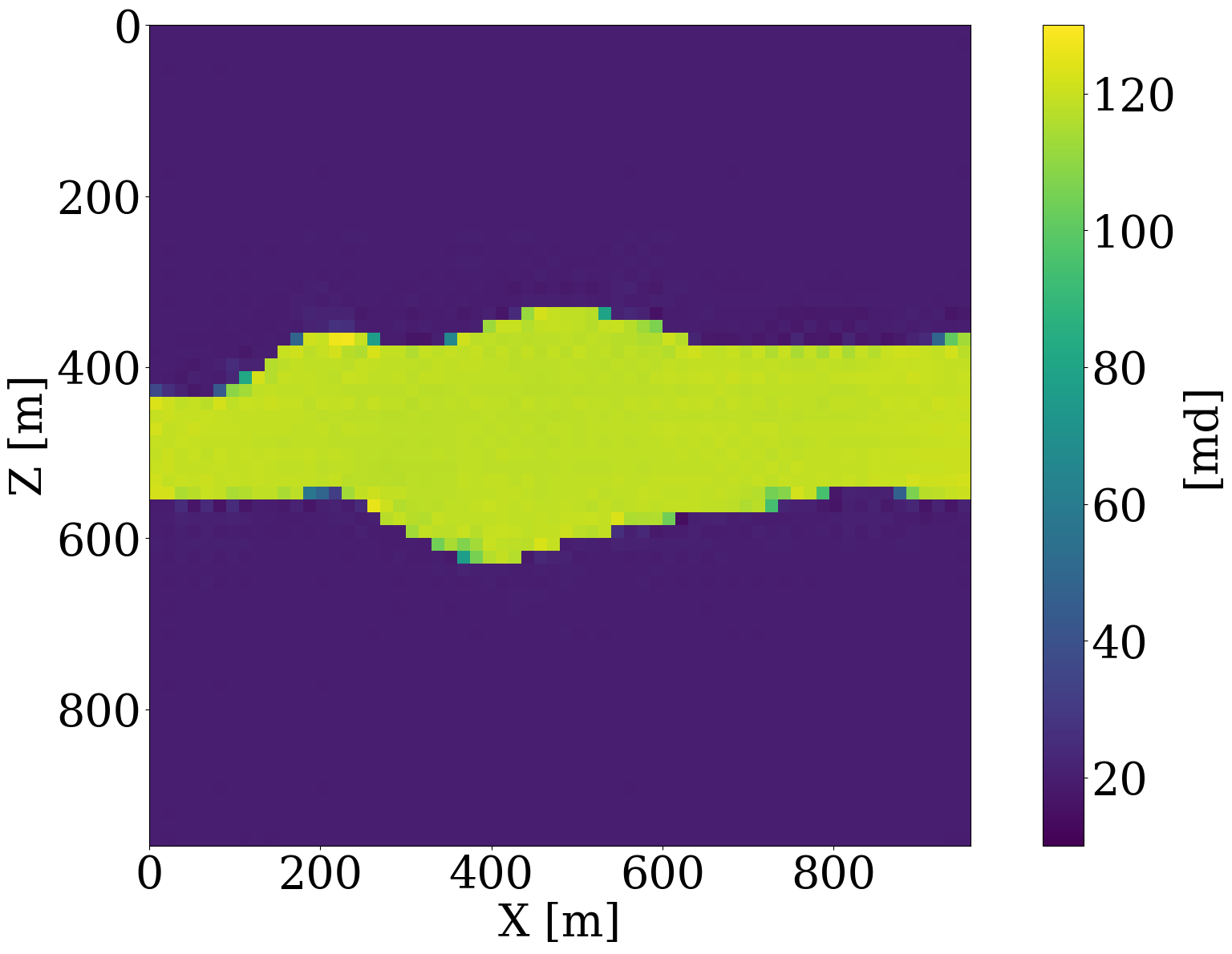}

}

}

\subcaption{\label{fig-combine-inv-jutul-nf}}
\end{minipage}%
\begin{minipage}[t]{0.25\linewidth}

{\centering 

\raisebox{-\height}{

\includegraphics{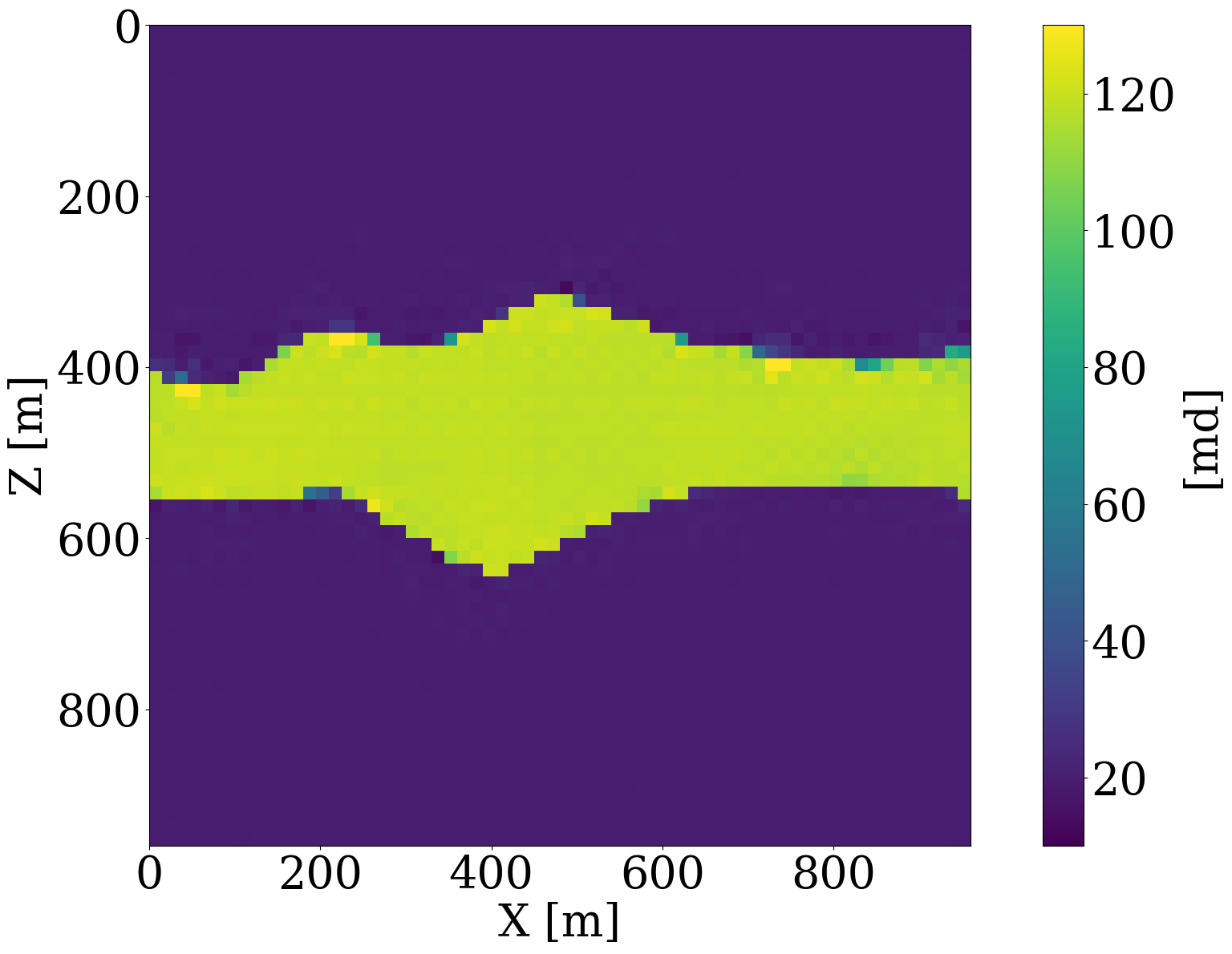}

}

}

\subcaption{\label{fig-combine-inv-fno-nf}}
\end{minipage}%
\newline
\begin{minipage}[t]{0.25\linewidth}

{\centering 

\raisebox{-\height}{

\includegraphics{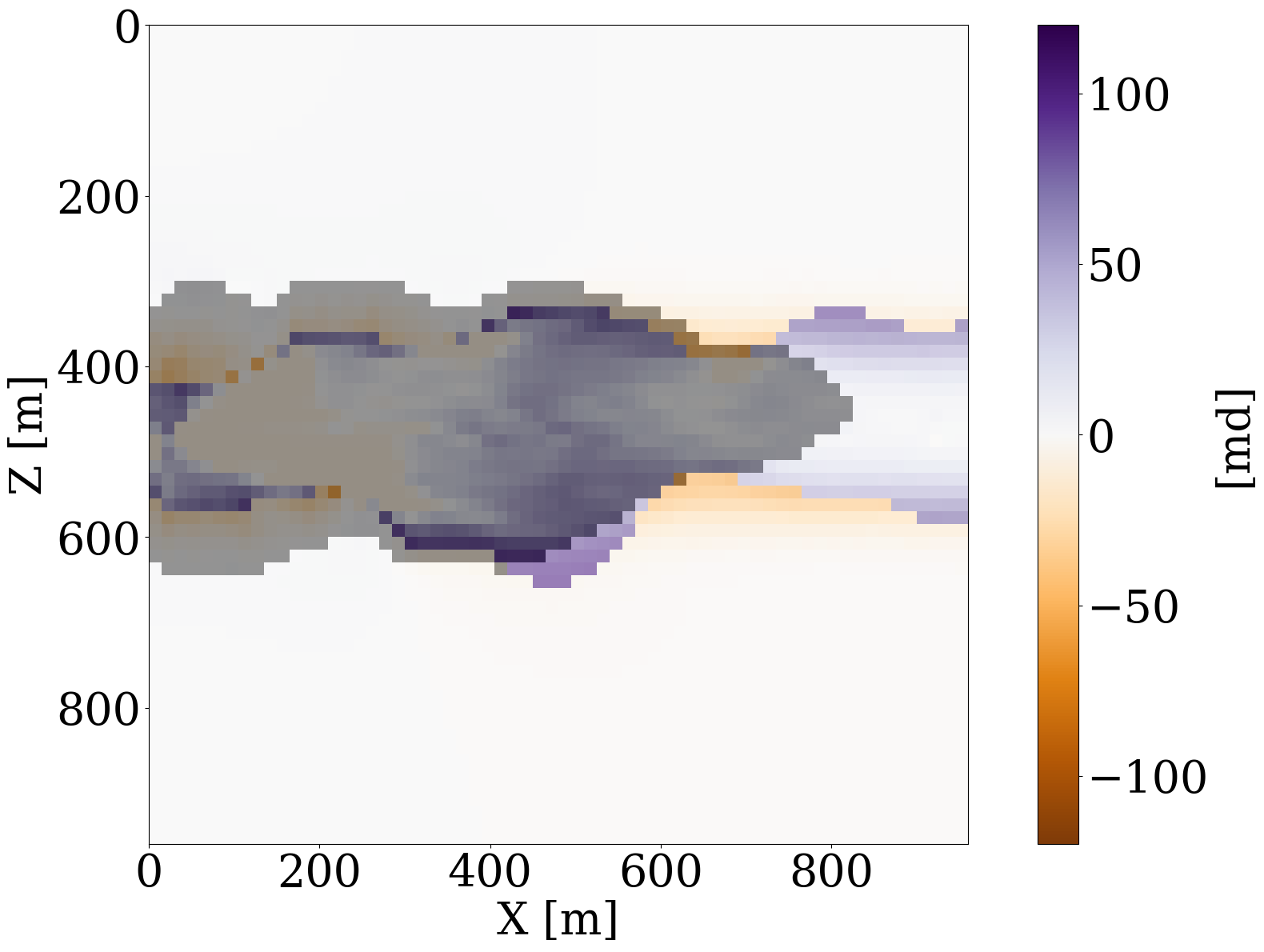}

}

}

\subcaption{\label{fig-combine-flow-inv-jutul}}
\end{minipage}%
\begin{minipage}[t]{0.25\linewidth}

{\centering 

\raisebox{-\height}{

\includegraphics{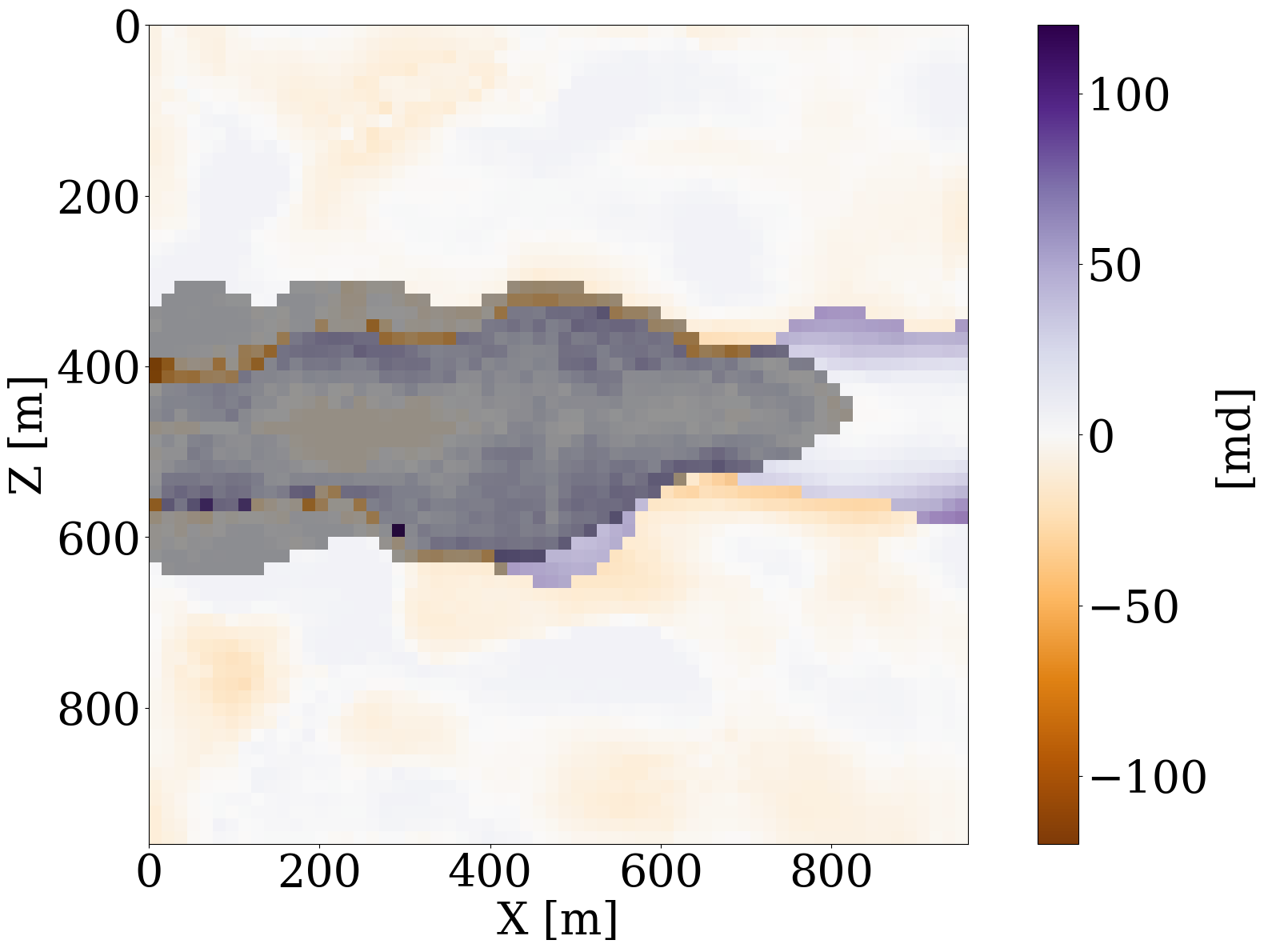}

}

}

\subcaption{\label{fig-diff-combine-flow-inv-fno}}
\end{minipage}%
\begin{minipage}[t]{0.25\linewidth}

{\centering 

\raisebox{-\height}{

\includegraphics{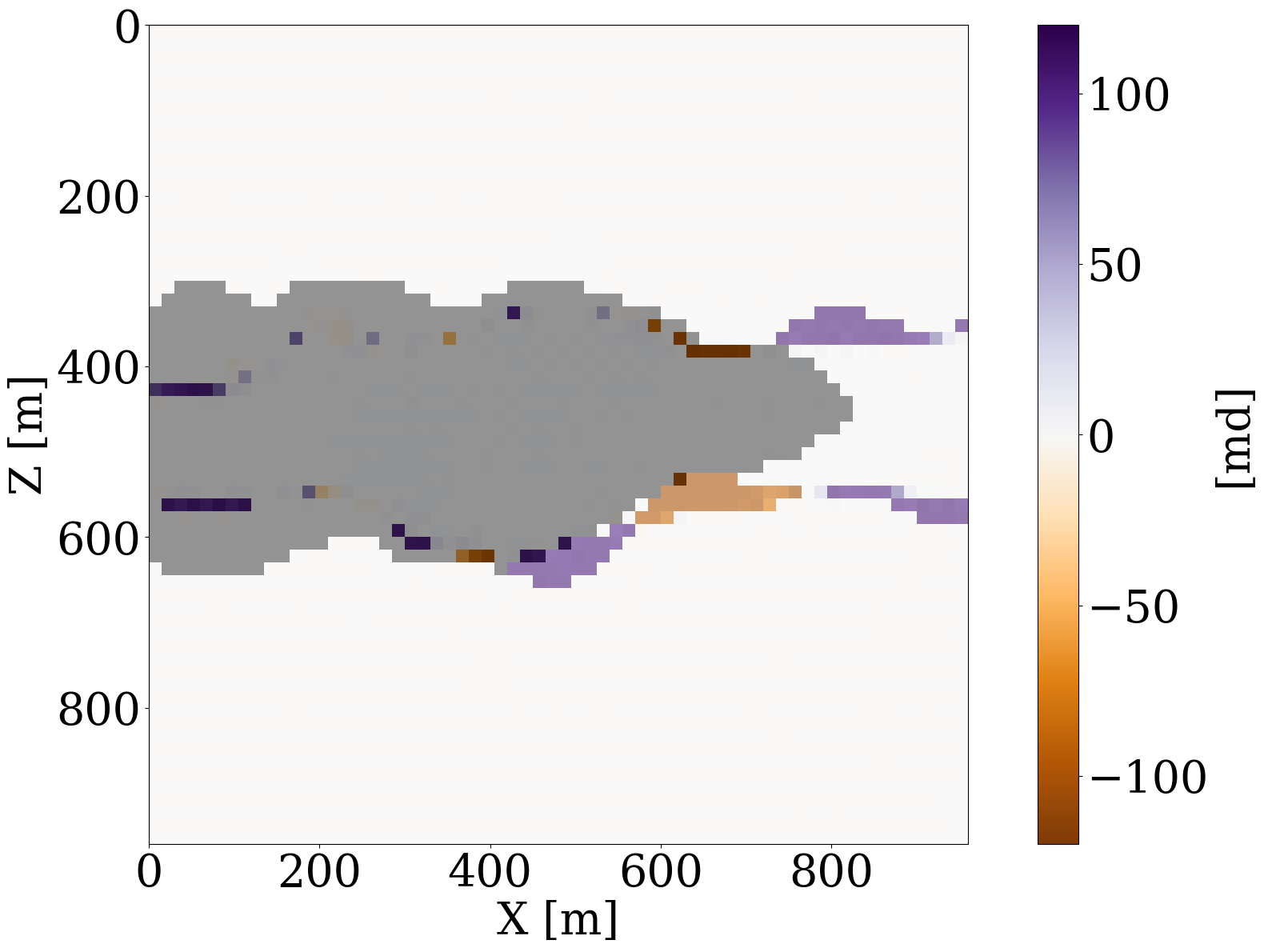}

}

}

\subcaption{\label{fig-diff-combine-flow-inv-jutul-nf}}
\end{minipage}%
\begin{minipage}[t]{0.25\linewidth}

{\centering 

\raisebox{-\height}{

\includegraphics{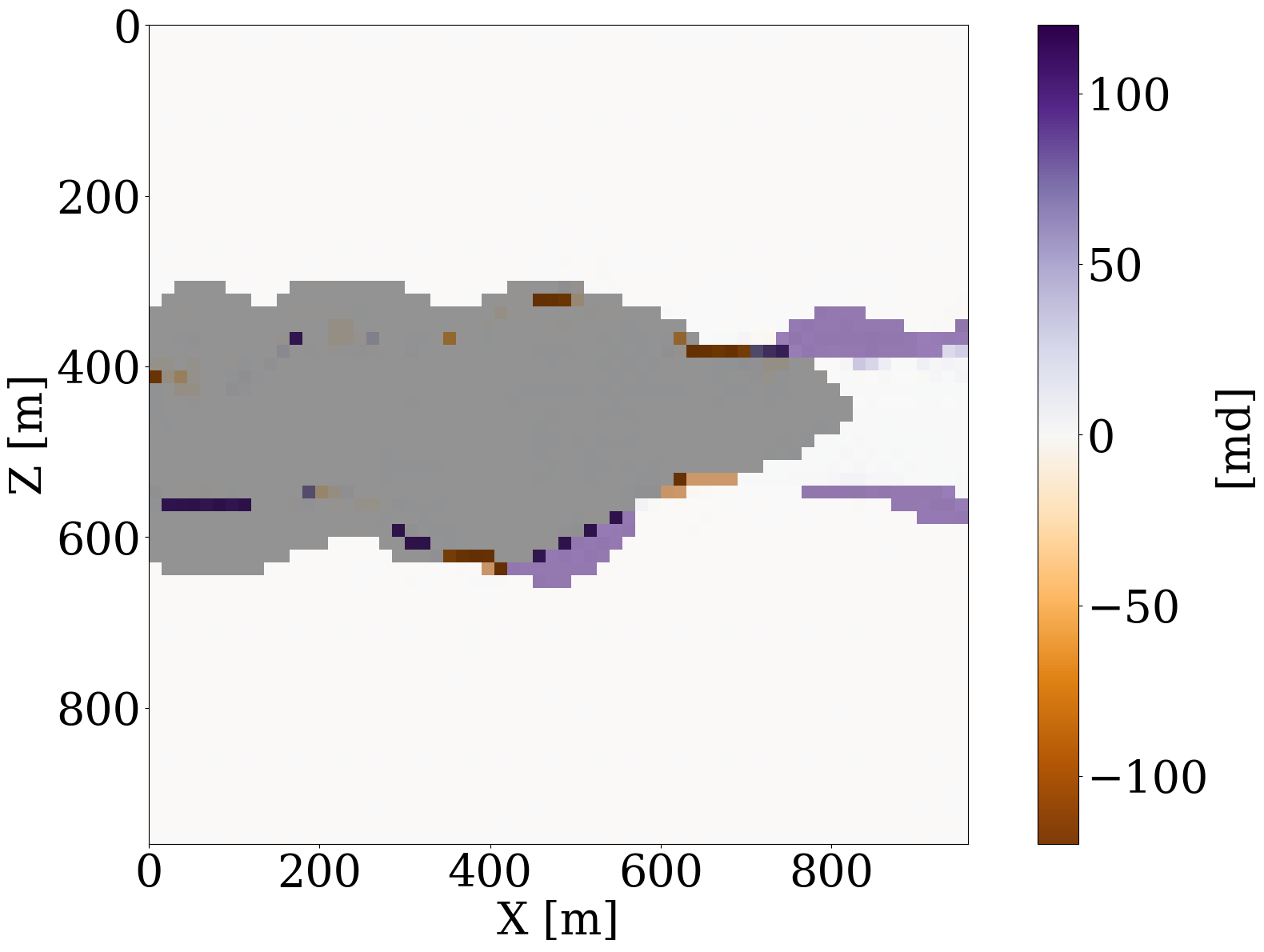}

}

}

\subcaption{\label{fig-diff-combine-flow-inv-fno-nf}}
\end{minipage}%
\newline
\begin{minipage}[t]{\linewidth}

{\centering 

\raisebox{-\height}{

\includegraphics{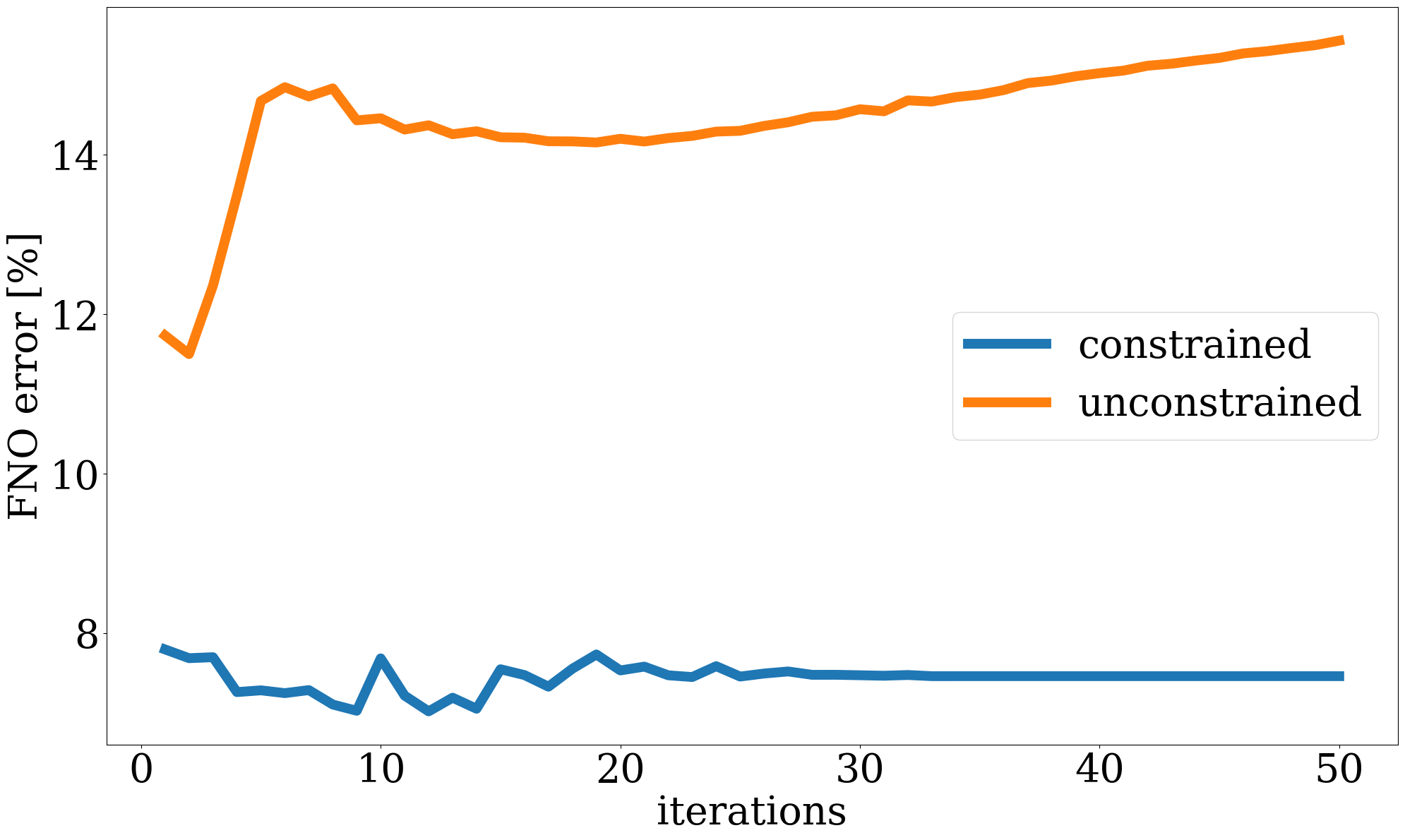}

}

}

\subcaption{\label{fig-combine-inv-fno-loss}}
\end{minipage}%

\caption{\label{fig-combine-inv}Joint permeability inversions from both
time-lapse seismic data and time-lapse well measurements. \emph{(a)}
Inversion result using PDE solvers. \emph{(b)} The same as \emph{(a)}
but for the FNO surrogate. \emph{(c)} The same as \emph{(a)} but with
the NF-based constraint. \emph{(d)} The same as \emph{(a)} but now for
the FNO surrogate with the NF-based constraint. \emph{(e)}-\emph{(h)}
The error of the permeability inversions in \emph{(a)}-\emph{(d)},
compared to the unseen ground truth shown in Figure~\ref{fig-true-perm}.
\emph{(i)} The FNO prediction errors as a function of the number of
iterations for \emph{(b)} and \emph{(d)}.}

\end{figure}

This objective function includes both the time-lapse seismic data misfit
from Equation~\ref{eq-inv-end2end2end} and the time-lapse well
measurement misfit from Equation~\ref{eq-inv-well} with a balancing term
\(\lambda\). While better choices can be made, we select this
\(\lambda\) in our numerical experiment to be \(10\), so that the
magnitudes of the two terms are relatively the same. The inversion
results and differences from the unseen ground truth permeability are
shown in Figure~\ref{fig-combine-inv}, where we again observe large
artifacts for the recovery when FNO surrogate is inverted without NF
constraints. This behavior is confirmed by the plot for the FNO error
curve as a function of the number of iterations. This error finally
reaches a value over \(15\%\).

We report quantitative measures for the permeability inversions for all
optimization methods and different types of observed data in
Table~\ref{tbl-snrs-perm} for the signal-to-noise ratios (S/Ns) and the
structural similarity index measure (SSIM, Wang et al. 2004). To avoid
undue influence of the null space for the permeability, we only
calculate the S/N and SSIM values based on the parts of the models that
are touched by CO\textsubscript{2} plume. From these values, following
observations can be made. First, the NF-constrained permeability
inversion are superior in both S/Ns and SSIMs, which demonstrates the
efficacy of the learned constraint. Second, by virtue of this NF
constraint, the results yielded by either the PDE solver or by the FNO
surrogate produce very similar S/Ns and SSIMs. This behavior reaffirms
that the trained FNO behavior is similar to the behavior yielded by PDE
solver when its inputs remain in-distribution, which is controlled by
the NF constraints.

\hypertarget{tbl-snrs-perm}{}
\begin{longtable}[]{@{}
  >{\raggedright\arraybackslash}p{(\columnwidth - 6\tabcolsep) * \real{0.4286}}
  >{\raggedright\arraybackslash}p{(\columnwidth - 6\tabcolsep) * \real{0.2041}}
  >{\raggedright\arraybackslash}p{(\columnwidth - 6\tabcolsep) * \real{0.2041}}
  >{\raggedright\arraybackslash}p{(\columnwidth - 6\tabcolsep) * \real{0.1633}}@{}}
\caption{\label{tbl-snrs-perm}S/N (in dB) and SSIM values of
permeability recovery.}\tabularnewline
\toprule()
\begin{minipage}[b]{\linewidth}\raggedright
Inversion method
\end{minipage} & \begin{minipage}[b]{\linewidth}\raggedright
Well measurement
\end{minipage} & \begin{minipage}[b]{\linewidth}\raggedright
Time-lapse seismic
\end{minipage} & \begin{minipage}[b]{\linewidth}\raggedright
Both
\end{minipage} \\
\midrule()
\endfirsthead
\toprule()
\begin{minipage}[b]{\linewidth}\raggedright
Inversion method
\end{minipage} & \begin{minipage}[b]{\linewidth}\raggedright
Well measurement
\end{minipage} & \begin{minipage}[b]{\linewidth}\raggedright
Time-lapse seismic
\end{minipage} & \begin{minipage}[b]{\linewidth}\raggedright
Both
\end{minipage} \\
\midrule()
\endhead
Unconstrained inversion with PDE solvers & 9.34 dB / 0.67 & 10.50 dB /
0.73 & 10.70 dB / 0.73 \\
Unconstrained inversion with FNO surrogates & 9.64 dB / 0.68 & 11.94 dB
/ 0.72 & 11.98 dB / 0.72 \\
Constrained inversion with PDE solvers & 12.2 dB / 0.77 & 14.18 dB /
0.80 & 15.20 dB / 0.85 \\
Constrained inversion with FNO surrogates & 11.06 dB / 0.74 & 14.16 dB /
0.81 & 14.92 dB / 0.83 \\
\bottomrule()
\end{longtable}

\hypertarget{co2-plume-estimation-and-forecast}{%
\subsubsection{\texorpdfstring{CO\textsubscript{2} plume estimation and
forecast}{CO2 plume estimation and forecast}}\label{co2-plume-estimation-and-forecast}}

\begin{figure}

\begin{minipage}[t]{0.20\linewidth}

{\centering 

\raisebox{-\height}{

\includegraphics{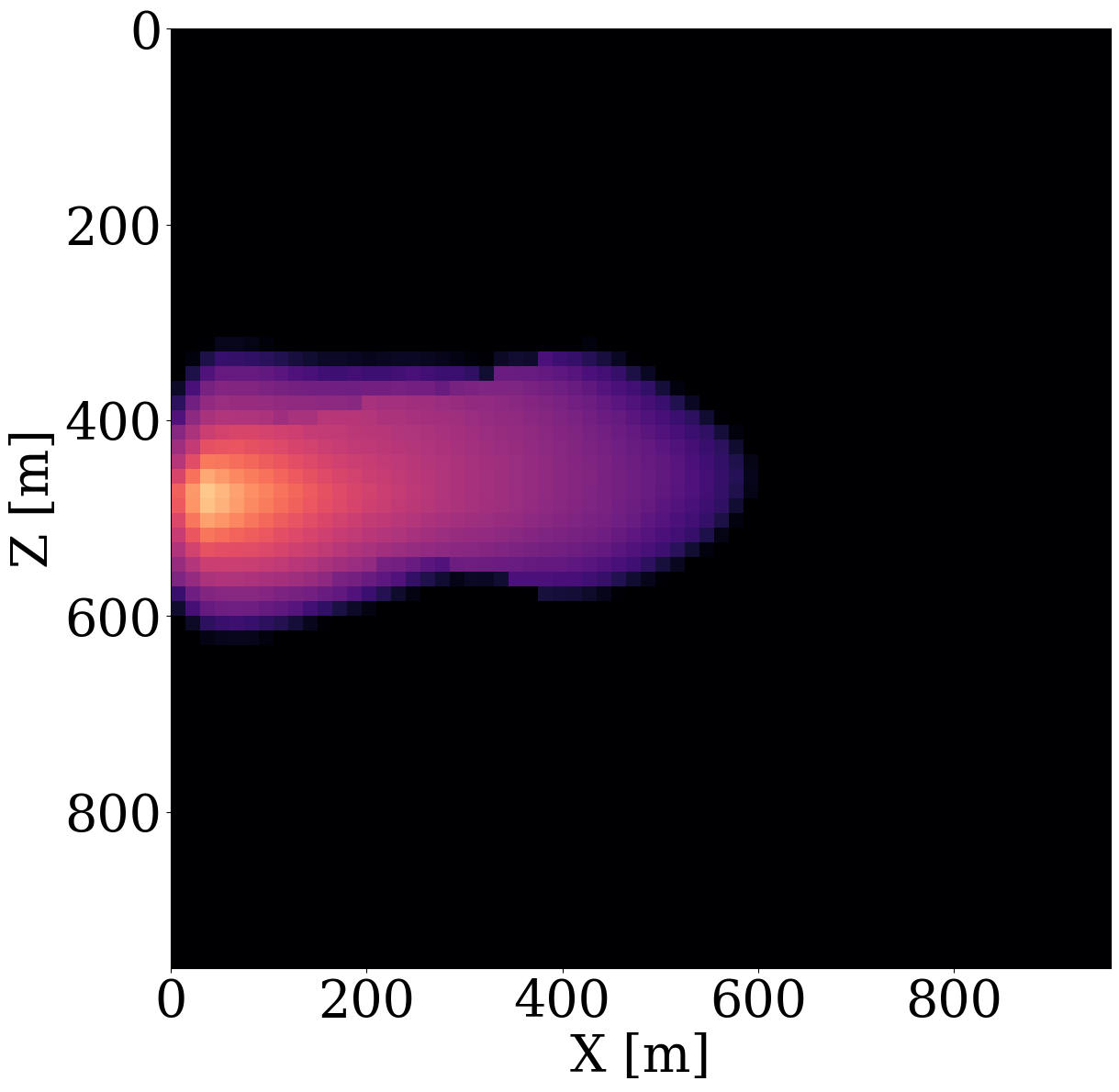}

}

}

\end{minipage}%
\begin{minipage}[t]{0.20\linewidth}

{\centering 

\raisebox{-\height}{

\includegraphics{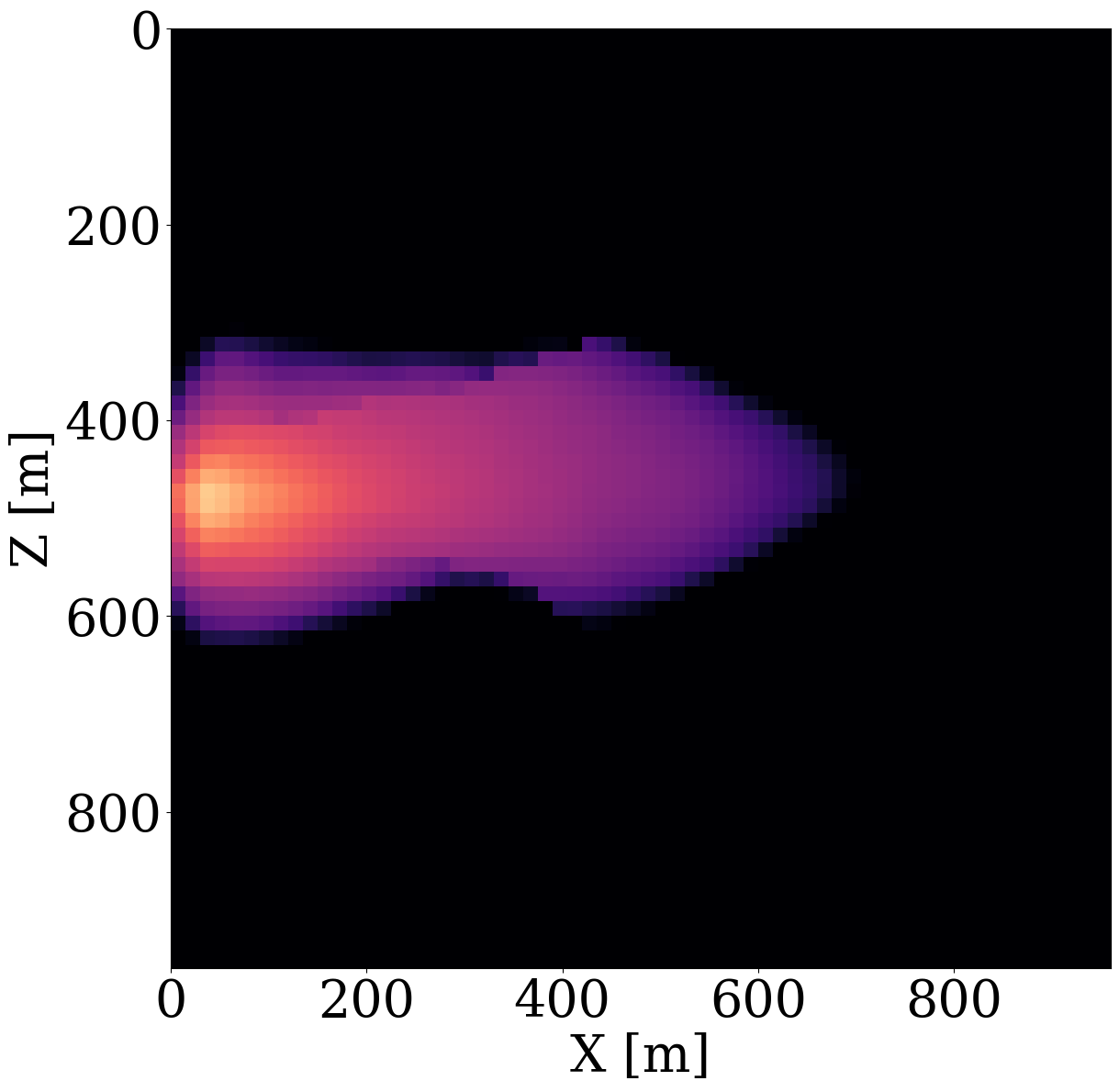}

}

}

\end{minipage}%
\begin{minipage}[t]{0.20\linewidth}

{\centering 

\raisebox{-\height}{

\includegraphics{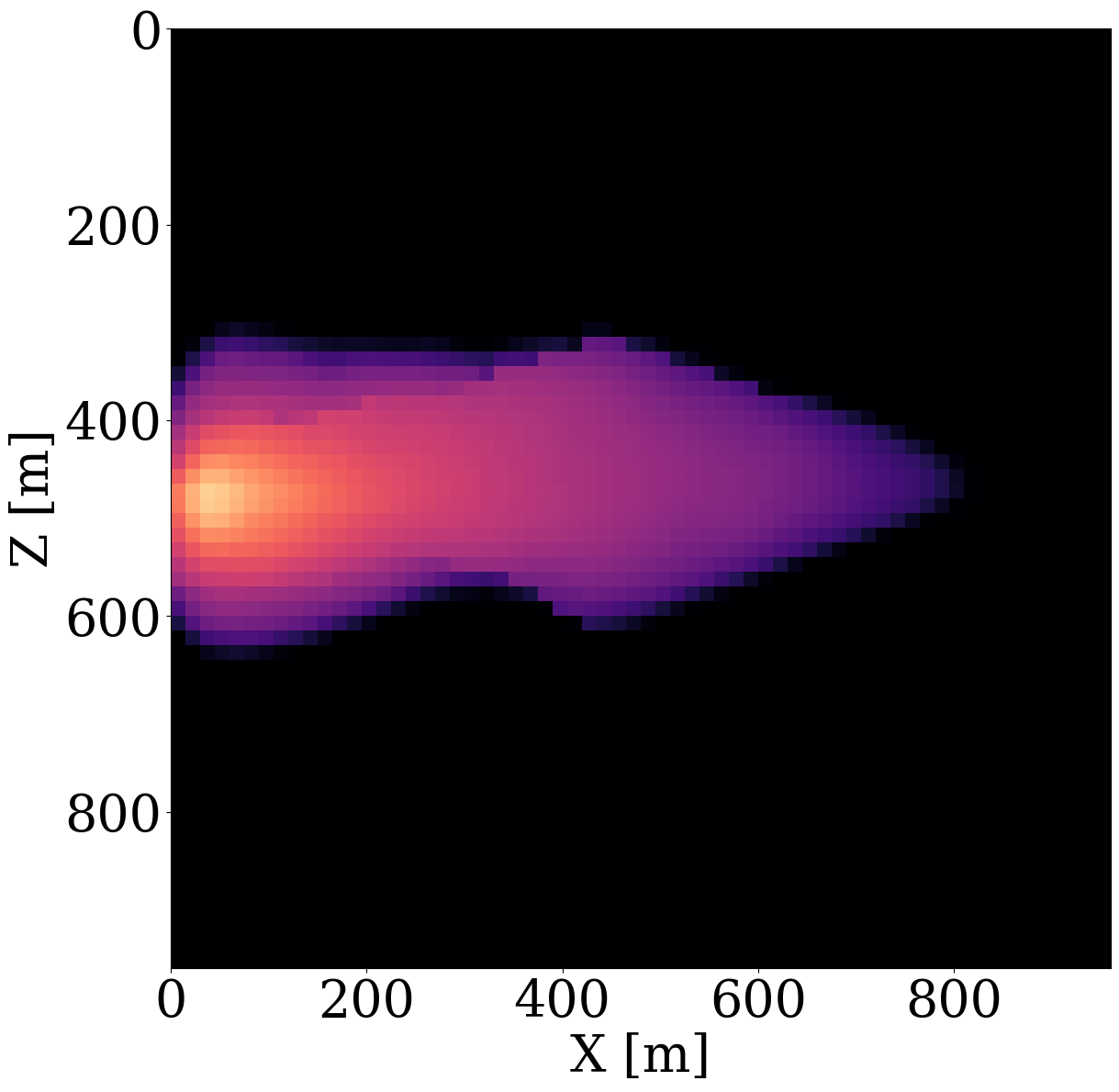}

}

}

\end{minipage}%
\begin{minipage}[t]{0.20\linewidth}

{\centering 

\raisebox{-\height}{

\includegraphics{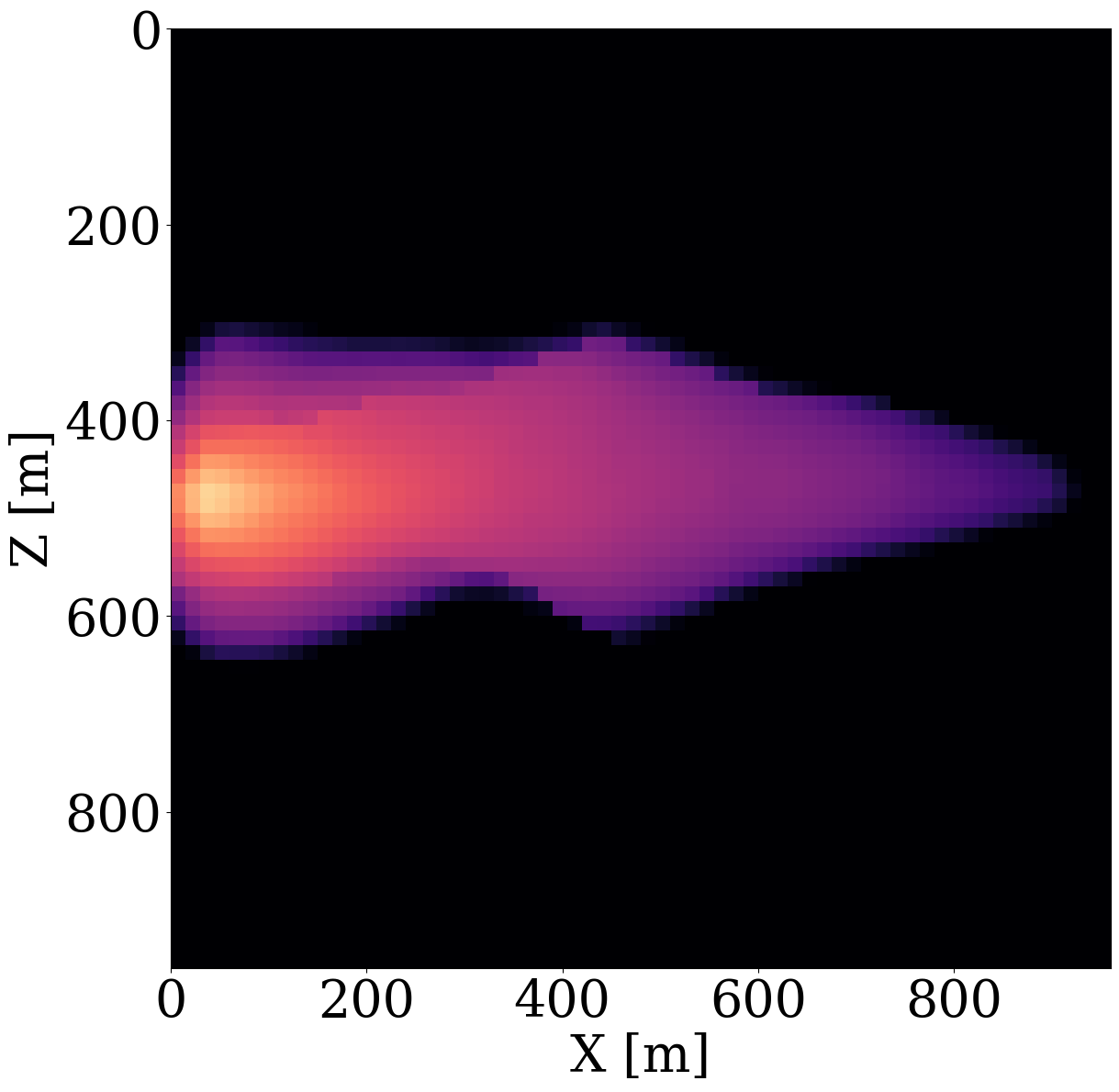}

}

}

\end{minipage}%
\begin{minipage}[t]{0.20\linewidth}

{\centering 

\raisebox{-\height}{

\includegraphics{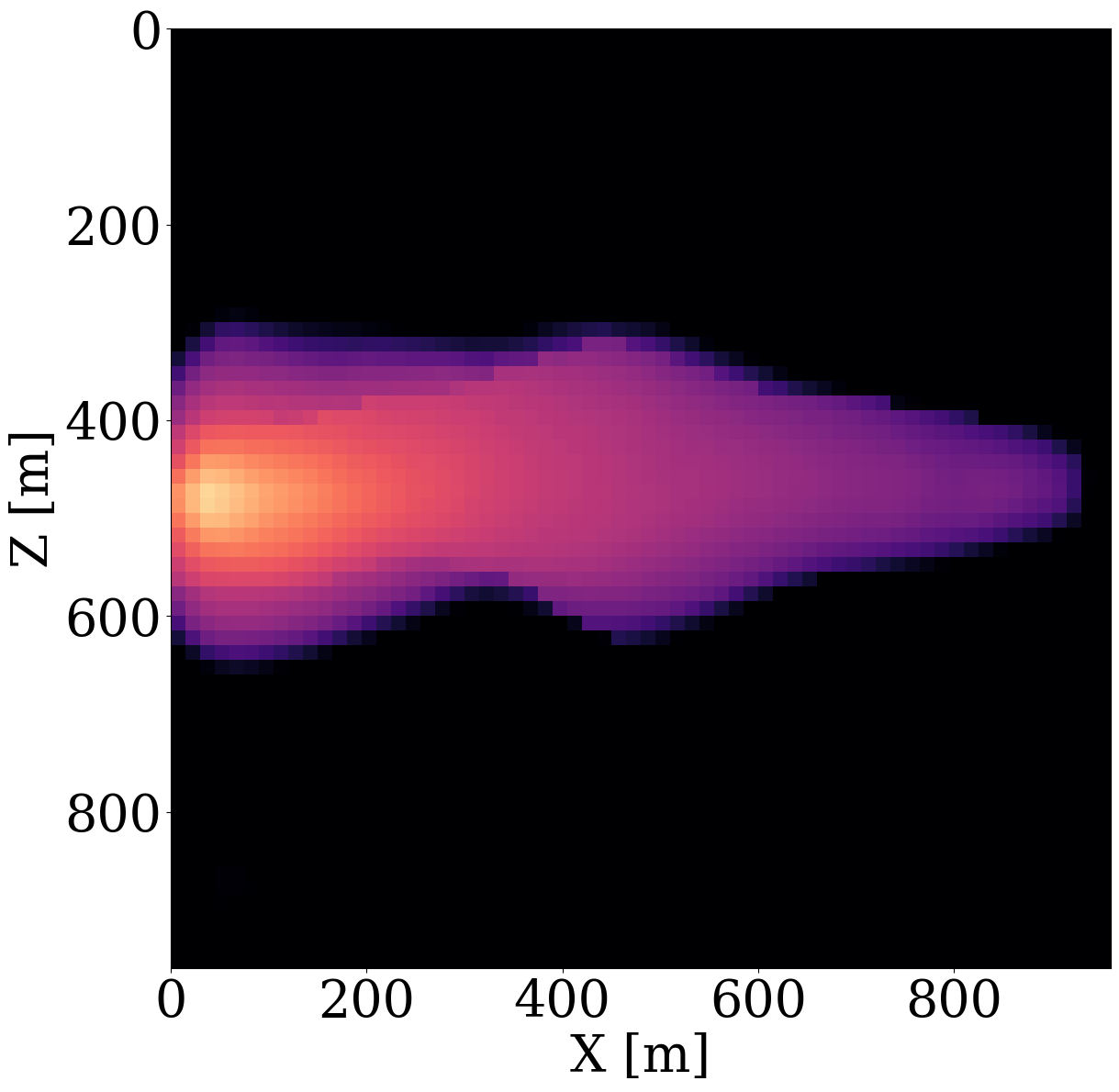}

}

}

\end{minipage}%
\newline
\begin{minipage}[t]{0.20\linewidth}

{\centering 

\raisebox{-\height}{

\includegraphics{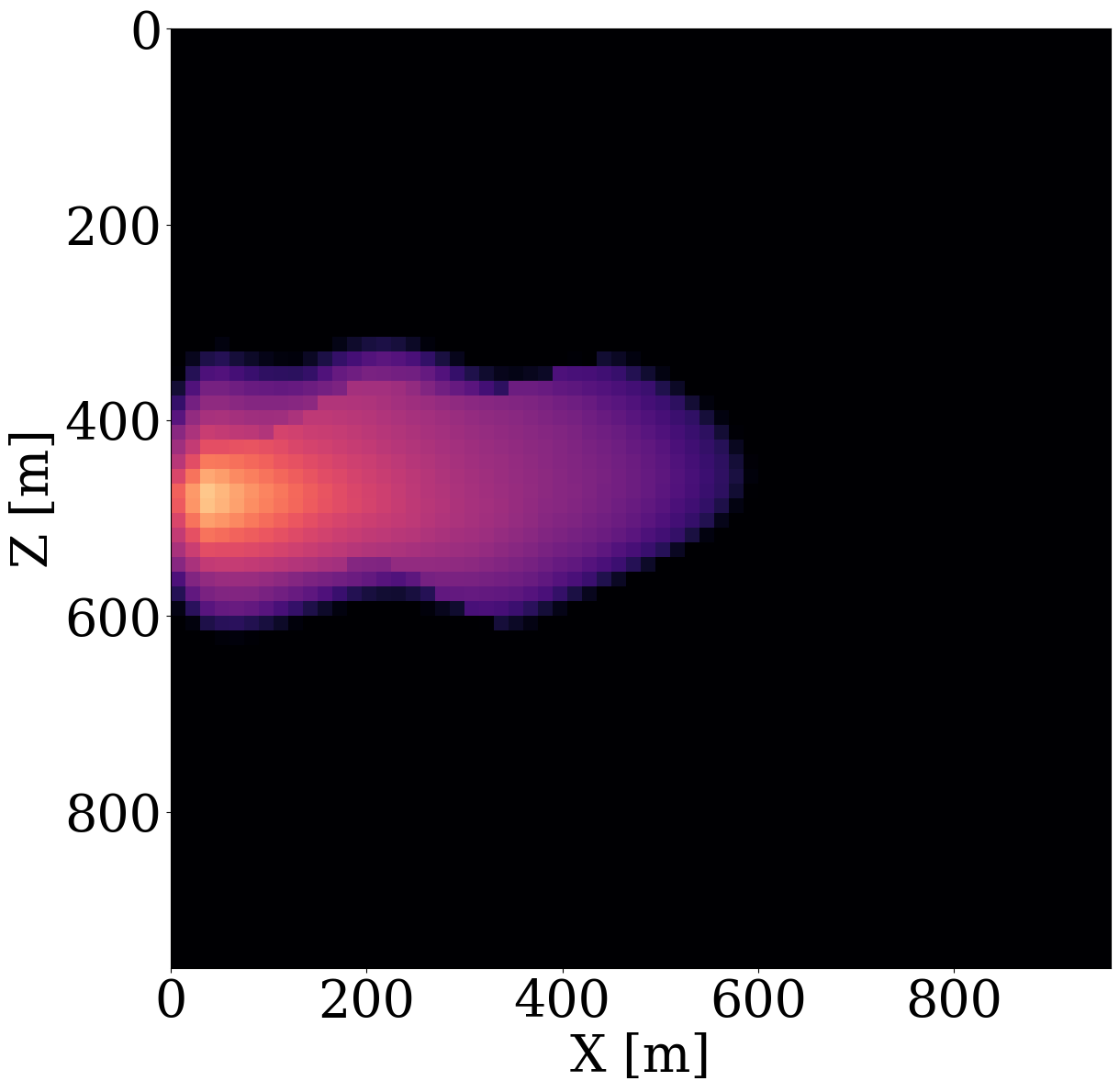}

}

}

\end{minipage}%
\begin{minipage}[t]{0.20\linewidth}

{\centering 

\raisebox{-\height}{

\includegraphics{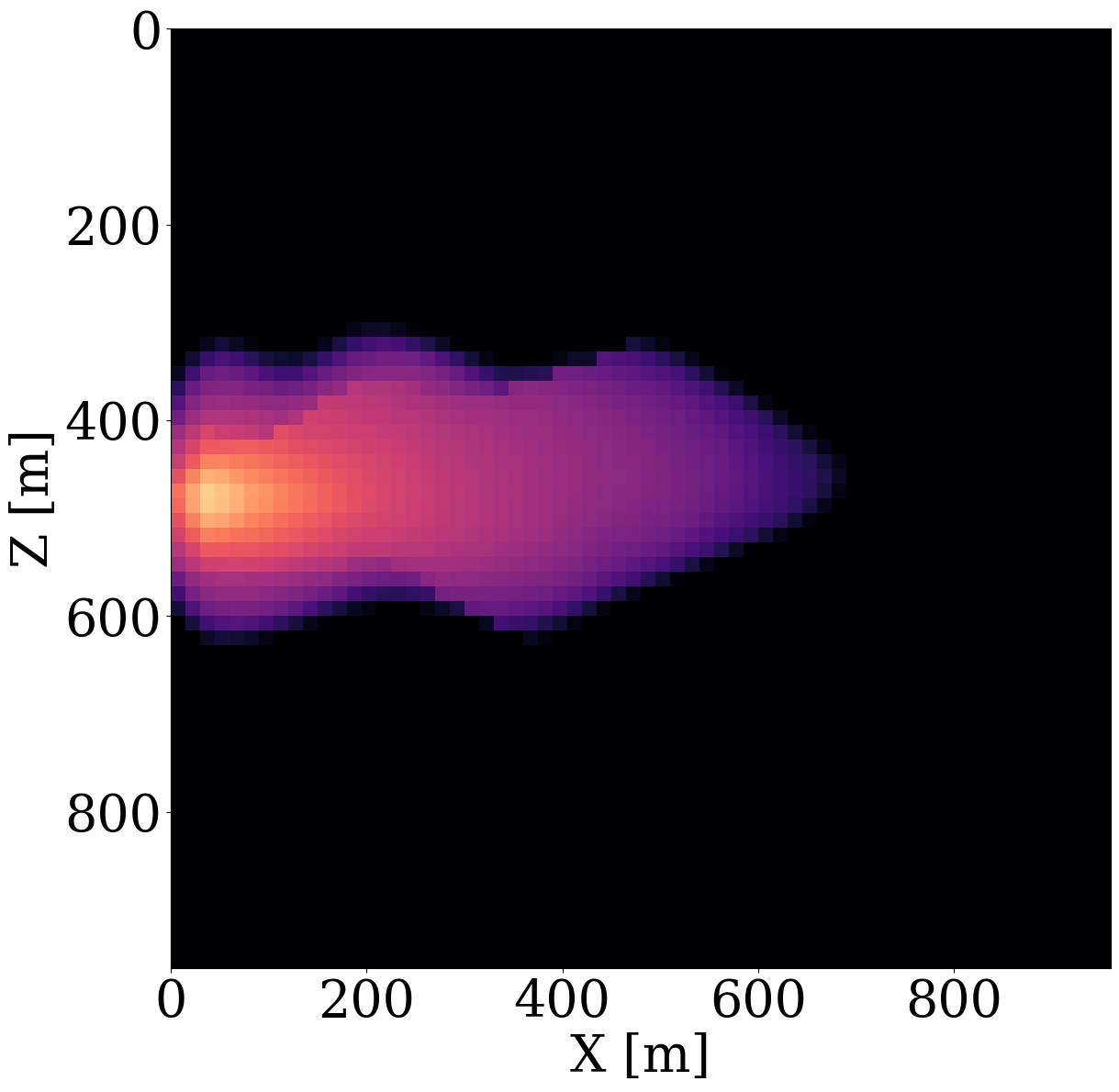}

}

}

\end{minipage}%
\begin{minipage}[t]{0.20\linewidth}

{\centering 

\raisebox{-\height}{

\includegraphics{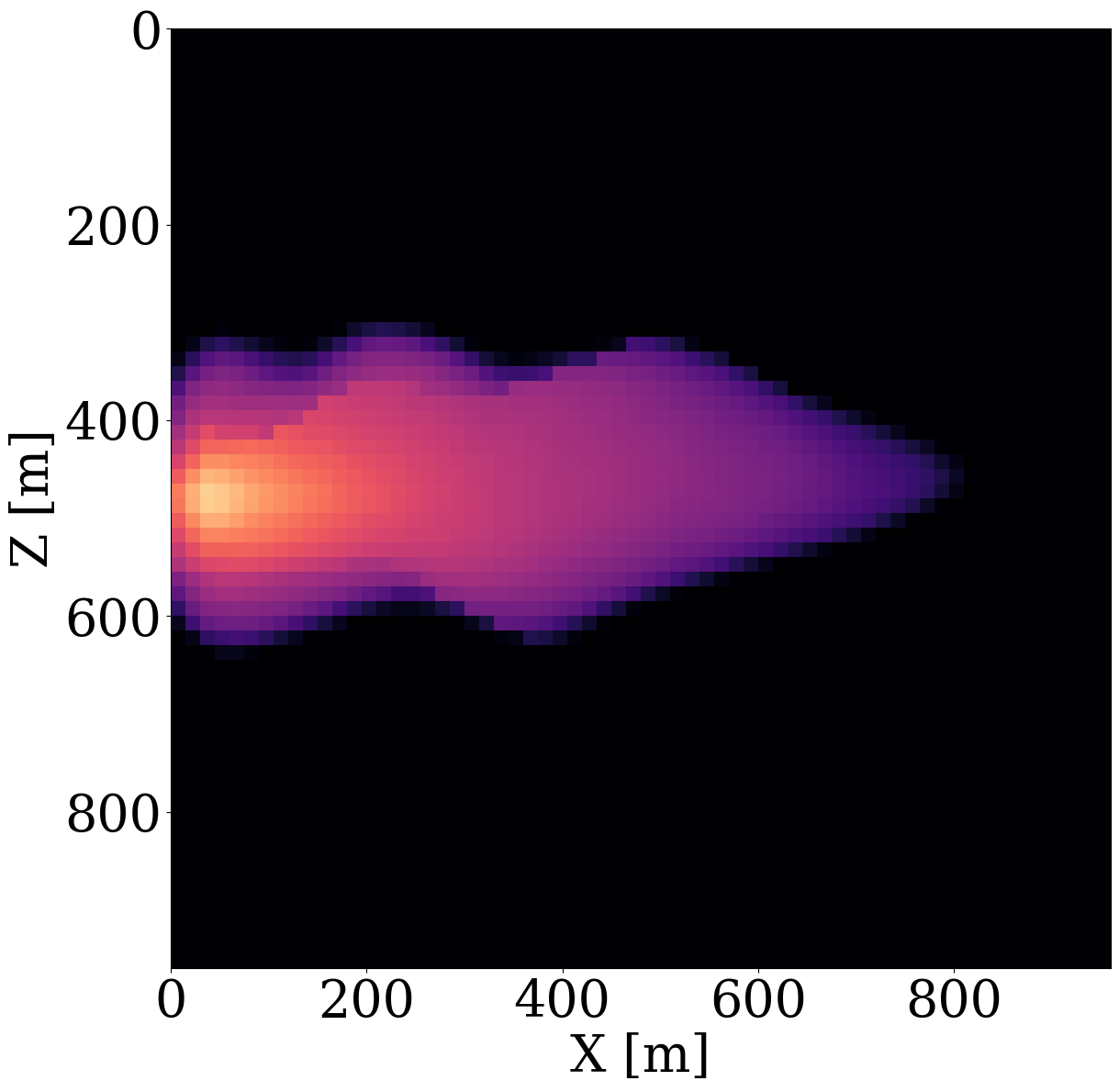}

}

}

\end{minipage}%
\begin{minipage}[t]{0.20\linewidth}

{\centering 

\raisebox{-\height}{

\includegraphics{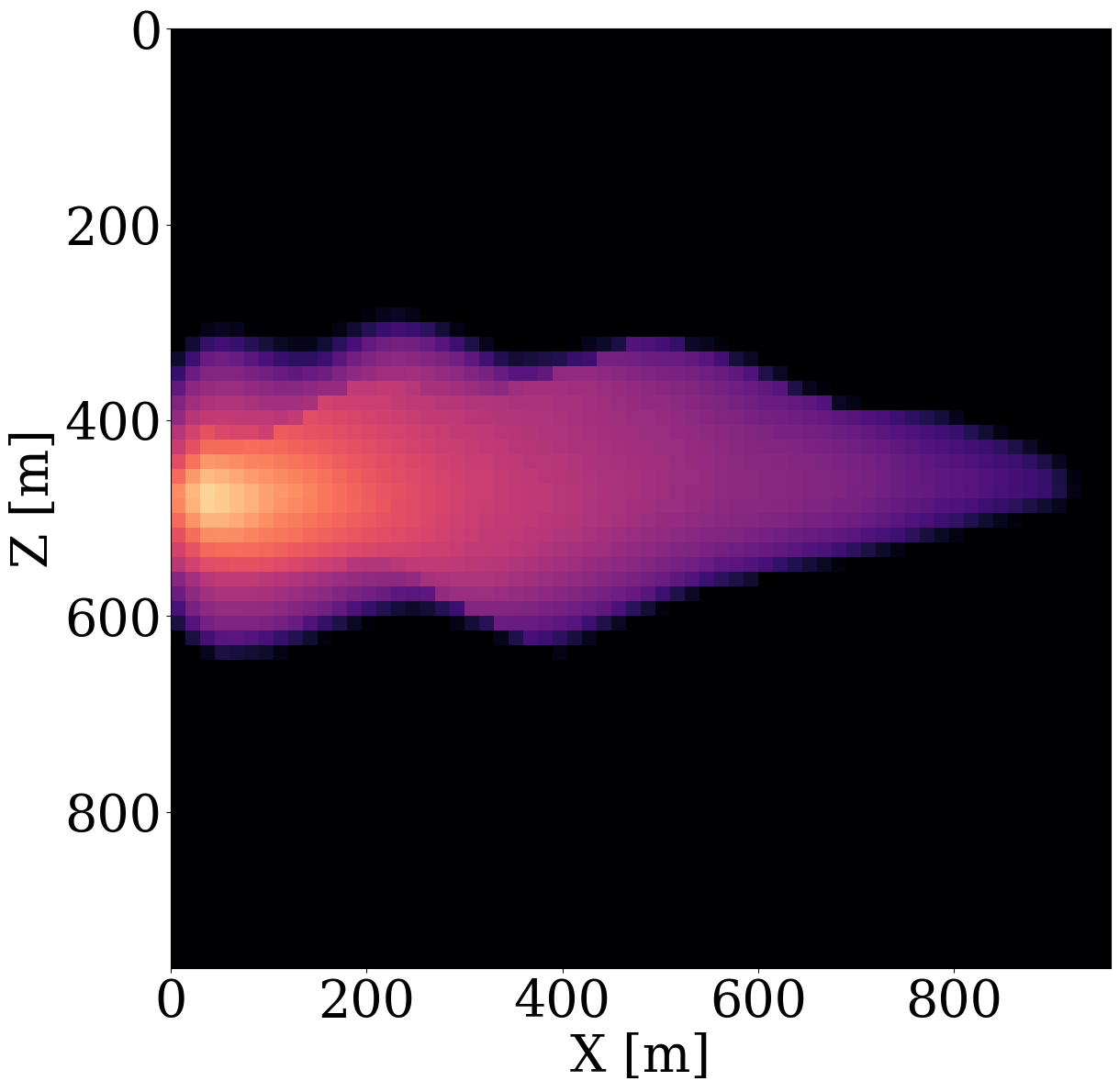}

}

}

\end{minipage}%
\begin{minipage}[t]{0.20\linewidth}

{\centering 

\raisebox{-\height}{

\includegraphics{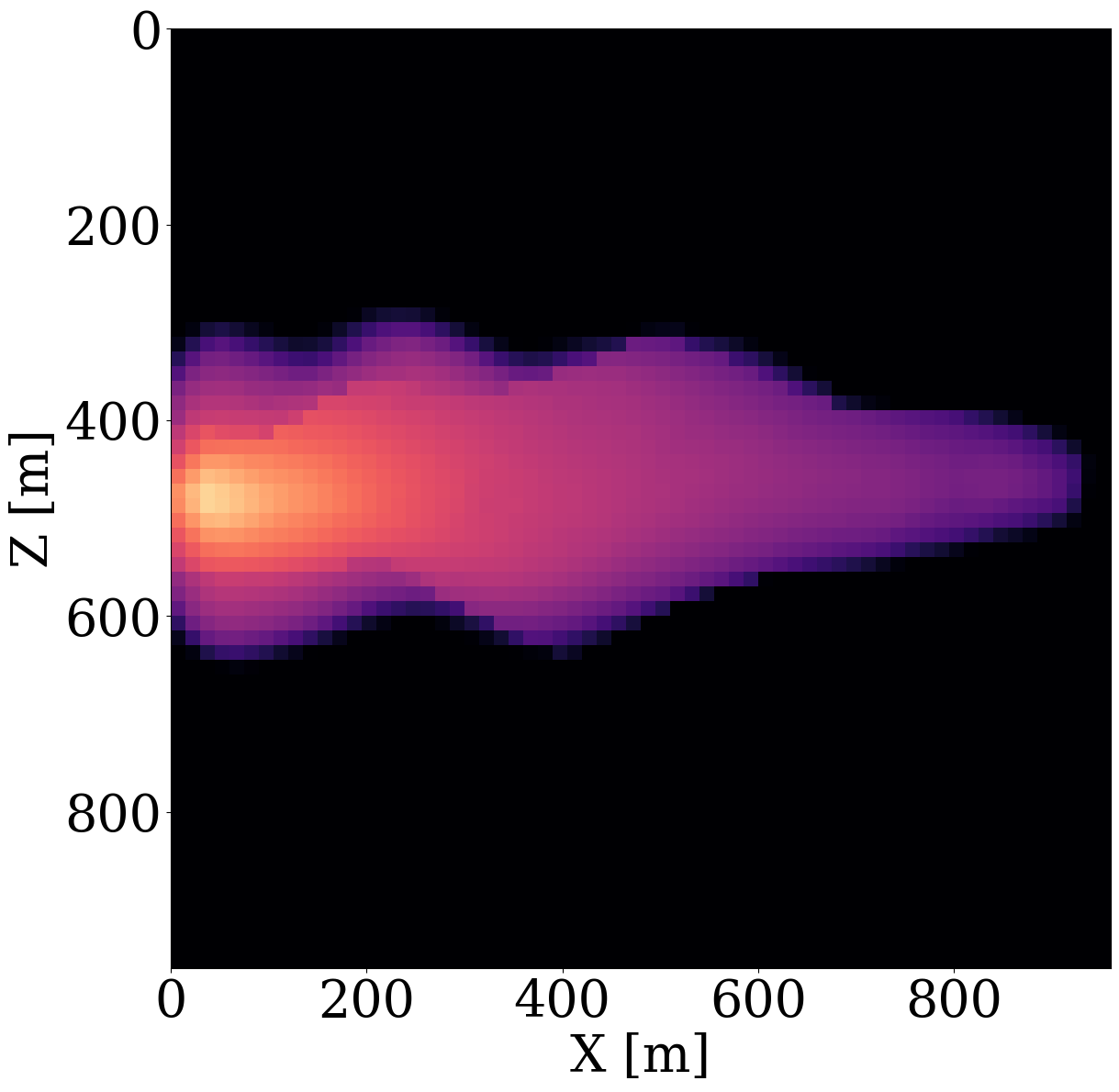}

}

}

\end{minipage}%
\newline
\begin{minipage}[t]{0.20\linewidth}

{\centering 

\raisebox{-\height}{

\includegraphics{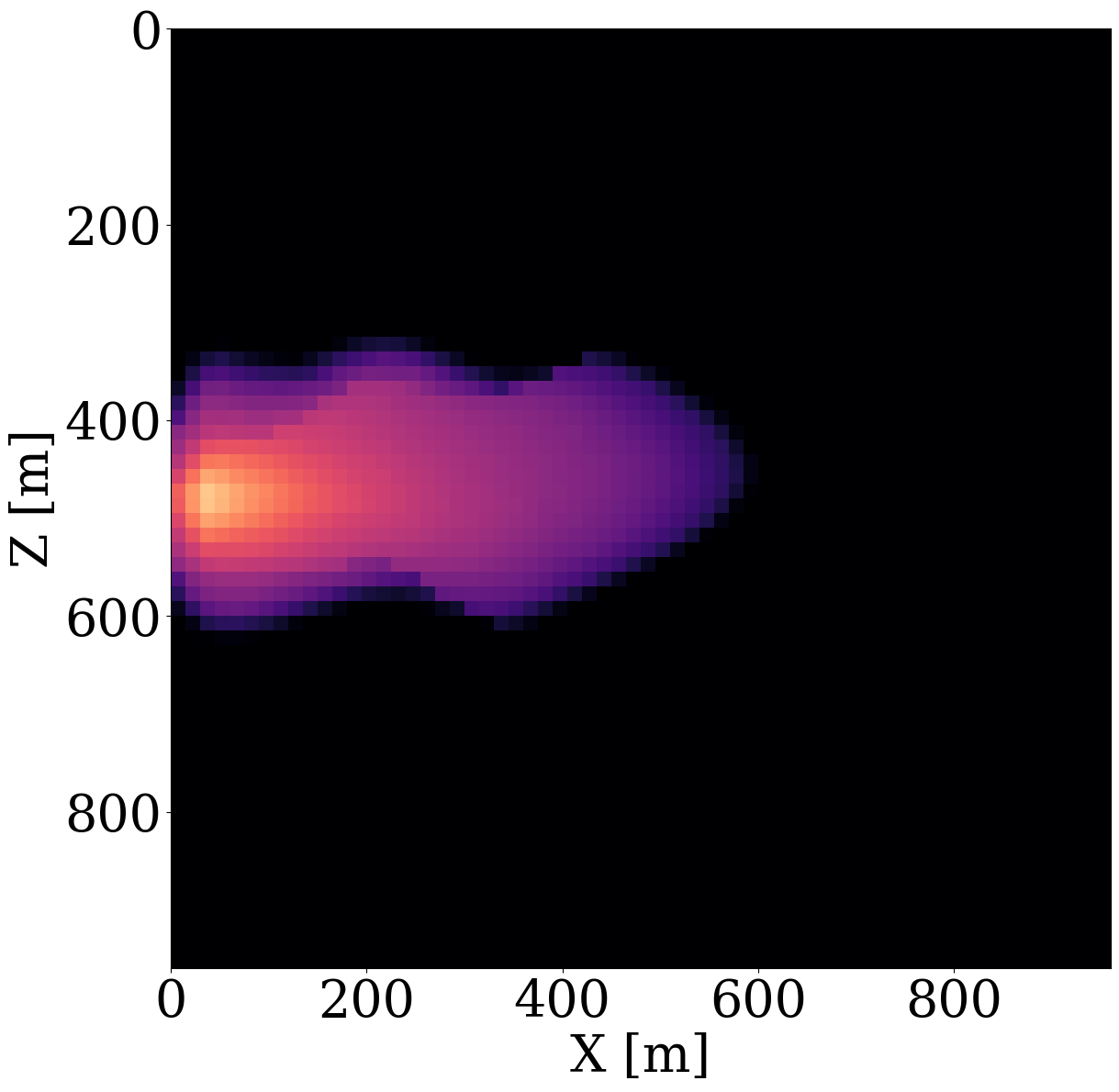}

}

}

\end{minipage}%
\begin{minipage}[t]{0.20\linewidth}

{\centering 

\raisebox{-\height}{

\includegraphics{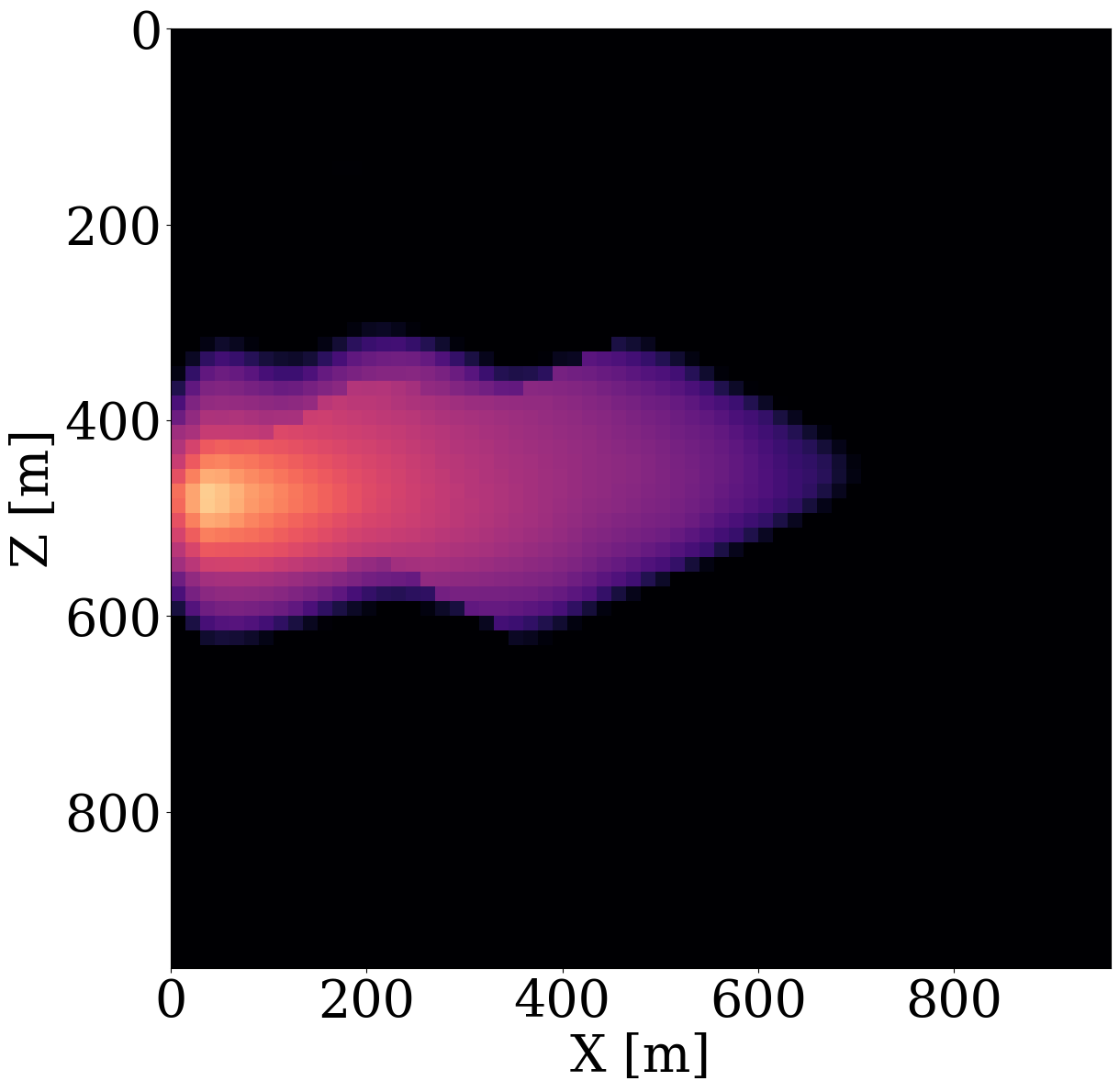}

}

}

\end{minipage}%
\begin{minipage}[t]{0.20\linewidth}

{\centering 

\raisebox{-\height}{

\includegraphics{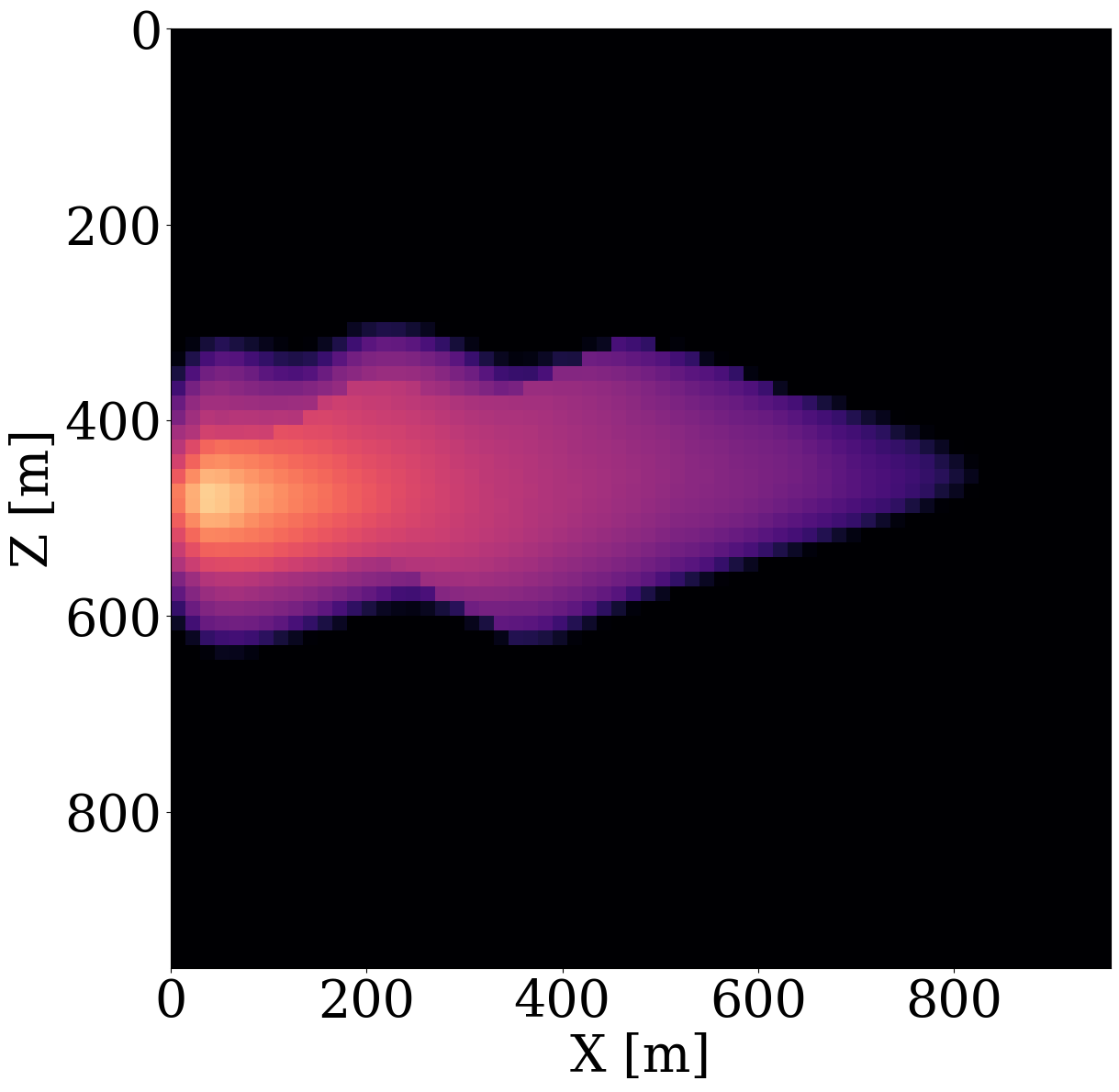}

}

}

\end{minipage}%
\begin{minipage}[t]{0.20\linewidth}

{\centering 

\raisebox{-\height}{

\includegraphics{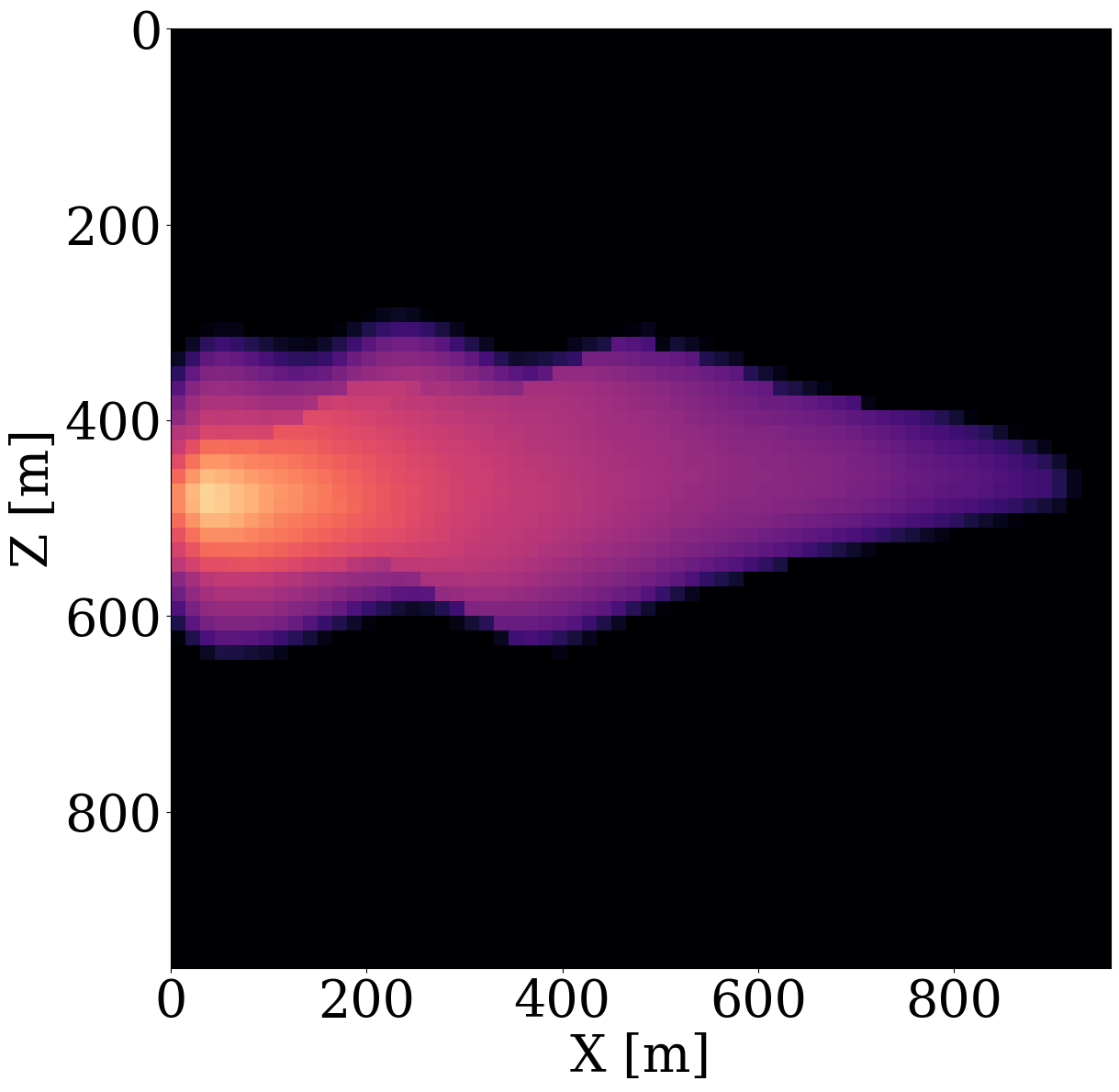}

}

}

\end{minipage}%
\begin{minipage}[t]{0.20\linewidth}

{\centering 

\raisebox{-\height}{

\includegraphics{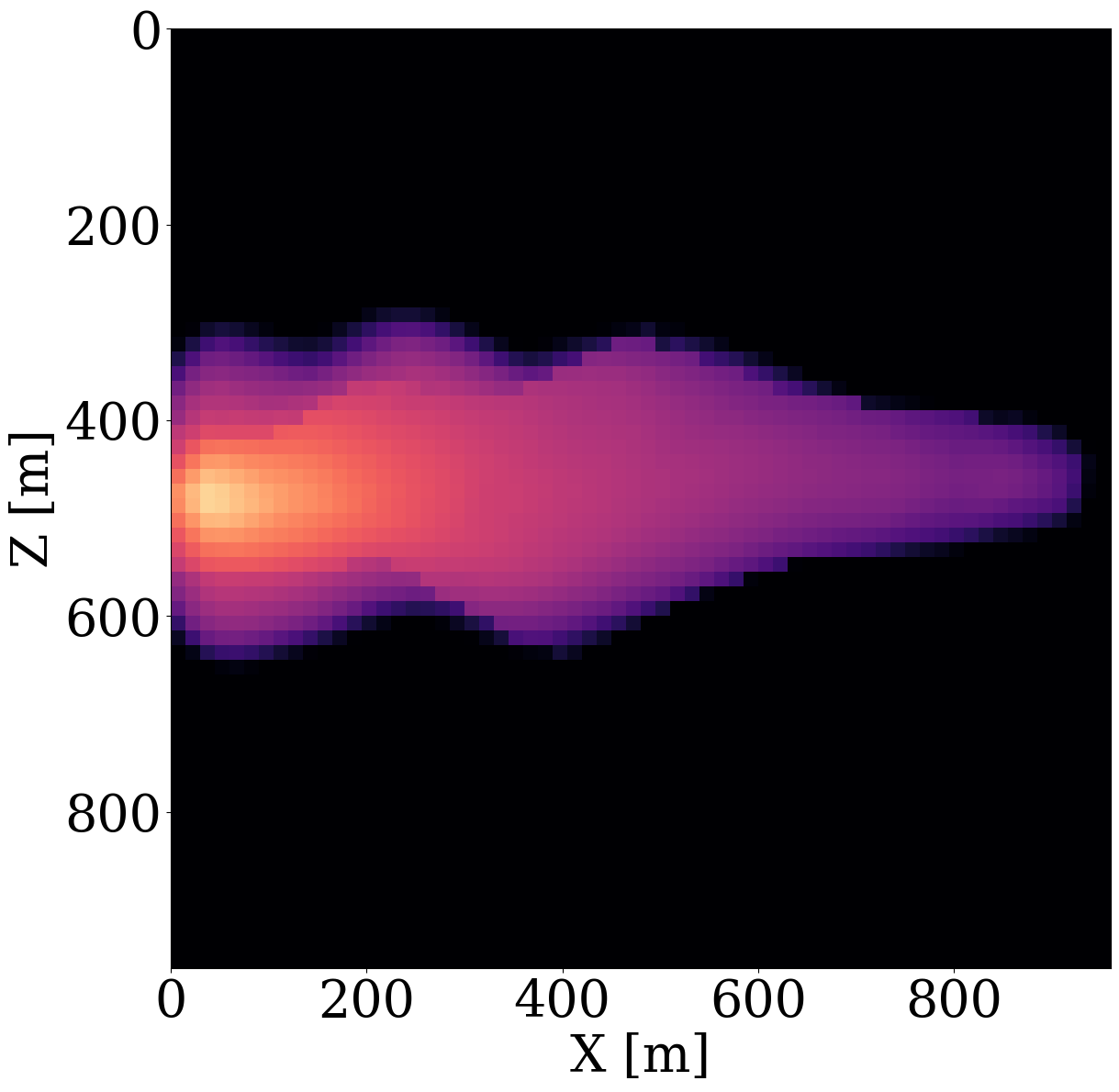}

}

}

\end{minipage}%
\newline
\begin{minipage}[t]{0.20\linewidth}

{\centering 

\raisebox{-\height}{

\includegraphics{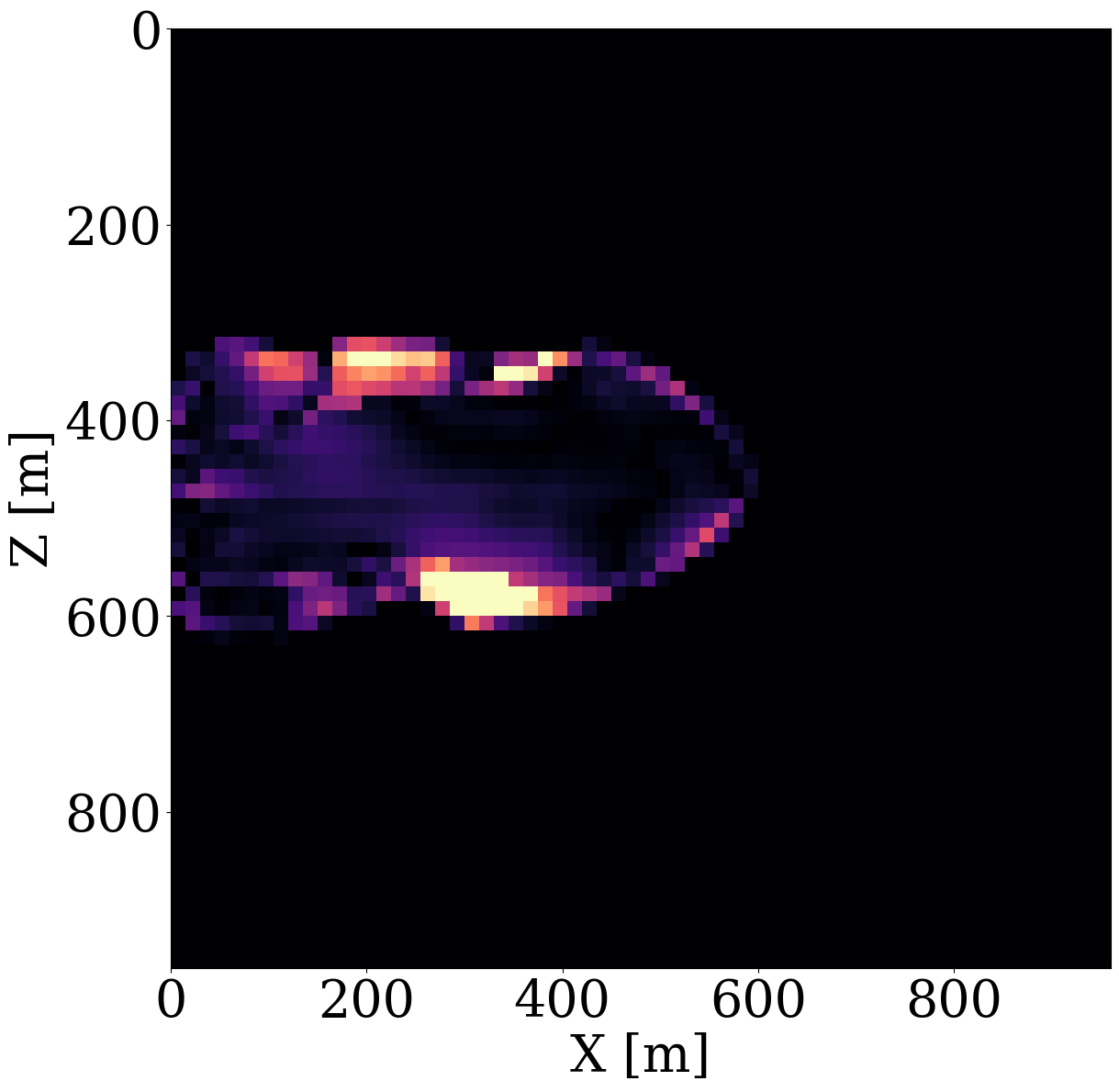}

}

}

\end{minipage}%
\begin{minipage}[t]{0.20\linewidth}

{\centering 

\raisebox{-\height}{

\includegraphics{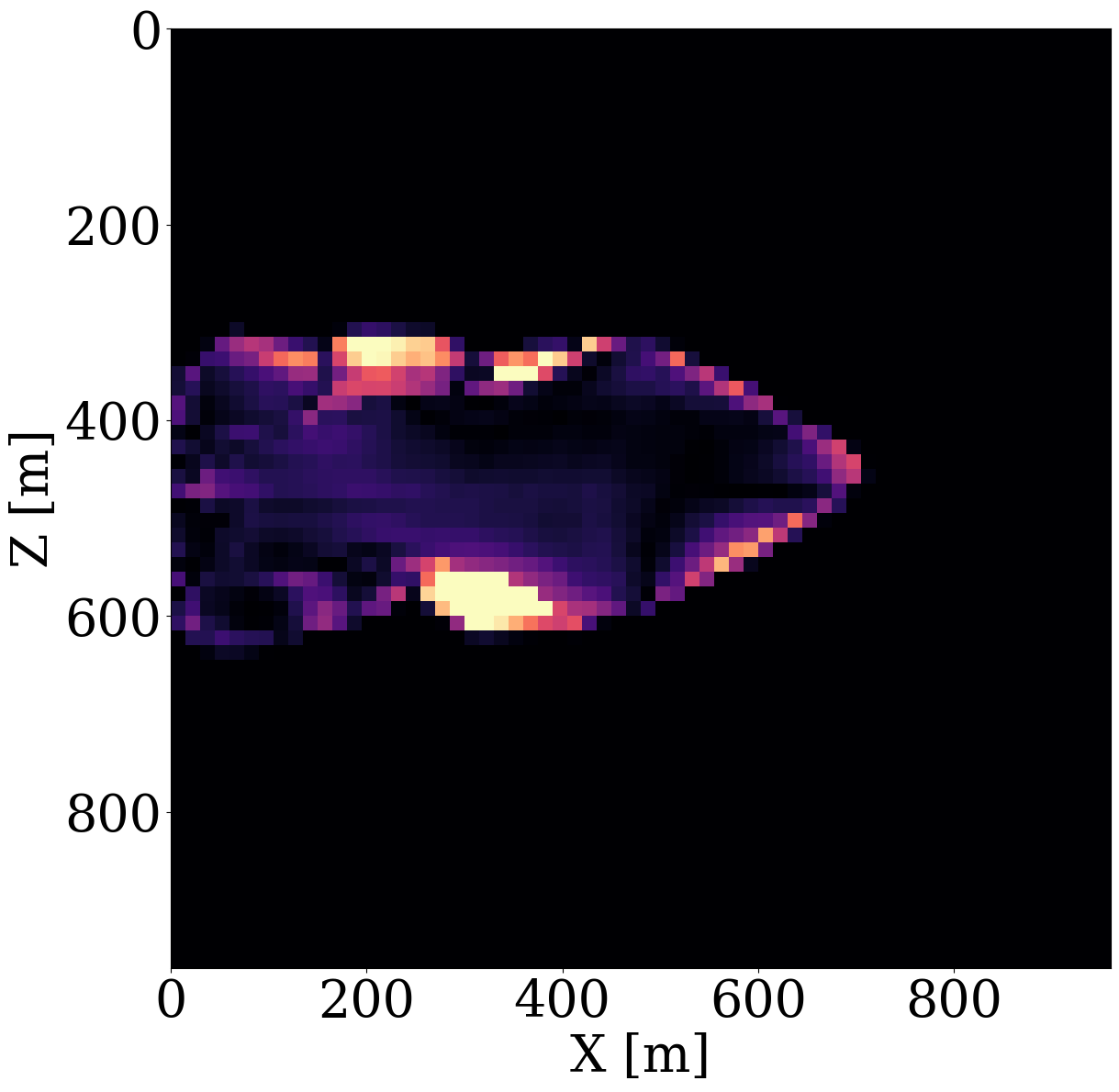}

}

}

\end{minipage}%
\begin{minipage}[t]{0.20\linewidth}

{\centering 

\raisebox{-\height}{

\includegraphics{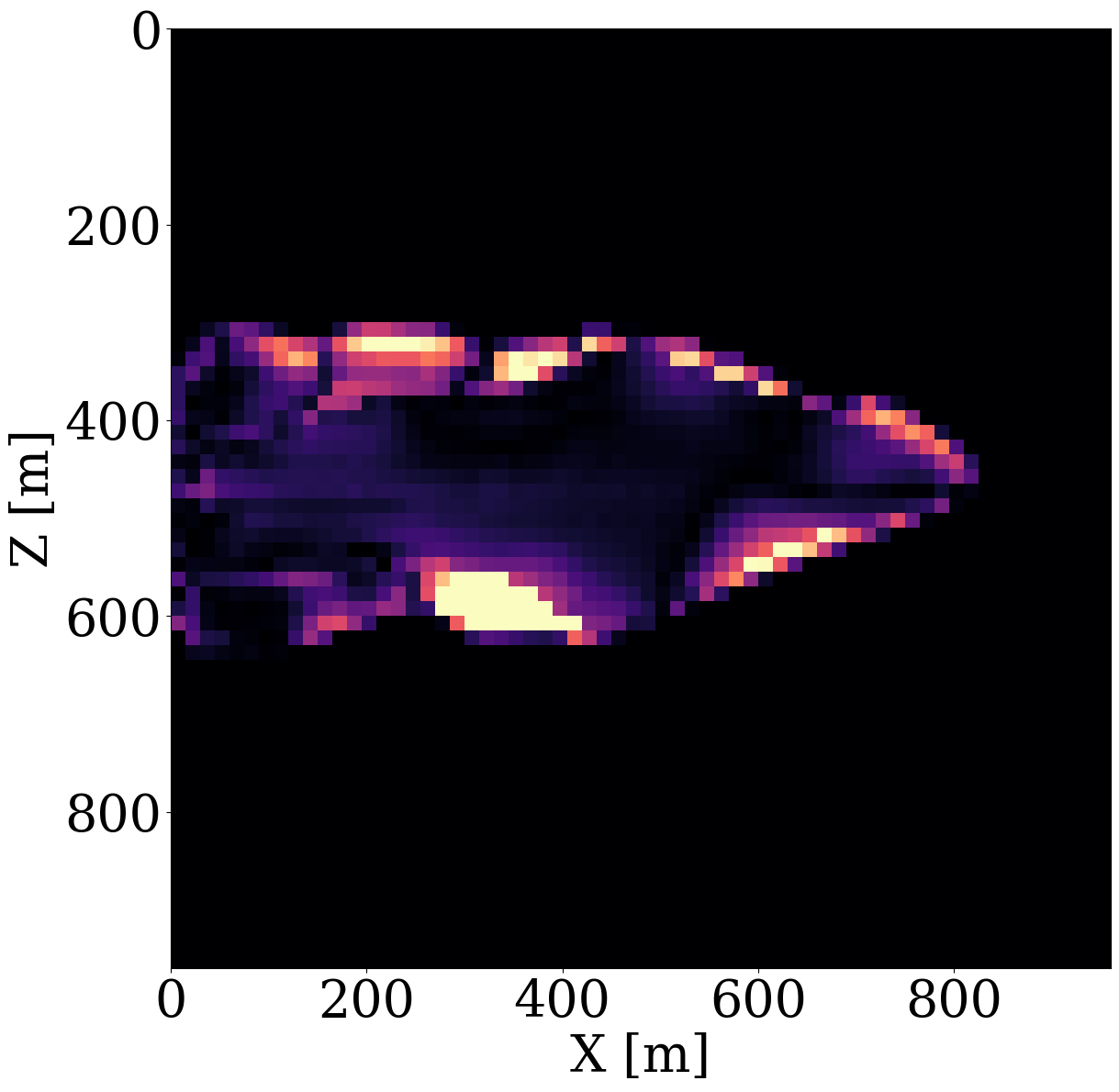}

}

}

\end{minipage}%
\begin{minipage}[t]{0.20\linewidth}

{\centering 

\raisebox{-\height}{

\includegraphics{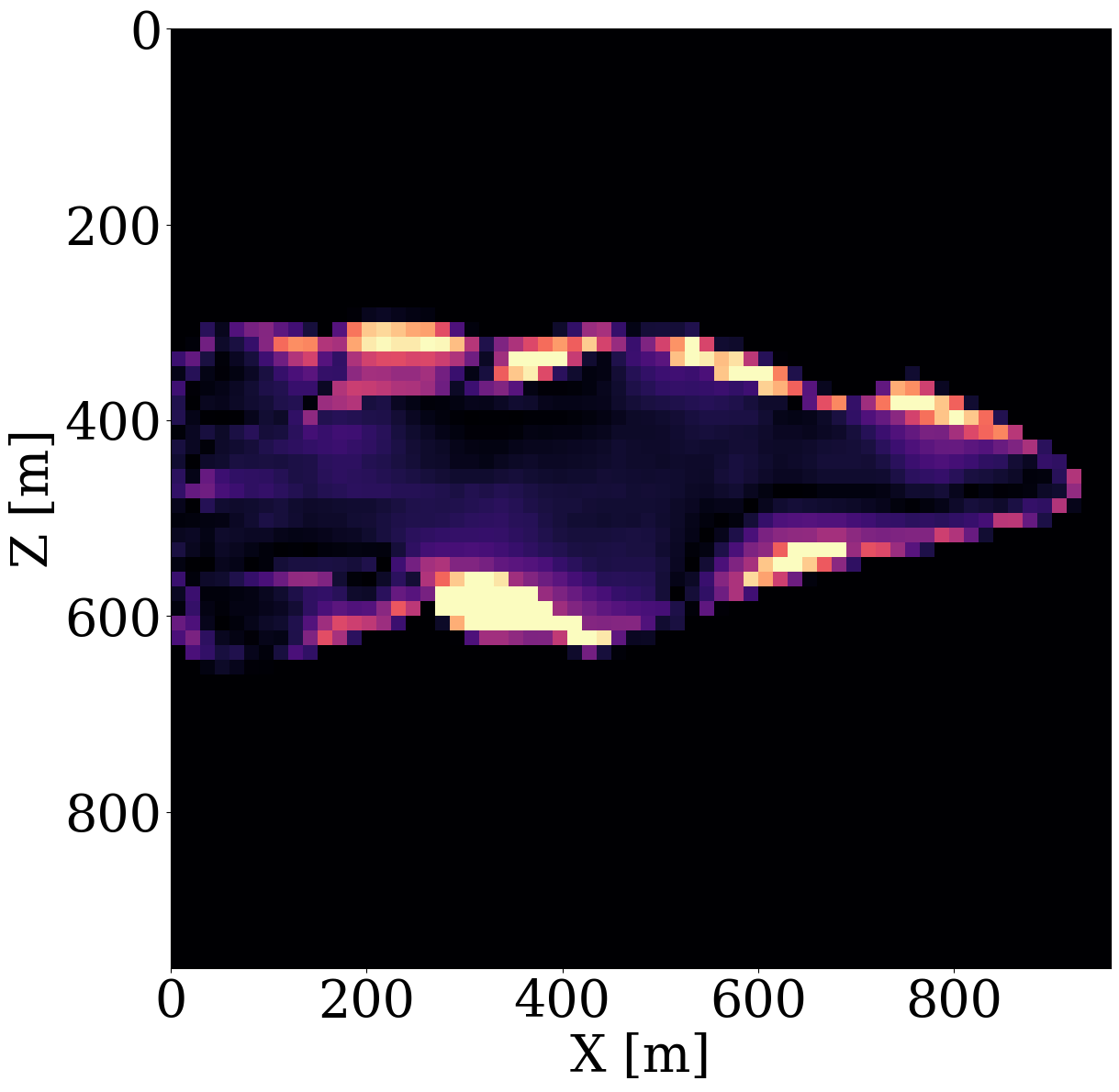}

}

}

\end{minipage}%
\begin{minipage}[t]{0.20\linewidth}

{\centering 

\raisebox{-\height}{

\includegraphics{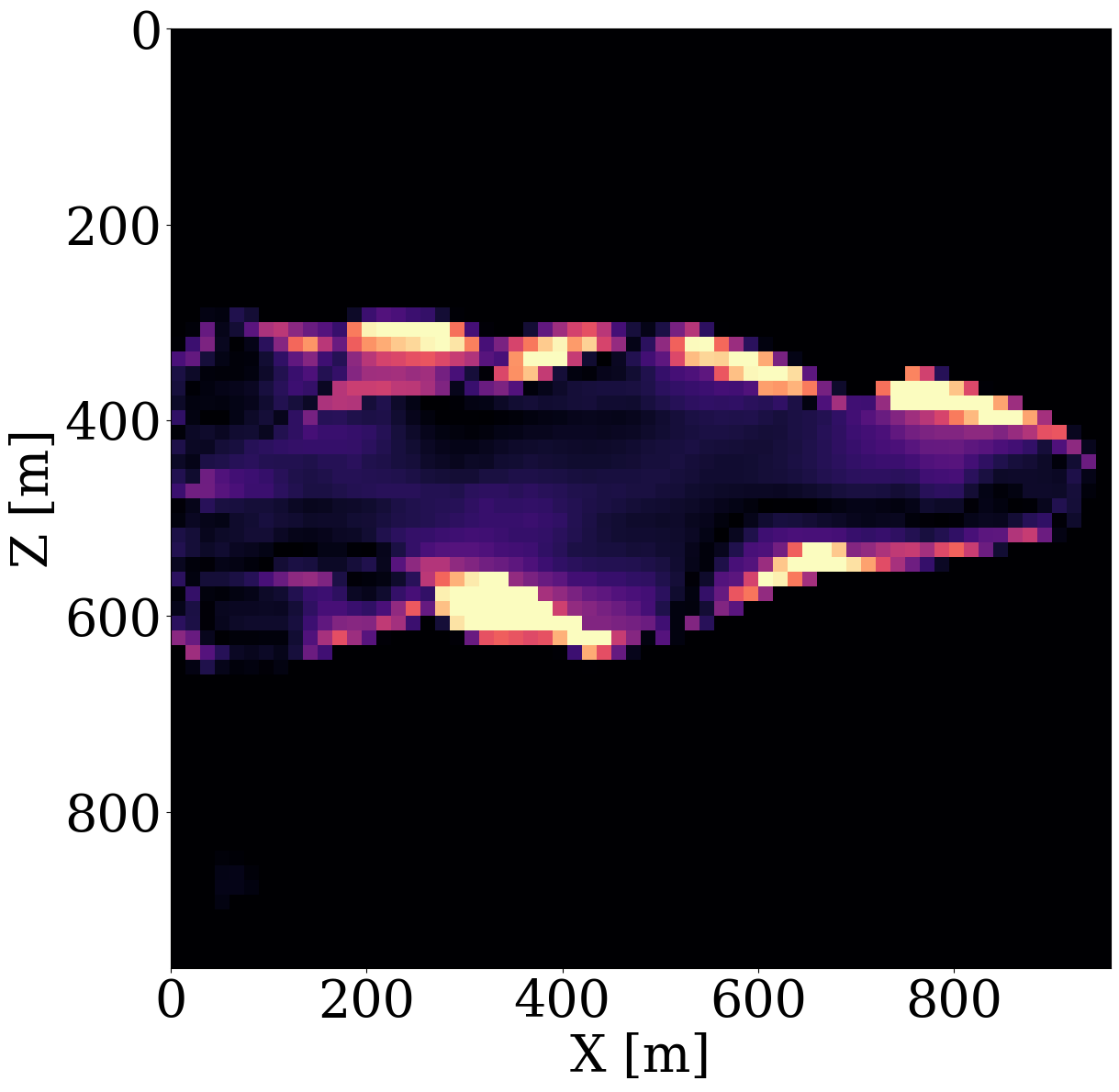}

}

}

\end{minipage}%
\newline
\begin{minipage}[t]{0.20\linewidth}

{\centering 

\raisebox{-\height}{

\includegraphics{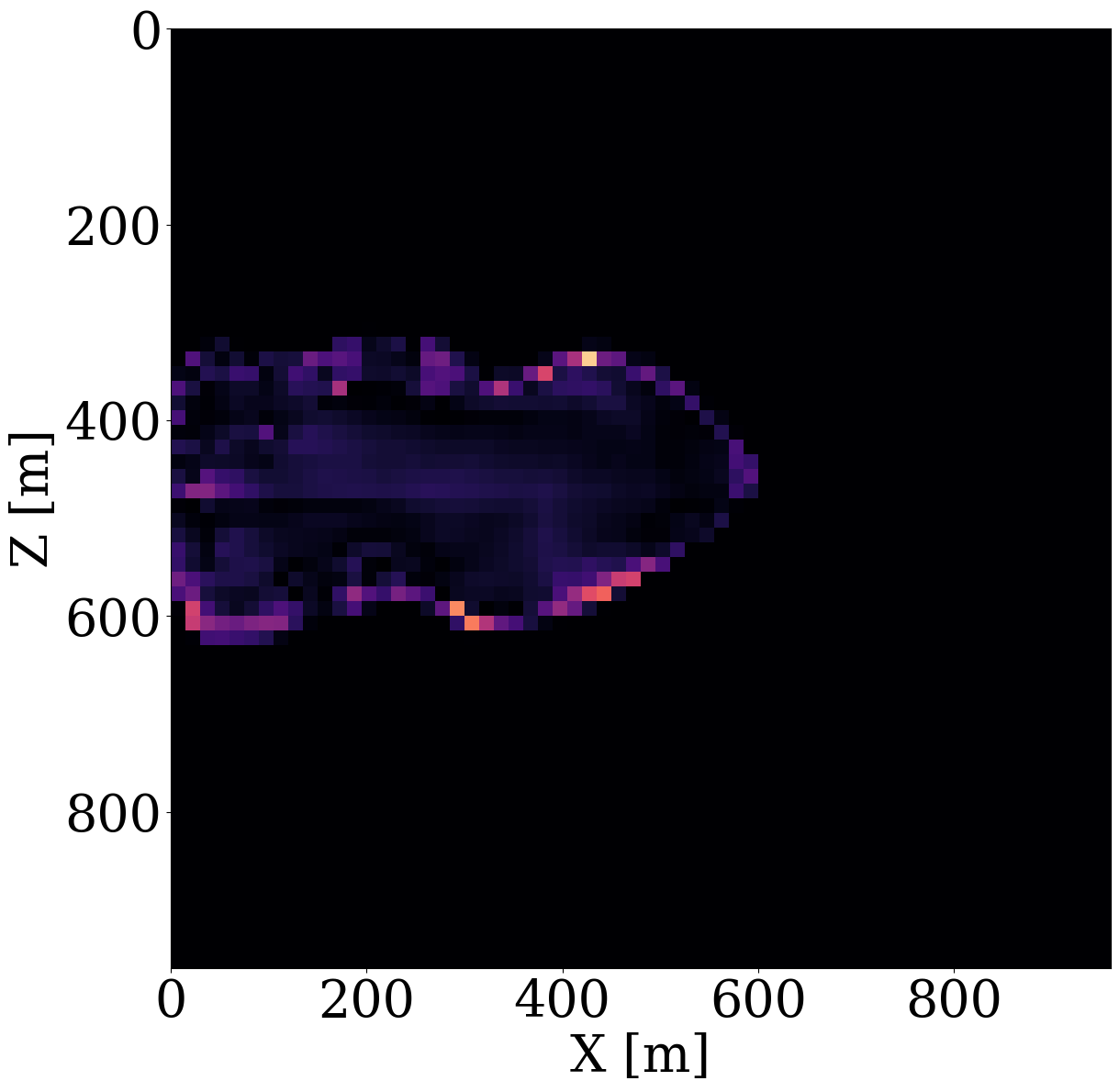}

}

}

\end{minipage}%
\begin{minipage}[t]{0.20\linewidth}

{\centering 

\raisebox{-\height}{

\includegraphics{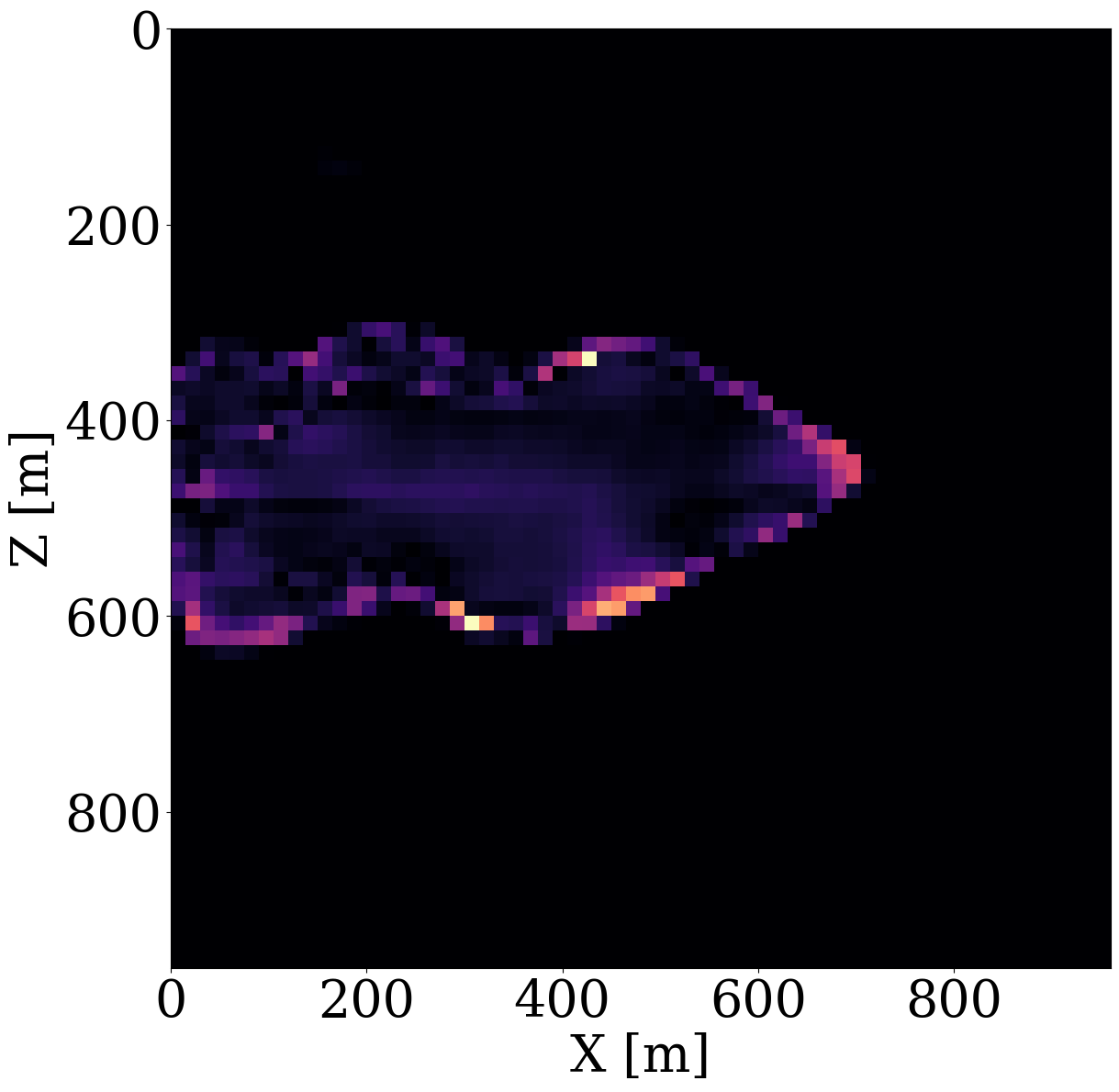}

}

}

\end{minipage}%
\begin{minipage}[t]{0.20\linewidth}

{\centering 

\raisebox{-\height}{

\includegraphics{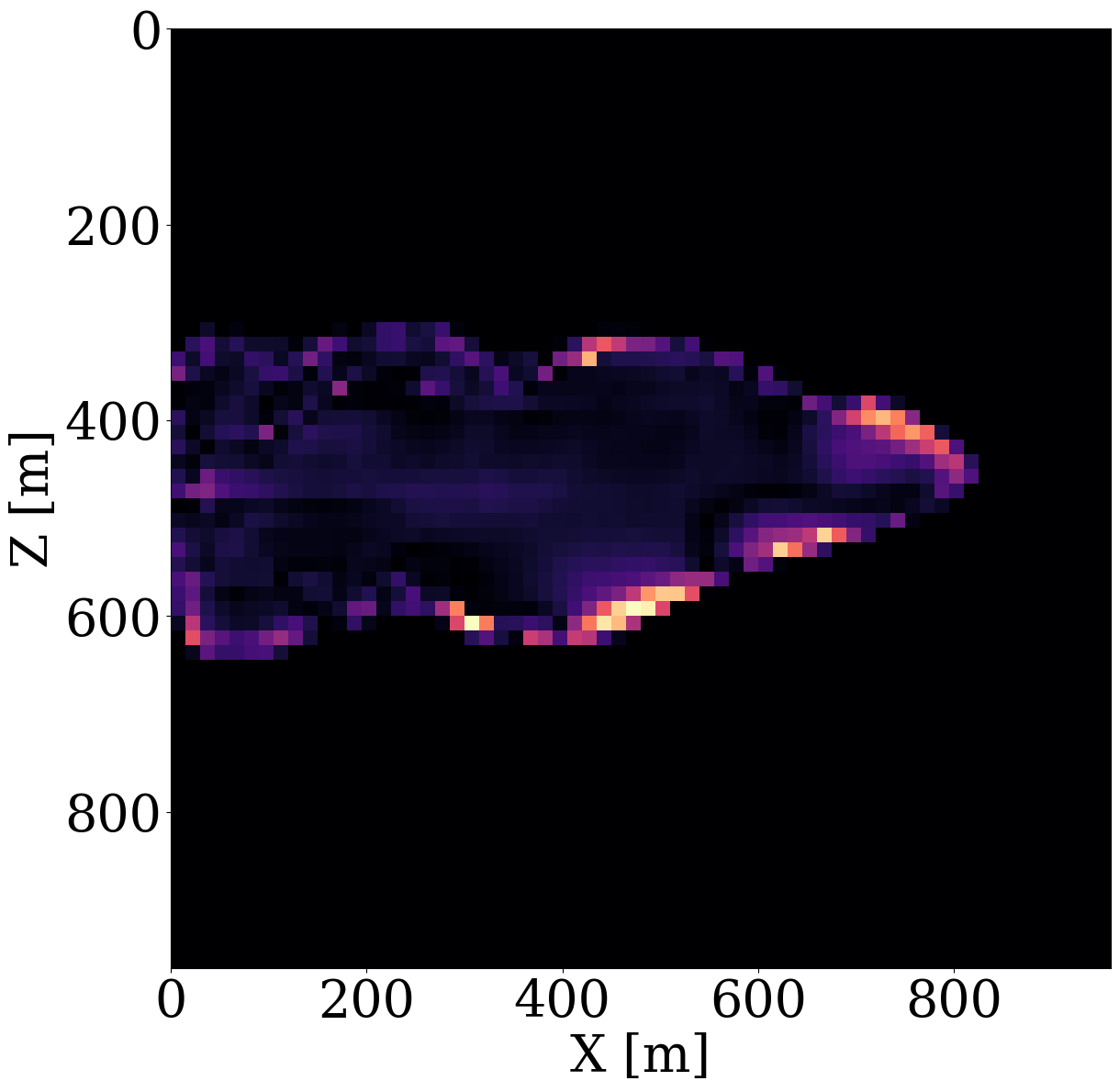}

}

}

\end{minipage}%
\begin{minipage}[t]{0.20\linewidth}

{\centering 

\raisebox{-\height}{

\includegraphics{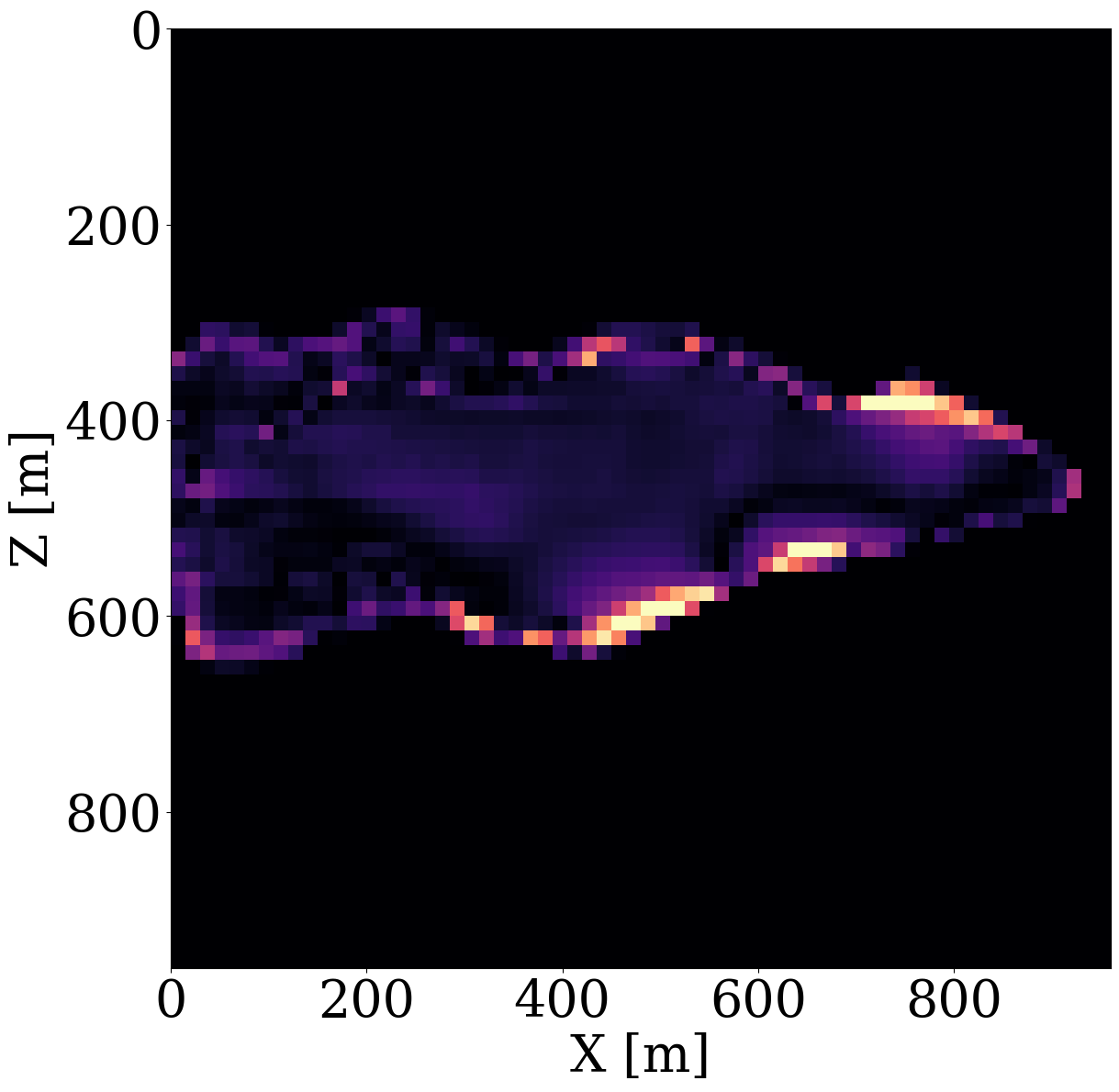}

}

}

\end{minipage}%
\begin{minipage}[t]{0.20\linewidth}

{\centering 

\raisebox{-\height}{

\includegraphics{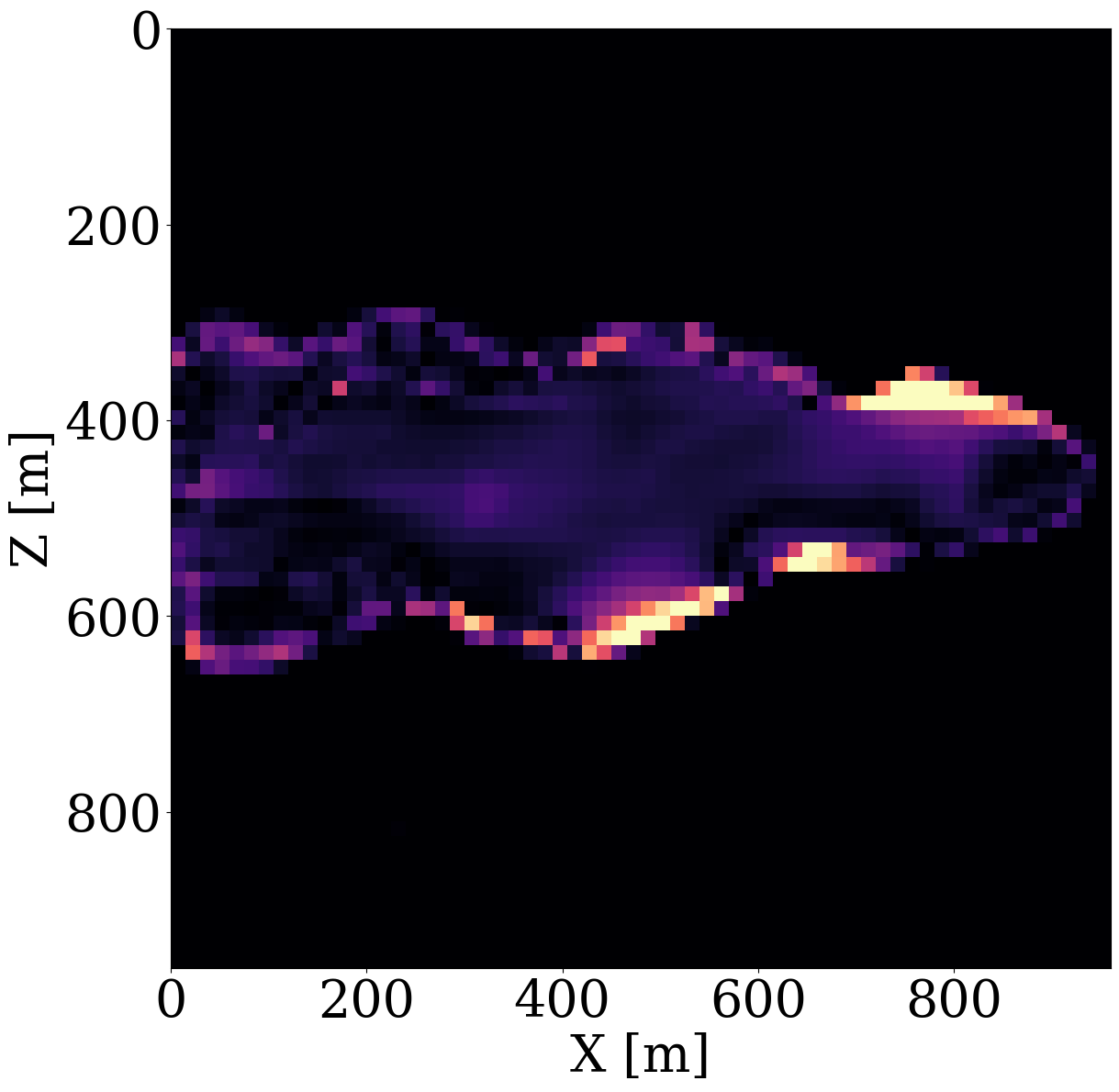}

}

}

\end{minipage}%
\newline
\begin{minipage}[t]{0.20\linewidth}

{\centering 

\raisebox{-\height}{

\includegraphics{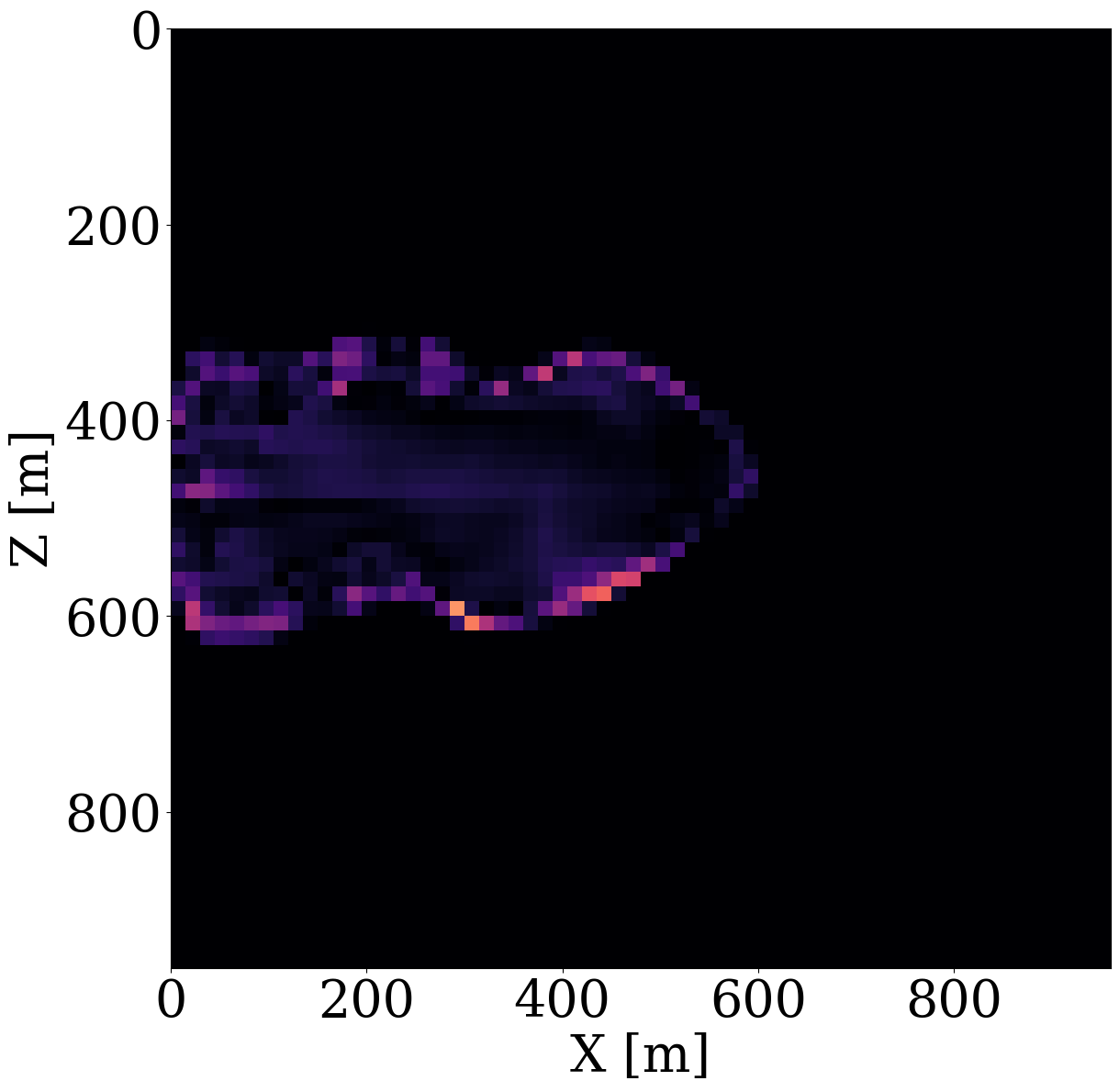}

}

}

\end{minipage}%
\begin{minipage}[t]{0.20\linewidth}

{\centering 

\raisebox{-\height}{

\includegraphics{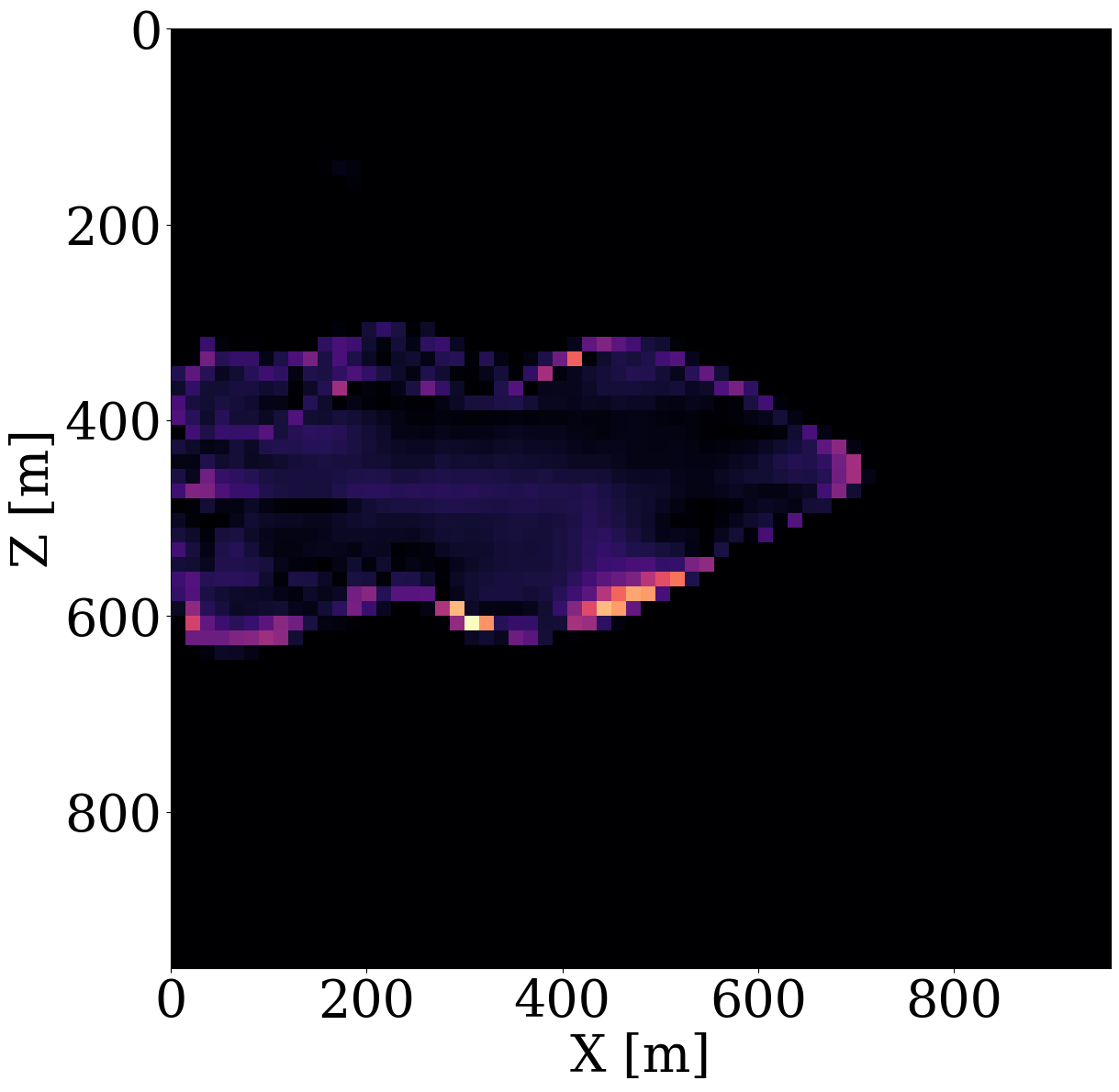}

}

}

\end{minipage}%
\begin{minipage}[t]{0.20\linewidth}

{\centering 

\raisebox{-\height}{

\includegraphics{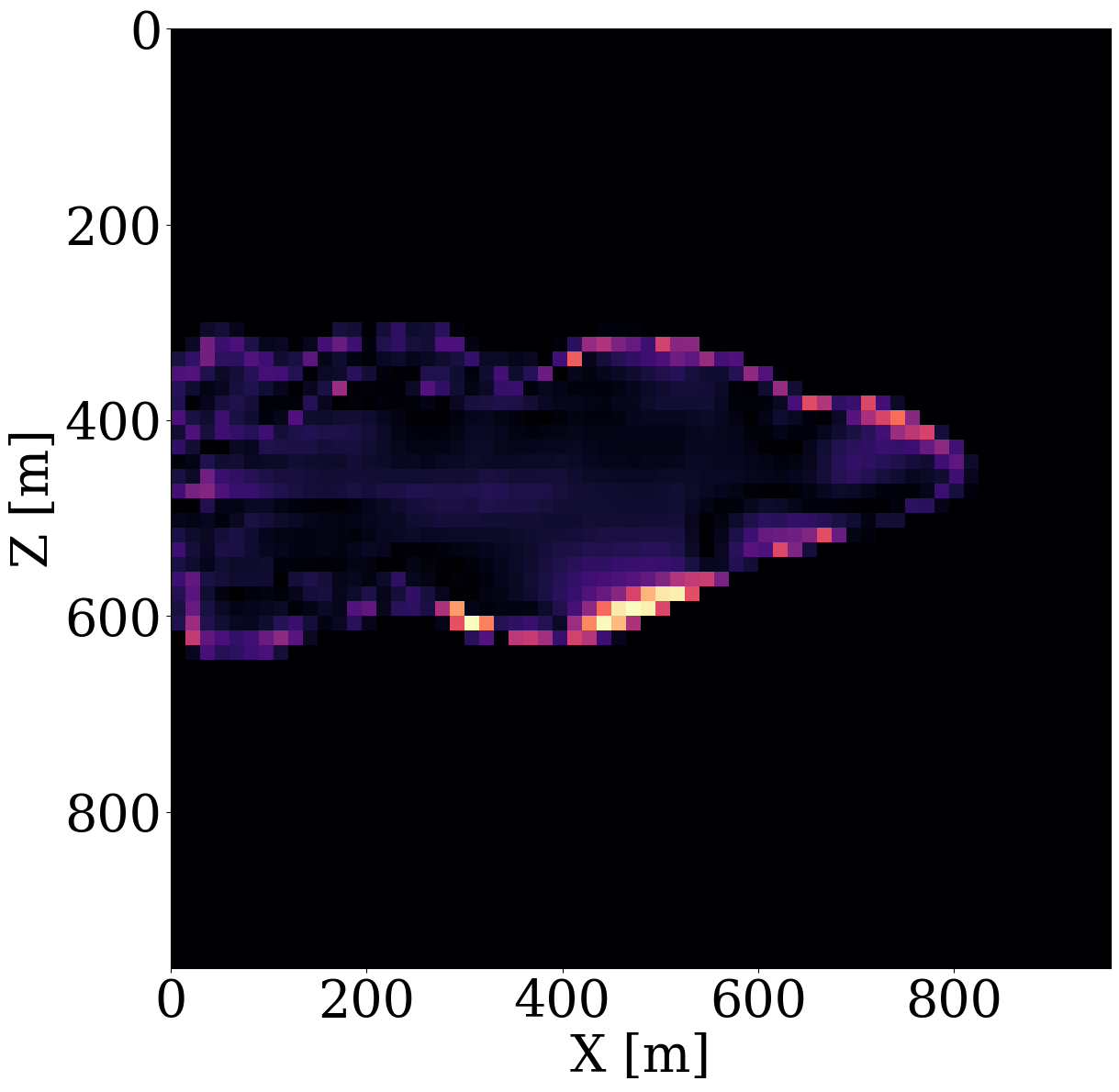}

}

}

\end{minipage}%
\begin{minipage}[t]{0.20\linewidth}

{\centering 

\raisebox{-\height}{

\includegraphics{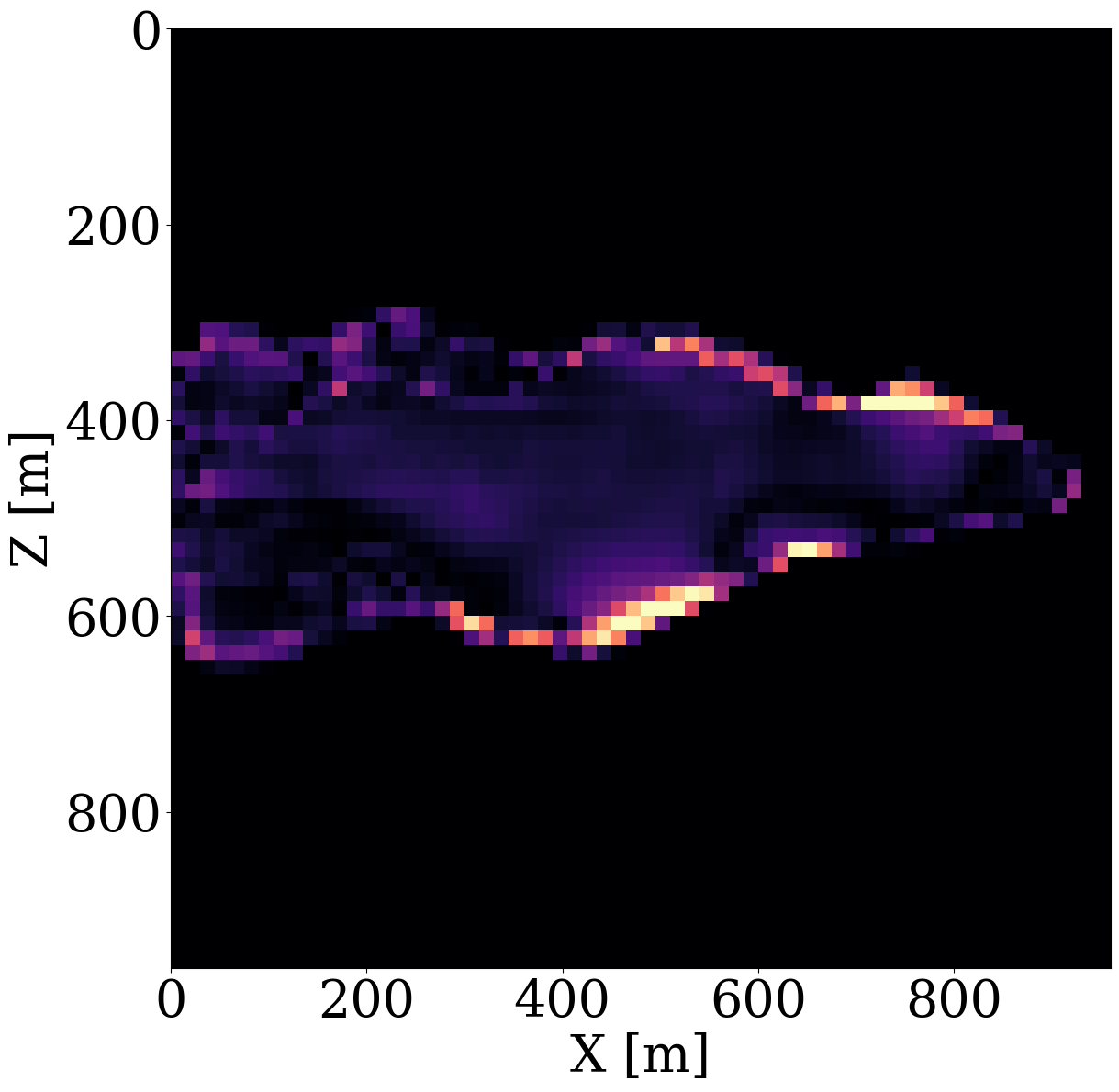}

}

}

\end{minipage}%
\begin{minipage}[t]{0.20\linewidth}

{\centering 

\raisebox{-\height}{

\includegraphics{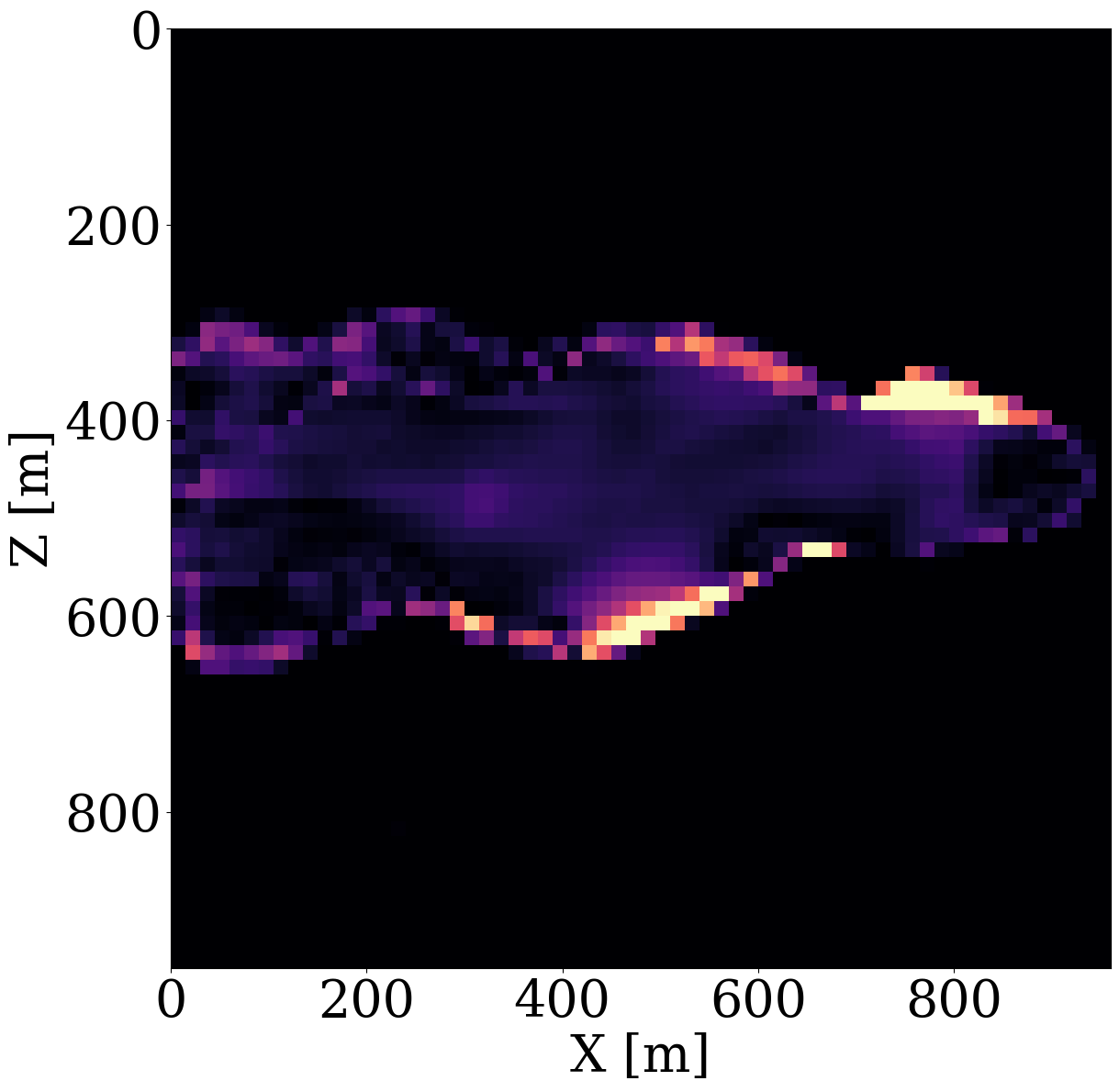}

}

}

\end{minipage}%

\caption{\label{fig-co2-all}CO\textsubscript{2} plume estimation and
forecast using FNO surrogates and NF constraints to invert different
modalities of observed data. The first three columns represent past
CO\textsubscript{2} saturations at day 400, 500, and 600 of the first
600 days of CO\textsubscript{2} saturation monitored either through the
well measurements or time-lapse data. The last two columns include
forecasts for the saturations at future days 700 and 800, where no
observed data is available. The first row shows the past and future
CO\textsubscript{2} estimates yielded by inverting well measurements
only. The second row is the same but now inverting time-lapse seismic
data. The third row is the same but now jointly inverting well
measurements and time-lapse seismic data. The fourth, fifth, and sixth
rows show \(5\times\) difference between the ground truth
CO\textsubscript{2} plume (first row of
Figure~\ref{fig-fno-train-plume}) and the first, second, third row,
respectively. The S/Ns for the first, the second, and the third rows are
15.26 dB,~20.14 dB, 20.46 dB, respectively.}

\end{figure}

While end-to-end permeability inversion from time-lapse data provides
novel access to this important fluid-flow property, the real interest in
monitoring GCS lies in determining where CO\textsubscript{2} plumes are
and will be in the foreseeable future, say of 100 and 200 days ahead. To
demonstrate the value of the proposed surrogates and of the use of
time-lapse seismic data, as opposed to time-lapse saturation data
measured at the wells only, we in Figure~\ref{fig-fno-train-plume}
juxtapose CO\textsubscript{2} predictions obtained from fluid-flow
simulations based on the inverted permeabilities in situations where
either well data is available (first row), or where time-lapse seismic
data is available (second row), or where both data modalities are
available (third row). These results are achieved by first inverting for
permeabilities using FNO surrogates and NF constraints, followed by
running the fluid-flow simulations for additional time steps given the
inverted permeabilities yielded by well-only
(Figure~\ref{fig-flow-inv-fno-nf}), time-lapse data
(Figure~\ref{fig-end2end-inv-fno-nf}), and both
(Figure~\ref{fig-combine-inv-fno-nf}). From these plots, we draw the
following two conclusion. First, the predicted CO\textsubscript{2}
plumes estimated from seismic data are significantly more accurate than
those obtained by inverting time-lapse saturations measured at the wells
only. As expected, there are large errors in the regions away from the
wells for the CO\textsubscript{2} plumes estimated from wells shown in
the fourth row of Figure~\ref{fig-co2-all}. Second, thanks to the
NF-constraint, the CO\textsubscript{2} predictions obtained with the
computationally beneficial surrogate approximation remain close to the
ground truth CO\textsubscript{2} plume plotted in the first row of
Figure~\ref{fig-fno-train-plume}, with only minor artifacts at the
edges. Third, using both seismic data and well measurements produces
CO\textsubscript{2} plume predictions with the smallest errors, while
the uplift of well measurements on top of seismic observations is modest
(comparing the second and the third rows of Figure~\ref{fig-co2-all}).
Finally, while the CO\textsubscript{2} plume estimates for the past
(monitored) vintages (i.e.~first three columns of the third row of
Figure~\ref{fig-co2-all}) are accurate, the near-future forecasts
without time-lapse well or seismic data (i.e.~last two columns of the
third row of Figure~\ref{fig-co2-all}) could be less accurate. This is
because the right-hand side and the bottom of the permeability model are
not touched yet by the CO\textsubscript{2} plume during the first 600
days. As a result, the error on the permeability recovery on the
right-hand side leads to the slightly larger errors on the
CO\textsubscript{2} plume forecast. Overall, these CO\textsubscript{2}
forecasts for the future 100 and 200 days match the general trend of the
CO\textsubscript{2} plume without any observed data despite minor
errors. A continuous monitoring system, where multiple modalities of
data are being acquired and inverted throughout the GCS project, could
allow for updating the reservoir permeability and forecasting the
CO\textsubscript{2} plume consistently.

\hypertarget{analysis-of-computational-gains}{%
\subsubsection{Analysis of computational
gains}\label{analysis-of-computational-gains}}

FNOs, and deep neural surrogates in general, have the potential to be
orders of magnitude faster than conventional PDE solvers (Z. Li et al.
2020), and this speed-up is generally problem-dependent. In our
numerical experiments, the PDE solver from
\href{https://github.com/sintefmath/Jutul.jl}{Jutul.jl} (Møyner et al.
2023; Møyner, Bruer, and Yin 2023; Yin, Bruer, and Louboutin 2023)
currently only supports CPUs and we find an average runtime for both the
forward and gradient on the \(64 \times 64\) model to be 10.6 seconds on
average on an 8-core Intel(R) Xeon(R) W-2245 CPU. The trained FNO,
implemented using modules from
\href{https://github.com/FluxML/Flux.jl}{Flux.jl} (Innes 2018), takes
16.4 milliseconds on average for both the forward and gradient. This
means that the trained FNO in our case provides \(646\times\) speed up
compared to conventional PDE solvers. The training of FNO takes about 4
hours on an NVIDIA T1000 8GB GPU. Given these numbers, we can calculate
the break-even point --- i.e., the point where using FNO surrogate
becomes cheaper in terms of the overall runtime, by the following
formula:

\begin{equation}\protect\hypertarget{eq-breakeven}{}{
\textrm{breakeven} = \frac{\textrm{generating training set time} + \textrm{training time}}{\textrm{PDE solver runtime}-\textrm{FNO runtime}}
\approx 3364.
}\label{eq-breakeven}\end{equation}

This means that after 3364 calls to the forward simulator and its
gradients, the computational savings gained from using the FNO surrogate
evaluations during the inversion process balances out the initial
upfront costs. These upfront costs include the generation of the
training dataset and the training of the FNO. Therefore, after this
break-even point of 3364 calls, the use of the FNO surrogate becomes
more cost-effective compared to the conventional PDE solver. Because the
trained FNO has the potential to generalize to different kinds of
inversion problems, and potentially also different GCS sites, 3364 calls
is justifiable in practice. However, we acknowledge that a more detailed
analysis on a more realistic 4D scale problem will be necessary to
understand the potential computational gains and tradeoffs of the
proposed methodology. For details on a high-performance computing
parallel implementation of FNOs, we refer to Grady et al. (2023) who
also conducted a realistic performance on large-scale 4D multiphase
fluid-flow problems. Even in cases where the computational advances are
perhaps challenging to justify, the use of FNOs has the additional
benefit by providing access to the gradient with respect to model
parameters (i.e.~permeability) through automatic differentiation. This
feature is important since it is an enabler for inversion problems that
involve complex PDE solvers for which gradients are often not readily
available, e.g. Gross and Mazuyer (2021). By training FNOs on
input-output pairs, ``gradient-free'' gradient-based inversion is made
possible in situations where the simulator does not support gradients.

\hypertarget{discussion-and-conclusions}{%
\subsection{Discussion and
conclusions}\label{discussion-and-conclusions}}

Monitoring of geological carbon storage is challenging because of
excessive computational needs and demands on data collection by drilling
monitor wells or by collecting time-lapse seismic data. To offset the
high computational costs of solving multiphase flow equations and to
improve permeability inversions from possibly multimodal time-lapse
data, we introduce the usage of trained Fourier neural operators (FNOs)
that act as surrogates for the fluid-flow simulations. We propose to do
this in combination with trained normalizing flows (NFs), which serve as
regularizers to keep the inversion and the accuracy of the FNOs in
check. Since the computational expense of FNO's online evaluation is
negligible compared to numerically expensive partial differential
equation solves, FNOs only incur upfront offline training costs. While
this obviously presents a major advantage, the approximation accuracy of
FNOs is, unfortunately, only guaranteed when its argument, the
permeability, is in distribution---i.e., is drawn from the same
distribution as the FNO was trained on. This creates a problem because
there is, thanks to the non-trivial null space of permeability
inversion, no guarantee the model iterates remain in-distribution. Quite
the opposite, our numerical examples show that these iterates typically
move out-of-distribution during the (early) iterations. This results in
large errors in the FNO and in rather poor inversion results for the
permeability.

To overcome this out-of-distribution dillema for the model iterates
during permeability inversion with FNOs, we propose adding learned
constraints, which ensure that model iterates remain in-distribution
during the inversion. We accomplish this by training a NF on the same
training set for the permeability used to train the FNO. After training,
the NF is capable of generating in-distribution samples for the
permeability from random realizations of standard Gaussian noise in the
latent space. We employ this learned ability by parameterizing the
unknown permeability in the latent space, which offers additional
control on whether the model iterates remain in-distribution during the
inversion. After establishing that out-of-distribution permeability
models can be mapped to in-distribution models by restricting the
\(\ell_2\)-norm of their latent representation, we introduce
permeability inversion as a constrained optimization problem where the
data misfit is minimized subject to a constraint on the \(\ell_2\)-norm
of the latent space. Compared to adding pretrained NFs as priors via
additative penalty terms, use of constraints ensures that model iterates
remain at all times in-distribution. We show that this holds as long as
the size of constraint set does not exceed the size of the
\(\ell_2\)-norm ball of the standard normal distribution. As a result,
we arrive at a computationally efficient continuation scheme, known as a
homotopy, during which the \(\ell_2\)-norm constraint is relaxed slowly,
so the data misfit objectives can be minimized while the model iterates
remain in distribution.

By means of a series of carefully designed numerical experiments, we
were able to establish the advocasy of combining learned surrogates and
constraints, yielding solutions to permeability inversion problems that
are close to solutions yielded by costly PDE-based methods. The examples
also clearly show the advantages of working with gradually relaxed
constraints where model iterates remain at all times in distribution
with the additional joint benefit of slowly building up the model while
bringing down the data misfit, an approach known to mitgate the effects
of local minima (Peters and Herrmann 2017; Esser et al. 2018; Peters,
Smithyman, and Herrmann 2019). Consequently, the quality of all
time-lapse inversions improved significantly without requiring
information that goes beyond having access to the training set of
permeability models.

While we applied the proposed method to gradient-based iterative
inversion, a similar approach can be used for other types of inversion
methods, including inference with Markov chain Monte Carlo methods for
uncertainty quantification (Lan, Li, and Shahbaba 2022). We also
envisage extensions of the proposed method to other physics-based
inverse problems (Heidenreich et al. 2014; Yang et al. 2023) and
simulation-based inference problems (Cranmer, Brehmer, and Louppe 2020),
where numerical simulations often form the computational bottleneck.

Despite the encouraging results from the numerical experiments, the
presented approach leaves room for improvements, which we will leave for
future work. For instance, the gradient with respect to the model
parameters (permeability) derived from the neural surrogate is not
guaranteed to be accurate---e.g.~close to the gradient yielded by the
adjoint-state method. As recent work by O'Leary-Roseberry et al. (2022)
has shown, this potential source of error can be addressed by training
neural surrogates on the simulator's gradient with respect to the model
parameters, provided it is available. Unfortunately, deriving gradients
of complex HPC implementatons of numerical PDE solvers is often
extremely challenging, explaining why this information is often not
available. Because our method solely relies on gradients of the
surrogate, which are readily available through algorithmic
differentiation, we only need access to numerical PDE solvers available
in legacy codes. While this approach may go at the expense of some
accuracy, this feature offers a distinct practical advantage. However,
as with many other machine learning approaches, our learned methods may
also suffer from time-lapse observations that are
out-of-distribution---i.e., produced by a permeability model that is
out-of-distribution. While this is a common problem in data-driven
methods, recent developments (Siahkoohi et al. 2023a) may remedy this
problem by applying latent space corrections, a solution that is
amenable to our approach. On the other hand, expanding the latent
space's \(\ell_2\) norm ball during inversion would allow NFs to
generate any out-of-distribution model parameter. However, in that case
the accuracy of the learned surrogate is not guaranteed. For such cases,
transitioning from the learned surrogate to the numerical solver during
later iterations may be advantageous and merits further study. The
choice for the size of the \(\ell_2\)-norm ball at the beginning and at
the end can also be further investigated (Aster, Borchers, and Thurber
2018).

While our paper primarily presents a proof of concept through a
relatively small 2D experiment, our inversion strategy is designed to
scale to large-scale 3D problems. NFs, with their inherent memory
efficiency due to invertibility, are already primed for extension to 3D
problems. For the learned surrogates, Grady et al. (2023) showcases
model-parallel FNOs, demonstrating success in simulating 4D multiphase
flow physics of over 2.6 billon variables. By combining these strengths,
we are optimistic scaling this inversion strategy to 3D.

To end on a positive note and forward looking note, we argue that the
presented approach makes a strong case for the inversion of multimodal
data, consisting of time-lapse well and seismic data. While inversions
from time-lapse saturation data collected from wells are feasible and
fall within the realm of reservoir engineering, their performance, as
expected, degrades away from the well. We argue that adding
active-source seismic provides essential fill-in away from the wells. As
such, it did not come to our surprise that joint inversion of multimodal
data resulted in the best permeability estimates. From our perspective,
our successfull combination of these often disjoint data modalities
holds future promise when addressing challenges that come with
monitoring and control of geological carbon storage and enhanced
geothermal systems.

\hypertarget{availability-of-data-and-materials}{%
\subsection{Availability of data and
materials}\label{availability-of-data-and-materials}}

The scripts to reproduce the experiments are available on the SLIM
GitHub page \url{https://github.com/slimgroup/FNO-NF.jl}.

\hypertarget{abbreviations}{%
\subsection{Abbreviations}\label{abbreviations}}

FNO: Fourier neural operator\\
GCS: geological carbon storage\\
NF: normalizing flow\\
PDE: partial differential equations

\hypertarget{acknowledgements}{%
\subsection{Acknowledgements}\label{acknowledgements}}

The authors would like to thank Philipp A. Witte at Microsoft, Ali
Siahkoohi at Rice University, and Olav Møyner at SINTEF for constructive
discussions. The authors gratefully acknowledge the contribution of
OpenAI's ChatGPT for refining sentence structure and enhancing the
overall readability of this manuscript.

\hypertarget{funding}{%
\subsection{Funding}\label{funding}}

This research was carried out with the support of Georgia Research
Alliance and partners of the ML4Seismic Center. This research was also
supported in part by the US National Science Foundation grant OAC
2203821 and the Department of Energy grant No.~DE-SC0021515.

\hypertarget{author-information}{%
\subsection{Author information}\label{author-information}}

\hypertarget{authors-and-affiliations}{%
\subsubsection{Authors and
Affiliations}\label{authors-and-affiliations}}

Georgia Institute of Technology, 756 West Peachtree Street NW, Atlanta,
GA 30308

Ziyi Yin, Rafael Orozco, Mathias Louboutin, and Felix J. Herrmann

\hypertarget{contributions}{%
\subsubsection{Contributions}\label{contributions}}

ZY concepted the general outline of the project, implemented the
algorithm, designed and conducted the numerical experiments, and wrote
the manuscript. RO helped concept the general outline of the project,
discussed the approach and helped design the numerical experiments. ML
helped concept the general outline of the project and discussed the
approach. FH concepted the general outline of the project, discussed the
approach, helped design the numerical experiments, edited the
manuscript, and supervised the project. All authors reviewed the results
and approved the final version of the manuscript.

\hypertarget{corresponding-author}{%
\subsubsection{Corresponding author}\label{corresponding-author}}

Correspondence to \href{mailto:ziyi.yin@gatech.edu}{Ziyi Yin}.

\hypertarget{ethics-declarations}{%
\subsection{Ethics declarations}\label{ethics-declarations}}

\hypertarget{competing-interests}{%
\subsubsection{Competing interests}\label{competing-interests}}

The authors declare that they have no competing interests.

\hypertarget{refs}{}
\begin{CSLReferences}{1}{0}
\leavevmode\vadjust pre{\hypertarget{ref-alexanderian2021optimal}{}}%

\subsection{References}
Alexanderian, Alen. 2021. {``Optimal Experimental Design for
Infinite-Dimensional Bayesian Inverse Problems Governed by PDEs: A
Review.''} \emph{Inverse Problems} 37 (4): 043001.

\leavevmode\vadjust pre{\hypertarget{ref-alnes2011results}{}}%
Alnes, Håvard, Ola Eiken, Scott Nooner, Glenn Sasagawa, Torkjell
Stenvold, and Mark Zumberge. 2011. {``Results from Sleipner Gravity
Monitoring: Updated Density and Temperature Distribution of the CO2
Plume.''} \emph{Energy Procedia} 4: 5504--11.

\leavevmode\vadjust pre{\hypertarget{ref-ardizzone2018analyzing}{}}%
Ardizzone, Lynton, Jakob Kruse, Sebastian Wirkert, Daniel Rahner, Eric W
Pellegrini, Ralf S Klessen, Lena Maier-Hein, Carsten Rother, and Ullrich
Köthe. 2018. {``Analyzing Inverse Problems with Invertible Neural
Networks.''} \emph{arXiv Preprint arXiv:1808.04730}.

\leavevmode\vadjust pre{\hypertarget{ref-arridge1999optical}{}}%
Arridge, Simon R. 1999. {``Optical Tomography in Medical Imaging.''}
\emph{Inverse Problems} 15 (2): R41.

\leavevmode\vadjust pre{\hypertarget{ref-arts2004monitoring}{}}%
Arts, Rob, Ola Eiken, Andy Chadwick, Peter Zweigel, L Van der Meer, and
B Zinszner. 2004. {``Monitoring of CO2 Injected at Sleipner Using
Time-Lapse Seismic Data.''} \emph{Energy} 29 (9-10): 1383--92.

\leavevmode\vadjust pre{\hypertarget{ref-asher2015review}{}}%
Asher, Michael J, Barry FW Croke, Anthony J Jakeman, and Luk JM Peeters.
2015. {``A Review of Surrogate Models and Their Application to
Groundwater Modeling.''} \emph{Water Resources Research} 51 (8):
5957--73.

\leavevmode\vadjust pre{\hypertarget{ref-asim2020invertible}{}}%
Asim, Muhammad, Max Daniels, Oscar Leong, Ali Ahmed, and Paul Hand.
2020. {``Invertible Generative Models for Inverse Problems: Mitigating
Representation Error and Dataset Bias.''} In \emph{International
Conference on Machine Learning}, 399--409. PMLR.

\leavevmode\vadjust pre{\hypertarget{ref-aster2018parameter}{}}%
Aster, Richard C, Brian Borchers, and Clifford H Thurber. 2018.
\emph{Parameter Estimation and Inverse Problems}. Elsevier.

\leavevmode\vadjust pre{\hypertarget{ref-avseth2010quantitative}{}}%
Avseth, Per, Tapan Mukerji, and Gary Mavko. 2010. \emph{Quantitative
Seismic Interpretation: Applying Rock Physics Tools to Reduce
Interpretation Risk}. Cambridge university press.

\leavevmode\vadjust pre{\hypertarget{ref-beck2014introduction}{}}%
Beck, Amir. 2014. \emph{Introduction to Nonlinear Optimization: Theory,
Algorithms, and Applications with MATLAB}. SIAM.

\leavevmode\vadjust pre{\hypertarget{ref-benitez2023fine}{}}%
Benitez, Jose Antonio Lara, Takashi Furuya, Florian Faucher, Xavier
Tricoche, and Maarten V de Hoop. 2023. {``Fine-Tuning Neural-Operator
Architectures for Training and Generalization.''} \emph{arXiv Preprint
arXiv:2301.11509}.

\leavevmode\vadjust pre{\hypertarget{ref-canchumuni2019history}{}}%
Canchumuni, Smith WA, Alexandre A Emerick, and Marco Aurelio C Pacheco.
2019. {``History Matching Geological Facies Models Based on Ensemble
Smoother and Deep Generative Models.''} \emph{Journal of Petroleum
Science and Engineering} 177: 941--58.

\leavevmode\vadjust pre{\hypertarget{ref-cao2003adjoint}{}}%
Cao, Yang, Shengtai Li, Linda Petzold, and Radu Serban. 2003. {``Adjoint
Sensitivity Analysis for Differential-Algebraic Equations: The Adjoint
DAE System and Its Numerical Solution.''} \emph{SIAM Journal on
Scientific Computing} 24 (3): 1076--89.

\leavevmode\vadjust pre{\hypertarget{ref-carcione2012cross}{}}%
Carcione, Jose M, Davide Gei, Stefano Picotti, and Alberto Michelini.
2012. {``Cross-Hole Electromagnetic and Seismic Modeling for CO2
Detection and Monitoring in a Saline Aquifer.''} \emph{Journal of
Petroleum Science and Engineering} 100: 162--72.

\leavevmode\vadjust pre{\hypertarget{ref-chandra2020surrogate}{}}%
Chandra, Rohitash, Danial Azam, Arpit Kapoor, and R Dietmar Müller.
2020. {``Surrogate-Assisted Bayesian Inversion for Landscape and Basin
Evolution Models.''} \emph{Geoscientific Model Development} 13 (7):
2959--79.

\leavevmode\vadjust pre{\hypertarget{ref-cowles1996markov}{}}%
Cowles, Mary Kathryn, and Bradley P Carlin. 1996. {``Markov Chain Monte
Carlo Convergence Diagnostics: A Comparative Review.''} \emph{Journal of
the American Statistical Association} 91 (434): 883--904.

\leavevmode\vadjust pre{\hypertarget{ref-cranmer2020frontier}{}}%
Cranmer, Kyle, Johann Brehmer, and Gilles Louppe. 2020. {``The Frontier
of Simulation-Based Inference.''} \emph{Proceedings of the National
Academy of Sciences} 117 (48): 30055--62.

\leavevmode\vadjust pre{\hypertarget{ref-de2022cost}{}}%
De Hoop, Maarten, Daniel Zhengyu Huang, Elizabeth Qian, and Andrew M
Stuart. 2022. {``The Cost-Accuracy Trade-Off in Operator Learning with
Neural Networks.''} \emph{arXiv Preprint arXiv:2203.13181}.

\leavevmode\vadjust pre{\hypertarget{ref-dinh2016density}{}}%
Dinh, Laurent, Jascha Sohl-Dickstein, and Samy Bengio. 2016. {``Density
Estimation Using Real Nvp.''} \emph{arXiv Preprint arXiv:1605.08803}.

\leavevmode\vadjust pre{\hypertarget{ref-esser2018total}{}}%
Esser, Ernie, Lluis Guasch, Tristan van Leeuwen, Aleksandr Y Aravkin,
and Felix J Herrmann. 2018. {``Total Variation Regularization Strategies
in Full-Waveform Inversion.''} \emph{SIAM Journal on Imaging Sciences}
11 (1): 376--406.

\leavevmode\vadjust pre{\hypertarget{ref-freifeld2009recent}{}}%
Freifeld, Barry M, Thomas M Daley, Susan D Hovorka, Jan Henninges, Jim
Underschultz, and Sandeep Sharma. 2009. {``Recent Advances in Well-Based
Monitoring of CO2 Sequestration.''} \emph{Energy Procedia} 1 (1):
2277--84.

\leavevmode\vadjust pre{\hypertarget{ref-FURRE20173916}{}}%
Furre, Anne-Kari, Ola Eiken, Håvard Alnes, Jonas Nesland Vevatne, and
Anders Fredrik Kiær. 2017. {``20 Years of Monitoring {CO2}-Injection at
Sleipner.''} \emph{Energy Procedia} 114: 3916--26.
https://doi.org/\url{https://doi.org/10.1016/j.egypro.2017.03.1523}.

\leavevmode\vadjust pre{\hypertarget{ref-golub1999tikhonov}{}}%
Golub, Gene H, Per Christian Hansen, and Dianne P O'Leary. 1999.
{``Tikhonov Regularization and Total Least Squares.''} \emph{SIAM
Journal on Matrix Analysis and Applications} 21 (1): 185--94.

\leavevmode\vadjust pre{\hypertarget{ref-goodfellow2014generative}{}}%
Goodfellow, Ian, Jean Pouget-Abadie, Mehdi Mirza, Bing Xu, David
Warde-Farley, Sherjil Ozair, Aaron Courville, and Yoshua Bengio. 2014.
{``Generative Adversarial Nets.''} \emph{Advances in Neural Information
Processing Systems} 27.

\leavevmode\vadjust pre{\hypertarget{ref-grady2023model}{}}%
Grady, Thomas J, Rishi Khan, Mathias Louboutin, Ziyi Yin, Philipp A
Witte, Ranveer Chandra, Russell J Hewett, and Felix J Herrmann. 2023.
{``Model-Parallel Fourier Neural Operators as Learned Surrogates for
Large-Scale Parametric PDEs.''} \emph{Computers \& Geosciences}, 105402.

\leavevmode\vadjust pre{\hypertarget{ref-griewank1989automatic}{}}%
Griewank, Andreas et al. 1989. {``On Automatic Differentiation.''}
\emph{Mathematical Programming: Recent Developments and Applications} 6
(6): 83--107.

\leavevmode\vadjust pre{\hypertarget{ref-gross2021geosx}{}}%
Gross, Herve, and Antoine Mazuyer. 2021. {``GEOSX: A Multiphysics,
Multilevel Simulator Designed for Exascale Computing.''} In \emph{SPE
Reservoir Simulation Conference}. OnePetro.

\leavevmode\vadjust pre{\hypertarget{ref-heidenreich2014surrogate}{}}%
Heidenreich, S, H Gross, MA Henn, C Elster, and M Bär. 2014. {``A
Surrogate Model Enables a Bayesian Approach to the Inverse Problem of
Scatterometry.''} In \emph{Journal of Physics: Conference Series},
490:012007. 1. IOP Publishing.

\leavevmode\vadjust pre{\hypertarget{ref-hennenfent2008new}{}}%
Hennenfent, Gilles, Ewout van den Berg, Michael P Friedlander, and Felix
J Herrmann. 2008. {``New Insights into One-Norm Solvers from the Pareto
Curve.''} \emph{Geophysics} 73 (4): A23--26.

\leavevmode\vadjust pre{\hypertarget{ref-hijazi2023pod}{}}%
Hijazi, Saddam, Melina Freitag, and Niels Landwehr. 2023.
{``POD-Galerkin Reduced Order Models and Physics-Informed Neural
Networks for Solving Inverse Problems for the Navier--Stokes
Equations.''} \emph{Advanced Modeling and Simulation in Engineering
Sciences} 10 (1): 1--38.

\leavevmode\vadjust pre{\hypertarget{ref-horvat2022intrinsic}{}}%
Horvat, Christian, and Jean-Pascal Pfister. 2022. {``Intrinsic
Dimensionality Estimation Using Normalizing Flows.''} \emph{Advances in
Neural Information Processing Systems} 35: 12225--36.

\leavevmode\vadjust pre{\hypertarget{ref-huang2022geophysical}{}}%
Huang, Lianjie. 2022. {``Geophysical Monitoring for Geologic Carbon
Storage.''}

\leavevmode\vadjust pre{\hypertarget{ref-innes2018flux}{}}%
Innes, Mike. 2018. {``Flux: Elegant Machine Learning with Julia.''}
\emph{Journal of Open Source Software} 3 (25): 602.

\leavevmode\vadjust pre{\hypertarget{ref-jansen2011adjoint}{}}%
Jansen, Jan Dirk. 2011. {``Adjoint-Based Optimization of Multi-Phase
Flow Through Porous Media--a Review.''} \emph{Computers \& Fluids} 46
(1): 40--51.

\leavevmode\vadjust pre{\hypertarget{ref-karniadakis2021physics}{}}%
Karniadakis, George Em, Ioannis G Kevrekidis, Lu Lu, Paris Perdikaris,
Sifan Wang, and Liu Yang. 2021. {``Physics-Informed Machine Learning.''}
\emph{Nature Reviews Physics} 3 (6): 422--40.

\leavevmode\vadjust pre{\hypertarget{ref-kingma2013auto}{}}%
Kingma, Diederik P, and Max Welling. 2013. {``Auto-Encoding Variational
Bayes.''} \emph{arXiv Preprint arXiv:1312.6114}.

\leavevmode\vadjust pre{\hypertarget{ref-kobyzev2020normalizing}{}}%
Kobyzev, Ivan, Simon JD Prince, and Marcus A Brubaker. 2020.
{``Normalizing Flows: An Introduction and Review of Current Methods.''}
\emph{IEEE Transactions on Pattern Analysis and Machine Intelligence} 43
(11): 3964--79.

\leavevmode\vadjust pre{\hypertarget{ref-kontolati2023learning}{}}%
Kontolati, Katiana, Somdatta Goswami, George Em Karniadakis, and Michael
D Shields. 2023. {``Learning in Latent Spaces Improves the Predictive
Accuracy of Deep Neural Operators.''} \emph{arXiv Preprint
arXiv:2304.07599}.

\leavevmode\vadjust pre{\hypertarget{ref-kovachki2021universal}{}}%
Kovachki, Nikola, Samuel Lanthaler, and Siddhartha Mishra. 2021. {``On
Universal Approximation and Error Bounds for Fourier Neural
Operators.''} \emph{Journal of Machine Learning Research} 22: Art--No.

\leavevmode\vadjust pre{\hypertarget{ref-kovachki2021neural}{}}%
Kovachki, Nikola, Zongyi Li, Burigede Liu, Kamyar Azizzadenesheli,
Kaushik Bhattacharya, Andrew Stuart, and Anima Anandkumar. 2021.
{``Neural Operator: Learning Maps Between Function Spaces.''}
\emph{arXiv Preprint arXiv:2108.08481}.

\leavevmode\vadjust pre{\hypertarget{ref-kruse2021hint}{}}%
Kruse, Jakob, Gianluca Detommaso, Ullrich Köthe, and Robert Scheichl.
2021. {``HINT: Hierarchical Invertible Neural Transport for Density
Estimation and Bayesian Inference.''} In \emph{Proceedings of the AAAI
Conference on Artificial Intelligence}, 35:8191--99. 9.

\leavevmode\vadjust pre{\hypertarget{ref-lan2022scaling}{}}%
Lan, Shiwei, Shuyi Li, and Babak Shahbaba. 2022. {``Scaling up Bayesian
Uncertainty Quantification for Inverse Problems Using Deep Neural
Networks.''} \emph{SIAM/ASA Journal on Uncertainty Quantification} 10
(4): 1684--1713.

\leavevmode\vadjust pre{\hypertarget{ref-lensink2022fully}{}}%
Lensink, Keegan, Bas Peters, and Eldad Haber. 2022. {``Fully Hyperbolic
Convolutional Neural Networks.''} \emph{Research in the Mathematical
Sciences} 9 (4): 60.

\leavevmode\vadjust pre{\hypertarget{ref-li2020coupled}{}}%
Li, Dongzhuo, Kailai Xu, Jerry M Harris, and Eric Darve. 2020.
{``Coupled Time-Lapse Full-Waveform Inversion for Subsurface Flow
Problems Using Intrusive Automatic Differentiation.''} \emph{Water
Resources Research} 56 (8): e2019WR027032.

\leavevmode\vadjust pre{\hypertarget{ref-li2012fast}{}}%
Li, Xiang, Aleksandr Y Aravkin, Tristan van Leeuwen, and Felix J
Herrmann. 2012. {``Fast Randomized Full-Waveform Inversion with
Compressive Sensing.''} \emph{Geophysics} 77 (3): A13--17.

\leavevmode\vadjust pre{\hypertarget{ref-li2020fourier}{}}%
Li, Zongyi, Nikola Kovachki, Kamyar Azizzadenesheli, Burigede Liu,
Kaushik Bhattacharya, Andrew Stuart, and Anima Anandkumar. 2020.
{``Fourier Neural Operator for Parametric Partial Differential
Equations.''} \emph{arXiv Preprint arXiv:2010.08895}.

\leavevmode\vadjust pre{\hypertarget{ref-li2021physics}{}}%
Li, Zongyi, Hongkai Zheng, Nikola Kovachki, David Jin, Haoxuan Chen,
Burigede Liu, Kamyar Azizzadenesheli, and Anima Anandkumar. 2021.
{``Physics-Informed Neural Operator for Learning Partial Differential
Equations.''} \emph{arXiv Preprint arXiv:2111.03794}.

\leavevmode\vadjust pre{\hypertarget{ref-lie2019introduction}{}}%
Lie, Knut-Andreas. 2019. \emph{An Introduction to Reservoir Simulation
Using MATLAB/GNU Octave: User Guide for the MATLAB Reservoir Simulation
Toolbox (MRST)}. Cambridge University Press.

\leavevmode\vadjust pre{\hypertarget{ref-liu1989limited}{}}%
Liu, Dong C, and Jorge Nocedal. 1989. {``On the Limited Memory BFGS
Method for Large Scale Optimization.''} \emph{Mathematical Programming}
45 (1-3): 503--28.

\leavevmode\vadjust pre{\hypertarget{ref-liu2023joint}{}}%
Liu, Mingliang, Divakar Vashisth, Dario Grana, and Tapan Mukerji. 2023.
{``Joint Inversion of Geophysical Data for Geologic Carbon Sequestration
Monitoring: A Differentiable Physics-Informed Neural Network Model.''}
\emph{Journal of Geophysical Research: Solid Earth} 128 (3):
e2022JB025372.

\leavevmode\vadjust pre{\hypertarget{ref-louboutin2018dae}{}}%
Louboutin, Mathias, Fabio Luporini, Michael Lange, Navjot Kukreja,
Philipp A. Witte, Felix J. Herrmann, Paulius Velesko, and Gerard J.
Gorman. 2019. {``Devito (V3.1.0): An Embedded Domain-Specific Language
for Finite Differences and Geophysical Exploration.''}
\emph{Geoscientific Model Development}.
\url{https://doi.org/10.5194/gmd-12-1165-2019}.

\leavevmode\vadjust pre{\hypertarget{ref-louboutin2022SEGais}{}}%
Louboutin, Mathias, Philipp A. Witte, Ali Siahkoohi, Gabrio Rizzuti,
Ziyi Yin, Rafael Orozco, and Felix J. Herrmann. 2022. {``Accelerating
Innovation with Software Abstractions for Scalable Computational
Geophysics.''} \url{https://doi.org/10.1190/image2022-3750561.1}.

\leavevmode\vadjust pre{\hypertarget{ref-JUDI}{}}%
Louboutin, Mathias, Philipp Witte, Ziyi Yin, Henryk Modzelewski, Kerim,
Carlos da Costa, and Peterson Nogueira. 2023. \emph{Slimgroup/JUDI.jl:
V3.2.3} (version v3.2.3). Zenodo.
\url{https://doi.org/10.5281/zenodo.7785440}.

\leavevmode\vadjust pre{\hypertarget{ref-louboutin2023learned}{}}%
Louboutin, Mathias, Ziyi Yin, Rafael Orozco, Thomas J. Grady II, Ali
Siahkoohi, Gabrio Rizzuti, Philipp A. Witte, Olav Møyner, Gerard J.
Gorman, and Felix J. Herrmann. 2023. {``Learned Multiphysics Inversion
with Differentiable Programming and Machine Learning.''} \emph{The
Leading Edge} 42 (7): 452--516.
\url{https://doi.org/10.1190/tle42070474.1}.

\leavevmode\vadjust pre{\hypertarget{ref-lu2019review}{}}%
Lu, Kuan, Yulin Jin, Yushu Chen, Yongfeng Yang, Lei Hou, Zhiyong Zhang,
Zhonggang Li, and Chao Fu. 2019. {``Review for Order Reduction Based on
Proper Orthogonal Decomposition and Outlooks of Applications in
Mechanical Systems.''} \emph{Mechanical Systems and Signal Processing}
123: 264--97.

\leavevmode\vadjust pre{\hypertarget{ref-lu2019deeponet}{}}%
Lu, Lu, Pengzhan Jin, and George Em Karniadakis. 2019. {``Deeponet:
Learning Nonlinear Operators for Identifying Differential Equations
Based on the Universal Approximation Theorem of Operators.''}
\emph{arXiv Preprint arXiv:1910.03193}.

\leavevmode\vadjust pre{\hypertarget{ref-lumley20104d}{}}%
Lumley, David. 2010. {``4D Seismic Monitoring of CO 2 Sequestration.''}
\emph{The Leading Edge} 29 (2): 150--55.

\leavevmode\vadjust pre{\hypertarget{ref-luporini2020architecture}{}}%
Luporini, Fabio, Mathias Louboutin, Michael Lange, Navjot Kukreja,
Philipp Witte, Jan Hückelheim, Charles Yount, Paul HJ Kelly, Felix J
Herrmann, and Gerard J Gorman. 2020. {``Architecture and Performance of
Devito, a System for Automated Stencil Computation.''} \emph{ACM
Transactions on Mathematical Software (TOMS)} 46 (1): 1--28.

\leavevmode\vadjust pre{\hypertarget{ref-mosser2019deepflow}{}}%
Mosser, Lukas, Olivier Dubrule, and Martin J Blunt. 2019. {``Deepflow:
History Matching in the Space of Deep Generative Models.''} \emph{arXiv
Preprint arXiv:1905.05749}.

\leavevmode\vadjust pre{\hypertarget{ref-JutulDarcy}{}}%
Møyner, Olav, Grant Bruer, and Ziyi Yin. 2023.
\emph{Sintefmath/JutulDarcy.jl: V0.2.3} (version v0.2.3). Zenodo.
\url{https://doi.org/10.5281/zenodo.7855628}.

\leavevmode\vadjust pre{\hypertarget{ref-Jutul}{}}%
Møyner, Olav, Martin Johnsrud, Halvor Møll Nilsen, Xavier Raynaud,
Kjetil Olsen Lye, and Ziyi Yin. 2023. \emph{Sintefmath/Jutul.jl: V0.2.6}
(version v0.2.6). Zenodo. \url{https://doi.org/10.5281/zenodo.7855605}.

\leavevmode\vadjust pre{\hypertarget{ref-nogues2011detecting}{}}%
Nogues, Juan P, Jan M Nordbotten, and Michael A Celia. 2011.
{``Detecting Leakage of Brine or CO2 Through Abandoned Wells in a
Geological Sequestration Operation Using Pressure Monitoring Wells.''}
\emph{Energy Procedia} 4: 3620--27.

\leavevmode\vadjust pre{\hypertarget{ref-nooner2007constraints}{}}%
Nooner, Scott L, Ola Eiken, Christian Hermanrud, Glenn S Sasagawa,
Torkjell Stenvold, and Mark A Zumberge. 2007. {``Constraints on the in
Situ Density of CO2 Within the Utsira Formation from Time-Lapse Seafloor
Gravity Measurements.''} \emph{International Journal of Greenhouse Gas
Control} 1 (2): 198--214.

\leavevmode\vadjust pre{\hypertarget{ref-nordbotten2011geological}{}}%
Nordbotten, Jan M, and Michael Anthony Celia. 2011. {``Geological
Storage of CO2: Modeling Approaches for Large-Scale Simulation.''} In
\emph{Geological Storage of CO 2: Modeling Approaches for Large-Scale
Simulation}. John Wiley; Sons.

\leavevmode\vadjust pre{\hypertarget{ref-o2022derivate}{}}%
O'Leary-Roseberry, Thomas, Peng Chen, Umberto Villa, and Omar Ghattas.
2022. {``Derivate Informed Neural Operator: An Efficient Framework for
High-Dimensional Parametric Derivative Learning.''} \emph{arXiv Preprint
arXiv:2206.10745}.

\leavevmode\vadjust pre{\hypertarget{ref-orozco2022memory}{}}%
Orozco, Rafael, Mathias Louboutin, and Felix J Herrmann. 2022. {``Memory
Efficient Invertible Neural Networks for 3D Photoacoustic Imaging.''}
\emph{arXiv Preprint arXiv:2204.11850}.

\leavevmode\vadjust pre{\hypertarget{ref-orozco2023MIDLanf}{}}%
Orozco, Rafael, Mathias Louboutin, Ali Siahkoohi, Gabrio Rizzuti,
Tristan van Leeuwen, and Felix J. Herrmann. 2023. {``Amortized
Normalizing Flows for Transcranial Ultrasound with Uncertainty
Quantification,''} March.
\url{https://openreview.net/forum?id=LoJG-lUIlk}.

\leavevmode\vadjust pre{\hypertarget{ref-orozco2023refining}{}}%
Orozco, Rafael, Ali Siahkoohi, Mathias Louboutin, and Felix J Herrmann.
2023. {``Refining Amortized Posterior Approximations Using
Gradient-Based Summary Statistics.''} \emph{arXiv Preprint
arXiv:2305.08733}.

\leavevmode\vadjust pre{\hypertarget{ref-orozco2021photoacoustic}{}}%
Orozco, Rafael, Ali Siahkoohi, Gabrio Rizzuti, Tristan van Leeuwen, and
Felix J. Herrmann. 2021. {``Photoacoustic Imaging with Conditional
Priors from Normalizing Flows.''}
\url{https://openreview.net/forum?id=woi1OTvROO1}.

\leavevmode\vadjust pre{\hypertarget{ref-adjoint}{}}%
---------. 2023. {``{Adjoint operators enable fast and amortized machine
learning based Bayesian uncertainty quantification}.''} In \emph{Medical
Imaging 2023: Image Processing}, edited by Olivier Colliot and Ivana
IÅ¡gum, 12464:124641L. International Society for Optics; Photonics;
SPIE. \url{https://doi.org/10.1117/12.2651691}.

\leavevmode\vadjust pre{\hypertarget{ref-pestourie2020active}{}}%
Pestourie, Raphaël, Youssef Mroueh, Thanh V Nguyen, Payel Das, and
Steven G Johnson. 2020. {``Active Learning of Deep Surrogates for PDEs:
Application to Metasurface Design.''} \emph{Npj Computational Materials}
6 (1): 164.

\leavevmode\vadjust pre{\hypertarget{ref-peters2017constraints}{}}%
Peters, Bas, and Felix J Herrmann. 2017. {``Constraints Versus Penalties
for Edge-Preserving Full-Waveform Inversion.''} \emph{The Leading Edge}
36 (1): 94--100.

\leavevmode\vadjust pre{\hypertarget{ref-peters2019projection}{}}%
Peters, Bas, Brendan R Smithyman, and Felix J Herrmann. 2019.
{``Projection Methods and Applications for Seismic Nonlinear Inverse
Problems with Multiple Constraints.''} \emph{Geophysics} 84 (2):
R251--69.

\leavevmode\vadjust pre{\hypertarget{ref-plessix2006review}{}}%
Plessix, R-E. 2006. {``A Review of the Adjoint-State Method for
Computing the Gradient of a Functional with Geophysical Applications.''}
\emph{Geophysical Journal International} 167 (2): 495--503.

\leavevmode\vadjust pre{\hypertarget{ref-powell1985radial}{}}%
Powell, Michael JD. 1985. {``Radial Basis Functions for Multivariable
Interpolation: A Review.''} \emph{Algorithms for the Approximation of
Functions and Data.}

\leavevmode\vadjust pre{\hypertarget{ref-qian2020lift}{}}%
Qian, Elizabeth, Boris Kramer, Benjamin Peherstorfer, and Karen Willcox.
2020. {``Lift \& Learn: Physics-Informed Machine Learning for
Large-Scale Nonlinear Dynamical Systems.''} \emph{Physica D: Nonlinear
Phenomena} 406: 132401.

\leavevmode\vadjust pre{\hypertarget{ref-rahman2022u}{}}%
Rahman, Md Ashiqur, Zachary E Ross, and Kamyar Azizzadenesheli. 2022.
{``U-No: U-Shaped Neural Operators.''} \emph{arXiv Preprint
arXiv:2204.11127}.

\leavevmode\vadjust pre{\hypertarget{ref-rasmussen2021open}{}}%
Rasmussen, Atgeirr Flø, Tor Harald Sandve, Kai Bao, Andreas Lauser,
Joakim Hove, Bård Skaflestad, Robert Klöfkorn, et al. 2021. {``The Open
Porous Media Flow Reservoir Simulator.''} \emph{Computers \& Mathematics
with Applications} 81: 159--85.

\leavevmode\vadjust pre{\hypertarget{ref-razavi2012review}{}}%
Razavi, Saman, Bryan A Tolson, and Donald H Burn. 2012. {``Review of
Surrogate Modeling in Water Resources.''} \emph{Water Resources
Research} 48 (7).

\leavevmode\vadjust pre{\hypertarget{ref-rezende2015variational}{}}%
Rezende, Danilo, and Shakir Mohamed. 2015. {``Variational Inference with
Normalizing Flows.''} In \emph{International Conference on Machine
Learning}, 1530--38. PMLR.

\leavevmode\vadjust pre{\hypertarget{ref-schilders2008model}{}}%
Schilders, Wilhelmus HA, Henk A Van der Vorst, and Joost Rommes. 2008.
\emph{Model Order Reduction: Theory, Research Aspects and Applications}.
Vol. 13. Springer.

\leavevmode\vadjust pre{\hypertarget{ref-sheriff1995exploration}{}}%
Sheriff, Robert E, and Lloyd P Geldart. 1995. \emph{Exploration
Seismology}. Cambridge university press.

\leavevmode\vadjust pre{\hypertarget{ref-siahkoohi2021Seglbe}{}}%
Siahkoohi, Ali, and Felix J. Herrmann. 2021. {``Learning by Example:
Fast Reliability-Aware Seismic Imaging with Normalizing Flows.''} In
\emph{First International Meeting for Applied Geoscience {\&} Energy},
1580--85. Society of Exploration Geophysicists; Expanded Abstracts.
\url{https://doi.org/10.1190/segam2021-3581836.1}.

\leavevmode\vadjust pre{\hypertarget{ref-siahkoohi2020ABIpto}{}}%
Siahkoohi, Ali, Gabrio Rizzuti, Mathias Louboutin, Philipp Witte, and
Felix J. Herrmann. 2021. {``Preconditioned Training of Normalizing Flows
for Variational Inference in Inverse Problems.''}
\url{https://openreview.net/pdf?id=P9m1sMaNQ8T}.

\leavevmode\vadjust pre{\hypertarget{ref-siahkoohi2023reliable}{}}%
Siahkoohi, Ali, Gabrio Rizzuti, Rafael Orozco, and Felix J Herrmann.
2023a. {``Reliable Amortized Variational Inference with Physics-Based
Latent Distribution Correction.''} \emph{Geophysics} 88 (3): R297--322.

\leavevmode\vadjust pre{\hypertarget{ref-doi:10.1190ux2fgeo2022-0472.1}{}}%
Siahkoohi, Ali, Gabrio Rizzuti, Rafael Orozco, and Felix J. Herrmann.
2023b. {``Reliable Amortized Variational Inference with Physics-Based
Latent Distribution Correction.''} \emph{Geophysics} 88 (3).
\url{https://doi.org/10.1190/geo2022-0472.1}.

\leavevmode\vadjust pre{\hypertarget{ref-stanimirovic2010accelerated}{}}%
Stanimirović, Predrag S, and Marko B Miladinović. 2010. {``Accelerated
Gradient Descent Methods with Line Search.''} \emph{Numerical
Algorithms} 54: 503--20.

\leavevmode\vadjust pre{\hypertarget{ref-tarantola1984inversion}{}}%
Tarantola, Albert. 1984. {``Inversion of Seismic Reflection Data in the
Acoustic Approximation.''} \emph{Geophysics} 49 (8): 1259--66.

\leavevmode\vadjust pre{\hypertarget{ref-tarantola2005inverse}{}}%
---------. 2005. \emph{Inverse Problem Theory and Methods for Model
Parameter Estimation}. SIAM.

\leavevmode\vadjust pre{\hypertarget{ref-wang2004image}{}}%
Wang, Zhou, Alan C Bovik, Hamid R Sheikh, and Eero P Simoncelli. 2004.
{``Image Quality Assessment: From Error Visibility to Structural
Similarity.''} \emph{IEEE Transactions on Image Processing} 13 (4):
600--612.

\leavevmode\vadjust pre{\hypertarget{ref-wen2022u}{}}%
Wen, Gege, Zongyi Li, Kamyar Azizzadenesheli, Anima Anandkumar, and
Sally M Benson. 2022. {``U-FNO---an Enhanced Fourier Neural
Operator-Based Deep-Learning Model for Multiphase Flow.''}
\emph{Advances in Water Resources} 163: 104180.

\leavevmode\vadjust pre{\hypertarget{ref-wen2023real}{}}%
Wen, Gege, Zongyi Li, Qirui Long, Kamyar Azizzadenesheli, Anima
Anandkumar, and Sally Benson. 2023. {``Real-Time High-Resolution CO2
Geological Storage Prediction Using Nested Fourier Neural Operators.''}
\emph{Energy \& Environmental Science}.

\leavevmode\vadjust pre{\hypertarget{ref-wen2023conditional}{}}%
Wen, Jeffrey, Rizwan Ahmad, and Philip Schniter. 2023. {``A Conditional
Normalizing Flow for Accelerated Multi-Coil MR Imaging.''} \emph{arXiv
Preprint arXiv:2306.01630}.

\leavevmode\vadjust pre{\hypertarget{ref-witte2022sciai4industry}{}}%
Witte, Philipp A, Russell J Hewett, Kumar Saurabh, AmirHossein Sojoodi,
and Ranveer Chandra. 2022. {``SciAI4Industry--Solving PDEs for
Industry-Scale Problems with Deep Learning.''} \emph{arXiv Preprint
arXiv:2211.12709}.

\leavevmode\vadjust pre{\hypertarget{ref-witte2023fast}{}}%
Witte, Philipp A, Tugrul Konuk, Erik Skjetne, and Ranveer Chandra. 2023.
{``Fast CO2 Saturation Simulations on Large-Scale Geomodels with
Artificial Intelligence-Based Wavelet Neural Operators.''}
\emph{International Journal of Greenhouse Gas Control} 126: 103880.

\leavevmode\vadjust pre{\hypertarget{ref-witte2018alf}{}}%
Witte, Philipp A., Mathias Louboutin, Navjot Kukreja, Fabio Luporini,
Michael Lange, Gerard J. Gorman, and Felix J. Herrmann. 2019. {``A
Large-Scale Framework for Symbolic Implementations of Seismic Inversion
Algorithms in Julia.''} \emph{Geophysics} 84 (3): F57--71.
\url{https://doi.org/10.1190/geo2018-0174.1}.

\leavevmode\vadjust pre{\hypertarget{ref-witte2022industry}{}}%
Witte, Philipp A, WA Redmond, Russell J Hewett, and Ranveer Chandra.
2022. {``Industry-Scale CO2 Flow Simulations with Model-Parallel Fourier
Neural Operators.''} In \emph{NeurIPS 2022 Workshop Tackling Climate
Change with Machine Learning}.

\leavevmode\vadjust pre{\hypertarget{ref-InvertibleNetworks}{}}%
Witte, Philipp, Mathias Louboutin, Rafael Orozco, grizzuti, Ali
Siahkoohi, Felix Herrmann, Bas Peters, Páll Haraldsson, and Ziyi Yin.
2023. \emph{Slimgroup/InvertibleNetworks.jl: V2.2.5} (version v2.2.5).
Zenodo. \url{https://doi.org/10.5281/zenodo.7850287}.

\leavevmode\vadjust pre{\hypertarget{ref-yang2023rapid}{}}%
Yang, Yan, Angela F Gao, Kamyar Azizzadenesheli, Robert W Clayton, and
Zachary E Ross. 2023. {``Rapid Seismic Waveform Modeling and Inversion
with Neural Operators.''} \emph{IEEE Transactions on Geoscience and
Remote Sensing} 61: 1--12.

\leavevmode\vadjust pre{\hypertarget{ref-JutulDarcyRules}{}}%
Yin, Ziyi, Grant Bruer, and Mathias Louboutin. 2023.
\emph{Slimgroup/JutulDarcyRules.jl: V0.2.5} (version v0.2.5). Zenodo.
\url{https://doi.org/10.5281/zenodo.7863970}.

\leavevmode\vadjust pre{\hypertarget{ref-yin2023derisking}{}}%
Yin, Ziyi, Huseyin Tuna Erdinc, Abhinav Prakash Gahlot, Mathias
Louboutin, and Felix J Herrmann. 2023. {``Derisking Geologic Carbon
Storage from High-Resolution Time-Lapse Seismic to Explainable Leakage
Detection.''} \emph{The Leading Edge} 42 (1): 69--76.
\url{https://doi.org/10.1190/tle42010069.1}.

\leavevmode\vadjust pre{\hypertarget{ref-yin2022SEGlci}{}}%
Yin, Ziyi, Ali Siahkoohi, Mathias Louboutin, and Felix J. Herrmann.
2022. {``Learned Coupled Inversion for Carbon Sequestration Monitoring
and Forecasting with Fourier Neural Operators.''}
\url{https://doi.org/10.1190/image2022-3722848.1}.

\leavevmode\vadjust pre{\hypertarget{ref-zhang2020seismic}{}}%
Zhang, Xin, and Andrew Curtis. 2020. {``Seismic Tomography Using
Variational Inference Methods.''} \emph{Journal of Geophysical Research:
Solid Earth} 125 (4): e2019JB018589.

\leavevmode\vadjust pre{\hypertarget{ref-zhang2021bayesian2}{}}%
---------. 2021. {``Bayesian Geophysical Inversion Using Invertible
Neural Networks.''} \emph{Journal of Geophysical Research: Solid Earth}
126 (7): e2021JB022320.

\leavevmode\vadjust pre{\hypertarget{ref-zhao2020bayesian}{}}%
Zhao, Xuebin, Andrew Curtis, and Xin Zhang. 2021. {``{Bayesian seismic
tomography using normalizing flows}.''} \emph{Geophysical Journal
International} 228 (1): 213--39.
\url{https://doi.org/10.1093/gji/ggab298}.

\leavevmode\vadjust pre{\hypertarget{ref-zhdanov2013electromagnetic}{}}%
Zhdanov, Michael S, Masashi Endo, Noel Black, Lee Spangler, Stacey
Fairweather, A Hibbs, GA Eiskamp, and R Will. 2013. {``Electromagnetic
Monitoring of CO2 Sequestration in Deep Reservoirs.''} \emph{First
Break} 31 (2).

\end{CSLReferences}

\end{document}